\documentclass{article}

\PassOptionsToPackage{numbers, compress}{natbib}

 \usepackage[main, final]{neurips_2026}

\usepackage{mathtools}
\usepackage{amsmath}
\usepackage{amssymb}
\usepackage{amsthm}
\usepackage{mathtools,etoolbox,physics}
\usepackage{xstring}
\usepackage{enumitem}
\usepackage[framemethod=tikz]{mdframed}
\usepackage{algorithm}
\usepackage{algpseudocode}
\usepackage{url}
\usepackage{graphicx}
\usepackage[utf8]{inputenc} 
\usepackage[T1]{fontenc}    
\usepackage[hidelinks]{hyperref}
\usepackage{url}            
\usepackage{booktabs}       
\usepackage{amsfonts}       
\usepackage{nicefrac}       
\usepackage{microtype}      
\usepackage{xcolor}         
\usepackage[capitalize,noabbrev]{cleveref}
\AddToHook{cmd/appendix/before}{\crefalias{section}{appendix}}
\usepackage{comment}

\usepackage{tikz}
\usepackage{pgfplots}
\pgfplotsset{compat=1.18}
\usetikzlibrary{calc}

\DeclareSymbolFont{extraup}{U}{zavm}{m}{n}
\DeclareMathSymbol{\varheart}{\mathalpha}{extraup}{86}
\DeclareMathSymbol{\vardiamond}{\mathalpha}{extraup}{87}
\DeclareUnicodeCharacter{2663}{$\clubsuit$}

\usepackage{aliascnt}

\newtheorem{theorem}{Theorem}[section]
\newtheorem{problem}{Problem}
\makeatletter
\newcommand{\newsharedtheorem}[3]{%
  \newaliascnt{#1}{theorem}%
  \newtheorem{#1}[#1]{#2}%
  \aliascntresetthe{#1}%
  \@addtoreset{#1}{section}%
  \AtEndPreamble{\crefname{#1}{#2}{#3}\Crefname{#1}{#2}{#3}}%
}
\makeatother
\newsharedtheorem{openquestion}{Open Problem}{Open Problems}
\newsharedtheorem{proposition}{Proposition}{Propositions}
\newsharedtheorem{lemma}{Lemma}{Lemmas}
\newsharedtheorem{corollary}{Corollary}{Corollaries}
\newsharedtheorem{example}{Example}{Examples}
\theoremstyle{definition}
\newsharedtheorem{definition}{Definition}{Definitions}
\newsharedtheorem{assumption}{Assumption}{Assumptions}
\theoremstyle{remark}
\newsharedtheorem{remark}{Remark}{Remarks}
\AtBeginDocument{}
\AtEndPreamble{%
  \crefname{theorem}{Theorem}{Theorems}\Crefname{theorem}{Theorem}{Theorems}%
  \crefname{metatheorem}{Meta Theorem}{Meta Theorems}\Crefname{metatheorem}{Meta Theorem}{Meta Theorems}%
  \crefname{problem}{Problem}{Problems}\Crefname{problem}{Problem}{Problems}%
}

\newenvironment{theoremcontinued}[1][]{%
  \par\medskip
  \noindent\textbf{Theorem~\ref{theorem:optimality-approximation} continued}%
  \if\relax\detokenize{#1}\relax\else\ (#1)\fi.%
  \itshape
}{%
  \par\medskip
}
\newcommand{\sepbound}{\hyperref[theorem:optimality-approximation-part1]{Theorem~\ref*{theorem:optimality-approximation}, Part~I}}
\newcommand{\sepconstruct}{\hyperref[theorem:optimality-approximation-part2]{Theorem~\ref*{theorem:optimality-approximation}, Part~II}}

\renewcommand{\epsilon}{\varepsilon}

\newcommand{\horizon}{T}
\newcommand{\nagents}{N}
\newcommand{\timestep}{t}
\newcommand{\nexttimestep}{\timestep+1}
\newcommand{\action}{a}
\newcommand{\state}{s}
\newcommand{\noise}{w}
\newcommand{\fleet}{\mu}
\newcommand{\reward}{r}

\newcommand{\policy}[1]{\pi_{#1}}

\newcommand{\statespace}{S}
\newcommand{\actionspace}{A}
\newcommand{\noisespace}{W}
\newcommand{\X}[1]{X_{#1}}

\newcommand{\probabilityspace}[1]{\mathcal{P}(#1)}
\newcommand{\reals}{\mathbb{R}}
\newcommand{\N}{\mathbb N}
\newcommand{\realsBar}{\bar{\reals}}
\newcommand{\nonnegativeRealsBar}{\realsBar_{\geq 0}}

\newcommand{\expectedvalue}[2]{\mathbb{E}_{#1}\left[#2\right]}
\newcommand{\dynamics}[3]{f_{#3}\ifstrempty{#1}{}{(#1, #2)}}

\newcommand{\noisedist}[1]{\xi_{#1}}
\newcommand{\empnoisedist}[1]{\hat{\noisedist{}}_{#1}}

\newcommand{\plan}[1]{#1}
\newcommand{\plans}[2]{\Gamma(#1\ifstrempty{#2}{}{,#2})}

\newcommand{\E}{\mathbb{E}}

\DeclareMathOperator{\supp}{\mathrm{supp}}

\DeclareMathOperator{\proj}{proj}

\DeclareMathOperator{\Id}{Id}

\renewcommand{\Id}[1]{\mathrm{Id}_{#1}}
\newcommand{\st}{\,|\,}

\newcommand{\costtogosmall}[1]{j_{#1}}
\newcommand{\stagecostsmall}[1]{g_{#1}}
\newcommand{\terminalcostsmall}{g_\horizon}
\newcommand{\refVar}{z}
\newcommand{\refPVar}{\nu}
\newcommand{\refSpace}{Z}

\newcommand{\pushforward}[1]{{#1}_{\#}}
\newcommand{\projectionFromTo}[2]{\proj^{#1}_{#2}}
\newcommand{\probabilitySA}[2]{\lambda^{#1}_{#2}}
\newcommand{\probabilitySAW}[2]{\rho^{#1}_{#2}}
\newcommand{\costtogo}[1]{J_{#1}}

\newcommand{\kantorovich}[4]{\mathcal{K}[#1]\ifstrempty{#2}{}{(#2\ifstrempty{#4}{\ifstrempty{#3}{}{,}#3}{\ldots,#4})}}

\newcommand{\distoverdist}[1]{\eta_{#1}}

\newcommand{\famalg}{\text{SALT}}
\newcommand{\algname}[1]{\text{SALT-{#1}}}

\newcommand{\dispersion}[1]{D_{#1}}      
\newcommand{\statemismatch}[1]{\varepsilon^f_{#1}}     
\newcommand{\popLip}[1]{\iota_{#1}}              
\newcommand{\septol}[1]{\theta_{#1}}             
\newcommand{\refProfile}[1]{\fleet^{\mathrm{ref}}_{#1}} 

\usepackage[utf8]{inputenc} 
\usepackage[T1]{fontenc}    
\usepackage{hyperref}       
\usepackage{url}            
\usepackage{booktabs}       
\usepackage{amsfonts}       
\usepackage{nicefrac}       
\usepackage{microtype}      
\usepackage{import}
\usepackage{xcolor}         

\usepackage{wrapfig}
\usepackage{pgfplots}
\usepackage{graphicx}
\usepackage{subcaption}
\usepackage{fontawesome5}
\usepackage{makecell}
\newcommand{\Cell}[7]{%
  \fill[fill={rgb,255:red,#3;green,#4;blue,#5},draw=white,line width=0.22pt]
    (#1,#2) rectangle ++(1,1);
  \node[font=\scriptsize,text=#6] at (#1+0.5,#2+0.5) {#7};
}
\newcommand{\FigurePanelHeight}{6.6cm}
\usepackage[T1]{fontenc}
\usepackage{pgfplots}
\usepgfplotslibrary{groupplots,fillbetween}
\pgfplotsset{compat=1.18}
\usepgfplotslibrary{statistics}

\definecolor{boxgreen}{RGB}{150,205,150}
\definecolor{vir1}{RGB}{68,1,84}
\definecolor{vir2}{RGB}{64,67,135}
\definecolor{vir3}{RGB}{41,120,142}
\definecolor{vir4}{RGB}{34,167,132}
\definecolor{vir5}{RGB}{121,209,81}
\definecolor{vir6}{RGB}{253,231,37}

\pgfplotsset{
  boxplotstyle/.style={
    boxplot/draw direction=y,
    boxplot/box extend=0.50,
    draw=black!65,
    line width=0.75pt,
    fill=vir3,
    mark=none,
    boxplot/every box/.style={draw=black!65, line width=0.75pt, fill=vir4},
    boxplot/every whisker/.style={draw=black, line width=0.75pt},
    boxplot/every median/.style={draw=black, line width=1.35pt},
  }
}
\pgfplotsset{
  colormap={myviridis}{
    rgb255(0cm)=(68,1,84);
    rgb255(13cm)=(71,44,122);
    rgb255(25cm)=(59,81,139);
    rgb255(38cm)=(44,113,142);
    rgb255(50cm)=(33,144,141);
    rgb255(63cm)=(39,173,129);
    rgb255(75cm)=(92,200,99);
    rgb255(88cm)=(170,220,50);
    rgb255(100cm)=(253,231,37)
  }
}
\usetikzlibrary{arrows.meta, positioning}
\usepackage{adjustbox}

\title{A Separation Principle for Cooperative Multi-Agent Reinforcement Learning}

\author{%
  Lucia Pezzetti\\
  ETH Z\"urich\\
  \texttt{lucia.pezzetti@ai.ethz.ch}\\
  \And
  Nicolas Lanzetti\\
   Caltech\\
  \texttt{lnicolas@caltech.edu}\\
  \And
  Antonio Terpin\\
  ETH Z\"urich\\
  \texttt{aterpin@ethz.ch}\\
  \And
  Florian D\"orfler\\
  ETH Z\"urich\\
  \texttt{dorfler@ethz.ch}
  \And
  Giorgia Ramponi\\
  University of Z\"urich\\
  \texttt{ramponi@ifi.uzh.ch}
}
\hypersetup{
  pdftitle    = {A Separation Principle for Cooperative Multi-Agent Reinforcement Learning},
  pdfauthor   = {Lucia Pezzetti, Nicolas Lanzetti, Antonio Terpin, Florian Dörfler, Giorgia Ramponi},
  pdfsubject  = {Cooperative multi-agent reinforcement learning},
  pdfkeywords = {multi-agent reinforcement learning, separation principle, optimal transport},
}

\begin{document}

\maketitle

\begin{abstract}
We study cooperative multi-agent reinforcement learning (MARL) problems in which agents 
evolve under decoupled noisy dynamics but share a population objective. 
A canonical example is a mobility operator that orchestrates a fleet of vehicles to meet customer demand.
The stochastic and time-varying nature of these problems calls for learning approaches, 
yet existing MARL algorithms do not scale: learning assignment and routing jointly over the 
full state-action space becomes prohibitive as the number of agents grows.
Formulating the problem over the space of probability measures, we prove a separation principle: 
the population cost-to-go function is upper bounded by an optimal transport problem whose transportation 
cost is the target-conditioned single-agent cost-to-go function. 
This simplifies the multi-agent problem to learning a single-agent policy and coordinating the agents via optimal transport.
Building on this principle, we introduce $\famalg$ (Separation-based Assignment and Learning via optimal Transport), 
a family of algorithms that learn a target-conditioned single-agent policy using any standard 
reinforcement learning algorithm, and then coordinate the agents via optimal transport. 
Since learning occurs at the single-agent level, $\famalg$ scales to fleet sizes where MARL 
baselines become intractable and transfers without retraining to fleet sizes unseen during training.
In a vehicle-delivery experiment on the southern Manhattan road network, where MARL baselines are computationally prohibitive, 
$\famalg$ improves travel time by $15$--$21\%$ over a shortest-path routing heuristic.
In stochastic grid worlds, $\famalg$ reduces average travel times by $23$--$61\%$ relative to MARL baselines 
and scales to population sizes where these methods fail.
\end{abstract}

\section{Introduction}
\label{section:introduction}

\begin{figure}[t]
    \centering
    \resizebox{\linewidth}{!}{\input{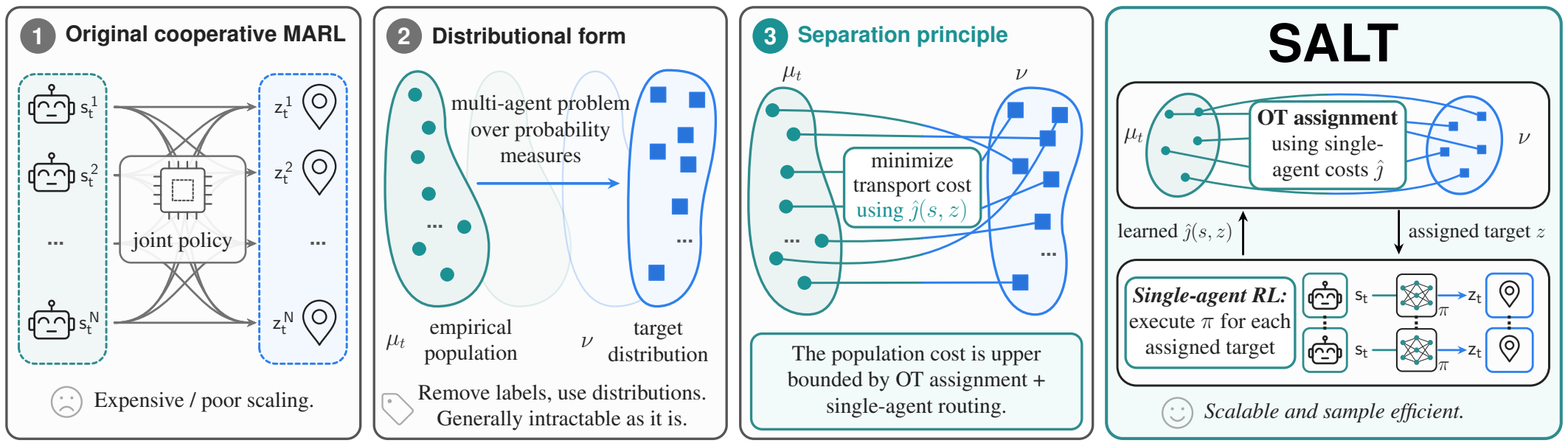}}
    \caption{Overview of $\famalg$. (1)~Cooperative MARL learns a joint fleet policy and scales poorly with $\nagents$. 
    (2)~Formulating the problem over probability measures replaces agent labels with an empirical population $\fleet_\timestep$ and target distribution $\refPVar$, but remains intractable. 
    (3)~The separation principle decouples assignment from routing, upper bounding the population cost-to-go by an optimal transport assignment with cost $\hat{j}(\state,\refVar)$. 
    $\famalg$ assigns targets through this optimal transport problem and routes agents with a shared single-agent policy $\pi$, yielding scalable and sample-efficient learning.}
    \label{fig:salt_overview}
    \vspace{-4pt}
\end{figure}

Consider a fleet of autonomous vehicles serving delivery requests in a city. 
At each decision time, the operator must simultaneously decide which vehicle should serve which request 
and how each vehicle should navigate to its assigned destination. 
These decisions are tightly coupled.
The best assignment depends on how long each vehicle will actually take to reach each target, while routing itself depends on the assignment. 
Assignments based only on nominal shortest paths 
can therefore be poor when travel times depend on traffic lights, congestion, and random delays.
Yet solving this full joint assignment-routing problem over the fleet quickly becomes computationally prohibitive.
Beyond delivery fleets, the same structure arises across a wide range of cooperative systems: inspection drones achieving coverage patterns~\cite{jing2020multi}, 
warehouse robots repositioning to incoming orders~\cite{lamballais2022dynamic}, emergency response units relocating to optimal positions~\cite{maxwell2010approximate}, 
ride-hailing platforms balancing supply and demand~\cite{li2019efficient}, and electric-vehicle fleets sharing charging infrastructure~\cite{ma2011decentralized}. 
In all these settings, the goal is to optimally coordinate a fleet of decoupled agents so that their collective terminal state matches or approximates a target configuration,
such as uniform coverage, a set of delivery locations, or a demand-weighted positioning pattern. 
These problems are typically solved in a receding-horizon fashion to account for changing conditions and request patterns and, 
thus, require modular approaches that generalize to different fleet sizes or target distributions.

Fleet assignment and routing are classically addressed via combinatorial solvers~\cite{zardini2022analysis, alonso2017demand} and, more recently, learned routing heuristics~\cite{kool2018attention}. Since the joint assignment-routing problem is combinatorial in the fleet size, these methods decouple the two steps using nominal travel costs, such as shortest-path distances, to solve the assignment problem. These heuristics, however, ignore the stochasticity of the costs and may favor targets that ``look close'' but are expensive to reach in expectation.
A natural learning-based alternative is multi-agent reinforcement learning (MARL) \cite{zhang2021multi, krnjaic2024scalable, lin2018efficient, shen2023multi}, often with centralized training and decentralized execution. 
Policy-gradient methods such as MAPPO \cite{yu2022surprisingeffectivenessppocooperative} and MADDPG \cite{lowe2017multi}, and value-decomposition methods such as VDN \cite{sunehag2018value}, QMIX \cite{rashid2020monotonic}, and QPLEX \cite{wang2021qplex} are effective in many settings \cite{papoudakis2021benchmarking, al2019deeppool}, but learn over a joint state-action space whose size grows exponentially with the number of agents~\cite{busoniu2008comprehensive}, limiting scalability and generalization to unseen fleet sizes. 
Value-decomposition methods gain tractability by imposing architectural assumptions on the joint value function, such as additivity or monotonicity, which are provably too restrictive even
for simple coordination tasks~\cite{son2019qtran, rashid2020monotonic}.
%
Mean-field reinforcement learning~\cite{yang2018mean, lauriere2022learning} instead
replaces pairwise interactions with an aggregate representation of the population. Although practical in large systems \cite{li2019efficient, wang2020joint}, its approximations can lose fidelity when modeling finite populations \cite{yardim2024mean}. 
Closer to our motivation, some methods learn latent subtasks and dynamically assign them among agents \cite{yang2022ldsa}, since in general cooperative MARL no such decomposition is available a priori.
We instead study a structured subclass, with decoupled dynamics and a terminal optimal transport cost, in which the assignment-routing separation can be exploited directly.

In this work, we seek a principled separation of population-level assignment and single-agent control (Figure~\ref{fig:salt_overview}).
Inspired by the ``deterministic'' separation principle for multi-agent control in \cite{terpin2024dynamic}, 
we formulate the cooperative MARL problem as an optimal control problem over the space of probability measures
where terminal deviation from a desired target measure is quantified via an optimal transport cost.
By construction, this representation is permutation invariant and captures both finite populations (as empirical measures) and a continuum of agents (as continuous measures).
In the deterministic setting, \citet{terpin2024dynamic} show that the population cost-to-go is exactly an optimal transport cost whose transportation cost is the single-agent cost-to-go, thus separating assignment from single-agent control.
This equality does not extend directly to the stochastic setting, where the population measure is random and future assignments can adapt to realized noise. 
Nevertheless, we show that a carefully designed separation principle, where noise is ``averaged out'', continues to upper bound the population cost-to-go. 
This analysis is non-trivial: the randomness of the population measure requires handling probability measures over probability measures, a modeling challenge in itself, compounded by delicate measurability issues.
Finally, \citet{terpin2024dynamic} provide a purely theoretical analysis, whereas we leverage our separation principle to design scalable algorithms for MARL. 

\noindent\textbf{Contributions.}
Our contributions are as follows:
\begin{enumerate}[leftmargin=16pt, topsep=0pt]
    \item We cast the cooperative multi-agent control problem with homogeneous agents, decoupled stochastic dynamics, and a terminal optimal transport formation cost as an optimal control problem over probability measures in the spirit of the work of~\citet{terpin2024dynamic}. This removes
    agent labels, makes the formulation permutation invariant, and applies to both finite populations and a continuum of agents.

    
    \item We prove a separation principle for this optimal control problem (\sepbound). The population cost-to-go is upper bounded by an optimal transport problem whose transportation cost is the target-conditioned single-agent cost-to-go. 
    This separates target allocation from routing, which can be solved independently of the fleet size.
    The upper bound becomes exact in the deterministic setting, recovering \cite[Corollary~4.2]{terpin2024dynamic}. 
    We then bound the gap between the two sides of \sepbound{} in terms of the noise of the single-agent dynamics (\cref{theorem:bq4:direct_bound}), and the loss of optimality under weak population interactions  (\cref{theorem:wc:performance}).
    Importantly, our proof is constructive and yields an implementable policy.

    \item This separation principle yields $\famalg$, a family of MARL algorithms that learn a shared target-conditioned single-agent policy and coordinate agents via optimal transport. 
    $\famalg$ is agnostic to the single-agent algorithm and generalizes to population sizes not seen during training.
    
    \item We validate $\famalg$ on a time-varying southern Manhattan road network with $2087$ intersections and $4255$ directed roads. 
    $\famalg$ trained with $\nagents = 20$ vehicles transfers without retraining to fleets of $1$--$100$ vehicles and reduces travel time by $15$--$21\%$ relative to a nominal shortest-path heuristic. 
    In controlled stochastic grid worlds, the $\famalg$ variants reduce the average travel time by $23$--$61\%$ relative to the standard MARL baselines, consistently across different single-agent reinforcement learning (RL) algorithms. 
    Additionally, $\famalg$ improves sample efficiency: with PPO~\cite{schulman2017proximalpolicyoptimizationalgorithms}, it reaches the same reach-rate milestones as MAPPO with up to $480\times$ fewer environment interactions, and scales to $\nagents=128$, while MAPPO fails at $\nagents=8$ within a $2\times10^9$-interaction budget.  
\end{enumerate}

\noindent\textbf{Notation.}
We work with Polish spaces, Borel functions, and Borel probability measures. For a Polish space $X$, $\probabilityspace{X}$ denotes the space of probability measures on $X$. We write $\delta_x$ for the Dirac measure at $x \in X$. The projection from $X \times Y$ onto $X$ is denoted by $\projectionFromTo{X\times Y}{X}$.
For $\mu \in \probabilityspace{X}$ and $F\colon X \to Y$, the pushforward of $\mu$ through $F$, $\pushforward{F}{\mu} \in \probabilityspace{Y}$, is defined by $\pushforward{F}{\mu}(B) = \mu(F^{-1}(B))$ for all Borel sets $B\subseteq Y$. 
A coupling between the probability measures $\mu \in \probabilityspace{X}$ and $\nu \in \probabilityspace{Y}$ is a joint probability measure with marginals $\mu$ and $\nu$; the set of couplings between $\mu$ and $\nu$ is $\plans{\mu}{\nu}$.
For a nonnegative cost $c\colon X \times Y \to \nonnegativeRealsBar$, the optimal transport cost between $\mu$ and $\nu$ is
\begin{equation*}
    \kantorovich{c}{\mu}{\nu}{} 
    \coloneqq
    \inf_{\plan{\gamma} \in \plans{\mu}{\nu}} 
    \int_{X \times Y}
    c(x,y)
    \dd \plan{\gamma}(x,y).
\end{equation*}
If $k\colon X\to \probabilityspace{Y}$ is a Markov kernel, $\mu\otimes k$ denotes the induced joint law on $X\times Y$. In particular, if $\pi(\dd x,\dd y)=\mu(\dd x)k(x;\dd y)$ is a disintegration, we write $\pi=\mu\otimes k^x$.

\section{Problem setting}
\label{section:problem_setting}
We consider cooperative MARL problems in which $\nagents$ homogeneous agents coordinate to reach a desired configuration over a finite horizon $\horizon$, while incurring agent-level stage costs and a population-level terminal cost.
Let $\statespace$, $\actionspace$, $\noisespace$, and $\refSpace$ denote the state, action, noise, and target spaces. 
While $\refSpace = \statespace$ for canonical target matching, we keep $\refSpace$ distinct 
to account for lower-dimensional targets or task-level descriptors, such as destinations when the state also includes velocity or battery level.
%
\subsection{Agent-based formulation}
Each agent $i$ has state $\state^i_\timestep\in\statespace$ and chooses action $\action_\timestep^i \in \actionspace$.
Its noisy dynamics are
$\state^i_{\timestep+1} = \dynamics{}{}{\timestep}(\state^i_\timestep, \action^i_\timestep, \noise^i_\timestep)$,
where 
$\noise^i_\timestep \sim \noisedist{\timestep}(\cdot \st \state^i_\timestep, \action^i_\timestep)$ 
and $\noisedist{\timestep}(\cdot \st \state^i_\timestep, \action^i_\timestep) \in \probabilityspace{\noisespace}$ gives the conditional law of the noise affecting an agent in state $\state$ taking action $\action$. 
We consider three noise models, which differ in how noise is correlated across agents: \textit{individual noise}, where noise is sampled independently across agents; \textit{local noise}, where agents with the same state-action pair share the same noise realization; and \textit{global noise}, where all agents share a common noise realization (see \cref{section:noise}).
Let $\boldsymbol{\state}_\timestep \coloneqq (\state_\timestep^1, \ldots, \state_\timestep^\nagents)$ and $\boldsymbol{\action}_\timestep \coloneqq  (\action_\timestep^1, \ldots, \action_\timestep^\nagents)$ denote the ordered states and actions of the agents.
We take the control cost to be the sum of agent-level stage costs 
$\stagecostsmall{\timestep} \colon
\statespace \times \actionspace \times \noisespace 
\to \nonnegativeRealsBar$ and a population-level terminal formation cost.
Given a transportation cost $\terminalcostsmall: \statespace \times \refSpace \to \nonnegativeRealsBar$, the terminal cost compares the final population distribution $\fleet_\horizon = \frac{1}{\nagents}\sum_{i=1}^\nagents \delta_{\state_\horizon^i}$ to a target configuration $\refPVar \in \probabilityspace{\refSpace}$ through the optimal transport cost
\begin{equation}
\label{eq:terminal_cost}
    \kantorovich{\terminalcostsmall}{\fleet_\horizon}{\refPVar}{} =
    \inf_{\plan{\gamma} \in \plans{\fleet_\horizon}{\refPVar}} 
    \int_{\statespace \times \refSpace} 
    \terminalcostsmall(\state, \refVar) 
    \dd\plan{\gamma}(\state, \refVar).
\end{equation} 
For $\terminalcostsmall(\state, \refVar)=\Vert \state- \refVar\Vert^p$ with $p\geq 1$, $\kantorovich{\terminalcostsmall}{\fleet_\horizon}{\refPVar}{}$ is the $p$-th power of the type-$p$ Wasserstein distance.
Since this is a fully observed cooperative problem, it can be cast as a multi-agent Markov decision process (MDP)~\cite{boutilier1996planning}, which reduces to a single-agent MDP over the joint action space. 
By standard MDP theory \cite{puterman1994markov, bertsekas1996stochastic}, we can restrict attention, without loss of optimality, to deterministic Markov policies $\policy{} = (\policy{\timestep})_{\timestep=0}^{\horizon-1}$, universally measurable on general Borel state and action spaces \cite[Ch.~8]{bertsekas1996stochastic}. 
We denote this class by $\Pi$. Throughout, we assume that, given the current states and actions, the noise is independent of the past, so the law of the noise sequence $(\boldsymbol{\noise}_\timestep)_{\timestep=0}^{\horizon-1}$, $\boldsymbol{\noise}_\timestep \coloneqq  (\noise_\timestep^1, \ldots, \noise_\timestep^\nagents)$, under which expectations are taken, is determined by the dynamics, the policy, and the noise model.

\begin{mdframed}[hidealllines=true,backgroundcolor=blue!5]
\begin{problem}
\label{problem:agent-level}
\vspace{.1cm}
Given initial states $\bar{\boldsymbol{\state}}_0 = (\bar\state_0^1, \ldots, \bar\state_0^\nagents)$ and a target distribution $\refPVar$, learn a policy $\policy{} \in \Pi$ minimizing the expected cumulative cost
\begin{displaymath}
\begin{aligned}
\costtogo{}^\nagents(\bar{\boldsymbol{\state}}_0, \refPVar) =
    \inf_{\policy{} \in \Pi} 
    \,\, & \expectedvalue{\substack{(\boldsymbol{\noise}_\timestep)_{\timestep=0}^{\horizon-1}\\}}{
    \frac{1}{\nagents}
    \sum_{i=1}^\nagents
    \sum_{t=0}^{\horizon-1}
    \stagecostsmall{\timestep}(\state_\timestep^i, \action_\timestep^i
    , \noise_\timestep^i) + 
    \kantorovich{\terminalcostsmall}{\fleet_\horizon}{\refPVar}{}}\\
     \mathrm{s.t.} \quad & \,\,
    \state_{\nexttimestep}^i = \dynamics{\state_\timestep^i}{\action_\timestep^i, \noise_\timestep^i}{\timestep},
    \qquad \boldsymbol{\state}_0 = \bar{\boldsymbol{\state}}_0.
\end{aligned}
\end{displaymath}
\end{problem}
\end{mdframed}

\subsection{Population-based formulation}
\label{section:problem_setting:population}
We generalize Problem~\ref{problem:agent-level} at the \textit{population level} by replacing the ordered state $\boldsymbol{\state}_\timestep$ with the empirical measure $\fleet_\timestep = \frac{1}{\nagents}\sum_{i=1}^\nagents \delta_{\state_\timestep^i} \in \probabilityspace{\statespace}$, which discards agent labels and makes the permutation invariance in the problem explicit.
We initially allow general population states $\fleet_\timestep\in\probabilityspace{\statespace}$ and we will later specialize this formulation back to the finite-agent setting.

\noindent\textbf{Population control.}
A population-level control must specify how the mass at each state is split among actions. Such a control is a \textit{state-action distribution} $\probabilitySA{}{\timestep}\equiv\probabilitySA{}{\timestep}(\fleet_\timestep)\in \probabilityspace{\statespace\times\actionspace}$ whose $\statespace$-marginal is the current population state: $\pushforward{\left(\projectionFromTo{\statespace\times\actionspace}{\statespace}\right)}{\probabilitySA{}{\timestep}} = \fleet_\timestep$.
%
%
For example, if all agents follow a common deterministic policy $\pi_\timestep\colon \statespace \to \actionspace$, then $\probabilitySA{}{\timestep} = \pushforward{(\Id{\statespace}, \pi_\timestep)}{\fleet_\timestep}$,
that is, all the mass that $\fleet_\timestep$ places at state $\state$ is assigned to the state-action pair
$(\state,\pi_\timestep(\state))$.

\noindent\textbf{Noise and population dynamics.}
When dynamics are deterministic, the next population state is the pushforward of $\probabilitySA{}{\timestep}$ under the dynamics, i.e., $\fleet_{\nexttimestep} = \pushforward{\dynamics{}{}{\timestep}}{\probabilitySA{}{\timestep}}$. 
The stochastic case instead requires more care, as the transition also depends on the specific noise realizations across the population.
In the agent-based formulation, each agent $i$ draws $\noise_\timestep^i \sim \noisedist{\timestep}(\cdot \st \state_\timestep^i, \action_\timestep^i)$ and transitions to $\dynamics{\state_\timestep^i}{\action_\timestep^i, \noise_\timestep^i}{\timestep}$. 
At the population level, these individual state-action dependent noise draws cannot simply be replaced by their underlying distribution $\noisedist{\timestep}(\cdot \st \state, \action)$.
Indeed, the next population state depends on the \textit{collection of noise draws actually realized} across all agents, which is itself a random object. 
We therefore represent the realized noise draws at each state-action pair by a state-action dependent \textit{empirical} noise distribution $\empnoisedist{\timestep}^{\state,\action}\in \probabilityspace{\noisespace}$.
The distinction between $\noisedist{\timestep}(\cdot \st \state, \action)$ and $\empnoisedist{\timestep}^{\state,\action}$ is crucial: the former is the single-agent noise law, while the latter collects the noise realizations that
drive the population transition.
Collecting the realized empirical kernels $\empnoisedist{\timestep}^{\state, \action}$ across all state-action pairs, we define the joint state-action-noise distribution $\probabilitySAW{}{\timestep}$ and the population dynamics as
\begin{equation*}
\probabilitySAW{}{\timestep} = \probabilitySA{}{\timestep} \otimes \empnoisedist{\timestep}^{\state, \action} \in \probabilityspace{\statespace\times\actionspace\times\noisespace},
\qquad
\fleet_{\nexttimestep} =  \pushforward{\dynamics{}{}{\timestep}}{\probabilitySAW{}{\timestep}},
\end{equation*}
where $\probabilitySAW{}{\timestep}$ records the fraction of the population experiencing each triple $(\state,\action,\noise)$.

\begin{example}
Consider three agents with state-action distribution
$\probabilitySA{}{\timestep} = \frac23\delta_{(0,\action)}
+ \frac13\delta_{(1,\action)}$, i.e., two agents are in state 0 and use action $\action$, and one agent is in state 1 and uses action $\action$.
Assume that the agents are subject to individual noise and that the noise realizations for the two agents at $(0,\action)$ are $0$ and $1$, while the noise realization for the agent at $(1,\action)$ is $1$. Then, $\empnoisedist{\timestep}^{0,\action} = \frac12\delta_0+\frac12\delta_1$ and $\empnoisedist{\timestep}^{1,\action} = \delta_1$. Accordingly, the state-action-noise distribution is $\probabilitySAW{}{\timestep} \coloneqq \probabilitySA{}{\timestep}\otimes\empnoisedist{\timestep}^{\state,\action} = \frac13\delta_{(0,\action,0)} + \frac13\delta_{(0,\action,1)}+\frac13\delta_{(1,\action,1)}$ and the next population distribution is $\fleet_{\nexttimestep}=\pushforward{(\dynamics{}{}{\timestep})}{\probabilitySAW{}{\timestep}} = \frac13\delta_{\dynamics{}{}{\timestep}(0,\action,0)} + \frac13\delta_{\dynamics{}{}{\timestep}(0,\action,1)} + \frac13\delta_{\dynamics{}{}{\timestep}(1,\action,1)}$.
\end{example}
\noindent\textbf{Cost.}
Since $\frac{1}{\nagents}\sum_{i=1}^\nagents\stagecostsmall{\timestep}(\state_\timestep^i, \action_\timestep^i, \noise_\timestep^i)=\int_{\statespace \times \actionspace \times \noisespace}\stagecostsmall{\timestep}(\state, \action, \noise)\dd\probabilitySAW{}{\timestep}(\state, \action, \noise)$, the cost of a given control $\probabilitySA{}{} = (\probabilitySA{}{\timestep})_{\timestep=0}^{\horizon-1}$ is
\begin{equation*}
    \costtogo{}^{\probabilitySA{}{}}(\bar{\fleet}_0, \refPVar) =
    \,\, 
    \expectedvalue{(\empnoisedist{\timestep})_{\timestep=0}^{\horizon-1}}
    {\sum_{t=0}^{\horizon-1}
    \int_{\statespace \times \actionspace \times \noisespace}
    {\stagecostsmall{\timestep}(\state, \action, \noise)}
    \dd\probabilitySAW{}{\timestep}(\state, \action, \noise) + 
    \kantorovich{\terminalcostsmall}{\fleet_\horizon}{\refPVar}{}},
\end{equation*}
where $\empnoisedist{\timestep} = (\empnoisedist{\timestep}^{\state, \action})_{(\state, \action) \in \statespace\times \actionspace}$ collects all realized noise kernels at time $\timestep$. 
The expectation is then taken over the
randomness of these profiles. This keeps finite-population fluctuations and noise
correlations explicit while reducing the population-level randomness to the single
object $\empnoisedist{\timestep}$. We refer the reader to Appendix~\ref{section:noise} for additional details and explicit construction for canonical noise models.
Since $\fleet_\timestep$ is random for $\timestep \geq 1$, an \textit{admissible population control} is a sequence of (universally measurable) feedback maps $\fleet \mapsto \probabilitySA{}{\timestep}(\fleet)$ with $\statespace$-marginal $\fleet$, applied before $\empnoisedist{\timestep}$ is realized. We write $\probabilitySA{}{\timestep}$ for $\probabilitySA{}{\timestep}(\fleet_\timestep)$.

\begin{mdframed}[hidealllines=true,backgroundcolor=blue!5]
\begin{problem}
\label{problem:population-level}
\vspace{.1cm}
Given an initial population distribution $\bar{\fleet}_0$ and a target distribution $\refPVar$, find an admissible population control $(\probabilitySA{}{\timestep})_{\timestep=0}^{\horizon-1}$ that minimizes the expected cumulative cost
\begin{displaymath}
\begin{aligned}
    \costtogo{}(\bar{\fleet}_0, \refPVar) =&
    \inf_{(\probabilitySA{}{\timestep})_{\timestep=0}^{\horizon-1}} 
    \,\, \expectedvalue{(\empnoisedist{\timestep})_{\timestep=0}^{\horizon-1}}
    {\sum_{t=0}^{\horizon-1}
    \int_{\statespace \times \actionspace \times \noisespace}
    {\stagecostsmall{\timestep}(\state, \action, \noise)}
    \dd\probabilitySAW{}{\timestep}(\state, \action, \noise) + 
    \kantorovich{\terminalcostsmall}{\fleet_\horizon}{\refPVar}{}}\\
     \mathrm{s.t.} \quad & \fleet_0 = \bar{\fleet}_0, \qquad
    \fleet_{\nexttimestep} = \pushforward{\dynamics{}{}{\timestep}}{\probabilitySAW{}{\timestep}}, \qquad \pushforward{(\projectionFromTo{\statespace \times \actionspace}{\statespace})}{ 
    \probabilitySA{}{\timestep}} =
    \fleet_\timestep,
    \qquad
    \probabilitySAW{}{\timestep} = \probabilitySA{}{\timestep} \otimes \empnoisedist{\timestep}^{\state, \action}.
\end{aligned}
\end{displaymath}
\end{problem}
\end{mdframed}

The population formulation removes the artificial labels from the state description and generalizes Problem~\ref{problem:agent-level}. 
Indeed, any finite-agent policy induces an empirical state-action distribution $\probabilitySA{}{\timestep} = \frac{1}{\nagents}\sum_{i=1}^\nagents \delta_{(\state_\timestep^i, \action_\timestep^i)}$ with $\statespace$-marginal $\fleet_\timestep$, 
thus every policy admissible for Problem~\ref{problem:agent-level} is feasible in Problem~\ref{problem:population-level} and $\costtogo{}(\bar\fleet_0, \refPVar) \leq \costtogo{}^\nagents(\bar{\boldsymbol{\state}}_0, \refPVar)$. 
The converse inequality does not hold in general. Problem~\ref{problem:population-level} allows for arbitrary population
state-action distributions, including those with more than $\nagents$ atoms, which are not realizable by $\nagents$ agents. 
However, we show in Corollary~\ref{corollary:agent_problem_upper_bound} that, for empirical target distributions $\refPVar^\nagents$ (i.e., $\nagents$ targets with equal mass), the separation bound derived for Problem~\ref{problem:population-level} 
directly upper bounds the agent-based cost-to-go $\costtogo{}^\nagents(\bar{\boldsymbol{\state}}_0, \refPVar^\nagents)$, yielding a deployable policy for Problem~\ref{problem:agent-level}.

\section{Separation principle}
\label{section:main_results}
%
In this section, we prove a separation principle for Problem~\ref{problem:population-level}, showing that the population cost-to-go function separates into optimal single-agent control toward a target and optimal target assignment across the population; see Theorem~\ref{theorem:optimality-approximation} below.
We then specialize the result to the finite-agent value $\costtogo{}^\nagents$ of Problem~\ref{problem:agent-level} and obtain a deployable policy that we use in~\cref{section:algorithm} for algorithm design.

\noindent\textbf{Target-conditioned single-agent problem.}
Fix a target $\refVar \in \refSpace$. For an agent in state $\state \in \statespace$ at time $\timestep$, define $\costtogosmall{\timestep}(\state, \refVar)$ as the optimal cost-to-go for reaching $\refVar$ over the remaining horizon:
\begin{equation}
\begin{aligned}\label{eq:dp:single_agent}
    \costtogosmall{\timestep}(\state, \refVar) &=
    \inf_{(\policy{\tau})_{\tau=\timestep}^{\horizon-1}}
    \expectedvalue{(\noise_{\tau})_{\tau=\timestep}^{\horizon-1}}{
    \sum_{\tau = \timestep}^{\horizon-1}
    \stagecostsmall{\tau}(\state_\tau, \action_\tau, \noise_\tau) +
    \terminalcostsmall(\state_\horizon, \refVar)
    },\\
    &\qquad\:\text{s.t. } \quad 
    \state_\timestep = \state, \quad 
    \state_{\tau + 1} = \dynamics{\state_\tau}{\action_\tau, \noise_\tau}{\tau}, \quad 
    \noise_\tau \sim \noisedist{\tau}(\cdot \st \state_\tau, \action_\tau).
\end{aligned}
\end{equation}
This is a standard single-agent stochastic optimal control problem, whose complexity is independent of the population size $\nagents$. Its Bellman recursion is
\begin{align}
\label{eq:bellman_recursion:single_agent}
    \costtogosmall{\timestep}(\state_\timestep, \refVar) &= 
    \inf_{\action_\timestep \in \actionspace}
    \expectedvalue{\noise_\timestep} {\stagecostsmall{\timestep}(\state_\timestep, \action_\timestep, \noise_\timestep) + 
    \costtogosmall{\nexttimestep}
    (\dynamics{\state_\timestep}{\action_\timestep, \noise_\timestep}{\timestep}, \refVar)},
    \quad 
    \costtogosmall{\horizon}(\state_\horizon, \refVar)= \terminalcostsmall(\state_\horizon, \refVar).
\end{align}

\noindent\textbf{A separation principle.}
For $\fleet_\timestep \in \probabilityspace{\statespace}$, let $\costtogo{\timestep}(\fleet_\timestep, \refPVar)$ denote the optimal expected cost at time $\timestep$, when the population $\fleet_\timestep$ aims to match the target distribution $\refPVar$. 
At $\timestep = \horizon$, $\costtogo{\horizon}(\fleet, \refPVar) = \kantorovich{\terminalcostsmall}{\fleet}{\refPVar}{}$.
Fix $\timestep < \horizon$. Since $\probabilityspace{\statespace}$ is a Polish space whenever $\statespace$ is Polish \cite[Theorem 15.15]{aliprantis06infinite}, the cost is additive over time, and the population dynamics are Markovian, standard dynamic programming yields the population-level Bellman recursion \cite[Ch.~8]{bertsekas1996stochastic}
\begin{equation}\label{eq:dp:population}
\costtogo{\timestep}(\fleet_\timestep, \refPVar)
=
\inf_{
\probabilitySA{}{\timestep}:\, \pushforward{(\projectionFromTo{\statespace\times\actionspace}{\statespace})}{\probabilitySA{}{\timestep}} = \fleet_\timestep
}
\E_{(\empnoisedist{\timestep})_{\timestep=0}^{\horizon-1}}\left[
\int_{\statespace\times\actionspace\times\noisespace}
\stagecostsmall{\timestep}(\state, \action, \noise)\,
\dd \probabilitySAW{}{\timestep}(\state, \action, \noise)
+ 
\costtogo{\nexttimestep}(\pushforward{\dynamics{}{}{\timestep}}{\probabilitySAW{}{\timestep}}, \refPVar)\right].
\end{equation}


\begin{mdframed}[hidealllines=true,backgroundcolor=blue!5]
\begin{theorem}[Separation principle, Part I: upper bound for the population cost-to-go]
\label{theorem:optimality-approximation}
\label{theorem:optimality-approximation-part1}
Consider Problem~\ref{problem:population-level}.
For all $\timestep\in\{0,\ldots,\horizon\}$, the population cost-to-go function \eqref{eq:dp:population} is upper bounded by an optimal transport problem whose transportation cost is the single-agent cost-to-go function $\costtogosmall{\timestep}$ in~\eqref{eq:dp:single_agent}:
\begin{equation}\label{eq:value_function_probability_space}
    \costtogo{\timestep}(\fleet_\timestep, \refPVar)
    \le
    \kantorovich{\costtogosmall{\timestep}}{\fleet_\timestep}{\refPVar}{}
    \;=\;
    \inf_{
        \plan{\gamma}\in \plans{\fleet_\timestep}{\refPVar}
    }
    \int_{\statespace\times\refSpace}
    \costtogosmall{\timestep}(\state, \refVar)
    \dd \plan{\gamma}(\state, \refVar).
\end{equation}
In particular, combined with the admissible control constructed in Part~II below, the same upper bound holds for the optimal cost $\costtogo{}(\bar\fleet_0, \refPVar)$ of Problem~\ref{problem:population-level}.
\end{theorem}
\end{mdframed}

\noindent\textbf{Discussion.}
The upper bound \eqref{eq:value_function_probability_space} separates the stochastic population control problem into target-conditioned single-agent control and population-level assignment, in which $\costtogosmall{\timestep}$ serves as the transport cost.
This separation generally comes at the price of an inequality because the two problems account for future noise differently.
In the optimal transport cost, $\kantorovich{\costtogosmall{\timestep}}{\fleet_\timestep}{\refPVar}{}$, mass is assigned to targets based on $\costtogosmall{\timestep}(\state, \refVar)$, which \emph{averages} over future noise realizations.
By contrast, the Bellman recursion~\eqref{eq:dp:population} evaluates the remaining cost $\costtogo{\nexttimestep}(\pushforward{\dynamics{}{}{\timestep}}{\probabilitySAW{}{\timestep}}, \refPVar)$ at the realized future population $\pushforward{\dynamics{}{}{\timestep}}{\probabilitySAW{}{\timestep}}$,
allowing the population-level assignment to \emph{adapt} to the observed transition.
The resulting minimization and expectation do not commute, which is the cause of the inequality~\eqref{eq:value_function_probability_space}.
At the same time, averaging over the noise is necessary for a meaningful separation, since it preserves the structure of a standard stochastic optimal control problem in~\eqref{eq:dp:single_agent}.
Appendix~\ref{section:bound_quantification_v4} quantifies the tightness of this inequality:
if $\costtogosmall{\tau+1}$ is $L_{\tau+1}$-Lipschitz in the state, the gap $\kantorovich{\costtogosmall{\timestep}}{\fleet_\timestep}{\refPVar}{}-\costtogo{\timestep}(\fleet_\timestep,\refPVar)$ in~\eqref{eq:value_function_probability_space} is bounded by $\sum_{\tau=\timestep}^{\horizon-1}L_{\tau+1}\dispersion{\tau}$,
where $\dispersion{\tau}$ is the worst-case expected distance between two (conditionally independent) next states generated from the same state-action pair.
The gap therefore depends linearly on the dispersion, and vanishes whenever dynamics are deterministic.
In particular, under deterministic dynamics, the next population state is fully determined by $(\fleet_\timestep, \probabilitySA{}{\timestep})$,  
the expectation in~\eqref{eq:dp:population} disappears, and the upper bound in \sepbound{} becomes exact, recovering the deterministic separation of~\cite{terpin2024dynamic}.
\begin{mdframed}[hidealllines=true,backgroundcolor=blue!5]
\begin{corollary}[Exact separation in the deterministic setting \protect{\cite[Corollary~4.2]{terpin2024dynamic}}]
\label{theorem:deterministic_exact}
\vspace{0.1cm}
When the dynamics are deterministic, the population cost-to-go function \eqref{eq:dp:population} is an optimal transport cost with the single-agent cost-to-go function as its transportation cost:
\begin{equation*}
        \costtogo{\timestep}(\fleet_\timestep, \refPVar)
        =
        \inf_{
            \plan{\gamma}\in \plans{\fleet_\timestep}{\refPVar}
        }
        \int_{\statespace\times\refSpace}
        \costtogosmall{\timestep}(\state, \refVar)
        \dd \plan{\gamma}(\state, \refVar).
    \end{equation*}
\end{corollary}
\end{mdframed}

\noindent\textbf{From a bound to a policy.}
For any $\varepsilon>0$, \sepconstruct{} constructs an explicit population-level control $(\probabilitySA{\varepsilon}{\timestep})_{\timestep=0}^{\horizon-1}$ whose cost exceeds the bound by at most $\varepsilon$.
Here, $\costtogo{}^{\probabilitySA{}{}}(\bar\fleet_0, \refPVar)$ denotes the expected cumulative cost of Problem~\ref{problem:population-level} when the population follows the control $(\probabilitySA{}{\timestep})_{\timestep=0}^{\horizon-1}$.

\begin{mdframed}[hidealllines=true,backgroundcolor=blue!5, nobreak=true]
\phantomsection
\label{thm:separation-part2}
\begin{theoremcontinued}[Separation principle, Part II: constructive separation-based control]\label{theorem:optimality-approximation-part2}
For every $\varepsilon > 0$ and each time step $\timestep$, let $\plan{\gamma}^\varepsilon_\timestep \in \plans{\fleet_\timestep}{\refPVar}$ be an $\frac{\varepsilon}{2\horizon}$-optimal coupling for the transport problem $\kantorovich{\costtogosmall{\timestep}}{\fleet_\timestep}{\refPVar}{}$ and let $\action_\timestep^\varepsilon:\statespace\times\refSpace\to\actionspace$ be an $\frac{\varepsilon}{2\horizon}$-optimal single-agent action for the Bellman recursion~\eqref{eq:bellman_recursion:single_agent}.
Define the separation-based state-action distribution $\probabilitySA{\varepsilon}{\timestep} \coloneqq \pushforward{(\projectionFromTo{\statespace\times\refSpace}{\statespace}, \action_\timestep^\varepsilon)}{\plan{\gamma}^\varepsilon_\timestep}$. Then,
    \begin{equation*}
        \costtogo{}^{\probabilitySA{\varepsilon}{}}(\bar\fleet_0, \refPVar)
        \le
        \inf_{
            \plan{\gamma}\in \plans{\bar\fleet_0}{\refPVar}
        }
        \int_{\statespace\times\refSpace}
        \costtogosmall{0}(\state, \refVar)
        \dd \plan{\gamma}(\state, \refVar)
        + \varepsilon.
    \end{equation*}
    In particular, if optimal couplings $\plan{\gamma}^\star_\timestep$ and optimal single-agent actions $\action_\timestep^\star$ exist and can be selected measurably, we can choose $\varepsilon=0$.
    \end{theoremcontinued}
\end{mdframed}
This construction has a natural interpretation. Assuming, for simplicity, that optimal couplings and actions exist (e.g., all spaces compact and all functions continuous; see the remark in Appendix~\ref{section:main_proof}), the coupling $\plan{\gamma}^\star_\timestep \in \plans{\fleet_\timestep}{\refPVar}$ optimally assigns the population mass at each state $\state$ to targets,
while $\action_\timestep^\star$ selects the optimal action for each state--target pair.
This is the core of our separation principle: the coupling $\plan{\gamma}^\star_\timestep$ indicates how to dispatch agents to targets, while $\action_\timestep^\star$ indicates how to route the single agent to the assigned target.
Part~II of Theorem~\ref{theorem:optimality-approximation} guarantees that the resulting population control has expected cost no greater than the transport upper bound established in Part~I.

\noindent\textbf{Robustness to weakly coupled dynamics.}
While \cref{theorem:optimality-approximation} assumes decoupled dynamics and stage costs, 
Appendix~\ref{section:weak_coupling} shows that the separation-based policy remains near-optimal when both depend weakly on the population.
In particular, we construct the separation-based policy of \sepconstruct{} using a decoupled surrogate and deploy it in the coupled system.
Let $\statemismatch{\tau}$ bound the distance between the next states the
surrogate and the coupled system produce from the same state, action, and noise,
let $\delta_\tau$ the expected difference between their stage costs, and let $L_\tau$ as above.
The loss of optimality is then at most $\sum_{\tau=0}^{\horizon-1}L_{\tau+1}\dispersion{\tau} + 2 \sum_{\tau=0}^{\horizon-1}(\delta_\tau+L_{\tau+1}\statemismatch{\tau})$.
The first term is the bound on the gap in~\eqref{eq:value_function_probability_space} derived above, while the second isolates the loss caused by population interactions, and vanishes with them.

\noindent\textbf{Implications for agent-based formulation $\costtogo{}^\nagents$.}
Theorem~\ref{theorem:optimality-approximation} is stated for the population formulation in Problem~\ref{problem:population-level}.
Corollary~\ref{corollary:agent_problem_upper_bound} bridges this to the original agent-based setting: for $\nagents$ agents and $\nagents$ target masses, the same transport surrogate $\kantorovich{\costtogosmall{\timestep}}{}{}{}$ upper bounds $\costtogo{\timestep}^{\nagents}$.
\begin{mdframed}[hidealllines=true,backgroundcolor=blue!5]
\begin{corollary}[Agent-based upper bound]
\label{corollary:agent_problem_upper_bound}
Let $\boldsymbol{\state}_\timestep = (\state_\timestep^1, \ldots, \state_\timestep^\nagents) \in \statespace^\nagents$ be the agent states at time $\timestep$, with empirical distribution $\fleet_\timestep^\nagents = \frac{1}{\nagents}\sum_{i=1}^\nagents\delta_{\state_\timestep^i}$. 
Let $\boldsymbol{\refVar}=(\refVar^1, \ldots, \refVar^\nagents) \in \refSpace^\nagents$ be fixed targets with empirical distribution $\refPVar^\nagents = \frac{1}{\nagents}\sum_{i=1}^\nagents\delta_{\refVar^i}$. 
Then, the cost-to-go function for the optimal control problem in Problem~\ref{problem:agent-level}, denoted by $\costtogo{\timestep}^\nagents(\boldsymbol{\state}_\timestep, \refPVar^\nagents)$, satisfies
\begin{equation}\label{eq:agent_cost_upper_bound}
    \costtogo{\timestep}^\nagents(\boldsymbol{\state}_\timestep, \refPVar^\nagents)
    \;\leq\;
    \kantorovich{\costtogosmall{\timestep}}{\fleet_\timestep^\nagents}{\refPVar^\nagents}{}.
\end{equation}
In particular, the inequality holds true for the cost-to-go $\costtogo{}^\nagents(\bar{\boldsymbol{\state}}_0, \refPVar^\nagents)$ in Problem~\ref{problem:agent-level}.
\end{corollary}
\end{mdframed}

We defer the proofs of \sepbound{}, \sepconstruct{}, and Corollary~\ref{corollary:agent_problem_upper_bound} to Appendix~\ref{section:main_proof}.

\section{SALT: Separation-based Assignment and Learning via optimal Transport}
\label{section:algorithm}

\begin{algorithm}[t]
\caption{$\famalg$: Separation-based Assignment and Learning via optimal Transport}
\label{alg:separation}
\begin{algorithmic}[1]
\Require Initial target-conditioned policy $\policy{}^{(0)}(\action \st \state,\refVar)$, cost-to-go function estimates $\hat{j}^{(0)}_\timestep(\state,\refVar)$, horizon $\horizon$, exploration schedule $(\alpha_k)_{k\geq 0}$
\For{episodes $k = 0,1,2,\dots$}
    \State Sample the episode initial population $\fleet_0^{(k)}$ and target distribution $\refPVar_k$.
    
    \State \textbf{Initial population-level assignment.} With probability $1-\alpha_k$, assign agents to targets via optimal transport with cost $\hat{j}^{(k)}_0$. Otherwise assign randomly.

    \State \textbf{Single-agent learning.} 
    Roll out the shared target-conditioned policy $\policy{}^{(k)}$ in parallel for all agents over the horizon $\horizon$, with the assigned targets kept fixed.

    \State Update $\policy{}^{(k+1)}$ and $\hat{j}^{(k+1)}_\timestep(\state,\refVar)$ with the chosen single-agent RL algorithm from the $\nagents$ trajectories collected in Step 4.
\EndFor

\State \textbf{Deployment (online-assignment variant).}
\For{$t=0, \dots, \horizon-1$}
    \State Assign agents to targets via optimal transport with cost $\hat{j}_\timestep$.
    \State Execute one step of $\policy{\timestep}$ with the current target assignments.
\EndFor
\end{algorithmic}
\end{algorithm}

Corollary~\ref{corollary:agent_problem_upper_bound} suggests using the upper bound
\begin{equation*}
    \kantorovich{\costtogosmall{\timestep}}{\fleet_\timestep}{\refPVar}{} \;=\;
    \inf_{
            \plan{\gamma}\in \plans{\fleet_\timestep}{\refPVar}
        }
        \int
        \costtogosmall{\timestep}(\state, \refVar)
        \dd \plan{\gamma}(\state, \refVar)
\end{equation*}
as a tractable surrogate for the agent-based cost-to-go $\costtogo{\timestep}^\nagents(\boldsymbol{\state}_\timestep, \refPVar^\nagents)$. 
The resulting learning paradigm, summarized in Algorithm~\ref{alg:separation}, has two phases: 
(i) learn the single-agent target-conditioned cost-to-go function $\hat{j}_\timestep$, 
and (ii) use it as the transport cost to assign agents to targets.
The choice of single-agent algorithm is entirely open: any method capable of learning a target-conditioned cost-to-go estimate $\hat{j}_\timestep$ 
(Q-learning \cite{watkins1992q}, DDQN \cite{hasselt2016deep}, PPO \cite{schulman2017proximalpolicyoptimizationalgorithms}, A2C~\cite{mnih2016asynchronous}, etc.) can be plugged in as the base learner.
We refer to this family of methods as $\famalg$, for Separation-based Assignment and Learning via optimal Transport. An instance using a single-agent RL algorithm is denoted $\algname{ALG}$; e.g., $\algname{PPO}$.

\noindent\textbf{Discussion.}
$\famalg$ replaces learning over the joint population state-action space with a target-conditioned single-agent policy $\policy{}(\cdot \st \state, \refVar)$ and cost-to-go estimate $\hat{j}(\state, \refVar)$ shared across all agents.
Thus, population-level coordination is handled only through the optimal transport assignment, while both learned functions depend only on a single agent's state--target pair.
This has three main benefits. (i) Scalability: learning occurs entirely at the target-conditioned single-agent level, avoiding the curse of dimensionality of the joint state-action space and generalizing naturally across population sizes. 
(ii) Trajectory pooling: the same single-agent model is updated from every agent's trajectory simultaneously. 
(iii) Modularity: the coordination layer requires only $\hat{j}(\state, \refVar)$ as its transport cost, making $\famalg$ compatible with any single-agent RL algorithm.

The time-indexed bound in \sepbound{} makes the assignment frequency a deployment choice. 
The transport assignment can be recomputed at every time step with $\hat j_\timestep$, as in Algorithm~\ref{alg:separation}, or solved once at $\timestep=0$ and kept fixed for the entire episode.
Both variants use the same target-conditioned learner and differ only in when the transport assignment is updated.

\noindent\textbf{Training details.}
%
At each episode $k$, we sample the initial population $\fleet_0^{(k)}$ and target distribution $\refPVar_k$ from the environment, 
and assign agents to targets by solving the optimal transport problem $\kantorovich{\hat{j}^{(k)}_0}{\fleet_0^{(k)}}{\refPVar_k}{}$
with the current initial-time estimate $\hat{j}^{(k)}_0(\state, \refVar)$ as cost.
For empirical population and target measures with $\nagents$ equally weighted samples, a coupling $\plan{\gamma}\in\plans{\fleet_\timestep}{\refPVar}$ 
can be represented as a doubly stochastic matrix (up to the factor $1/\nagents$). 
By the Birkhoff--von Neumann theorem an optimal coupling can be taken to be a permutation (see the proof of Corollary~\ref{corollary:agent_problem_upper_bound}),
and thus the optimal transport problem reduces to a \emph{one-to-one assignment} problem. 
In particular, given a cost matrix $C_{ij}=\hat j^{(k)}_\timestep(\state^i,\refVar^j)$, 
we solve $\min_{\sigma\in S_\nagents}\sum_{i=1}^\nagents \hat j^{(k)}_\timestep(\state^i,\refVar^{\sigma(i)})$, where $S_\nagents$ is the set of permutations of $\{1,\dots,\nagents\}$, 
and assign target $\refVar^{\sigma(i)}$ to agent $i$. 
Such an assignment is computed exactly in $O(\nagents^3)$ with the Hungarian algorithm, or approximately using entropic regularization~\cite{cuturi2013sinkhorn}.
During training we apply this construction at $\timestep=0$ with probability $1-\alpha_k$, else we assign targets uniformly at random to ensure broad coverage of $(\state, \refVar)$ pairs 
and prevent the cost-to-go estimate from overfitting to its own assignments. 
We then keep the assignments fixed for the entire episode. 
Each agent therefore faces a single-agent target-conditioned task with terminal cost $\terminalcostsmall(\state_\horizon, \refVar^i)$,
ensuring that the collected transitions are consistent with the cost-to-go $\costtogosmall{\timestep}(\state, \refVar^i)$ being learned.
Once targets are assigned, agents roll out the shared target-conditioned policy in parallel, 
producing $\nagents$ trajectories that are pooled to update the shared policy and cost-to-go estimate using the chosen single-agent algorithm. 

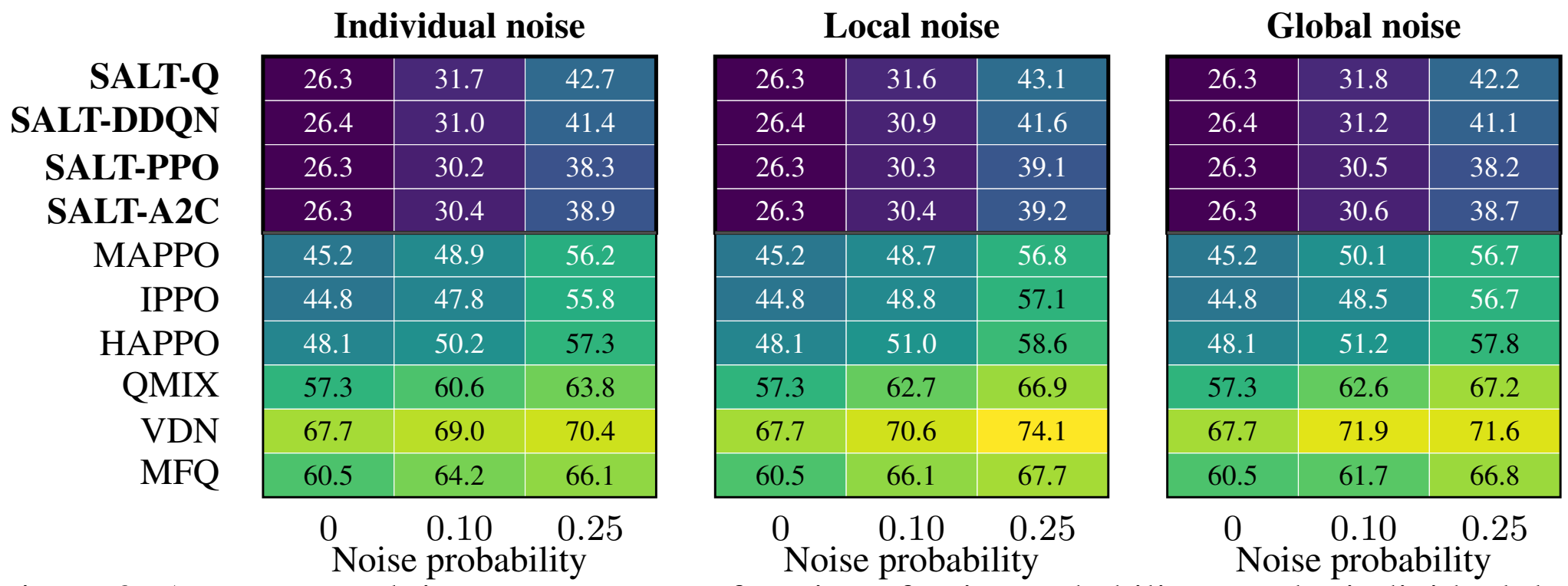
\begin{figure}[t]
    \begin{tikzpicture}[x=1.13cm,y=0.38cm]
  \node[anchor=east,font=\small\bfseries] at (-0.22,9.5) {SALT-Q};
  \node[anchor=east,font=\small\bfseries] at (-0.22,8.5) {SALT-DDQN};
  \node[anchor=east,font=\small\bfseries] at (-0.22,7.5) {SALT-PPO};
  \node[anchor=east,font=\small\bfseries] at (-0.22,6.5) {SALT-A2C};
  \node[anchor=east,font=\small] at (-0.22,5.5) {MAPPO};
  \node[anchor=east,font=\small] at (-0.22,4.5) {IPPO};
  \node[anchor=east,font=\small] at (-0.22,3.5) {HAPPO};
  \node[anchor=east,font=\small] at (-0.22,2.5) {QMIX};
  \node[anchor=east,font=\small] at (-0.22,1.5) {VDN};
  \node[anchor=east,font=\small] at (-0.22,0.5) {MFQ};

  \node[anchor=south,font=\bfseries\small] at (1.500,10.1) {Individual noise};
  \Cell{0.000}{9}{68}{1}{84}{white}{26.3}
  \Cell{1.000}{9}{72}{40}{120}{white}{31.7}
  \Cell{2.000}{9}{48}{106}{142}{white}{42.7}
  \Cell{0.000}{8}{68}{1}{84}{white}{26.4}
  \Cell{1.000}{8}{72}{36}{117}{white}{31.0}
  \Cell{2.000}{8}{50}{100}{142}{white}{41.4}
  \Cell{0.000}{7}{68}{1}{84}{white}{26.3}
  \Cell{1.000}{7}{72}{31}{112}{white}{30.2}
  \Cell{2.000}{7}{59}{82}{139}{white}{38.3}
  \Cell{0.000}{6}{68}{1}{84}{white}{26.3}
  \Cell{1.000}{6}{72}{32}{113}{white}{30.4}
  \Cell{2.000}{6}{57}{85}{140}{white}{38.9}
  \Cell{0.000}{5}{42}{119}{142}{white}{45.2}
  \Cell{1.000}{5}{35}{138}{141}{white}{48.9}
  \Cell{2.000}{5}{40}{174}{128}{white}{56.2}
  \Cell{0.000}{4}{43}{117}{142}{white}{44.8}
  \Cell{1.000}{4}{37}{132}{142}{white}{47.8}
  \Cell{2.000}{4}{38}{173}{129}{white}{55.8}
  \Cell{0.000}{3}{37}{133}{142}{white}{48.1}
  \Cell{1.000}{3}{33}{144}{141}{white}{50.2}
  \Cell{2.000}{3}{47}{180}{124}{black}{57.3}
  \Cell{0.000}{2}{47}{180}{124}{black}{57.3}
  \Cell{1.000}{2}{76}{194}{108}{black}{60.6}
  \Cell{2.000}{2}{112}{207}{87}{black}{63.8}
  \Cell{0.000}{1}{165}{219}{54}{black}{67.7}
  \Cell{1.000}{1}{186}{222}{40}{black}{69.0}
  \Cell{2.000}{1}{205}{225}{29}{black}{70.4}
  \Cell{0.000}{0}{76}{194}{108}{black}{60.5}
  \Cell{1.000}{0}{119}{209}{83}{black}{64.2}
  \Cell{2.000}{0}{144}{215}{67}{black}{66.1}
  \draw[black,line width=0.55pt] (0.000,0) rectangle (3.000,10);
  \draw[black,line width=0.95pt] (0.000,6) rectangle (3.000,10);
  \draw[black!70,line width=0.75pt] (0.000,6) -- (3.000,6);
  \node[anchor=north,font=\small] at (0.500,-0.1) {$0$};
  \node[anchor=north,font=\small] at (1.500,-0.1) {$0.10$};
  \node[anchor=north,font=\small] at (2.500,-0.1) {$0.25$};
  \node[overlay,anchor=north,font=\small] at (1.500,-0.8) {Noise probability};

  \node[anchor=south,font=\bfseries\small] at (4.950,10.1) {Local noise};
  \Cell{3.450}{9}{68}{1}{84}{white}{26.3}
  \Cell{4.450}{9}{72}{40}{120}{white}{31.6}
  \Cell{5.450}{9}{46}{109}{142}{white}{43.1}
  \Cell{3.450}{8}{68}{1}{84}{white}{26.4}
  \Cell{4.450}{8}{72}{35}{116}{white}{30.9}
  \Cell{5.450}{8}{50}{101}{142}{white}{41.6}
  \Cell{3.450}{7}{68}{1}{84}{white}{26.3}
  \Cell{4.450}{7}{72}{31}{112}{white}{30.3}
  \Cell{5.450}{7}{57}{86}{140}{white}{39.1}
  \Cell{3.450}{6}{68}{1}{84}{white}{26.3}
  \Cell{4.450}{6}{72}{32}{113}{white}{30.4}
  \Cell{5.450}{6}{56}{88}{140}{white}{39.2}
  \Cell{3.450}{5}{42}{119}{142}{white}{45.2}
  \Cell{4.450}{5}{35}{137}{142}{white}{48.7}
  \Cell{5.450}{5}{44}{177}{126}{white}{56.8}
  \Cell{3.450}{4}{43}{117}{142}{white}{44.8}
  \Cell{4.450}{4}{35}{137}{142}{white}{48.8}
  \Cell{5.450}{4}{46}{179}{124}{black}{57.1}
  \Cell{3.450}{3}{37}{133}{142}{white}{48.1}
  \Cell{4.450}{3}{31}{148}{140}{white}{51.0}
  \Cell{5.450}{3}{58}{186}{118}{black}{58.6}
  \Cell{3.450}{2}{47}{180}{124}{black}{57.3}
  \Cell{4.450}{2}{101}{203}{94}{black}{62.7}
  \Cell{5.450}{2}{155}{217}{60}{black}{66.9}
  \Cell{3.450}{1}{165}{219}{54}{black}{67.7}
  \Cell{4.450}{1}{208}{225}{28}{black}{70.6}
  \Cell{5.450}{1}{253}{231}{37}{black}{74.1}
  \Cell{3.450}{0}{76}{194}{108}{black}{60.5}
  \Cell{4.450}{0}{144}{215}{67}{black}{66.1}
  \Cell{5.450}{0}{165}{219}{54}{black}{67.7}
  \draw[black,line width=0.55pt] (3.450,0) rectangle (6.450,10);
  \draw[black,line width=0.95pt] (3.450,6) rectangle (6.450,10);
  \draw[black!70,line width=0.75pt] (3.450,6) -- (6.450,6);
  \node[anchor=north,font=\small] at (3.950,-0.1) {$0$};
  \node[anchor=north,font=\small] at (4.950,-0.1) {$0.10$};
  \node[anchor=north,font=\small] at (5.950,-0.1) {$0.25$};
  \node[overlay,anchor=north,font=\small] at (4.950,-0.8) {Noise probability};

  \node[anchor=south,font=\bfseries\small] at (8.400,10.1) {Global noise};
  \Cell{6.900}{9}{68}{1}{84}{white}{26.3}
  \Cell{7.900}{9}{72}{41}{121}{white}{31.8}
  \Cell{8.900}{9}{49}{104}{142}{white}{42.2}
  \Cell{6.900}{8}{68}{1}{84}{white}{26.4}
  \Cell{7.900}{8}{72}{37}{118}{white}{31.2}
  \Cell{8.900}{8}{51}{98}{141}{white}{41.1}
  \Cell{6.900}{7}{68}{1}{84}{white}{26.3}
  \Cell{7.900}{7}{72}{32}{113}{white}{30.5}
  \Cell{8.900}{7}{59}{81}{139}{white}{38.2}
  \Cell{6.900}{6}{68}{1}{84}{white}{26.3}
  \Cell{7.900}{6}{72}{33}{115}{white}{30.6}
  \Cell{8.900}{6}{58}{84}{140}{white}{38.7}
  \Cell{6.900}{5}{42}{119}{142}{white}{45.2}
  \Cell{7.900}{5}{33}{144}{141}{white}{50.1}
  \Cell{8.900}{5}{42}{176}{127}{white}{56.7}
  \Cell{6.900}{4}{43}{117}{142}{white}{44.8}
  \Cell{7.900}{4}{35}{136}{142}{white}{48.5}
  \Cell{8.900}{4}{44}{177}{126}{white}{56.7}
  \Cell{6.900}{3}{37}{133}{142}{white}{48.1}
  \Cell{7.900}{3}{31}{149}{139}{white}{51.2}
  \Cell{8.900}{3}{50}{182}{122}{black}{57.8}
  \Cell{6.900}{2}{47}{180}{124}{black}{57.3}
  \Cell{7.900}{2}{99}{203}{95}{black}{62.6}
  \Cell{8.900}{2}{160}{218}{57}{black}{67.2}
  \Cell{6.900}{1}{165}{219}{54}{black}{67.7}
  \Cell{7.900}{1}{226}{228}{24}{black}{71.9}
  \Cell{8.900}{1}{221}{227}{24}{black}{71.6}
  \Cell{6.900}{0}{76}{194}{108}{black}{60.5}
  \Cell{7.900}{0}{88}{199}{101}{black}{61.7}
  \Cell{8.900}{0}{155}{217}{60}{black}{66.8}
  \draw[black,line width=0.55pt] (6.900,0) rectangle (9.900,10);
  \draw[black,line width=0.95pt] (6.900,6) rectangle (9.900,10);
  \draw[black!70,line width=0.75pt] (6.900,6) -- (9.900,6);
  \node[anchor=north,font=\small] at (7.400,-0.1) {$0$};
  \node[anchor=north,font=\small] at (8.400,-0.1) {$0.10$};
  \node[anchor=north,font=\small] at (9.400,-0.1) {$0.25$};
  \node[overlay,anchor=north,font=\small] at (8.400,-0.8) {Noise probability};
\end{tikzpicture}
    \caption{Average travel time to targets as a function of noise probability $p$ under individual, local, and global noise. 
    Cells show the pooled arithmetic mean over $50$ evaluation rollouts from each of $5$ independently trained policies. 
    Figure~\ref{fig:gridworld_combined_metrics} shows the corresponding distributions. 
    Brighter cells indicate worse relative performance. $\famalg$ variants consistently outperform standard MARL baselines.}
    \label{fig:TravelTimesBenchmark}
\end{figure}
\section{Experimental results}
\label{section:results}
We evaluate $\famalg$ against MARL baselines in stochastic grid worlds, and test scalability and fleet-size transfer in a Manhattan urban logistics task. 
More ablations and experiment details are in Appendix~\ref{app:additional_experiments}.
The code for our experiments is available at \href{https://github.com/lucia-pezzetti/SALT}{\texttt{https://github.com/lucia-pezzetti/SALT}}. 

\subsection{Stochastic grid-world experiments}
\noindent\textbf{Separation leads to better performance.}
\label{section:results:benchmarking}
%
We first evaluate $\famalg$ on a stochastic grid-world environment. 
A population of $\nagents=7$ homogeneous agents must reach $\nagents$ target locations within a horizon $\horizon = 100$ on an $11\times 30$ grid. 
Agents start in the first column, while targets are uniformly sampled from the last ten columns. 
At each step, the intended action is replaced uniformly by one of the other four actions with probability $p$.
We evaluate the \textit{individual}, \textit{local}, and \textit{global} noise models of Section~\ref{section:problem_setting}, 
according to which replacement draws are independent across agents, shared among agents with the same state-action pair, or shared across the population.
The objective is to minimize average travel time and terminal mismatch to the target distribution. 
We instantiate $\famalg$ with Q-learning \cite{watkins1992q} ($\algname{Q}$), DDQN \cite{hasselt2016deep} ($\algname{DDQN}$), PPO \cite{schulman2017proximalpolicyoptimizationalgorithms} ($\algname{PPO}$), and A2C~\cite{mnih2016asynchronous} ($\algname{A2C}$), and compare them 
against six MARL baselines: MAPPO \cite{yu2022surprisingeffectivenessppocooperative}, IPPO \cite{dewitt2020independent}, HAPPO \cite{kuba2022trust}, VDN \cite{sunehag2018value}, QMIX \cite{rashid2020monotonic}, and mean-field Q-learning (MFQ) \cite{yang2018mean}. 
At evaluation, $\famalg$ uses the online-assignment variant in Algorithm~\ref{alg:separation}, recomputing the transport assignment at every time step. 
All methods are trained with $5$ seeds and evaluated on $50$ evaluation seeds per trained policy. 
%
%
Figure~\ref{fig:TravelTimesBenchmark} shows that $\famalg$ consistently outperforms the MARL baselines across all noise levels and noise models. 
The four $\famalg$ variants perform similarly, suggesting that the gain comes from the separation structure rather than from the particular single-agent RL algorithm.
Complementary metrics, as well as the complete training, evaluation, and uncertainty details, are reported in Appendix~\ref{appendix:results:performance}.

\noindent\textbf{Separation leads to better scalability.}
\label{section:results:scalability}
\begin{figure}[t]
\centering
    \pgfplotsset{
  colormap={myviridis}{
    rgb255(0cm)=(68,1,84);
    rgb255(13cm)=(71,44,122);
    rgb255(25cm)=(59,81,139);
    rgb255(38cm)=(44,113,142);
    rgb255(50cm)=(33,144,141);
    rgb255(63cm)=(39,173,129);
    rgb255(75cm)=(92,200,99);
    rgb255(88cm)=(170,220,50);
    rgb255(100cm)=(253,231,37)
  }
}

\pgfplotsset{
  heatnumbers/.style={
    scatter,
    only marks,
    mark=none,
    point meta=explicit symbolic,
    nodes near coords={\strut\pgfplotspointmeta},
    every node near coord/.append style={
      font=\scriptsize,
      text=white,
      anchor=center,
      inner sep=0pt,
      outer sep=0pt,
      align=center,
    },
  },
  heatnumbers dark/.style={
    heatnumbers,
    every node near coord/.append style={
      text=black,
    },
  },
}

\begin{tikzpicture}

\begin{groupplot}[
  group style={
    group size=2 by 1,
    horizontal sep=1.5mm,
  },
  colormap name=myviridis,
  point meta min=0,
  point meta max=1.6,
  ymin=-0.5, ymax=4.5,
  ytick={0,1,2,3,4},
  yticklabels={70\%,80\%,90\%,95\%,99\%},
  y dir=normal,
  enlargelimits=false,
  scale only axis,
  unbounded coords=discard,
  tick style={draw=none},
  axis line style={black, line width=0.8pt},
  axis on top,
  clip=true,
  xlabel={Agents / grid size},
  ylabel={Reach-rate milestone},
  title style={font=\small, yshift=-0.8ex},
  xlabel style={font=\small},
  ylabel style={font=\small},
  ticklabel style={font=\footnotesize},
  xticklabel style={align=center,font=\footnotesize},
]

\nextgroupplot[
  title={SALT-PPO (ours)},
  title style={font=\small\bfseries, yshift=-0.8ex},
  width=0.6\linewidth,
  height=0.23\linewidth,
  xmin=-0.5, xmax=7.5,
  xtick={0,1,2,3,4,5,6,7},
  xticklabels={
    {\shortstack{$N\!=\!2$\\10$\times$10}},
    {\shortstack{$N\!=\!4$\\14$\times$14}},
    {\shortstack{$N\!=\!6$\\17$\times$17}},
    {\shortstack{$N\!=\!8$\\20$\times$20}},
    {\shortstack{$N\!=\!16$\\28$\times$28}},
    {\shortstack{$N\!=\!32$\\40$\times$40}},
    {\shortstack{$N\!=\!64$\\57$\times$57}},
    {\shortstack{$N\!=\!128$\\80$\times$80}}
  },
]

\addplot[
  matrix plot*,
  mesh/cols=8,
  point meta=explicit,
  shader=flat corner,
  draw=none,
] coordinates {
  (0,0) [0.000344576] (1,0) [0.001053696]  (2,0) [0.0019166208] (3,0) [0.003321856]  (4,0) [0.011153408] (5,0) [0.03842048]  (6,0) [0.147554304] (7,0) [0.55574528]
  (0,1) [0.000404992] (1,1) [0.0011855872] (2,1) [0.002193408]  (3,1) [0.003694592]  (4,1) [0.012730368] (5,1) [0.04276224]  (6,1) [0.162729984] (7,1) [0.62980096]
  (0,2) [0.000510464] (1,2) [0.0014651392] (2,2) [0.0025772544] (3,2) [0.004395008]  (4,2) [0.015167488] (5,2) [0.0507904]   (6,2) [0.190746624] (7,2) [0.72679424]
  (0,3) [0.000650752] (1,3) [0.0017647616] (2,3) [0.0030081024] (3,3) [0.0050176]    (4,3) [0.018321408] (5,3) [0.0602112]   (6,3) [0.230436864] (7,3) [0.86048768]
  (0,4) [0.00233472]  (1,4) [0.0326918144] (2,4) [0.0253260288] (3,4) [0.016797696]  (4,4) [0.085471232] (5,4) [0.17547264]  (6,4) [0.445464576] (7,4) [1.54075136]
};

\addplot+[
  heatnumbers,
] table[x=x,y=y,meta=label]
  {figures/reach_rate_labels_salt_white.dat};

\addplot+[
  heatnumbers dark,
] table[x=x,y=y,meta=label]
  {figures/reach_rate_labels_salt_black.dat};

\nextgroupplot[
  title={MAPPO},
  title style={font=\small, yshift=-0.8ex},
  width=0.29\linewidth,
  height=0.23\linewidth,
  xmin=-0.5, xmax=3.5,
  ylabel={},
  yticklabels={},
  xtick={0,1,2,3},
  xticklabels={
    {\shortstack{$N\!=\!2$\\10$\times$10}},
    {\shortstack{$N\!=\!4$\\14$\times$14}},
    {\shortstack{$N\!=\!6$\\17$\times$17}},
    {\shortstack{$N\!=\!8$\\20$\times$20}}
  },
]

\addplot[
  matrix plot*,
  mesh/cols=4,
  point meta=explicit,
  shader=flat corner,
  draw=none,
] coordinates {
  (0,0) [0.007297536] (1,0) [0.1594607616] (2,0) [0.8635865088] (3,0) [nan]
  (0,1) [0.008282624] (1,1) [0.181113856]  (2,1) [1.0313796096] (3,1) [nan]
  (0,2) [0.00896768]  (1,2) [0.20694016]   (2,2) [1.2390614016] (3,2) [nan]
  (0,3) [0.011582464] (1,3) [0.2638483456] (2,3) [1.452662784]  (3,3) [nan]
  (0,4) [0.11273728]  (1,4) [nan]          (2,4) [nan]          (3,4) [nan]
};

\addplot+[
  heatnumbers,
] table[x=x,y=y,meta=label]
  {figures/reach_rate_labels_mappo_white.dat};

\addplot+[
  heatnumbers dark,
] table[x=x,y=y,meta=label]
  {figures/reach_rate_labels_mappo_black.dat};

\node[font=\normalsize] at (axis cs:1,4) {--};
\node[font=\normalsize] at (axis cs:2,4) {--};
\node[font=\normalsize] at (axis cs:3,4) {--};

\node[font=\normalsize] at (axis cs:3,0) {--};
\node[font=\normalsize] at (axis cs:3,1) {--};
\node[font=\normalsize] at (axis cs:3,2) {--};
\node[font=\normalsize] at (axis cs:3,3) {--};

\end{groupplot}

\end{tikzpicture}
    \vspace{-1.3\baselineskip}
    \caption{Training environment interactions required by $\algname{PPO}$ and MAPPO to attain each reach-rate milestone ($70$--$99\%$) for $10$ consecutive batches 
    as population size $\nagents$ increases (grid area scaled proportionally to maintain target density). 
    Each cell reports the mean over $5$ training seeds, a dash (--) indicates that no seed attained the milestone within the training budget.}
    \label{fig:ReachRateMilestones}
\end{figure}
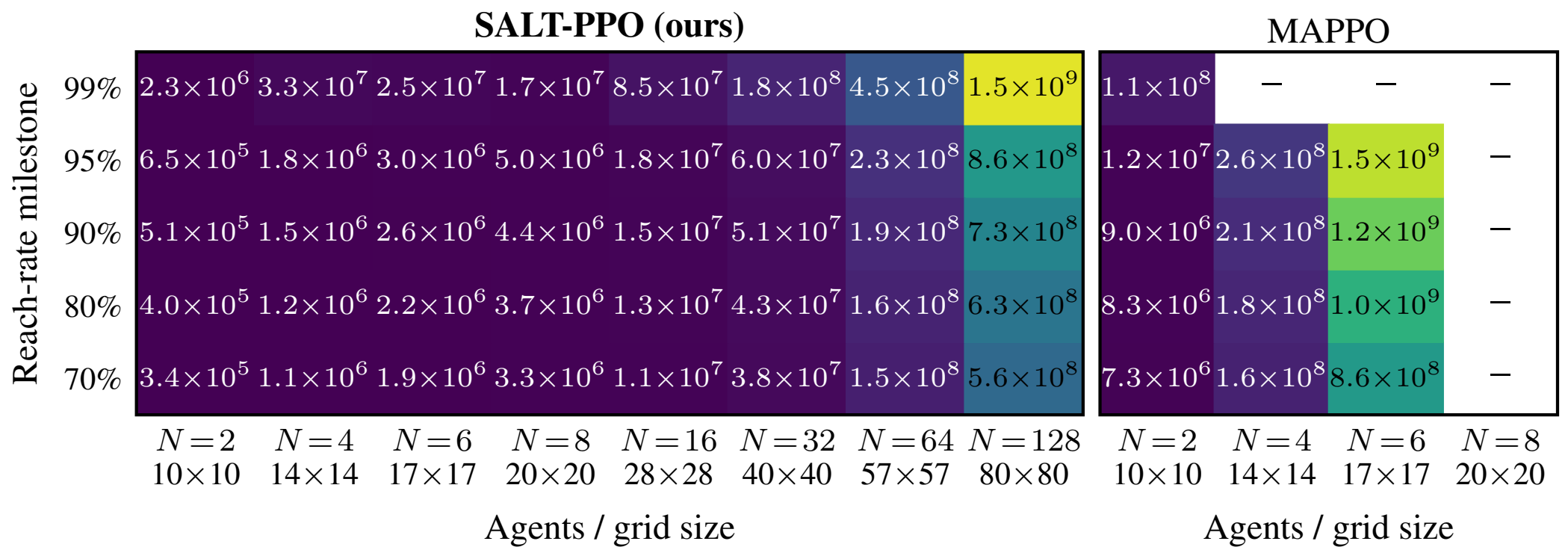
We next study sample complexity as the number of agents increases from $\nagents = 2$ to $\nagents = 128$. 
We use the same grid-world dynamics with individual noise, and increase grid area to keep the target density roughly constant and isolate the effects of scaling from the complexity of the task. 
During training, agents start from independently sampled rows of the first column, and targets are sampled uniformly across the grid (see Appendix~\ref{appendix:results:scaling}).
We select PPO as the single-agent learner in $\famalg$, yielding $\algname{PPO}$, and compare it against MAPPO.
We report the number of environment interactions required to attain reach-rate milestones of $70\%$, $80\%$, $90\%$, $95\%$, and $99\%$, up to a maximum training budget of $2\times10^9$ interactions.
Figure~\ref{fig:ReachRateMilestones} shows that $\algname{PPO}$ attains the same milestones with up to $480\times$ fewer interactions than MAPPO, 
and reaches the $99\%$ milestone for populations of up to $\nagents=128$ agents, while MAPPO fails to reach the $70\%$ milestone at $\nagents=8$ within the configured budget.
This gap directly reflects their different scalability properties. MAPPO learns over a joint multi-agent state-action space, whose size grows exponentially with $\nagents$, 
whereas $\algname{PPO}$ learns a target-conditioned policy that is independent of $\nagents$.

\subsection{Towards a realistic time-varying stochastic urban logistics task}
\label{section:results:manhattan}

\begin{figure}[t]
    \centering
    \begin{subfigure}[t]{0.245\textwidth}
        \centering
            \includegraphics[height=\FigurePanelHeight,
                width=\linewidth,
                keepaspectratio
            ]{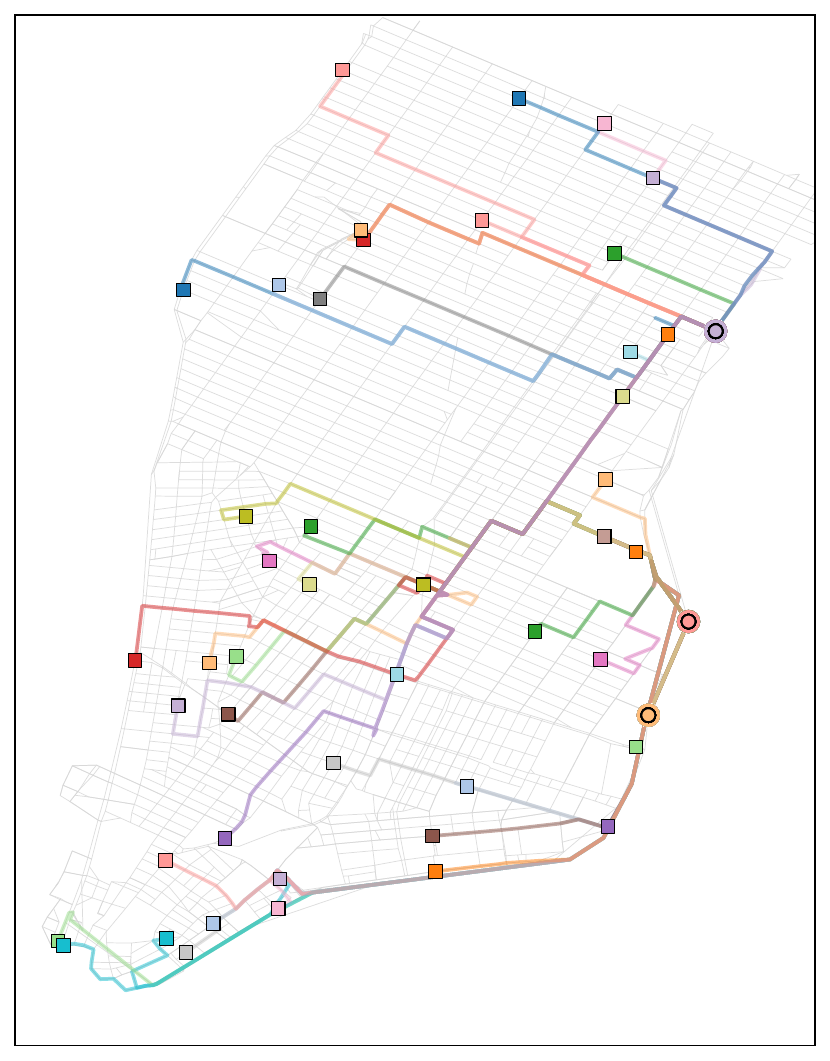}
    \end{subfigure}
    \hfill
    \begin{subfigure}[t]{0.725\textwidth}
        \centering
        \begin{tikzpicture}
\begin{axis}[
  width=0.995\linewidth,
  height=0.45\linewidth,
  xmin=0.5,
  xmax=5.5,
  ymin=-52,
  ymax=68,
  axis lines=box,
  axis line style={black, line width=0.75pt},
  tick align=outside,
  tick pos=left,
  tick style={black, line width=0.75pt},
  xtick={1,2,3,4,5},
  xticklabels={1,5,10,20,100},
  minor x tick num=1,
  minor tick style={draw=none},
  ytick={-40,-20,0,20,40,60},
  yticklabels={-40\%,-20\%,0\%,20\%,40\%,60\%},
  xlabel={Number of agents},
  ylabel={Relative improvement},
  label style={font=\small},
  tick label style={font=\small},
  xmajorgrids=true,
  xminorgrids=true,
  grid style={draw=gray!35, solid, line width=0.35pt},
  ymajorgrids=false,
]
\foreach \yy in {-40,-20,20,40,60}{%
  \addplot[forget plot, draw=gray!35, dashed, line width=0.35pt]
    coordinates {(0.5,\yy) (5.5,\yy)};
}
\addplot[forget plot, draw=black, dashed, line width=0.55pt]
  coordinates {(0.5,0) (5.5,0)};

\addplot[boxplotstyle,
  boxplot prepared={
    draw position=1,
    lower whisker=-45.056668,
    lower quartile=0.000000,
    median=14.845996,
    upper quartile=30.696553,
    upper whisker=63.335937
  }
] coordinates {};

\addplot[boxplotstyle,
  boxplot prepared={
    draw position=2,
    lower whisker=-7.833956,
    lower quartile=13.416065,
    median=21.213559,
    upper quartile=28.012151,
    upper whisker=48.222901
  }
] coordinates {};

\addplot[boxplotstyle,
  boxplot prepared={
    draw position=3,
    lower whisker=0.013008,
    lower quartile=15.631962,
    median=21.060194,
    upper quartile=26.106960,
    upper whisker=40.971681
  }
] coordinates {};

\addplot[boxplotstyle,
  boxplot prepared={
    draw position=4,
    lower whisker=6.442881,
    lower quartile=17.435774,
    median=21.224713,
    upper quartile=25.000285,
    upper whisker=36.163655
  }
] coordinates {};

\addplot[boxplotstyle,
  boxplot prepared={
    draw position=5,
    lower whisker=14.088794,
    lower quartile=19.528262,
    median=21.325891,
    upper quartile=23.248151,
    upper whisker=28.268776
  }
] coordinates {};
\end{axis}
\end{tikzpicture}
    \end{subfigure}
    \vspace{-\baselineskip}
    \caption{Relative travel-time improvement of $\algname{Q}$ over static shortest-path routing on the time-varying Manhattan road network, as a function of fleet size. 
    Boxplots pool $500$ evaluation instances from each of $8$ training seeds. 
    Boxes span the interquartile range, center lines show medians, and whiskers follow the Tukey $1.5$-interquartile-range rule. 
    The dashed line indicates zero improvement. 
    $\algname{Q}$ is trained with $20$ vehicles and evaluated without retraining on fleets of $1$--$100$ vehicles.}
    \vspace{-\baselineskip}
     \label{fig:manhattan_travel_times}
\end{figure}
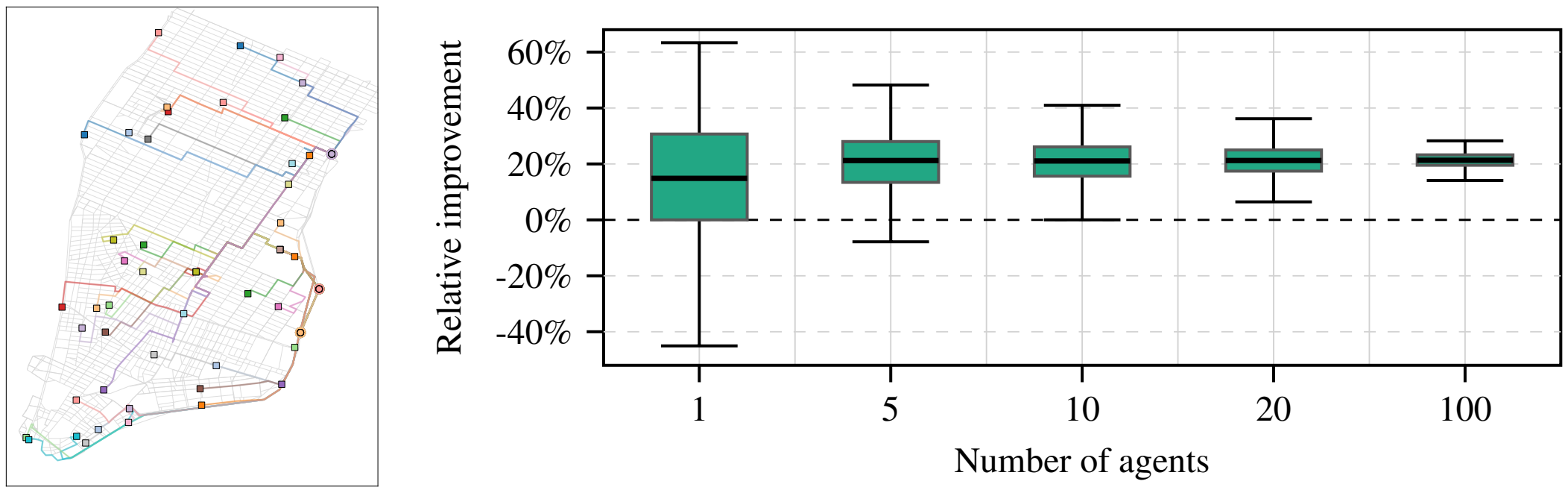

A fleet of delivery vehicles must serve requests across Manhattan while jointly minimizing total travel time. 
This task couples vehicle-to-request assignment with time-efficient routing through a stochastic, time-varying network and is precisely the setting that motivated this work. 
$\famalg$ separates the problem into a coordination step (assigning vehicles to requests via optimal transport) and a routing step (navigating each vehicle to its current target via a shared target-conditioned learned policy). 
As motivated in \cref{section:algorithm}, assignments remain fixed during each training episode to learn the target-conditioned subproblem, and are recomputed at every routing step during evaluation.
The environment is a directed southern Manhattan road graph extracted from OpenStreetMap \cite{OpenStreetMap}, with $2087$ intersections and $4255$ directed roads (see Figure~\ref{fig:manhattan_travel_times}).
Travel times depend on road length, road type, traffic-light cycles of $90$ seconds (with a road-dependent green-phase duration and a random offset), and stochastic congestion effects.
Vehicles start from three depot locations in random proportions and serve delivery requests sampled uniformly across the city.
Further environment details are provided in Appendix~\ref{appendix:results:manhattan}.
We instantiate $\famalg$ with tabular Q-learning ($\algname{Q}$) and compare it against a static baseline that assigns vehicles once at $t=0$ using nominal shortest-path distances and then follows the resulting paths. 
This standard operational heuristic ignores traffic-light timing and stochastic congestion, since incorporating them exactly requires a much larger spatiotemporal routing problem.
We exclude the deep MARL baselines of Section~\ref{section:results:benchmarking} because, as demonstrated, they become computationally prohibitive on a graph of this scale and are tied to a fixed training fleet size. 
$\algname{Q}$ policies are trained with $\nagents = 20$ agents and evaluated, without retraining, on fleets of $1$, $5$, $10$, $20$, and $100$ vehicles.

Figure~\ref{fig:manhattan_travel_times} reports the relative improvement in travel time over the static shortest-path baseline as a function of fleet size. 
$\algname{Q}$ consistently outperforms the shortest-path baseline, with median improvements ranging from $14.8\%$ ($\nagents=1$) to $21.3\%$ ($\nagents=100$), 
reflecting its ability to anticipate traffic-light timing and adapt to stochastic congestion. 
Performance remains stable across fleet sizes, confirming that $\famalg$ generalizes naturally to fleet sizes unseen at training time. 
The residual variance arises from a natural bias of distributional training: the policy concentrates on high-probability origin--destination pairs, improving sample efficiency at the cost of reduced accuracy on rare assignments.
Additional ablations regarding $\famalg$ fleet-size transferability and reassignment variants of the shortest-path baseline are reported in \cref{appendix:results:pair-frequency} and \cref{appendix:results:manhattan}, respectively.

\section{Conclusion and limitations}
We established a separation principle for cooperative MARL and used it to design $\famalg$, a family of algorithms that learn a single-agent policy and use optimal transport to coordinate agents. $\famalg$ outperforms existing MARL baselines in grid worlds, both in terms of performance and scalability. In the southern Manhattan road network, where MARL baselines are impractical, $\famalg$ improves travel time by $15$--$21\%$ over a shortest-path heuristic and generalizes to population sizes not seen \mbox{during training.}
Our work does not address several important challenges, which we leave to future work.
From a theoretical perspective, our separation principle requires decoupled dynamics and an optimal transport terminal cost, ruling out settings with strong interactions, collisions, or non-transport objectives. Our bound on the optimal cost is also not tight in general.
From a practical perspective, the target-conditioned single-agent problem must generalize across the target space, which is challenging when targets are high-dimensional or sampled from distributions far from the training one.

\begin{ack}
This publication was made possible by the ETH AI Center doctoral fellowship to Lucia Pezzetti. This work was also supported as a part of NCCR Automation,
a National Centre of Competence in Research, funded by the Swiss National
Science Foundation (grant number 51NF40 225155).
Nicolas Lanzetti was in part supported by the PIMCO Fellows Program, by the National Science Foundation under Grant \#CCF-2326609, and by the Resnick Sustainability Institute.
\end{ack}

\bibliographystyle{unsrtnat}
\bibliography{refs}

\newpage
\appendix

\section{Noise models and averaging property}
\label{section:noise}
This appendix develops the measure-theoretic framework used to handle stochasticity in the multi-agent system. We first introduce the relevant general notation, and illustrate how common noise models fit into this framework through a series of examples. We conclude with Lemma~\ref{lemma:noiseexp}, which will be a key technical tool used in the proof of \sepbound{} in Appendix~\ref{section:main_proof}.
Throughout, the space of probability measures on a Polish space $X$, $\probabilityspace{X}$, is endowed with the topology of weak convergence and its Borel $\sigma$-algebra. Since $X$ is a Polish space, $\probabilityspace{X}$ is itself a Polish space~\cite[Theorem~15.15]{aliprantis06infinite}.
The product measure of two independent random variables with marginals $\mu \in \probabilityspace{X}$ and $\nu \in \probabilityspace{Y}$ is the unique measure $\mu \otimes \nu \in \probabilityspace{X\times Y}$ \cite[Theorem B, Section 35]{halmos1976measure} such that $(\mu \otimes \nu)(A \times B) = \mu(A)\, \nu(B)$ for all Borel sets $A\subseteq X,\, B \subseteq Y$.

\subsection{General framework}
In multi-agent systems, randomness enters at the \textit{individual} level, through the noise realizations for each individual agent. At the \textit{population} level, these realizations induce randomness in the population distribution and therefore in the next population state.
To capture this in a unified way, we work with distributions over noise distributions. Concretely, following the notation in Section~\ref{section:problem_setting:population}, let $\probabilitySA{}{\timestep} \in \probabilityspace{\statespace\times\actionspace}$ denote the joint state-action distribution of the population at time $\timestep$, and denote by $\noisedist{\timestep}(\cdot \st \state, \action) \in \probabilityspace{\noisespace}$ the \emph{nominal kernel}, i.e., the law of the noise experienced by an agent in state $\state$ taking action $\action$. For agents sharing the same state-action pair $(\state, \action)$, their noise realizations form the \emph{realized kernel} $\empnoisedist{\timestep}^{\state, \action} \in \probabilityspace{\noisespace}$. Unlike the nominal kernel $\noisedist{\timestep}(\cdot \st \state, \action)$, which is a fixed object, $\empnoisedist{\timestep}^{\state, \action}$ is random, fluctuating across episodes depending on the specific noise draws of the agents in that bin.
Therefore, we can talk about the distribution of $\empnoisedist{\timestep}^{\state, \action}$, which we denote by $\distoverdist{\timestep}^{\state, \action} \in \probabilityspace{\probabilityspace{\noisespace}}$, and collect these across all state-action pairs into the joint object $\distoverdist{\timestep} \in \probabilityspace{\probabilityspace{\noisespace}^{\statespace\times \actionspace}}$.
Finally, to describe the full population transition in this stochastic setting, it is not enough to track where agents are and what actions they take; one also needs to account for the noise they experience. This motivates the introduction of the joint state-action-noise distribution, defined as the gluing
\begin{equation*}
    \probabilitySAW{}{\timestep} = \probabilitySA{}{\timestep}(\fleet_\timestep) \otimes \empnoisedist{\timestep}^{\state, \action} \in \probabilityspace{\statespace\times\actionspace\times\noisespace}.
\end{equation*}
Intuitively, $\probabilitySAW{}{\timestep}(\{(\state,\action,\noise)\})$ is the fraction of agents that are at state $\state$, use action $\action$, and experience the noise realization $\noise$.
Since the noise realizations are random, $\empnoisedist{\timestep}^{\state, \action}$ is random and, consequently, so is $\probabilitySAW{}{\timestep}$. 
At this point, the next population is a pushforward of $\probabilitySAW{}{\timestep}$ under the system dynamics $\dynamics{}{}{\timestep}$: 
\begin{equation*}
    \fleet_{\nexttimestep} = \pushforward{\dynamics{}{}{\timestep}}{\probabilitySAW{}{\timestep}}.
\end{equation*}
Different noise models below correspond to different distributional assumptions on $\empnoisedist{\timestep}^{\state, \action}$, and hence on $\probabilitySAW{}{\timestep}$.
The objects introduced above are summarized in Table~\ref{table:noise_formulation}.

\begin{table}[t]
\centering
\renewcommand{\arraystretch}{1.2}
\caption{Notation for Appendices~\ref{section:noise}--~\ref{app:additional_experiments}.}
\label{table:noise_formulation}
\begin{tabular}{@{}p{0.08\textwidth} p{0.3\textwidth} p{0.5\textwidth}@{}}
\toprule
\textbf{Object} & \textbf{Definition} & \textbf{Intuition} \\ \midrule
 $\empnoisedist{\timestep}^{\state,\action}$ 
& Random element of $\probabilityspace{\noisespace}$ 
& Distribution of noise experienced by agents in $(\state,\action)$ at time $\timestep$ \\[6pt]
$\distoverdist{\timestep}^{\state,\action}$ 
& Distribution of $\empnoisedist{\timestep}^{\state,\action}$ 
& Describes the probability distribution associated with the noise realizations \\[6pt]
$\distoverdist{\timestep}$
& Probability measure in $\probabilityspace{\probabilityspace{\noisespace}^{\statespace\times \actionspace}}$
& Collects all $\distoverdist{\timestep}^{\state,\action}$ across bins, encoding population-level randomness \\[6pt]
$\probabilitySAW{}{\timestep}$
& $\probabilitySA{}{\timestep} \otimes \empnoisedist{\timestep}^{\state, \action}$
& Joint state-action-noise distribution \\
\bottomrule
\end{tabular}
\end{table}

The gluing $\probabilitySAW{}{\timestep} = \probabilitySA{}{\timestep} \otimes \empnoisedist{\timestep}^{\state, \action}$ is well-defined per realization only if the realized field is a measurable kernel. We impose this as a standing assumption. It holds directly for the individual and global models, while the only subtle case is discussed in Example~\ref{example:local_noise}.

\begin{assumption}[Well-posedness of the realized noise field]
\label{assumption:measurable_field}
Fix a time step $\timestep$ and an admissible state-action distribution $\probabilitySA{}{\timestep}$. The realized noise field, defined on a probability space $(\Omega, \mathcal F, \mathbb P)$, admits a version for which there exists a Borel set $D_\timestep \subseteq \statespace \times \actionspace$ with $\probabilitySA{}{\timestep}(D_\timestep) = 1$ such that
\begin{equation*}
    \bigl((\state, \action), \omega\bigr) \mapsto \empnoisedist{\timestep}^{\state, \action}(\omega)(B)
\end{equation*}
is jointly measurable on $D_\timestep \times \Omega$ for every Borel set $B \subseteq \noisespace$.
\end{assumption}

In particular, almost surely, $(\state, \action) \mapsto \empnoisedist{\timestep}^{\state, \action}$ is a Borel kernel on $D_\timestep$, so $\probabilitySAW{}{\timestep}$ is a well-defined random element of $\probabilityspace{\statespace\times\actionspace\times\noisespace}$, and expectations over the field can be interchanged with $\probabilitySA{}{\timestep}$-integrals by Tonelli's theorem.

We additionally impose the following consistency condition on the realized kernel.
\begin{assumption}
\label{assumption:barycenter_condition}
Fix a time step $\timestep$ and an admissible state-action distribution $\probabilitySA{}{\timestep}$, together with its realized noise field. There exists a Borel set $E_\timestep \subseteq \statespace\times\actionspace$ with $\probabilitySA{}{\timestep}(E_\timestep) = 1$ such that for all $(\state, \action)\in E_\timestep$ and for every bounded Borel measurable $\varphi: \noisespace \to \reals$
\begin{equation}
\label{eq:test_functions}
\expectedvalue{\distoverdist{\timestep}^{\state, \action}}{
\int_\noisespace \varphi(\noise) 
\dd\empnoisedist{\timestep}^{\state, \action}(\noise)
}
=
\int_\noisespace \varphi(\noise)
\dd\noisedist{\timestep}(\noise \st \state, \action).
\end{equation}
\end{assumption}

Condition \eqref{eq:test_functions} should be read as an unbiased or barycenter condition between the realized kernel $\empnoisedist{\timestep}^{\state, \action}$ and the nominal kernel $\noisedist{\timestep}(\cdot \st \state, \action)$. More precisely, the assumption requires that these two descriptions agree
after averaging over the realization of $\empnoisedist{\timestep}^{\state, \action}$.
As we will show below, Assumption~\ref{assumption:barycenter_condition} is satisfied by several standard realized-kernel constructions. Specifically, in Proposition~\ref{proposition:averagingproperty} we show that it holds true for the noise models presented in Examples~\ref{example:individual_noise}--\ref{example:global_noise}.
More generally, Assumption~\ref{assumption:barycenter_condition} is satisfied whenever the realized kernel is chosen from a family of probability kernels whose mixture is the nominal kernel
$\noisedist{\timestep}(\cdot \st \state, \action)$.

What \eqref{eq:test_functions} excludes are biased realized-kernel constructions, as we show in the next example. 

\begin{example}
Let $\noisespace=\{0,1\}$ and
$\noisedist{\timestep}=\frac12\delta_0+\frac12\delta_1$. Suppose that we artificially bias all noise realizations towards $1$, so that
\begin{equation*}
\empnoisedist{\timestep}^{\state,\action}=\delta_1
\end{equation*}
almost surely. Then
\begin{equation*}
\expectedvalue{\distoverdist{\timestep}^{\state,\action}}{\int_\noisespace \noise\dd\empnoisedist{\timestep}^{\state,\action}(\noise)}=1
\neq
\frac12
=
\int_\noisespace \noise\dd \noisedist{\timestep}(\noise\st \state, \action).
\end{equation*}
Hence, \eqref{eq:test_functions} fails.
\end{example} 

More generally, \eqref{eq:test_functions} of~\cref{assumption:barycenter_condition} fails whenever the realized population-level kernel systematically overweights some noise realizations relative to the nominal kernel used in the single-agent Bellman equation.
In the remainder of this section, we specialize this general framework to the individual, local, and global noise models considered in this work (Appendix~\ref{app:canonical-noise-models}). 
Finally, in Appendix~\ref{app:averaging-realized-noise-kernels}, we prove a key property of all these noise models: we establish, in Lemma~\ref{lemma:noiseexp}, that the expected cost under the realized kernel $\empnoisedist{\timestep}^{\state, \action}$ equals the cost under the nominal kernel $\noisedist{\timestep}(\cdot \st \state, \action)$. Ultimately, this unbiasedness property, verified for all three noise models in Proposition~\ref{proposition:averagingproperty}, is what allows us to eliminate randomness from the backward recursion in Appendix~\ref{section:main_proof} and recover the single-agent Bellman equation.

\subsection{Noise models}
\label{app:canonical-noise-models}

We start with the case where the noise is i.i.d.\ over all agents with the same state-action pair. 
We start with the case of finitely many agents: 

\begin{example}[Individual noise, finitely many agents]
\label{example:individual_noise} 
Consider $\nagents$ finite agents, each with its ``own'' noise sample
\begin{equation*}
    \noise_\timestep^i \sim \noisedist{\timestep}(\cdot \st \state_\timestep^i, \action_\timestep^i).
\end{equation*}
We assume noise draws to be independent across agents and across state-action pairs. 
Given the state-action distribution $\probabilitySA{}{\timestep}$, its support $\supp{\probabilitySA{}{\timestep}}$ is finite (at most $\nagents$ atoms, since $\probabilitySA{}{\timestep} = \frac{1}{\nagents}\sum_{i=1}^\nagents \delta_{(\state_\timestep^i, \action_\timestep^i)}$). For each bin $(\state, \action) \in \supp{\probabilitySA{}{\timestep}}$, let 
\begin{equation*}
    \nagents_{\state, \action} \coloneqq \nagents\cdot \probabilitySA{}{\timestep}(\{(\state, \action)\}) \ge 1
\end{equation*}
be the number of agents in that bin. This is a strictly positive integer, determined by $\probabilitySA{}{\timestep}$. 
Define the empirical measure map
\begin{align*}
    \Psi^{\state, \action}_{\nagents_{\state, \action}}\colon 
    \noisespace^{\nagents_{\state, \action}} &\to \probabilityspace{\noisespace}\\
    \noise^{\state, \action} \,\,\,\,
    &\mapsto 
    \frac{1}{\nagents_{\state, \action}}
    \sum_{i = 1}^{\nagents_{\state, \action}}
    \delta_{\noise^{\state,\action,i}},
\end{align*}
where $\noise^{\state, \action} = (\noise^{\state,\action,i})_{i \in [\nagents_{\state, \action}]}$ collects the noise draws of the agents in bin $(\state,\action)$. The realized noise distribution for the state-action pair $(\state, \action)$ is
\begin{equation*}
    \empnoisedist{\timestep}^{\state, \action}=
    \Psi^{\state, \action}_{\nagents_{\state, \action}} \left(
    \noise^{\state, \action}
    \right)
    \in \probabilityspace{\noisespace},
\end{equation*}
with distribution
\begin{equation*}
    \distoverdist{\timestep}^{\state, \action}=
\pushforward{\Psi^{\state, \action}_{\nagents_{\state, \action}}}
    {\left(\bigotimes_{i\in[\nagents_{\state, \action}]} \noisedist{\timestep}(\cdot \st \state, \action)\right)}
    \in \probabilityspace{\probabilityspace{\noisespace}}.
\end{equation*}
Since $\supp{\probabilitySA{}{\timestep}}$ is finite and noise realizations are independent across state-action pairs, the joint law is the finite product
\begin{equation*}
    \distoverdist{\timestep} =
    \bigotimes_{(\state, \action) \in \supp{\probabilitySA{}{\timestep}}}
    \distoverdist{\timestep}^{\state, \action}
    \in \probabilityspace{\probabilityspace{\noisespace}^{\supp{\probabilitySA{}{\timestep}}}}.
\end{equation*}
We extend $\distoverdist{\timestep}$ to all state-action pairs by setting, for $(\state, \action) \notin\supp{\probabilitySA{}{\timestep}}$, $\empnoisedist{\timestep}^{\state, \action} = \delta_{\noise^0}$, for an arbitrarily chosen $\noise^0 \in \noisespace$.
Clearly, the specific choice of $\noise^0$ is inconsequential: since $\probabilitySA{}{\timestep}$ assigns zero mass to these bins, they do not contribute to $\probabilitySA{}{\timestep}$-weighted integrals, which are exactly the type of integrals in our problem setting.
The map $(\state, \action) \mapsto \distoverdist{\timestep}^{\state, \action}$ is Borel measurable: it takes finitely many distinct values (one per element of $\supp\probabilitySA{}{\timestep}$, plus the constant extension outside), and the preimage of each value is a finite union of singletons or the complement of $\supp\probabilitySA{}{\timestep}$, all of which are Borel sets.  
\end{example}


\begin{example}[Individual noise, infinitely many agents]
\label{example:indiv_noise_inf_agents}
In the limit case of infinitely many agents, the two-level randomization (each agent drawing a private noise sample, and the collection of noise realizations across the population) collapses into a single draw from the deterministic coupling
\begin{equation*}
    \probabilitySAW{}{\timestep}=\probabilitySA{}{\timestep} \otimes \noisedist{\timestep}(\cdot \st \state, \action).
\end{equation*}
Intuitively, the population is so numerous that its empirical profile of noise realizations matches the underlying noise kernel itself.
Rigorously, in the case of finite state and action spaces, we can assume that the number of agents in each state-action pair $(\state, \action)$ grows to infinity and Varadarajan’s theorem \cite[Theorem 11.4.1.]{dudley2002} in the space $\probabilityspace{\noisespace}$ guarantees
\begin{equation*}
    \empnoisedist{\timestep}^{\state, \action} \overset{\text{a.s.}}{\longrightarrow} \noisedist{\timestep}(\cdot \st \state, \action), \quad \text{as } \nagents_{\state, \action} \to \infty.
\end{equation*}
Consequently, the laws of the realized kernels $\distoverdist{\timestep}^{\state, \action}$ converge weakly to $\delta_{\noisedist{\timestep}(\cdot \st \state, \action)}$ \cite[Theorem~9.2.1, Proposition~9.3.5]{dudley2002}.
We therefore set $\empnoisedist{\timestep}^{\state, \action} = \noisedist{\timestep}(\cdot \st \state, \action)$ almost surely and
\begin{equation*}
    \distoverdist{\timestep}^{\state, \action}=\delta_{\noisedist{\timestep}(\cdot \st \state, \action)}
    \in \probabilityspace{\probabilityspace{\noisespace}}.
\end{equation*}
%
For general (possibly uncountable) Polish spaces $\statespace$ and $\actionspace$, we construct the joint law via Kolmogorov's extension theorem \cite[Theorems 36.1 and 36.2]{billingsley1980}. Specifically, for each finite subset $J \subset \statespace\times\actionspace$, define a probability measure
\begin{equation*}
    \distoverdist{\timestep}^J=\bigotimes_{(\state, \action) \in J}
    \delta_{\noisedist{\timestep}(\cdot \st \state, \action)} 
    \in \probabilityspace{\probabilityspace{\noisespace}^J}.
\end{equation*}
These finite-dimensional marginals are consistent (the marginal of $\distoverdist{\timestep}^J$ onto any $J' \subset J$ equals $\distoverdist{\timestep}^{J'}$, since products of Dirac masses marginalize correctly). Thus, Kolmogorov’s extension theorem guarantees that there exists a unique probability measure
\begin{equation*}
    \distoverdist{\timestep}=\bigotimes_{(\state, \action) \in \statespace \times \actionspace}
    \delta_{\noisedist{\timestep}(\cdot \st \state, \action)},
\end{equation*}
and the realized joint state-action-noise distribution collapses to the deterministic coupling $\probabilitySAW{}{\timestep}=\probabilitySA{}{\timestep} \otimes \noisedist{\timestep}(\cdot \st \state, \action)$.
\end{example}


\begin{example}[Local noise]
\label{example:local_noise} 
    Suppose the noise realizations are shared among agents with the same state-action pair.
    For each fixed $(\state, \action) \in \statespace \times \actionspace$ we draw 
    \begin{equation*}
    \noise_\timestep^{\state, \action} \sim \noisedist{\timestep}(\cdot \st \state, \action),
    \end{equation*}
    independently across state-action pairs.
    Then, all agents in bin $(\state, \action)$ experience the same noise realization $\noise_\timestep^i = \noise_\timestep^{\state, \action}$.
    In this framework, we define the empirical measure map
    \begin{align*}
        \Psi^{\state, \action}\colon \noisespace &\to \probabilityspace{\noisespace} \\
        \noise^{\state, \action}_\timestep &\mapsto
        \delta_{\noise^{\state, \action}_\timestep}. 
    \end{align*}
    The realized noise distribution for bin $(\state, \action)$ collapses to
    \begin{equation*}
        \empnoisedist{\timestep}^{\state, \action}=\delta_{\noise^{\state, \action}_\timestep},
    \end{equation*}
    with distribution
    \begin{equation*}
        \distoverdist{\timestep}^{\state, \action}=\pushforward{\Psi^{\state, \action}}
        {\noisedist{\timestep}(\cdot \st \state, \action)}
        \in \probabilityspace{\probabilityspace{\noisespace}}.
    \end{equation*}
    Since noise samples are independent across state-action pairs, the joint law is constructed by the same Kolmogorov extension argument as in 
    Example~\ref{example:indiv_noise_inf_agents} (applied with $\distoverdist{\timestep}^{\state, \action}$ in place of $\delta_{\noisedist{\timestep}(\cdot \st \state, \action)}$). 
    For every finite $J \subset \statespace \times \actionspace$, the marginal $\distoverdist{\timestep}^J = \bigotimes_{(\state, \action)\in J} \distoverdist{\timestep}^{\state, \action}$ is consistent, 
    yielding the unique extension\footnote{Assumption~\ref{assumption:measurable_field} holds whenever $\probabilitySA{}{\timestep}$ has countable support, including every finite population. 
    When $\probabilitySA{}{\timestep}$ is atomless, $\distoverdist{\timestep}$ still exists, but a continuum of independent draws admits no jointly measurable version with nondegenerate coordinates on an ordinary product probability space \cite{judd1985law,sun2006exact} and so
    $\probabilitySA{}{\timestep}\otimes\empnoisedist{\timestep}^{\state,\action}$ cannot be formed pathwise. In that case, we do not construct $\probabilitySAW{}{\timestep}$ but take it as the primitive disturbance, requiring only that its state-action marginal is $\probabilitySA{}{\timestep}$ almost 
    surely and that $\mathbb E[\probabilitySAW{}{\timestep}]=\probabilitySA{}{\timestep}\otimes\noisedist{\timestep}$. On a rich Fubini extension, an enlargement of the product space that retains the Fubini property and is rich enough to carry the continuum of independent noise draws itself, the exact law of large numbers of Sun~\cite{sun2006exact} then gives $\probabilitySAW{}{\timestep} =\probabilitySA{}{\timestep}\otimes\noisedist{\timestep}$ almost surely on the atomless part, while atoms keep their shared draws \cite{sun2006exact}. 
    We record this only to show that the formulation is consistent, indeed, the separation proof uses just the marginal and average properties above, so its conclusions also hold in this formulation.}
    \begin{equation*}
        \distoverdist{\timestep} =
        \bigotimes_{(\state, \action) \in \statespace \times \actionspace}
        \distoverdist{\timestep}^{\state, \action}
        \in \probabilityspace{\probabilityspace{\noisespace}^{\statespace \times \actionspace}}.
    \end{equation*}
\end{example}


\begin{example}[Global noise]
\label{example:global_noise}
    Consider the case in which all agents are subject to the same noise draw $\noise_\timestep \sim \noisedist{\timestep} \in \probabilityspace{\noisespace}$, regardless of their state and action.
    Since every state-action pair $(\state, \action)$ experiences the same noise realization, we have $\empnoisedist{\timestep}^{\state, \action} = \delta_{\noise_\timestep}$ for all $(\state, \action)$.
    We then define the diagonal embedding 
    \begin{align*}
        \Phi\colon \noisespace &\to \probabilityspace{\noisespace}^{\statespace\times\actionspace} \\
        \noise &\mapsto
        \left( (\state, \action) \mapsto \delta_{\noise}
        \right). 
    \end{align*}
    The entire random family $\left( \empnoisedist{\timestep}^{\state, \action} \right)_{(\state, \action)\in \statespace\times\actionspace} = \Phi(\noise_\timestep)$ is a $\probabilityspace{\noisespace}^{\statespace\times\actionspace}$-valued random variable and its law is
    \begin{equation*}
        \distoverdist{\timestep} =
        \pushforward{\Phi}{\noisedist{\timestep}} 
        \in \probabilityspace{\probabilityspace{\noisespace}^{\statespace \times \actionspace}}.
    \end{equation*}
    In words, $\noisedist{\timestep} \in \probabilityspace{\noisespace}$ is pushed forward through $\Phi$ to a measure on $\probabilityspace{\noisespace}^{\statespace \times \actionspace}$, concentrated on the diagonal $\{\Phi(\noise)\colon \noise \in \noisespace\}$.
We refer the reader to Remark~\ref{remark:well_posed_global} for a discussion on the well-posedness of $\Phi$.
\end{example}

\begin{remark}[Well-posedness of Example~\ref{example:global_noise}]
\label{remark:well_posed_global}
    We verify here that the map $\Phi \colon \noisespace \to \probabilityspace{\noisespace}^{\statespace\times\actionspace}$ is well-defined and a Borel measurable map between Polish spaces, so that the pushforward $\distoverdist{\timestep} = \pushforward{\Phi}{\noisedist{\timestep}}$ is well-defined. 
    
    \textbf{Step 1: Each coordinate map is continuous.} We start by noting that each coordinate map $\noise \mapsto \delta_\noise \colon \noisespace \to \probabilityspace{\noisespace}$ is continuous with respect to the weak topology on $\probabilityspace{\noisespace}$ \cite[Theorem~15.8]{aliprantis06infinite}.
    
    \textbf{Step 2: $\Phi$ is Borel measurable.} To show that the full map $\Phi$ is Borel measurable we consider two cases depending on whether $\statespace\times\actionspace$ is countable or uncountable.
    In the countable case, $\probabilityspace{\noisespace}^{\statespace\times\actionspace}$ is a countable product of Polish spaces, which is again a Polish space \cite[Proposition~3.3]{kechris1995classical} under the product topology. A map into a countable product is measurable if and only if each coordinate projection composed with it is measurable, which holds by Step 1. So $\Phi$ is Borel measurable. 
    If, instead, $\statespace \times \actionspace$ is uncountable, then $\probabilityspace{\noisespace}^{\statespace\times\actionspace}$ might fail to be Polish. However, this does not pose issues as the image of $\Phi$ is contained in the diagonal subset 
    \begin{equation*}
        \mathcal{D}=
        \{ \Phi(\noise) \colon \noise \in \noisespace \}=
        \{ ( (\state, \action) \mapsto \delta_\noise) \colon \noise \in \noisespace\} 
        \subset 
        \probabilityspace{\noisespace}^{\statespace \times \actionspace}.
    \end{equation*}
    The map $\Phi \colon \noisespace \to \mathcal D$ is a homeomorphism: both $\Phi$ and its inverse $\Phi^{-1} \colon \mathcal{D} \to \noisespace$ (which evaluates any coordinate and then inverts $\noise \mapsto \delta_\noise$) are continuous. 
    Hence, $\mathcal D$ is homeomorphic to $\noisespace$, which is Polish. In particular, $\mathcal D$ is a Polish space in the subspace topology inherited from $\probabilityspace{\noisespace}^{\statespace\times\actionspace}$. Since $\Phi \colon \noisespace\to \mathcal D$ is continuous between Polish spaces it is Borel measurable, and $\distoverdist{\timestep} = \pushforward{\Phi}{\noisedist{\timestep}}$ is a well-defined Borel probability measure on $\mathcal D$ and so also on $\probabilityspace{\noisespace}^{\statespace\times\actionspace}$.
    We finally observe that $\distoverdist{\timestep}$ is fully concentrated on $\mathcal{D}$, so all subsequent integrals only involve elements of $\mathcal D$ and the uncountability of $\statespace \times\actionspace$ does not pose an issue.
\end{remark}

We now show that these canonical noise models satisfy Assumption~\ref{assumption:barycenter_condition}.

\begin{proposition}[The noise models satisfy Assumption~\ref{assumption:barycenter_condition}]
\label{proposition:averagingproperty}
For each of the noise models in Examples~\ref{example:individual_noise}--\ref{example:global_noise}, the realized kernel $\empnoisedist{\timestep}^{\state, \action}$ and its distribution $\distoverdist{\timestep}$ satisfy Assumption~\ref{assumption:barycenter_condition}.
That is, there exists a Borel set $E_\timestep \subseteq \statespace\times\actionspace$ with $\probabilitySA{}{\timestep}(E_\timestep) = 1$ such that for all $(\state, \action)\in E_\timestep$ and for every bounded Borel measurable $\varphi: \noisespace \to \reals$ simultaneously,
\begin{equation*}
\expectedvalue{\distoverdist{\timestep}^{\state, \action}}{
\int_\noisespace \varphi(\noise) 
\dd\empnoisedist{\timestep}^{\state, \action}(\noise)
}
=
\int_\noisespace \varphi(\noise)
\dd\noisedist{\timestep}(\noise \st \state, \action).
\end{equation*}
\end{proposition}

\begin{proof}
We prove the statement for each noise model separately.

\textit{Individual noise (Example~\ref{example:individual_noise}).} Fix $(\state, \action) \in \supp{\probabilitySA{}{\timestep}}$, hence $\nagents_{\state, \action} \ge 1$. By construction,
\begin{equation*}
    \empnoisedist{\timestep}^{\state,\action}=\frac{1}{\nagents_{\state, \action}}
    \sum_{i = 1}^{\nagents_{\state, \action}}
    \delta_{\noise_\timestep^{\state,\action,i}}
\end{equation*}
where $\noise_\timestep^{\state,\action,i}$ are i.i.d. with law $\noisedist{\timestep}(\cdot \st \state, \action)$. Hence, for any bounded measurable $\varphi \colon \noisespace \to \reals$,

\begin{equation*}
\int_\noisespace \varphi(\noise) 
\dd\empnoisedist{\timestep}^{\state, \action}(\noise)
=
\frac{1}{\nagents_{\state, \action}}
\sum_{i = 1}^{\nagents_{\state, \action}} 
\varphi(\noise_\timestep^{\state,\action,i})
\end{equation*}
and therefore, by linearity of expectation and since each
$\noise_\timestep^{\state,\action,i}$ has law $\noisedist{\timestep}(\cdot \st \state, \action)$,
\begin{equation*}
\expectedvalue{\distoverdist{\timestep}^{\state, \action}}{
\int_\noisespace \varphi(\noise) 
\dd\empnoisedist{\timestep}^{\state, \action}(\noise)
}
=
\frac{1}{\nagents_{\state, \action}}
\sum_{i = 1}^{\nagents_{\state, \action}} 
\expectedvalue{\noisedist{\timestep}(\cdot \st \state, \action)}{\varphi(\noise)}
=
\int_\noisespace \varphi(\noise)
\dd\noisedist{\timestep}(\noise \st \state, \action).
\end{equation*}
Thus \eqref{eq:test_functions} holds on $E_\timestep = \supp{\probabilitySA{}{\timestep}}$, which clearly satisfies $\probabilitySA{}{\timestep}(E_\timestep)=1$.

\textit{Infinite-population model (Example~\ref{example:indiv_noise_inf_agents}).} 
By construction $\empnoisedist{\timestep}^{\state,\action}=\noisedist{\timestep}(\cdot \st \state, \action)$ almost surely for every $(\state, \action) \in \statespace\times \actionspace$. Therefore, for every bounded measurable $\varphi$
\begin{equation*}
    \expectedvalue{\distoverdist{\timestep}^{\state, \action}}{\int_\noisespace
    \varphi(\noise) 
    \dd\empnoisedist{\timestep}^{\state, \action}} =
    \int_\noisespace \varphi(\noise)
    \dd\noisedist{\timestep}(\noise \st \state, \action).
\end{equation*}
Thus, \eqref{eq:test_functions} holds with $E_\timestep=\statespace\times\actionspace$.

\textit{Local noise (Example~\ref{example:local_noise}).}
For each fixed $(\state, \action)$, a single draw $\noise_\timestep^{\state,\action} \sim \noisedist{\timestep}(\cdot \st \state, \action)$ is taken, independently across state-action pairs, and the realized noise distribution is
$\empnoisedist{\timestep}^{\state,\action}=\delta_{\noise_\timestep^{\state,\action}}$.
So, for every bounded measurable $\varphi$, the inner integral equals $\varphi(\noise_\timestep^{\state,\action})$ and, since $\noise_\timestep^{\state,\action}\sim\noisedist{\timestep}(\cdot \st \state, \action)$,
\begin{equation*}
\expectedvalue{\distoverdist{\timestep}^{\state, \action}}{\int_\noisespace \varphi(\noise)
\dd\empnoisedist{\timestep}^{\state,\action}(\noise)}
=
\expectedvalue{\noisedist{\timestep}(\cdot \st \state, \action)}{\varphi(\noise)}
=
\int_\noisespace \varphi(\noise)
\dd\noisedist{\timestep}(\noise \st \state, \action).
\end{equation*}
Thus, \eqref{eq:test_functions} holds again with $E_\timestep=\statespace\times\actionspace$.

\textit{Global-noise (Example~\ref{example:global_noise}).} This case is identical to the local noise one, with the same draw $\noise_\timestep$ shared across all $(\state,\action)$.
\end{proof}

\subsection{Averaging property of realized noise kernels}
\label{app:averaging-realized-noise-kernels} 

We now show that even if $\empnoisedist{\timestep}^{\state, \action}$ is random, the expected cost computed under its law $\distoverdist{\timestep}$ equals the cost computed directly under the nominal kernel $\noisedist{\timestep}(\cdot \st \state, \action)$. In other words, replacing the realized kernel $\empnoisedist{\timestep}^{\state, \action}$ by the nominal kernel $\noisedist{\timestep}(\cdot \st \state, \action)$ is unbiased in expectation. In particular, this lemma will allow us to remove randomness from the backward recursion in the proof of \sepbound{}.

\begin{lemma}[Averaging realized kernels]
\label{lemma:noiseexp}
Let $\statespace$, $\actionspace$, $\noisespace$ be Polish spaces and fix a time step $\timestep$. Let $\probabilitySA{}{\timestep}$ denote the joint state-action distribution and $\noisedist{\timestep} : \statespace \times \actionspace \to \probabilityspace{\noisespace}$ be a Borel stochastic kernel. For each $(\state, \action)$, let $\empnoisedist{\timestep}^{\state, \action}$ be a $\probabilityspace{\noisespace}$-valued random variable with distribution $\distoverdist{\timestep}^{\state, \action} \in \probabilityspace{\probabilityspace{\noisespace}}$, such that Assumption~\ref{assumption:barycenter_condition} holds and $(\state, \action) \mapsto \distoverdist{\timestep}^{\state, \action}$ is a Borel stochastic kernel.
%
Then, for every nonnegative lower semianalytic $c : \statespace \times \actionspace \times \noisespace \to \nonnegativeRealsBar$ (in particular, for every nonnegative Borel measurable $c$),
\begin{equation}
\label{eq:noiseexp}
\int_{\statespace \times \actionspace}
\expectedvalue{\distoverdist{\timestep}^{\state, \action}}
{\int_\noisespace c(\state,\action,\noise) 
\dd\empnoisedist{\timestep}^{\state, \action}(\noise)} 
\dd\probabilitySA{}{\timestep}
(\state, \action)
=
\int_{\statespace \times \actionspace}
\int_\noisespace
c(\state,\action,\noise)
\dd\noisedist{\timestep}(\noise \st \state,\action)
\dd\probabilitySA{}{\timestep}
(\state, \action).
\end{equation}
\end{lemma}

\begin{proof}
For $n\in\N$, define
\begin{equation*}
c_n(\state,\action,\noise) \coloneqq \min\{c(\state,\action,\noise),n\}.
\end{equation*}
By construction, each $c_n$ is bounded and lower semianalytic, and $c_n\uparrow c$ pointwise.

To start, we prove that the two integrands in \eqref{eq:noiseexp} are (universally) measurable.
For the left-hand side, for each $n$, the map $(\state, \action, \zeta) \mapsto \int_\noisespace c_n(\state, \action, \noise) \dd\zeta(\noise)$ is lower semianalytic on $\statespace\times\actionspace\times \probabilityspace{\noisespace}$: the evaluation map $q(\dd\noise \st \state, \action, \zeta) \coloneqq \zeta(\dd\noise)$ is a Borel stochastic kernel (since $\zeta \mapsto \zeta(B)$ is Borel measurable for every Borel set $B \subseteq \noisespace$ \cite[Proposition~7.28]{bertsekas1996stochastic}), and integration of a lower semianalytic function against a Borel stochastic kernel preserves lower semianalyticity \cite[Proposition~7.48]{bertsekas1996stochastic}.
Since $(\state, \action) \mapsto \distoverdist{\timestep}^{\state, \action}$ is a Borel stochastic kernel, a second application of \cite[Proposition~7.48]{bertsekas1996stochastic} yields lower semianalyticity, hence universal measurability, of
\begin{equation*}
(\state,\action)\mapsto\expectedvalue{\distoverdist{\timestep}^{\state, \action}}{\int_\noisespace c_n(\state,\action,\noise)\dd\empnoisedist{\timestep}^{\state, \action}(\noise)}.
\end{equation*}
Letting $n \uparrow\infty$, and using that a pointwise increasing limit of universally measurable functions is universally measurable, we obtain the measurability of
\begin{equation*}
    (\state, \action) \mapsto
    \expectedvalue{\distoverdist{\timestep}^{\state, \action}}
    {\int_\noisespace c(\state,\action,\noise)
    \dd\empnoisedist{\timestep}^{\state, \action}(\noise)}.
\end{equation*}
For the right-hand side, instead, the measurability of
\begin{equation*}
    (\state, \action) \mapsto
    \int_\noisespace
c(\state,\action,\noise)
\dd\noisedist{\timestep}(\noise \st \state,\action),
\end{equation*}
follows from the same argument, applied with the Borel stochastic kernel $\noisedist{\timestep}$ \cite[Proposition~7.48]{bertsekas1996stochastic}.

We now prove the equality~\eqref{eq:noiseexp}.
Fix $n$ and $(\state,\action) \in E_\timestep$, and consider the bounded test function $\varphi_{\state,\action}(\noise) \coloneqq c_n(\state,\action,\noise)$, viewed as a function of $\noise$ alone for fixed $(\state,\action)$.
Assumption~\ref{assumption:barycenter_condition} is stated for bounded \emph{Borel} test functions, while $\varphi_{\state,\action}$ is only lower semianalytic, hence universally measurable; we extend \eqref{eq:test_functions} to $\varphi_{\state,\action}$ as follows. Let $m \coloneqq \noisedist{\timestep}(\cdot \st \state, \action)$. Since $\varphi_{\state,\action}$ is universally measurable, it is measurable with respect to the $m$-completion of the Borel $\sigma$-algebra, so there exist a bounded Borel function $\varphi'$ and a Borel set $\mathcal N \subseteq \noisespace$ with $m(\mathcal N) = 0$ and $\{\varphi_{\state,\action} \neq \varphi'\} \subseteq \mathcal N$. Applying \eqref{eq:test_functions} to the bounded Borel function $\boldsymbol 1_{\mathcal N}$ gives $\expectedvalue{\distoverdist{\timestep}^{\state,\action}}{\empnoisedist{\timestep}^{\state,\action}(\mathcal N)} = m(\mathcal N) = 0$, and since $\empnoisedist{\timestep}^{\state,\action}(\mathcal N) \geq 0$, we conclude $\empnoisedist{\timestep}^{\state,\action}(\mathcal N) = 0$ almost surely. Consequently, $\int \varphi_{\state,\action} \dd\empnoisedist{\timestep}^{\state,\action} = \int \varphi' \dd\empnoisedist{\timestep}^{\state,\action}$ almost surely and $\int \varphi_{\state,\action} \dd m = \int \varphi' \dd m$; applying \eqref{eq:test_functions} to $\varphi'$ then yields \eqref{eq:test_functions} for $\varphi_{\state,\action}$ itself. This gives
\begin{equation*}
\expectedvalue{\distoverdist{\timestep}^{\state, \action}}{\int_\noisespace
c_n(\state,\action,\noise) 
\dd\empnoisedist{\timestep}^{\state, \action}(\noise)}
=
\int_\noisespace c_n(\state,\action,\noise)
\dd\noisedist{\timestep}(\noise \st \state, \action)
\end{equation*}
for all $(\state, \action) \in E_\timestep$. Since $\probabilitySA{}{\timestep}(E_\timestep) = 1$, integrating both sides with respect to $\probabilitySA{}{\timestep}$ gives
\begin{align*}
\int_{\statespace\times \actionspace}
\expectedvalue{\distoverdist{\timestep}^{\state, \action}}{
\int_\noisespace c_n(\state,\action,\noise)
\dd\empnoisedist{\timestep}^{\state, \action}(\noise)}
\dd\probabilitySA{}{\timestep}(\state, \action)
&=
\int_{\statespace\times \actionspace}
\int_\noisespace c_n(\state,\action,\noise)
\dd\noisedist{\timestep}(\noise \st \state, \action)
\dd\probabilitySA{}{\timestep}(\state, \action).
\end{align*}
Finally, let $n\to\infty$. Since $c_n\uparrow c$, the monotone convergence theorem yields \eqref{eq:noiseexp}.
\end{proof}

\section{Preliminary lemmata}
\label{appendix:preliminary_lemmata}

We now collect the auxiliary results required for the main proof in Appendix~\ref{section:main_proof}.

\noindent\textbf{Measurability convention.}
Throughout the appendices, primitive data such as the dynamics, stage costs, and terminal cost
are assumed to be Borel measurable, as in the main text. Since every nonnegative Borel
measurable function is lower semianalytic, these data also satisfy the analytic-measurability
assumptions used in the auxiliary results below. We use lower semianalytic functions because
the dynamic programming recursion involves pointwise infima over the action space, and such
infima need not be Borel measurable on general Polish spaces; see~\cite{bertsekas1996stochastic} for a detailed discussion. 
Accordingly, value functions and pointwise Bellman infima are interpreted as lower semianalytic, hence universally measurable, functions. This is sufficient for the integrals and measurable-selection arguments used in the proof
of Theorem~\ref{theorem:optimality-approximation}. For this reason, the new auxiliary lemmata below (Lemmas~\ref{lemma:infexp}, \ref{lemma:global_to_pointwise}, and~\ref{lemma:marginalinf}) state their regularity requirements explicitly in terms of lower semianalytic costs: this is the regularity actually available when they are invoked (the costs there involve the value functions $\costtogosmall{\timestep}$). The results recalled from \cite{terpin2024dynamic} are stated and proved there under the standing assumption that all spaces are Polish and all maps are Borel; their proofs (change-of-variables, gluing, and disintegration arguments) apply verbatim when the transportation cost is only lower semianalytic, hence universally measurable (all integrals remain well-defined by nonnegativity), and we use them in this form.
We work at this level of generality to simultaneously capture systems with finite and continuous state and action spaces. Nevertheless, when all spaces are finite, measurability is automatic and all measurability arguments can therefore be skipped. 

We begin by recalling from~\cite{terpin2024dynamic} a result that lets us ``move'' the pushforward from the set of couplings into the transportation cost, and vice versa.

\begin{lemma}[Pushforward and optimal transport \protect{\cite[Lemma 6.1]{terpin2024dynamic}}]
\label{lemma:pushandOT}
Given a transportation cost $\costtogosmall{\nexttimestep}: \statespace \times \refSpace \to \nonnegativeRealsBar$, a map $\dynamics{}{}{\timestep}\colon \statespace \times \actionspace \times \noisespace \to \statespace$, and a probability measure $\probabilitySAW{}{\timestep} \in \probabilityspace{\statespace \times \actionspace \times \noisespace}$, it holds that

\begin{multline*}\label{equation:pushforward:stability:1}
\inf_{\pushforward
{(\projectionFromTo{\statespace \times \actionspace \times \noisespace \times \refSpace}{\statespace \times \actionspace \times \noisespace})}
{\plan{\gamma}} 
=
\probabilitySAW{}{\timestep}}
\int_{\statespace \times \actionspace \times \noisespace \times \refSpace}
\costtogosmall{\nexttimestep} \circ (\dynamics{}{}{\timestep}\times \Id{\refSpace})
\dd\plan{\gamma}
=
\inf_{\pushforward
{(\projectionFromTo{\statespace \times \refSpace}{\statespace})}
{\plan{\tilde{\gamma}}} 
=
\pushforward{\dynamics{}{}{\timestep}}{\probabilitySAW{}{\timestep}}}
\int_{\statespace \times \refSpace}
\costtogosmall{\nexttimestep}
\dd\plan{\tilde{\gamma}},
\end{multline*}
where $\Id{\refSpace}$ is the identity map over $\refSpace$.
\end{lemma}

In our setting, this lemma tells us that evaluating the cost after applying the system dynamics is equivalent to pushing forward the distribution first: we can ``apply the dynamics first'' and then evaluate the population-level cost without changing the result.

Next, we recall how to merge additive costs into a single transport objective.

\begin{lemma}[Sum of optimal transport discrepancies~\protect{\cite[Lemma 6.2]{terpin2024dynamic}}]
\label{lemma:sumOTdiscr}
Given transportation costs $c_1\colon \X{} \to \nonnegativeRealsBar$, $c_2\colon \X{} \times \refSpace \to \nonnegativeRealsBar$, and probability measures $\probabilitySAW{}{} \in \probabilityspace{\X{}}$, $\refPVar \in \probabilityspace{\refSpace}$, it holds that


\begin{equation*}
\expectedvalue{\probabilitySAW{}{}}{c_1}
+ 
\kantorovich{c_2}{\probabilitySAW{}{}}{\refPVar}{}
=
\kantorovich{c}{\probabilitySAW{}{}}{\refPVar}{},
\end{equation*}
with $c: \X{} \times \refSpace \to \nonnegativeRealsBar$ defined as
$
c(x, \refVar) 
=
c_1(x) 
+ 
c_2(x, \refVar).
$
\end{lemma}

This result allows us to collapse stage costs and terminal transport costs into a unified expression, which is crucial for dynamic programming formulations.

We now introduce stochasticity which, in our setting, will appear due to the stochastic dynamics.
Through a standard argument, we show that we can safely exchange expectation over random noise with the infimum over couplings, at the price of an inequality. 

\begin{lemma}[Infimum and expectation]
\label{lemma:infexp}
Let $\X{}$, $\noisespace$, $\refSpace$ be Polish spaces, $\probabilitySA{}{} \in \probabilityspace{\X{}}$, and $\refPVar \in \probabilityspace{\refSpace}$. Let $\empnoisedist{} = (\empnoisedist{}^{x})_{x \in \X{}}$ be a random family of probability measures $\empnoisedist{}^{x} \in \probabilityspace{\noisespace}$, defined on a common probability space and such that, almost surely, $x \mapsto \empnoisedist{}^{x}$ is a Borel measurable kernel on a set of full $\probabilitySA{}{}$-measure (cf.\ Assumption~\ref{assumption:measurable_field}). Denote its distribution by $\distoverdist{}$.
Let $c : \X{} \times \noisespace \times \refSpace \to \nonnegativeRealsBar$ be measurable. Whenever the integrand of the left-hand side below fails to be measurable in $\empnoisedist{}$, the outer expectation is understood; in our applications, it is universally measurable (see the proof of Theorem~\ref{theorem:optimality-approximation}).
Then
\begin{multline*}
  \expectedvalue{\distoverdist{}}
  {\inf_{\plan{\gamma} \in \plans{\probabilitySA{}{}}{\refPVar}}
    \int_{\X{} \times \refSpace{}}
    \int_\noisespace
    c(x,\noise,\refVar) 
    \dd\empnoisedist{}^x(\noise)
    \dd\plan{\gamma}(x,\refVar)}\\
  \le
  \inf_{\plan{\gamma} \in \plans{\probabilitySA{}{}}{\refPVar}}
  \expectedvalue{\distoverdist{}}
  {\int_{\X{} \times \refSpace{}} 
  \int_\noisespace
    c(x,\noise,\refVar) 
    \dd\empnoisedist{}^x(\noise)
  \dd\plan{\gamma}(x,\refVar)}.
\end{multline*}
\end{lemma}

\begin{proof}
Fix any $\plan{\bar\gamma} \in \plans{\probabilitySA{}{}}{\refPVar}$. Pointwise in $\empnoisedist{}$,
\begin{equation*}
  \inf_{\plan{\gamma} \in \plans{\probabilitySA{}{}}{\refPVar}}
    \int_{\X{} \times \refSpace{}}
    \int_\noisespace
    c(x,\noise,\refVar)
    \dd\empnoisedist{}^x(\noise)
    \dd\plan{\gamma}(x,\refVar)
  \le
  \int_{\X{} \times \refSpace{}}
    \int_\noisespace
    c(x,\noise,\refVar)
    \dd\empnoisedist{}^x(\noise)
    \dd\plan{\bar\gamma}(x,\refVar).
\end{equation*}
Taking expectations over $\empnoisedist{}$ preserves the inequality (also when the left-hand side is interpreted as an outer expectation, by its monotonicity). Since the resulting inequality holds for every coupling $\plan{\bar\gamma}$, taking the infimum over $\plan{\bar\gamma} \in \plans{\probabilitySA{}{}}{\refPVar}$ yields the claim.
\end{proof}

With this inequality, randomness can be ``pushed inside'' the transport problem, a step that will later allow us to handle noisy transitions.

The next tool, recalled from~\cite{terpin2024dynamic}, refines our handling of couplings: it shows how an optimizer can be disintegrated with respect to one marginal, yielding conditional distributions.

\begin{lemma}[Disintegration of the optimizer \protect{\cite[Lemma 6.4]{terpin2024dynamic}}]
\label{lemma:disintegration}
Given a transportation cost $c: \statespace{}\times \actionspace \times \refSpace \to \nonnegativeRealsBar$ and probability measures $\mu \in \probabilityspace{\statespace}$, $\nu \in \probabilityspace{\refSpace}$, it holds that
\begin{multline}\label{eq:lemma:disintegration}
\inf_{\pushforward{\left(\projectionFromTo{\statespace \times \actionspace \times \refSpace}{\statespace \times \refSpace}\right)}{\plan{\gamma}} \in \plans{\mu}{\nu}}
\int_{\statespace\times \actionspace\times \refSpace}
c(\state,\action,\refVar)
\dd\plan{\gamma}(\state,\action,\refVar)
\\=
\inf_{\plan{\gamma'} \in \plans{\mu}{\nu}} 
\inf_{
\{\alpha^{\state, \refVar}\} \in \Lambda(\actionspace)
}
\int_{\statespace\times \refSpace}
\int_{\actionspace}
c(\state, \action, \refVar)
\dd\alpha^{\state, \refVar}(\action)
\dd\plan{\gamma'}(\state,\refVar),
\end{multline}
where $
\Lambda(\actionspace) 
\coloneqq
\{
\{\alpha^{\state, \refVar}\}_{(\state,\refVar) \in \statespace \times \refSpace}
\subseteq \probabilityspace{\actionspace} 
\st
\statespace\times \refSpace \ni (\state, \refVar) \mapsto \alpha^{\state, \refVar} \in \probabilityspace{\actionspace} \text{ Borel}
\}.
$
\end{lemma}

This separation will allow us to express the population-level policy as a distribution over conditional agent-level actions.

Next, we study more in detail the right-hand side in~\eqref{eq:lemma:disintegration}, showing that the infimum over the family $\{\alpha^{\state, \refVar}\} \in \Lambda(\actionspace)$ can be ``pushed'' into the integral.

\begin{lemma}[Reduction to pointwise infimum over deterministic policies]
\label{lemma:global_to_pointwise}
Let $\statespace$, $\actionspace$, $\refSpace$ be Polish spaces and let $c \colon \statespace\times \actionspace\times\refSpace \to \nonnegativeRealsBar$ be lower semianalytic.
Let $\fleet \in \probabilityspace{\statespace}$ and $\refPVar \in \probabilityspace{\refSpace}$.
Define
\begin{equation*}
    \Lambda(\actionspace) \coloneqq \bigl\{\{\alpha^{\state,\refVar}\}_{(\state,\refVar) \in \statespace \times \refSpace} \subseteq
\probabilityspace{\actionspace} \;\big|\; \statespace \times \refSpace \ni (\state,\refVar) \mapsto \alpha^{\state,\refVar} \in
\probabilityspace{\actionspace} \text{ is Borel measurable}\bigr\}.
\end{equation*} 
Then, the function $(\state, \refVar) \mapsto \inf_{\action \in \actionspace}c(\state, \action, \refVar)$ is universally measurable, and for every coupling $\plan{\gamma} \in \plans{\fleet}{\refPVar}$,
\begin{equation}
\label{eq:global_to_pointwise}
    \inf_{\{\alpha^{\state, \refVar}\} \in \Lambda(\actionspace)}
    \int_{\statespace\times\refSpace}
    \int_\actionspace
    c(\state, \action, \refVar)
    \dd\alpha^{\state, \refVar}(\action)
    \dd\plan{\gamma}(\state, \refVar)
    =
    \int_{\statespace\times\refSpace}
    \inf_{\action\in \actionspace}
    c(\state, \action, \refVar)
    \dd\plan{{\gamma}}(\state, \refVar).
\end{equation}
\end{lemma}

\begin{proof}
We start by assessing the measurability of the pointwise infimum. Since $c$ is lower semianalytic and $\actionspace$ is a Polish space, the function $(\state, \refVar) \mapsto \inf_{\action \in \actionspace}c(\state, \action, \refVar)$ is lower semianalytic by \cite[Proposition~7.47]{bertsekas1996stochastic}, hence universally measurable. In particular, the right-hand side of \eqref{eq:global_to_pointwise} is a well-defined element of $[0, +\infty]$.

We then prove \eqref{eq:global_to_pointwise} by showing ``$\le$'' and ``$\ge$'' separately.

\noindent\textbf{Lower bound ``$\ge$''.}
For any $\{\alpha^{\state,\refVar}\} \in \Lambda(\actionspace)$ and every $(\state, \refVar) \in \statespace\times\refSpace$, since $\alpha^{\state,\refVar} \in \probabilityspace{\actionspace}$ is a probability measure 
\begin{equation*}
    \int_\actionspace c(\state,\action,\refVar) \dd\alpha^{\state,\refVar}(\action) \geq
    \inf_{\action \in \actionspace}
    c(\state,\action,\refVar).
\end{equation*}
Integrating both sides with respect to $\plan{\gamma}$ and taking the infimum over all $\{\alpha^{\state,\refVar}\} \in \Lambda(\actionspace)$ leads to
\begin{equation*}
    \inf_{\{\alpha^{\state,\refVar}\} \in \Lambda(\actionspace)}
    \int_{\statespace \times \refSpace} 
    \int_\actionspace 
    c(\state, \action, \refVar) 
    \dd\alpha^{\state,\refVar}(\action)
    \dd\plan{\gamma}(\state,\refVar)
    \geq
    \int_{\statespace \times \refSpace}
    \inf_{\action \in \actionspace}
    c(\state,\action,\refVar) 
    \dd\plan{\gamma}(\state,\refVar).
\end{equation*}


\noindent\textbf{Upper bound ``$\le$''.} Let $\varepsilon > 0$. Since $c \ge 0$ is lower semianalytic and $\actionspace$ is a Polish space, \cite[Proposition~7.50]{bertsekas1996stochastic} guarantees the existence of a universally measurable $\varepsilon$-optimal selector $\action^\varepsilon \colon \statespace \times \refSpace \to \actionspace$ satisfying
\begin{equation*}
    c(\state, \action^\varepsilon(\state, \refVar), \refVar)
    \le 
    \inf_{\action\in \actionspace}
    c(\state, \action, \refVar) + \varepsilon
    \quad\forall (\state, \refVar) \in \statespace\times \refSpace.
\end{equation*}
Now, since $\plan{{\gamma}}$ is a $\sigma$-finite probability measure on $\statespace\times \refSpace$, by \cite[Lemma~7.27]{bertsekas1996stochastic} there exists a Borel measurable map $b^\varepsilon \colon \statespace\times \refSpace \to \actionspace$ with $b^\varepsilon = \action^\varepsilon$ $\plan{{\gamma}}$-almost everywhere, so that
\begin{equation*}
    c(\state, b^\varepsilon(\state, \refVar), \refVar)
    \le 
    \inf_{\action\in \actionspace}
    c(\state, \action, \refVar) + \varepsilon
    \quad \plan{{\gamma}}\text{-almost everywhere}.
\end{equation*}
The Dirac embedding $\actionspace\to \probabilityspace{\actionspace}$ defined by $\action \mapsto \delta_\action$ is continuous in the weak topology \cite[Theorem~15.8]{aliprantis06infinite}, hence Borel measurable.
Therefore, the composition $(\state, \refVar) \mapsto \delta_{b^\varepsilon(\state, \refVar)}$ is Borel measurable, and so is the family $\{\delta_{b^\varepsilon(\state, \refVar)}\}_{(\state, \refVar)} \in \Lambda(\actionspace)$.
Thus, 
\begin{align*}
    \inf_{\{\alpha^{\state, \refVar} \} \in \Lambda(\actionspace)}
    \int_{\statespace\times \refSpace}
    \int_\actionspace
    c(\state, \action, \refVar)
    \dd\alpha^{\state, \refVar}(\action)
    \dd\plan{{\gamma}}(\state, \refVar)
    &\le
    \int_{\statespace\times \refSpace}
    c(\state, b^\varepsilon(\state, \refVar), \refVar)
    \dd\plan{\gamma}(\state, \refVar)\\
    &\le
    \int_{\statespace\times \refSpace}
    \inf_{\action\in \actionspace}
    c(\state, \action, \refVar)
    \dd\plan{\gamma}(\state, \refVar)
    + \varepsilon.
\end{align*}
Since $\varepsilon > 0$ is arbitrary, we conclude 
\begin{equation*}
    \inf_{\{\alpha^{\state, \refVar} \} \in \Lambda(\actionspace)}
    \int_{\statespace\times \refSpace}
    \int_\actionspace
    c(\state, \action, \refVar)
    \dd\alpha^{\state, \refVar}(\action)
    \dd\plan{\gamma}(\state, \refVar)
    \le
    \int_{\statespace\times \refSpace}
    \inf_{\action\in \actionspace}
    c(\state, \action, \refVar)
    \dd\plan{\gamma}(\state, \refVar).
\end{equation*}
This completes the proof of the upper bound and, with it, the proof of the lemma. 
\end{proof}

This lemma will allow us to relate the dynamic programming over probability measures with the standard dynamic programming over the state and action spaces. 

Finally, we present a lemma that upper bounds the transport cost when replacing ``full couplings'' with ``marginal plans''. This inequality will be used to simplify the optimization by relaxing the coupling constraints, enabling us to separate noise from the population-level allocation.

\begin{lemma}[Suboptimality of marginal plans]
\label{lemma:marginalinf}
Let $\X{}$, $\noisespace$, $\refSpace$ be Polish spaces. Given a lower semianalytic transportation cost
$c: \X{}\times \noisespace \times \refSpace \to \nonnegativeRealsBar$ and probability measures
$\probabilitySAW{}{} \in \probabilityspace{\X{} \times \noisespace}$,
$\refPVar \in \probabilityspace{\refSpace}$, let $\probabilitySAW{}{} = \probabilitySA{}{}\otimes q^x$
be the disintegration of $\probabilitySAW{}{}$ with respect to its first marginal
$\probabilitySA{}{} \coloneqq \pushforward{(\projectionFromTo{\X{}\times\noisespace}{\X{}})}{\probabilitySAW{}{}}$,
where $q^x \in \probabilityspace{\noisespace}$ is a (deterministic) Borel kernel
(we use the symbol $q$ rather than $\empnoisedist{}$ to stress that no randomness is involved here).
Then,
\begin{equation}\label{eq:lemma:marginalinf}
\inf_{\plan{\gamma}\in \plans{\probabilitySAW{}{}}{\refPVar}}
\int_{\X{}\times \noisespace \times \refSpace}
c(x,\noise,\refVar)
\dd\plan{\gamma}(x,\noise,\refVar)
\le
\inf_{\plan{\gamma_1} \in \plans{\probabilitySA{}{}}{\refPVar}} 
\int_{\X{}\times \refSpace}
\int_{\noisespace}
c(x, \noise, \refVar)
\dd q^{x}(\noise)
\dd\plan{\gamma_1}(x,\refVar).
\end{equation}
\end{lemma}
\begin{proof}
    Consider the disintegration of $\probabilitySAW{}{}$ with respect to its first marginal $\probabilitySA{}{} = \pushforward{(\projectionFromTo{\X{} \times \noisespace}{\X{}})}{\probabilitySAW{}{}} \in \probabilityspace{\X{}}$,  
    $\probabilitySAW{}{} = \probabilitySA{}{} \otimes q^x$.
    Let $\plan{\gamma_1} \in \plans{\probabilitySA{}{}}{\refPVar}$ and consider the gluing
    $
    \plan{\bar{\gamma}} = \plan{\gamma_1 \otimes q^{x}} \in \probabilityspace{\X{} \times \refSpace \times \noisespace}.
    $ 
    To ``reorder'' marginals, consider the permutation 
    \begin{equation*}
    \sigma: \X{} \times \refSpace \times \noisespace \to \X{} \times \noisespace \times \refSpace, 
    \quad
    \sigma(x, \refVar, \noise) = (x, \noise, \refVar),
    \end{equation*}
    and set $
    \plan{\gamma} = 
    \pushforward{\sigma}{\plan{(\gamma_1 \otimes q^{x})}}
    \in \probabilityspace{\X{} \times \noisespace \times \refSpace}.
    $

    We now show that $\plan{\gamma}$ is a coupling between $\probabilitySAW{}{}$ and $\refPVar$ and, thus, a valid upper bound for the infimum in~\eqref{eq:lemma:marginalinf}.
    We check each marginal by testing against an arbitrary bounded measurable function.
    Let $\varphi\colon \X{} \times \noisespace \to \reals$ be bounded and measurable, we have
    \begin{align*}
    \int_{\X{} \times \noisespace}
    \varphi(x,\noise)&
    \dd\left(
    \pushforward
      {\left(\projectionFromTo{\X{} \times \noisespace \times \refSpace}{\X{} \times \noisespace}\right)}
      \gamma
      \right)
      (x,\noise)\\
    &=
    \int_{\X{} \times \noisespace \times \refSpace}
    \varphi\left(\projectionFromTo{\X{} \times \noisespace \times \refSpace}{\X{} \times \noisespace}(x,\noise,\refVar)\right)
    \dd\gamma(x,\noise,\refVar)\\
    &=
    \int_{\X{}\times \noisespace \times \refSpace}
    \varphi\bigl(\projectionFromTo{\X{} \times \noisespace \times \refSpace}{\X{} \times \noisespace}
    (x,\noise,\refVar)\bigr)
    \dd\left(\pushforward{\sigma}{(\gamma_{1}\otimes q^{x})}\right)
    (x,\noise,\refVar)\\
    &=
    \int_{\X{} \times \refSpace \times \noisespace}
    \varphi\left(\projectionFromTo{\X{} \times \noisespace \times \refSpace}{\X{} \times \noisespace}(\sigma(x,\refVar,\noise))\right)
    \dd(\gamma_1\otimes q^{x})(x,\refVar,\noise)\\
    &=
    \int_{\X{}\times \refSpace}\int_{\noisespace}
    \varphi(x,\noise)
    \dd q^{x}(\noise)
    \dd\gamma_1(x,\refVar)\\
    &=
    \int_{\X{}}\int_{\noisespace}
    \varphi(x,\noise)
    \dd q^{x}(\noise)
    \dd\probabilitySA{}{}(x).
    \end{align*}
    For every bounded and measurable $\varphi$, since $q^{x}$ is the disintegration of $\probabilitySAW{}{}$ on $\noisespace$ given $x$, and $\gamma_1 \in \plans{\probabilitySA{}{}}{\refPVar}$ has first marginal $\probabilitySA{}{}$, Fubini--Tonelli \cite[Theorem 18.3]{billingsley1980} gives
    \begin{multline*}
    \int_{\X{}}\int_{\noisespace}
    \varphi(x,\noise)
    \dd q^{x}(\noise)
    \dd\probabilitySA{}{}(x)
    =
    \int_{\X{}\times \noisespace}
    \varphi(x,\noise)
    \dd(\probabilitySA{}{}\otimes q^{x})(x,\noise)
    =
    \int_{\X{}\times \noisespace}
    \varphi(x, \noise)
    \dd\probabilitySAW{}{}(x, \noise).
    \end{multline*}
    Hence $\pushforward
            {(\projectionFromTo{\X{} \times \noisespace \times \refSpace}
            {\X{} \times \noisespace})}
            \gamma
            =
            \probabilitySAW{}{}$.
    Similarly, for the marginal over $\refSpace$, let $\psi\colon \refSpace \to \reals$ be bounded and measurable. Then, 
    \begin{align*}
    \int_{\refSpace}
    \psi(\refVar)
    &\dd\left(\pushforward{\left(\projectionFromTo{\X{} \times \noisespace \times \refSpace}{\refSpace}\right)}{\gamma}\right)(\refVar)
    \\
    &=
    \int_{\X{} \times \noisespace \times \refSpace}
    \psi\left(\projectionFromTo{\X{} \times \noisespace \times \refSpace}{\refSpace}(x,\noise,\refVar)\right)
    \dd\gamma(x,\noise,\refVar)\\
    &=
    \int_{\X{}\times \noisespace\times \refSpace}
    \psi\left(\projectionFromTo{\X{} \times \noisespace \times \refSpace}{\refSpace}(x,\noise,\refVar)\right)
    \dd\left(\pushforward{\sigma}{(\gamma_1\otimes q^{x})}\right)
    (x,\noise,\refVar)\\
    &=
    \int_{\X{} \times \refSpace \times \noisespace}
    \psi\left(\projectionFromTo{\X{} \times \noisespace \times \refSpace}{\refSpace}(\sigma(x,\refVar,\noise))\right)
    \dd(\gamma_1\otimes q^{x})(x,\refVar,\noise)\\
    &=
    \int_{\X{}\times\refSpace}\int_{\noisespace}
    \psi(\refVar)
    \dd q^{x}(\noise)
    \dd\gamma_1(x,\refVar).
    \end{align*}
    Since $\int_{\noisespace}\dd q^{x}(\noise)=1$ for all $x$ and $\refPVar$ is the marginal over $\refSpace$ of $\gamma_1 \in \plans{\probabilitySA{}{}}{\refPVar}$, this reduces to
    \begin{equation*}
    \int_{\X{}\times\refSpace}
    \psi(\refVar)
    \dd\gamma_1(x,\refVar)
    =
    \int_\refSpace\psi(\refVar)\dd\refPVar(\refVar).
    \end{equation*}
    Combining the two chains of equalities, $\int_\refSpace \psi \dd\bigl(\pushforward{(\projectionFromTo{\X{} \times \noisespace \times \refSpace}{\refSpace})}{\gamma}\bigr) = \int_\refSpace \psi \dd\refPVar$ for every bounded measurable $\psi$, and hence $\pushforward
            {(\projectionFromTo{\X{} \times \noisespace \times \refSpace}{\refSpace})}
            \gamma
            =
            \refPVar$.
    Overall, we have proven that, for all $\plan{\gamma_1} \in \plans{\probabilitySA{}{}}{\refPVar}$, $\plan{\gamma} = 
    \pushforward{\sigma}{\plan{(\gamma_1 \otimes q^{x})}} \in \plans{\probabilitySAW{}{}}{\refPVar}$.
    
    We are now ready to prove~\eqref{eq:lemma:marginalinf}. Fix $\plan{\gamma_1} \in \plans{\probabilitySA{}{}}{\refPVar}$. The coupling $\plan{\gamma} =
    \pushforward{\sigma}{\plan{(\gamma_1 \otimes q^{x})}} \in \plans{\probabilitySAW{}{}}{\refPVar}$ is a valid candidate for the left-hand side. Thus, we obtain the upper bound
    \begin{align*}
    \inf_{\plan{\tilde{\gamma}}\in\plans{\probabilitySAW{}{}}{\refPVar}}
    \int_{\X{} \times \noisespace \times \refSpace}
    c(x,\noise,\refVar)
    \dd\plan{\tilde{\gamma}}(x,\noise,\refVar)
    &\le 
    \int_{\X{} \times \noisespace \times \refSpace}
    c(x,\noise,\refVar)
    \dd\plan{\gamma}(x,\noise,\refVar)\\
    &=
    \int_{\X{}\times \noisespace \times \refSpace}
    c(x,\noise,\refVar)
    \dd(\pushforward{\sigma}{(\gamma_1\otimes q^{x})})
    (x,\noise,\refVar)
    \\
    &=
    \int_{\X{}\times \refSpace\times \noisespace }
    c(x,\noise,\refVar)
    \dd(\gamma_1\otimes q^{x})
    (x,\refVar,\noise)
    \\
    &=
    \int_{\X{} \times \refSpace}
    \int_{\noisespace}c(x,\noise,\refVar)
    \dd q^{x}(\noise)
    \dd\plan{\gamma_1}(x,\refVar),
    \end{align*}
    where the last equality uses the disintegration of $\gamma_1 \otimes q^{x}$ and Tonelli's theorem.
    Since $\plan{\gamma_1}\in\plans{\probabilitySA{}{}}{\refPVar}$ is arbitrary, we may take infima to obtain
    \begin{equation*}
    \inf_{\gamma\in\plans{\probabilitySAW{}{}}{\refPVar}}
    \int c(x, \noise, \refVar)
    \dd\plan{\gamma}(x, \noise, \refVar)
    \le
    \inf_{\gamma_1\in\plans{\probabilitySA{}{}}{\refPVar}}
    \int_{\X{} \times \refSpace}
    \int_{\noisespace}c(x,\noise,\refVar)
    \dd q^{x}(\noise)
    \dd\plan{\gamma_1}(x,\refVar),
    \end{equation*}
    which is the desired statement.
\end{proof}

Through the following example, we show that the inequality~\eqref{eq:lemma:marginalinf} is generally strict.
\begin{example}[Strictness of Lemma~\ref{lemma:marginalinf}]
\label{example:noise_coupling}
Let $\X{}=\noisespace=\refSpace=\reals$. Set $\probabilitySA{}{}=\mathcal N(0,1)$ on $\X{}$. For each $x\in \X{}$, let $q = q^x=\mathcal N(0,1)$ on $\noisespace$, independent of $x$. Then, the product measure $\probabilitySAW{}{}=\probabilitySA{}{}\otimes q$ is the law of a pair $(x,\noise)$ with $x, \noise \overset{\text{i.i.d.}}{\sim} \mathcal N(0,1)$. Let $\refPVar=\mathcal N(0,1)$ on $\refSpace$ and define the cost
\[
c(x,\noise,\refVar)=(\noise-\refVar)^2.
\]

To evaluate the left-hand side of~\eqref{eq:lemma:marginalinf}, we aim to solve
\begin{equation*}
    \inf_{\plan{\gamma} \in \plans{\probabilitySAW{}{}}{\refPVar}} 
    \int_{\reals^3} c(x, \noise, \refVar) 
    \dd\plan{\gamma}(x, \noise, \refVar).
\end{equation*}
To do so, we show that there exists a plan $\plan{\tilde{\gamma}}\in \plans{\probabilitySAW{}{}}{\refPVar}$ that attains a cost of $0$, which, by non-negativity of the cost, implies optimality. 
Consider $T: \reals^2 \to \reals^3$ defined by $T(x, \noise) = (x, \noise, \noise)$ and set $\plan{\tilde{\gamma}} = \pushforward{T}{\probabilitySAW{}{}}$. By construction, its $(x, \noise)$-marginal is $\probabilitySAW{}{}$ and its $\refVar$-marginal is given by $\noise \sim \mathcal N(0,1) = \refPVar$, hence $\plan{\tilde{\gamma}} \in \plans{\probabilitySAW{}{}}{\refPVar}$. Under this coupling
\begin{equation*}
    \int_{\reals^3} (\noise - \refVar)^2 \dd\plan{\tilde{\gamma}}(x, \noise, \refVar) =
    \int_{\reals^2} (\noise - \noise)^2 \dd\probabilitySAW{}{}(x, \noise) = 
    0
\end{equation*}
and so 
\begin{equation*}
    \inf_{\plan{\gamma} \in \plans{\probabilitySAW{}{}}{\refPVar}} 
    \int_{\reals^3} c(x,\noise,\refVar) \dd\plan{\gamma}(x, \noise, \refVar) =
    0.
\end{equation*}

We now consider the right-hand side in~\eqref{eq:lemma:marginalinf}.
For any $\gamma_1\in\plans{\probabilitySA{}{}}{\refPVar}$, we evaluate 
\begin{equation*}
    \int_{\X{} \times \refSpace}
    \int_\noisespace
    c(x, \noise, \refVar)
    \dd q^x (\noise)
    \dd\plan{\gamma_1} (x, \refVar).
\end{equation*}
Since the disintegration of $\probabilitySAW{}{}$ is $q = \mathcal N(0,1)$ (in particular, it is independent of $x$), we have
\begin{equation*}
    \int_\reals c(x,\noise,\refVar) \dd q(\noise)
    =
    \int_\reals (\noise - \refVar)^2 
    \dd q(\noise)
    = 
    \int_\reals \noise^2 \dd q(\noise)
    - 2\refVar \int_\reals \noise \dd q(\noise)
    + \refVar^2 =
    1 + \refVar^2,
\end{equation*}
and so
\begin{align*}
\inf_{\gamma_1\in\plans{\probabilitySA{}{}}{\refPVar}}
\int_{\reals^2}\!
\int_\reals c(x,\noise,\refVar)
\dd q(\noise)
\dd\gamma_1(x,\refVar)
&=
\inf_{\gamma_1\in\plans{\probabilitySA{}{}}{\refPVar}}
\int_{\reals^2}
1+\refVar^2
\dd\gamma_1(x,\refVar)
\\
&=
1+\int_\reals
\refVar^2
\dd\refPVar(\refVar)
\\
&=
2,
\end{align*}
where the last equality follows from the fact that the second marginal of $\plan{\gamma_1}$ is $\refPVar = \mathcal N(0,1)$ and thus its second moment is 1. 
Clearly, the left- and right-hand sides do not agree, and the inequality~\eqref{eq:lemma:marginalinf} is strict in general.    
\end{example}

Example~\ref{example:noise_coupling} shows that the inequality of Lemma~\ref{lemma:marginalinf}, and hence the bound of \sepbound{}, is generally strict. Informally, the gap corresponds to the value of being able to match noise realizations directly to targets, rather than averaging over them. In control problems where noise is small relative to the nominal cost (as in the low-noise regime of our grid-world experiments), we expect this gap to be small.

\section{Proofs of Section~\ref{section:main_results}}
\label{section:main_proof}

We now prove the main results presented in Section~\ref{section:main_results}.
Throughout the following appendices, we write for the set of admissible state-action distributions at $\fleet \in \probabilityspace{\statespace}$
\begin{equation*}
    M(\fleet) \coloneqq
    \left\{ \probabilitySA{}{} \in
    \probabilityspace{\statespace\times\actionspace} \colon
    \pushforward{(\projectionFromTo{\statespace\times\actionspace}{\statespace})}{\probabilitySA{}{}}
    = \fleet \right\},
\end{equation*}
that is, the state-action distributions whose $\statespace$-marginal is $\fleet$.
An admissible population control in the sense of Problem~\ref{problem:population-level} thus satisfies $\probabilitySA{}{\timestep}(\fleet) \in M(\fleet)$ for every $\fleet$.
Recall the single-agent problem and the dynamic programming for the cost-to-go function in \eqref{eq:dp:single_agent}:
\begin{align*}
    \costtogosmall{\horizon}(\state_\horizon, \refVar) &= \terminalcostsmall(\state_\horizon, \refVar);\\
    \costtogosmall{\timestep}(\state_\timestep, \refVar) &= 
    \inf_{\action_\timestep \in \actionspace}
    \expectedvalue{\noise_\timestep} {\stagecostsmall{\timestep}(\state_\timestep, \action_\timestep, \noise_\timestep) + 
    \costtogosmall{\nexttimestep}
    (\dynamics{\state_\timestep}{\action_\timestep, \noise_\timestep}{\timestep}, \refVar)}.
\end{align*}

\subsection{Proof of Theorem~\ref{theorem:optimality-approximation}}

We aim to show that the population-level cost-to-go function in Problem~\ref{problem:population-level} is upper bounded by an optimal transport problem whose cost is the single-agent cost-to-go function in \eqref{eq:dp:single_agent}.
Concretely, we show by backward induction that
\begin{equation*}
    \costtogo{\timestep}(\fleet_\timestep, \refPVar) 
    \le
    \inf_{
    \plan{\gamma}\in \plans{\fleet_\timestep}{\refPVar}
    }
    \int
    \costtogosmall{\timestep}(\state, \refVar)
    \dd\plan{\gamma}(\state, \refVar).
\end{equation*}
We first prove the theorem under the field-based setting of Assumptions~\ref{assumption:measurable_field} and~\ref{assumption:barycenter_condition}. The constructions in Examples~\ref{example:individual_noise}--\ref{example:global_noise} and Proposition~\ref{proposition:averagingproperty} cover individual noise with finitely or infinitely many agents, global noise, and local noise on countably supported state-action distributions. For atomless local noise, the footnote in Example~\ref{example:local_noise} gives the corresponding aggregate random-measure construction and explains how the same proof carries over. In the field-based setting, Lemma~\ref{lemma:noiseexp} allows us to replace the realized noise kernels by the nominal kernel after taking expectations.
Throughout, Assumption~\ref{assumption:measurable_field} makes the gluing $\probabilitySAW{}{\timestep} = \probabilitySA{}{\timestep} \otimes \empnoisedist{\timestep}^{\state,\action}$ a well-defined random measure and justifies interchanging expectations over the realized field with $\probabilitySA{}{\timestep}$-integrals. All integrands below are lower semianalytic (see Appendix~\ref{appendix:preliminary_lemmata}), so all expectations are well-defined.

We start with the proof of Part I of the theorem. 

\begin{proof}
The proof proceeds by induction, in three steps. First, we use the definition of the cost-to-go function and the induction hypothesis at time $\nexttimestep$. Second, we express the conditional expectations over noise realizations in terms of the averaged single-agent cost. Finally, we reorganize the infima over couplings by means of Lemma~\ref{lemma:disintegration}, which allows us to recover the desired distributional dynamic program and conclude the induction.

The base case of the induction is straightforward. By definition $\costtogo{\horizon}(\fleet_\horizon, \refPVar) = \kantorovich{\terminalcostsmall}{\fleet_\horizon}{\refPVar}{}$. Since $\costtogosmall{\horizon}(\state, \refVar) = \terminalcostsmall(\state, \refVar)$, we have $\kantorovich{\costtogosmall{\horizon}}{\fleet_\horizon}{\refPVar}{} = \kantorovich{\terminalcostsmall}{\fleet_\horizon}{\refPVar}{} = \costtogo{\horizon}(\fleet_\horizon, \refPVar)$.
Fix $\timestep<\horizon$, using the induction hypothesis at time $\nexttimestep$, i.e., 
\begin{equation*}
    \costtogo{\nexttimestep}(\pushforward{\dynamics{}{}{\timestep}}{\probabilitySAW{}{\timestep}}, \refPVar) 
    \le
    \inf_{\plan{\gamma} \in \plans{\pushforward{\dynamics{}{}{\timestep}}{\probabilitySAW{}{\timestep}}}{\refPVar}}
    \int_{\statespace\times\refSpace} 
    \costtogosmall{\nexttimestep}(\state^+, \refVar)
    \dd\plan{\gamma},
\end{equation*}
the backward recursion for the cost-to-go function at time $\timestep$ gives

\begin{align}
\label{eq:inductionstep}
&\costtogo{\timestep}(\fleet_\timestep, \refPVar)
=
\inf_{
\pushforward{(\projectionFromTo{\statespace\times\actionspace}{\statespace})}{\probabilitySA{}{\timestep}}=\fleet_\timestep
}
\expectedvalue{\distoverdist{\timestep}}{
\expectedvalue{\probabilitySAW{}{\timestep}}{\stagecostsmall{\timestep}(\state, \action, \noise)}
+
\costtogo{\nexttimestep}(\pushforward{\dynamics{}{}{\timestep}}{\probabilitySAW{}{\timestep}}, \refPVar)} \nonumber
\\
&\quad\leq
\inf_{
\pushforward{(\projectionFromTo{\statespace\times\actionspace}{\statespace})}{\probabilitySA{}{\timestep}}=\fleet_\timestep
}
\expectedvalue{\distoverdist{\timestep}}{
\underbrace{
\int_{\statespace\times\actionspace\times \noisespace}
\stagecostsmall{\timestep}(\state, \action, \noise)
\dd\probabilitySAW{}{\timestep} +
\inf_{\plan{\gamma} \in \plans{\pushforward{\dynamics{}{}{\timestep}}{\probabilitySAW{}{\timestep}}}{\refPVar}}
\int_{\statespace\times\refSpace}
\costtogosmall{\nexttimestep}(\state^+, \refVar)
\dd\plan{\gamma}}_{(\ast)}}.
\end{align}

We now focus on the term denoted by $(\ast)$. We analyze $(\ast)$ pointwise for each realization of $\empnoisedist{\timestep}$ before taking expectations.
By Lemma~\ref{lemma:pushandOT}, we can ``push'' the dynamics into the transportation cost, so that
\begin{align*}
(\ast)
&=
\int_{\statespace\times\actionspace\times \noisespace}
\stagecostsmall{\timestep}(\state, \action, \noise)
\dd\probabilitySAW{}{\timestep} +
\inf_{\plan{\gamma'} \in \plans{\probabilitySAW{}{\timestep}}{\refPVar}}
\int_{\statespace\times\actionspace\times\noisespace\times\refSpace} 
\costtogosmall{\nexttimestep}(\dynamics{\state}{\action, \noise}{\timestep}, \refVar)
\dd\plan{\gamma'}.
\end{align*}
Here we use that the coupling correspondence in Lemma~\ref{lemma:pushandOT} preserves the target marginal: since $\dynamics{}{}{\timestep}\times \Id{\refSpace}$ acts as the identity on the $\refSpace$-coordinate, it restricts to a bijection between the couplings with $\refSpace$-marginal $\refPVar$, so imposing $\plan{\gamma'}\in \plans{\probabilitySAW{}{\timestep}}{\refPVar}$ (with $\refPVar$ fixed) is equivalent to the corresponding fixed-$\refPVar$ constraint on $\pushforward{\dynamics{}{}{\timestep}}{\probabilitySAW{}{\timestep}}$.
Since $\probabilitySAW{}{\timestep}$ is the first marginal of $\plan{\gamma'}$, Lemma~\ref{lemma:sumOTdiscr}, with $\X{} = \statespace \times \actionspace \times \noisespace$, $c_1 = \stagecostsmall{\timestep}$ and $c_2 = \costtogosmall{\nexttimestep} \circ (\dynamics{}{}{\timestep}\times \Id{\refSpace})$, allows us to merge the two costs into a single transport objective on $\statespace \times \actionspace \times \noisespace \times \refSpace$. We therefore have
\begin{align*}
(\ast) &=
\inf_{\plan{\gamma'} \in \plans{\probabilitySAW{}{\timestep}}{\refPVar}}
\int_{\statespace\times\actionspace\times\noisespace\times\refSpace} 
\stagecostsmall{\timestep}(\state, \action, \noise) +
\costtogosmall{\nexttimestep}(\dynamics{\state}{\action, \noise}{\timestep}, \refVar)
\dd\plan{\gamma'}.
\end{align*}
We now use Lemma~\ref{lemma:marginalinf} with $\X{} = \statespace\times\actionspace$, $\probabilitySAW{}{\timestep} = \probabilitySA{}{\timestep} \otimes \empnoisedist{\timestep}^{\state, \action}$, and $c(\state, \action, \noise, \refVar) = \stagecostsmall{\timestep}(\state, \action, \noise) +
\costtogosmall{\nexttimestep}(\dynamics{\state}{\action, \noise}{\timestep}, \refVar)$, to replace the couplings of the full state-action-noise distribution $\probabilitySAW{}{\timestep}$ with couplings of its $(\statespace\times\actionspace)$-marginal $\probabilitySA{}{\timestep}$. This yields the inequality
\begin{align*}
(\ast) \leq
\inf_{\plan{\tilde{\gamma}} \in \plans{\probabilitySA{}{\timestep}}{\refPVar}}
\int_{\statespace\times\actionspace\times\refSpace} 
\int_{\noisespace}
\stagecostsmall{\timestep}(\state, \action, \noise) +
\costtogosmall{\nexttimestep}(\dynamics{\state}{\action, \noise}{\timestep}, \refVar)
\dd\empnoisedist{\timestep}^{\state, \action} \dd\plan{\tilde{\gamma}}.
\end{align*}
This relaxation removes the dependence on the realized noise from the optimal assignment, thus restricting it to depend only on $(\state, \action)$.

We now plug the upper bound obtained for $(\ast)$ inside the expectation over $\empnoisedist{\timestep}^{\state, \action}$, as in \eqref{eq:inductionstep}.
Since $\probabilitySA{}{\timestep}$ is fixed and the infimum is now over couplings $\plan{\tilde{\gamma}} \in \plans{\probabilitySA{}{\timestep}}{\refPVar}$ that do not depend on $\empnoisedist{\timestep}^{\state, \action}$, Lemma~\ref{lemma:infexp} applies, allowing us to interchange expectation and infimum and yielding the upper bound
\begin{align*}
\expectedvalue{\distoverdist{\timestep}}{
(\ast)
}
&\leq
\expectedvalue{\distoverdist{\timestep}}{
\inf_{\plan{\tilde{\gamma}} \in \plans{\probabilitySA{}{\timestep}}{\refPVar}}
\int_{\statespace\times\actionspace\times\refSpace}
\int_{\noisespace}
\stagecostsmall{\timestep}(\state, \action, \noise) +
\costtogosmall{\nexttimestep}(\dynamics{\state}{\action, \noise}{\timestep}, \refVar)
\dd\empnoisedist{\timestep}^{\state,\action} \dd\plan{\tilde{\gamma}}
}
\\
&\leq
\inf_{\plan{\tilde{\gamma}} \in \plans{\probabilitySA{}{\timestep}}{\refPVar}}
\expectedvalue{\distoverdist{\timestep}}{
\int_{\statespace\times\actionspace\times\refSpace}
\int_{\noisespace}
\stagecostsmall{\timestep}(\state, \action, \noise) +
\costtogosmall{\nexttimestep}(\dynamics{\state}{\action, \noise}{\timestep}, \refVar)
\dd\empnoisedist{\timestep}^{\state,\action} \dd\plan{\tilde{\gamma}}}
\\
&=
\inf_{\plan{\tilde{\gamma}} \in \plans{\probabilitySA{}{\timestep}}{\refPVar}}
\int_{\statespace\times\actionspace\times\refSpace}
\expectedvalue{\distoverdist{\timestep}^{\state, \action}}{\int_{\noisespace}
\stagecostsmall{\timestep}(\state, \action, \noise) +
\costtogosmall{\nexttimestep}(\dynamics{\state}{\action, \noise}{\timestep}, \refVar)
\dd\empnoisedist{\timestep}^{\state,\action}} \dd\plan{\tilde{\gamma}},
\end{align*}
where the last equality follows from Tonelli's theorem, which applies by nonnegativity of the costs and joint measurability of the realized field (Assumption~\ref{assumption:measurable_field}). Informally, the second inequality is precisely where we transition from ``choose the assignment after seeing the realized noise'' to ``choose one assignment against average noise''.
%
%
We now focus on the inner expectation. In particular, consider a fixed $\plan{\tilde{\gamma}} \in \plans{\probabilitySA{}{\timestep}}{\refPVar}$.
We can disintegrate the coupling with respect to its $(\statespace\times\actionspace)$-marginal $\probabilitySA{}{\timestep}$ as $\plan{\tilde{\gamma}} = \probabilitySA{}{\timestep}\otimes \kappa_{\state, \action}$ where $\kappa_{\state, \action}$ is a (conditional) probability over $\refSpace$. By non-negativity of all functions, Tonelli's theorem gives 
\begin{align*}
\int_{\statespace\times\actionspace\times\refSpace}
&\expectedvalue{\distoverdist{\timestep}^{\state, \action}}{\int_{\noisespace}
\stagecostsmall{\timestep}(\state, \action, \noise) +
\costtogosmall{\nexttimestep}(\dynamics{\state}{\action, \noise}{\timestep}, \refVar)
\dd\empnoisedist{\timestep}^{\state,\action}} \dd\plan{\tilde{\gamma}}
\\
&=
\int_{\statespace\times\actionspace}
\expectedvalue{\distoverdist{\timestep}^{\state, \action}}{\int_{\noisespace}\int_\refSpace
\stagecostsmall{\timestep}(\state, \action, \noise) +
\costtogosmall{\nexttimestep}(\dynamics{\state}{\action, \noise}{\timestep}, \refVar)
\dd\kappa_{\state, \action}
\dd\empnoisedist{\timestep}^{\state,\action}} \dd\probabilitySA{}{\timestep}.
\end{align*}
We can apply Lemma~\ref{lemma:noiseexp} (treating $(\state, \action)$ as parameters) to the function 
\begin{equation*}
    \tilde{c}(\state, \action, \noise)
    \coloneqq
    \int_\refSpace \stagecostsmall{\timestep}(\state, \action, \noise) +
\costtogosmall{\nexttimestep}(\dynamics{\state}{\action, \noise}{\timestep}, \refVar) \dd\kappa_{\state, \action}(\refVar).
\end{equation*}
Note, in particular, that Lemma~\ref{lemma:noiseexp} can be applied since $\kappa_{\state, \action}$ is fixed and not a function of the noise distribution $\empnoisedist{\timestep}^{\state,\action}$. Thus, we get 
\begin{align*}
    \int_{\statespace\times\actionspace}
    &\expectedvalue{\distoverdist{\timestep}^{\state, \action}}{\int_{\noisespace}\int_{\refSpace}
    \stagecostsmall{\timestep}(\state, \action, \noise) +
    \costtogosmall{\nexttimestep}(\dynamics{\state}{\action, \noise}{\timestep}, \refVar)
    \dd\kappa_{\state, \action}
    \dd\empnoisedist{\timestep}^{\state,\action}
    }
    \dd \probabilitySA{}{\timestep}
    \\
    &=
    \int_{\statespace\times\actionspace}
    \expectedvalue{\noise}{\int_\refSpace
    \stagecostsmall{\timestep}(\state, \action, \noise) +
    \costtogosmall{\nexttimestep}(\dynamics{\state}{\action, \noise}{\timestep}, \refVar)
    \dd\kappa_{\state, \action}}
    \dd \probabilitySA{}{\timestep}
    \\
    &=
    \int_{\statespace\times\actionspace\times\refSpace}
    \expectedvalue{\noise}{
    \stagecostsmall{\timestep}(\state, \action, \noise) +
    \costtogosmall{\nexttimestep}(\dynamics{\state}{\action, \noise}{\timestep}, \refVar)}\,
    \dd \plan{\tilde{\gamma}}
\end{align*}
where the last step follows again from Tonelli's theorem and the definition of $\tilde\gamma$.
Combined with the previous chain of inequalities, we therefore have
\begin{align*}
\expectedvalue{\distoverdist{\timestep}}{
(\ast)
}
&\leq
\inf_{\plan{\tilde{\gamma}} \in \plans{\probabilitySA{}{\timestep}}{\refPVar}}
\int_{\statespace\times\actionspace\times\refSpace} 
\expectedvalue{\noise}{
\stagecostsmall{\timestep}(\state, \action, \noise) +
\costtogosmall{\nexttimestep}(\dynamics{\state}{\action, \noise}{\timestep}, \refVar)
}
\dd\plan{\tilde{\gamma}}.
\end{align*}
We can now plug back our bound in \eqref{eq:inductionstep} to get 
\begin{align*}
\costtogo{\timestep}(\fleet_\timestep, \refPVar) &\leq
\inf_{
\pushforward{(\projectionFromTo{\statespace\times\actionspace}{\statespace})}{\probabilitySA{}{\timestep}}=\fleet_\timestep
}
\inf_{\plan{\tilde{\gamma}} \in \plans{\probabilitySA{}{\timestep}}{\refPVar}}
\int_{\statespace\times\actionspace\times\refSpace} 
\expectedvalue{\noise}{
\stagecostsmall{\timestep}(\state, \action, \noise) +
\costtogosmall{\nexttimestep}(\dynamics{\state}{\action, \noise}{\timestep}, \refVar)
}
\dd\plan{\tilde{\gamma}}
\\
&=
\inf_{
\pushforward{(\projectionFromTo{\statespace\times\actionspace\times \refSpace}{\statespace \times \refSpace})}{\plan{\tilde{\gamma}}}\in \plans{\fleet_\timestep}{\refPVar}
}
\int_{\statespace\times\actionspace\times\refSpace} 
\expectedvalue{\noise}{
\stagecostsmall{\timestep}(\state, \action, \noise) +
\costtogosmall{\nexttimestep}(\dynamics{\state}{\action, \noise}{\timestep}, \refVar)
}
\dd\plan{\tilde{\gamma}}
\\
&\eqqcolon (\star).
\end{align*}
Now, we can use Lemma~\ref{lemma:disintegration} to disintegrate the optimizer in $(\star)$, which gives 
\begin{align*}
(\star)
&=
\inf_{\substack{
\plan{\tilde{\gamma}}\in \plans{\fleet_\timestep}{\refPVar} \\ \alpha^{\state, \refVar} \in \Lambda(\actionspace)
}}
\int_{\statespace\times\refSpace} 
\int_\actionspace
\expectedvalue{\noise}{
\stagecostsmall{\timestep}(\state, \action, \noise) +
\costtogosmall{\nexttimestep}(\dynamics{\state}{\action, \noise}{\timestep}, \refVar)
}
\dd\alpha^{\state, \refVar}
\dd\plan{\tilde{\gamma}}
\end{align*}
The integrand is lower semianalytic: $\stagecostsmall{\timestep}$, $\terminalcostsmall$, and $\dynamics{}{}{\timestep}$ are Borel, and, by backward induction, $\costtogosmall{\nexttimestep}$ is lower semianalytic, since lower semianalyticity is preserved by composition with Borel maps \cite[Lemma~7.30]{bertsekas1996stochastic}, integration against the Borel kernel $\noisedist{\timestep}$ \cite[Proposition~7.48]{bertsekas1996stochastic}, and pointwise infima over $\actionspace$ \cite[Proposition~7.47]{bertsekas1996stochastic}. Therefore, we can apply Lemma~\ref{lemma:global_to_pointwise}, which gives
\begin{align*}
(\star)
&=
\inf_{
\plan{\tilde{\gamma}}\in \plans{\fleet_\timestep}{\refPVar}
}
\int_{\statespace\times\refSpace} 
\inf_{
\action \in \actionspace
}
\expectedvalue{\noise}{
\stagecostsmall{\timestep}(\state, \action, \noise) +
\costtogosmall{\nexttimestep}(\dynamics{\state}{\action, \noise}{\timestep}, \refVar)
}
\dd\plan{\tilde{\gamma}}.
\end{align*}
At this point, we recognize that the term in the integral coincides with the Bellman recursion for $\costtogosmall{\timestep}$:
\begin{equation*}
    \costtogosmall{\timestep}(\state_\timestep, \refVar) = 
    \inf_{\action_\timestep \in \actionspace}
    \expectedvalue{\noise_\timestep} {\stagecostsmall{\timestep}(\state_\timestep, \action_\timestep, \noise_\timestep) + 
    \costtogosmall{\nexttimestep}
    (\dynamics{\state_\timestep}{\action_\timestep, \noise_\timestep}{\timestep}, \refVar)}.
\end{equation*}
Hence,
\begin{equation*}
    \costtogo{\timestep}(\fleet_\timestep, \refPVar)
    \leq
    (\star)
    =
    \inf_{
    \plan{\tilde{\gamma}}\in \plans{\fleet_\timestep}{\refPVar}
    }
    \int_{\statespace\times\refSpace} 
    \costtogosmall{\timestep}(\state, \refVar)
    \dd\plan{\tilde{\gamma}}.
\end{equation*}
This concludes the inductive step and the proof of the theorem.
\end{proof}

Two comments on our proof are in order. 
First, while the proof follows the inductive strategy of the proof of the deterministic separation
in \cite{terpin2024dynamic}, there is a key difference: 
%
the next population state is random and depends on the realized empirical
noise kernel \(\empnoisedist{\timestep}\). This requires various additional steps (and preliminary lemmata) in the proof. All these steps trivialize in the absence of noise and, thus, our proof recovers the deterministic separation in \cite{terpin2024dynamic} when the dynamics are deterministic.
Second, our result is an upper bound on the cost-to-go function. 
Intuitively, this conservatism originates from the following fact: The population state-action distribution can, in principle, choose future couplings \emph{after} observing this realization, while our upper bound posits that couplings are chosen \emph{before} the noise is realized, averaging over the kernel
\(\noisedist{\timestep}(\cdot \st \state,\action)\). This is precisely where Lemmas~\ref{lemma:infexp} and~\ref{lemma:marginalinf} yield inequalities rather than equalities. 
In the deterministic case, all these steps trivialize and we obtain equality as in~\cite{terpin2024dynamic}.


We now prove Part II of the theorem. 

\begin{proof}
For each time step $\timestep$, consider the function
\begin{equation*}
    \kantorovich{\costtogosmall{\timestep}}{\fleet_\timestep}{\refPVar}{}
    =
    \inf_{
        \plan{\gamma}\in \plans{\fleet_\timestep}{\refPVar}
    }
    \int_{\statespace\times\refSpace}
    \costtogosmall{\timestep}(\state, \refVar)
    \dd\plan{\gamma}.
\end{equation*}
Let $\varepsilon > 0$. At each time $\timestep$, we can construct a ``separation-based'' state-action distribution $\probabilitySA{\varepsilon}{\timestep}$ as follows:  given a population distribution $\fleet_\timestep$, choose an $\frac{\varepsilon}{2\horizon}$-optimal coupling $\plan{\gamma_\timestep^\varepsilon} \in \plans{\fleet_\timestep}{\refPVar}$ satisfying
\begin{equation}\label{eq:thm:partii:eps_coupling}
    \int_{\statespace\times\refSpace}
    \costtogosmall{\timestep}(\state, \refVar)
    \dd \plan{\gamma_\timestep^\varepsilon}
    \leq
    \kantorovich{\costtogosmall{\timestep}}{\fleet_\timestep}{\refPVar}{} +
    \frac{\varepsilon}{2\horizon},
\end{equation}
and a universally measurable $\frac{\varepsilon}{2\horizon}$-optimal action selector $\action_\timestep^\varepsilon\colon \statespace\times\refSpace \to \actionspace$ satisfying, for all $(\state, \refVar)$,
\begin{equation}\label{eq:thm:partii:dp_ground_space}
    \expectedvalue{\noise}{
    \stagecostsmall{\timestep}(\state, \action_\timestep^\varepsilon(\state, \refVar), \noise) +
    \costtogosmall{\nexttimestep}(\dynamics{\state}{\action_\timestep^\varepsilon(\state, \refVar), \noise}{\timestep}, \refVar)
    }
    \leq
    \costtogosmall{\timestep}(\state, \refVar) +
    \frac{\varepsilon}{2\horizon}.
\end{equation}
Formally, we have 
\begin{equation*}
    \probabilitySA{\varepsilon}{\timestep}
    \coloneqq
    \pushforward{\left(
        \projectionFromTo{\statespace\times\refSpace}{\statespace}{},
        \action_\timestep^\varepsilon
    \right)}{
        \plan{\gamma_\timestep^\varepsilon} 
    }.
\end{equation*}
In words, this state-action distribution assigns agents to targets through $\plan{\gamma_\timestep^\varepsilon}$ and controls the individual agents through $\action_\timestep^\varepsilon$.
Since the first marginal of $\plan{\gamma_\timestep^\varepsilon}$ is
$\fleet_\timestep$, we have $\probabilitySA{\varepsilon}{\timestep}\in M(\fleet_\timestep)$.
In the rest of the proof, we will show that the resulting control sequence satisfies the bound, 
\begin{equation}\label{eq:thm:partii:eps_optimal}
    \costtogo{0}^{\varepsilon}(\fleet_0, \refPVar)
    \coloneqq
    \costtogo{0}^{\probabilitySA{\varepsilon}{0},\ldots,\probabilitySA{\varepsilon}{\horizon-1}}(\fleet_0, \refPVar)
    \leq
    \kantorovich{\costtogosmall{0}}{\fleet_0}{\refPVar}{}
    +
    \varepsilon.
\end{equation}

Before doing so, we verify well-posedness of all the objects. The $\frac{\varepsilon}{2\horizon}$-optimal universally measurable selector $\action_\timestep^\varepsilon$ exists by \cite[Proposition~7.50]{bertsekas1996stochastic}, since the Bellman integrand in \eqref{eq:thm:partii:dp_ground_space} is lower semianalytic and nonnegative (see the induction argument in the proof of Part~I).
For the existence of the $\frac{\varepsilon}{2\horizon}$-optimal coupling we work as follows. Since $\costtogosmall{\timestep}$ is lower semianalytic (by backward induction on \eqref{eq:dp:single_agent}), the map $\plan{\gamma} \mapsto \int_{\statespace\times\refSpace} \costtogosmall{\timestep }\dd\plan{\gamma}$ is lower semianalytic by~\cite[Corollary~7.48.1]{bertsekas1996stochastic} (jointly in $(\fleet, \plan{\gamma})$, as it does not depend on $\fleet$).
Further, the set $\{(\fleet, \plan{\gamma}): \plan{\gamma}\in\plans{\fleet}{\refPVar}\}$ is closed.
%
Indeed, fixing $\refPVar\in\probabilityspace{\refSpace}$, define
\begin{equation*}
\mathcal G_\refSpace
\coloneqq
\{(\fleet,\plan{\gamma})\in \probabilityspace{\statespace}\times \probabilityspace{\statespace\times\refSpace}:\plan{\gamma}\in\plans{\fleet}{\refPVar}\}.
\end{equation*}
The maps $F_\statespace(\plan{\gamma})\coloneqq \pushforward{(\projectionFromTo{{\statespace\times\refSpace}}{{\statespace}})}{\plan{\gamma}}$ and $F_\refSpace(\plan{\gamma})\coloneqq \pushforward{(\projectionFromTo{{\statespace\times\refSpace}}{{\refSpace}})}{\plan{\gamma}}$ are continuous for the weak topologies, being the pushforward under a continuous map. Hence,
\begin{equation*}
\mathcal G_\refSpace
=
\{(\fleet,\plan{\gamma}):F_\statespace(\plan{\gamma})=\fleet\}
\cap
\{(\fleet,\plan{\gamma}):F_\refSpace(\plan{\gamma})=\refPVar\}.
\end{equation*}
The first set is the preimage of the diagonal in
$\probabilityspace{\statespace}\times\probabilityspace{\statespace}$ under the continuous map $(\fleet,\plan{\gamma})\mapsto(\fleet,F_\statespace(\plan{\gamma}))$, and the second is the preimage of
the closed singleton $\{\refPVar\}$ under the continuous map
$(\fleet,\plan{\gamma})\mapsto F_\refSpace(\plan{\gamma})$. Since these spaces are Polish, the diagonal
and singleton are closed. By continuity of $F_\statespace$ and $F_\refSpace$, the set $\mathcal G_\refSpace$ is closed, hence Borel and in particular analytic.
Thus, we can apply \cite[Proposition~7.50]{bertsekas1996stochastic} to obtain the existence of a universally measurable map 
\begin{equation*}
    \fleet \mapsto \plan{\gamma_\timestep^\varepsilon(\fleet)} \in \plans{\fleet}{\refPVar} \quad \text{with } \int_{\statespace\times\refSpace} \costtogosmall{\timestep} \dd\plan{\gamma_\timestep^\varepsilon(\fleet)} \leq \kantorovich{\costtogosmall{\timestep}}{\fleet_\timestep}{\refPVar}{} + \frac{\varepsilon}{2\horizon}.
\end{equation*}
This shows the existence of a universally measurable selector $\plan{\gamma_\timestep^\varepsilon(\fleet)}$. In particular, for fixed $\fleet$, $\plan{\gamma_\timestep^\varepsilon(\fleet)}$ is a Borel probability measure, and $\probabilitySA{\varepsilon}{\timestep}$ is a well-defined Borel probability measure: it is the pushforward of $\plan{\gamma_\timestep^\varepsilon(\fleet)}$ under the universally measurable map $(\projectionFromTo{\statespace\times\refSpace}{\statespace}, \action_\timestep^\varepsilon)$, which coincides $\plan{\gamma_\timestep^\varepsilon(\fleet)}$-almost everywhere with a Borel map~\cite[Lemma~7.27]{bertsekas1996stochastic}.

We now turn our attention to the desired bound \eqref{eq:thm:partii:eps_optimal}.
We now show by backward induction the more general bound 
\begin{equation}\label{eq:cost_seppolicy}
    \costtogo{\timestep}^{\varepsilon}(\fleet_\timestep, \refPVar)
    \coloneqq
    \costtogo{\timestep}^{\probabilitySA{\varepsilon}{\timestep},\ldots,\probabilitySA{\varepsilon}{\horizon-1}}(\fleet_\timestep, \refPVar)
    \leq
    \kantorovich{\costtogosmall{\timestep}}{\fleet_\timestep}{\refPVar}{} + \frac{\horizon - \timestep}{\horizon}\varepsilon,
\end{equation}
which, evaluated at time $t=0$, readily implies~\eqref{eq:thm:partii:eps_optimal}. 
The base case $\timestep=\horizon$ holds exactly by definition of terminal cost and $\costtogosmall{\horizon}=\terminalcostsmall$:
\begin{equation*}
    \costtogo{\horizon}^{\varepsilon}(\fleet_\horizon, \refPVar) 
    =
    \kantorovich{\terminalcostsmall}{\fleet_\horizon}{\refPVar}{}
    =
    \inf_{
        \plan{\gamma}\in \plans{\fleet_\horizon}{\refPVar}
    }
    \int_{\statespace\times\refSpace}
    \terminalcostsmall(\state, \refVar)
    \dd \plan{\gamma}
    =
    \kantorovich{\costtogosmall{\horizon}}{\fleet_\horizon}{\refPVar}{}.
\end{equation*}
Now fix $\timestep < \horizon$ and assume \eqref{eq:cost_seppolicy} holds at $\nexttimestep$.
By the policy-cost recursion (which uses the independence of the noise across time) and the induction hypothesis at $\nexttimestep$ applied inside the expectation, with $\probabilitySAW{\varepsilon}{\timestep} = \probabilitySA{\varepsilon}{\timestep} \otimes \empnoisedist{\timestep}^{\state,\action}$,
\begin{equation}
\label{eq:tilde_J_recursion}
    \costtogo{\timestep}^{\varepsilon}(\fleet_\timestep, \refPVar)
    \leq
    \expectedvalue{\distoverdist{\timestep}}{
    \int_{\statespace\times\actionspace\times\noisespace}
    \stagecostsmall{\timestep}(\state, \action, \noise)
    \dd \probabilitySAW{\varepsilon}{\timestep}
    +
    \kantorovich{\costtogosmall{\nexttimestep}}{\fleet_{\nexttimestep}}{\refPVar}{}
    } +
    \frac{\horizon - \timestep -1}{\horizon}\varepsilon.
\end{equation}
Next, we define 
\begin{equation*}
    \plan{\bar\gamma_\timestep^\varepsilon}
    \coloneqq
    \pushforward{(\dynamics{}{}{\timestep},\Id{\refSpace})}{(\pushforward{\sigma}{\plan{\tilde\gamma_\timestep^\varepsilon}})}
    \in
    \probabilityspace{\statespace\times\refSpace}
\end{equation*}
where 
\begin{equation*}
    \plan{\tilde\gamma_\timestep^\varepsilon}
    \coloneqq
    \pushforward{(\Id{\statespace\times\refSpace},\action_\timestep^\varepsilon)}{\plan{\gamma_\timestep^\varepsilon}}
    \otimes 
    \empnoisedist{\timestep}^{\state, \action_\timestep^\varepsilon(\state, \refVar)}
    \in
    \probabilityspace{\statespace\times\refSpace\times\actionspace\times\noisespace}.
\end{equation*}
Where we use the permutation
    \begin{equation*}
    \sigma: \statespace{}\times\refSpace \times\actionspace \times \noisespace \to \statespace \times \actionspace \times \noisespace \times \refSpace,
    \quad
    \sigma(\state,\refVar,\action,\noise) = (\state,\action, \noise, \refVar),
    \end{equation*}
to ``re-order'' the marginals. (The gluing is well-posed: $(\state, \refVar) \mapsto \empnoisedist{\timestep}^{\state, \action_\timestep^\varepsilon(\state, \refVar)}$ is a universally measurable kernel, as the composition of the field of Assumption~\ref{assumption:measurable_field} with the universally measurable map $(\Id{\statespace}, \action_\timestep^\varepsilon)$.)
Informally, $\plan{\bar\gamma_\timestep^\varepsilon}$ is the law of $(\dynamics{\state}{\action_\timestep^\varepsilon(\state, \refVar), \noise}{\timestep}, \refVar)$ where $(\state, \refVar) \sim \plan{\gamma_\timestep^\varepsilon}$ and $\noise \sim \empnoisedist{\timestep}^{\state, \action_\timestep^\varepsilon(\state, \refVar)}$.
We now show $\plan{\bar\gamma_\timestep^\varepsilon} \in \plans{\fleet_{\nexttimestep}}{\refPVar}$. Specifically, the second marginal is $\refPVar$, since the $\refVar$-coordinate is unchanged by $\sigma$ and by $(\dynamics{}{}{\timestep}, \Id{\refSpace})$, and $\pushforward{(\projectionFromTo{\statespace\times\refSpace}{\refSpace})}{\plan{\gamma_\timestep^\varepsilon}} = \refPVar$.
For the first marginal, the $(\statespace\times\actionspace\times\noisespace)$-marginal of $\pushforward{\sigma}{\plan{\tilde\gamma_\timestep^\varepsilon}}$ equals $\probabilitySAW{\varepsilon}{\timestep}$, since the noise kernel depends on $(\state, \refVar)$ only through the state-action pair $(\state, \action_\timestep^\varepsilon(\state, \refVar))$, whose law under $\pushforward{(\Id{\statespace\times\refSpace},\action_\timestep^\varepsilon)}{\plan{\gamma}^\varepsilon_\timestep}$ is $\probabilitySA{\varepsilon}{\timestep}$ by definition. Hence the first marginal of $\plan{\bar\gamma_\timestep^\varepsilon}$ is $\pushforward{\dynamics{}{}{\timestep}}{\probabilitySAW{\varepsilon}{\timestep}} = \fleet_{\nexttimestep}$, completing the proof of $\plan{\bar\gamma_\timestep^\varepsilon} \in \plans{\fleet_{\nexttimestep}}{\refPVar}$.
By definition of the infimum, we then obtain 
\begin{align*}
    \kantorovich{\costtogosmall{\nexttimestep}}{\fleet_{\nexttimestep}}{\refPVar}{}
    \leq
    \int_{\statespace\times\refSpace}
    \costtogosmall{\nexttimestep}(\state, \refVar)
    \dd \plan{\bar\gamma_\timestep^\varepsilon}
    =
    \int_{\statespace\times\refSpace}
    \int_\noisespace
    \costtogosmall{\nexttimestep}(\dynamics{\state}{\action_\timestep^\varepsilon(\state, \refVar), \noise}{\timestep}, \refVar)
    \dd \empnoisedist{\timestep}^{\state, \action_\timestep^\varepsilon(\state, \refVar)}
    \dd \plan{\gamma_\timestep^\varepsilon}.
\end{align*}
Taking expectations over the realized noise field and applying Tonelli's theorem and Lemma~\ref{lemma:noiseexp} as in the proof of Part~I (with $\probabilitySA{\varepsilon}{\timestep}$ in place of $\probabilitySA{}{\timestep}$), we get
\begin{align*}
    &\expectedvalue{\distoverdist{\timestep}}{\int_{\statespace\times\actionspace\times\noisespace}
    \stagecostsmall{\timestep}(\state, \action, \noise)
    \dd\probabilitySAW{\varepsilon}{\timestep} +
    \kantorovich{\costtogosmall{\nexttimestep}}{\fleet_{\nexttimestep}}{\refPVar}{}
    }
    \\
    &\quad\quad\leq
    \expectedvalue{\distoverdist{\timestep}}{
    \int_{\statespace\times\refSpace}
    \int_\noisespace
    \stagecostsmall{\timestep}(\state, \action_\timestep^\varepsilon(\state, \refVar), \noise)
    +
    \costtogosmall{\nexttimestep}(\dynamics{\state}{\action_\timestep^\varepsilon(\state, \refVar), \noise}{\timestep}, \refVar)
    \dd \empnoisedist{\timestep}^{\state, \action_\timestep^\varepsilon(\state, \refVar)}
    \dd \plan{\gamma_\timestep^\varepsilon}
    }
    \\
    &\quad\quad=
    \int_{\statespace\times\refSpace}
    \expectedvalue{\noise}{\stagecostsmall{\timestep}(\state, \action_\timestep^\varepsilon(\state, \refVar), \noise) + \costtogosmall{\nexttimestep}(\dynamics{\state}{\action_\timestep^\varepsilon(\state, \refVar), \noise}{\timestep}, \refVar)}
    \dd\plan{\gamma_\timestep^\varepsilon}.
\end{align*}
Then, we use the $\frac{\varepsilon}{2\horizon}$-optimality of $\action_\timestep^\varepsilon$ in~\eqref{eq:thm:partii:dp_ground_space} and of $\plan{\gamma_\timestep^\varepsilon}$ in~\eqref{eq:thm:partii:eps_coupling} to get 
\begin{multline*}
    \int_{\statespace\times\refSpace}
    \underbrace{\expectedvalue{\noise}{\stagecostsmall{\timestep}(\state, \action_\timestep^\varepsilon(\state, \refVar), \noise) + \costtogosmall{\nexttimestep}(\dynamics{\state}{\action_\timestep^\varepsilon(\state, \refVar), \noise}{\timestep}, \refVar)}}_{\leq \costtogosmall{\timestep}(\state, \refVar) + \varepsilon/(2\horizon)}
    \dd\plan{\gamma_\timestep^\varepsilon}
    \\
    \leq
    \int_{\statespace\times\refSpace}
    \costtogosmall{\timestep}(\state, \refVar)
    \dd\plan{\gamma_\timestep^\varepsilon} +
    \frac{\varepsilon}{2\horizon}\leq
    \kantorovich{\costtogosmall{\timestep}}{\fleet_\timestep}{\refPVar}{} +
    \frac{\varepsilon}{2\horizon} + \frac{\varepsilon}{2\horizon}.
\end{multline*}
We can now combine this bound with~\eqref{eq:tilde_J_recursion} and get 
\begin{equation*}
    \costtogo{\timestep}^{\varepsilon}(\fleet_\timestep, \refPVar)
    \leq
    \kantorovich{\costtogosmall{\timestep}}{\fleet_\timestep}{\refPVar}{} +
    \frac{\varepsilon}{\horizon} +
    \frac{\horizon-\timestep-1}{\horizon}\varepsilon
    =
    \kantorovich{\costtogosmall{\timestep}}{\fleet_\timestep}{\refPVar}{} +
    \frac{\horizon-\timestep}{\horizon}\varepsilon.
\end{equation*}
This concludes the inductive step.
The bound at $\timestep=0$ gives $\costtogo{0}^{\varepsilon}(\fleet_0, \refPVar) \leq \kantorovich{\costtogosmall{0}}{\fleet_0}{\refPVar}{} + \varepsilon$.\qedhere
\end{proof}

\begin{remark}[Choosing $\varepsilon = 0$ under stronger regularity]
The proof of Part II only requires universally measurable
\(\varepsilon\)-optimal coupling selectors, which exist under the lower semianalytic assumptions used throughout Appendix~\ref{appendix:preliminary_lemmata}. Under the stronger assumption that the cost-to-go functions $\costtogosmall{\timestep}:\statespace\times \refSpace\to\nonnegativeRealsBar$
are lower semicontinuous and the spaces are compact, the infimum in $\kantorovich{\costtogosmall{\timestep}}{\fleet}{\refPVar}{}$ is attained for
each fixed $(\fleet,\refPVar)$ by the standard existence theorem for optimal
couplings~\cite[Theorem~4.1]{villani2009optimal}.
This pointwise existence is, however, not sufficient to choose $\varepsilon = 0$ in \sepconstruct{}: the construction requires \emph{measurable selectors} of exact optimizers. These exist when all spaces are compact, $\stagecostsmall{\timestep}, \terminalcostsmall, \dynamics{}{}{\timestep}$ are continuous, and $\noisedist{\timestep}$ is weakly continuous: then $\costtogosmall{\timestep}$ and the Bellman integrand are lower semicontinuous by backward induction, and exact Borel minimizing selectors exist \cite[Proposition~7.33]{bertsekas1996stochastic} (see also \cite[Theorem~18.19]{aliprantis06infinite}); the same argument applies to the coupling problem, where $\fleet \mapsto \plans{\fleet}{\refPVar}$ is a compact-valued continuous correspondence.
\end{remark}

\subsection{Proof of Corollary~\ref{corollary:agent_problem_upper_bound}}

\begin{proof}
We construct, for every $\varepsilon>0$, an admissible separation policy whose cost is within $\frac{(\horizon - \timestep)\varepsilon}{\horizon}$ of $\kantorovich{\costtogosmall{\timestep}}{\fleet_\timestep^\nagents}{\refPVar^\nagents}{}$. Since $\varepsilon>0$ is arbitrary, and $\costtogo{\timestep}^\nagents$ is the infimum of the expected cost over all admissible policies, the claim follows.
We start by giving an $\varepsilon$-optimal separation policy $\pi^\varepsilon_{\mathrm{sep},\timestep}$ which attains the upper bound. We then show that this policy is admissible, from which we directly obtain the desired inequality.

To define $\pi^\varepsilon_{\mathrm{sep},\timestep}$ we proceed as follows. 
First, since $\fleet_\timestep^\nagents$ and $\refPVar^\nagents$ are both empirical measures over $\nagents$ atoms, the optimal transport problem $\kantorovich{\costtogosmall{\timestep}}{\fleet_\timestep^\nagents}{\refPVar^\nagents}{}$ is a finite-dimensional transportation linear program. We work at the level of agent and target indices, i.e., we replace $\fleet_\timestep^\nagents$ and $\refPVar^\nagents$ by the uniform measures on $\{1,\ldots,\nagents\}$ and transport the cost $c_{ij} = \costtogosmall{\timestep}(\state_\timestep^i,\refVar^j)$. With this identification, every coupling between $\fleet_\timestep^\nagents$ and $\refPVar^\nagents$ can be represented by a matrix $\Gamma = (\gamma_{ij})_{i,j =1}^\nagents$ satisfying
$\gamma_{ij} \ge 0$, $\sum_{i=1}^\nagents \gamma_{ij}= \sum_{j=1}^\nagents \gamma_{ij} = \frac{1}{\nagents}$; equivalently, $\nagents\Gamma$ is doubly stochastic.
By the Birkhoff--von Neumann theorem \cite{birkhoff1946tres} applied to $\nagents\Gamma$, every such matrix is a convex combination of $\frac{1}{\nagents}$-scaled permutation matrices; since the objective is linear, the minimum is attained by a $\frac{1}{\nagents}$-scaled permutation matrix. Thus,
\begin{equation*}
    \kantorovich{\costtogosmall{\timestep}}{\fleet_\timestep^\nagents}{\refPVar^\nagents}{}
    =
    \frac{1}{\nagents}
    \min_{\sigma_\timestep \in S_\nagents} 
    \sum_{i=1}^\nagents
    \costtogosmall{\timestep}(\state_\timestep^i, \refVar^{\sigma_\timestep(i)}),
\end{equation*}
where $S_\nagents$ denotes the group of permutations of $\{1, \ldots, \nagents\}$.
Since $S_\nagents$ is finite and each map $\boldsymbol{\state} \mapsto \sum_{i=1}^\nagents \costtogosmall{\timestep}(\state^i, \refVar^{\sigma(i)})$ is universally measurable, a minimizing permutation $\sigma_\timestep^\ast \coloneqq \sigma_\timestep^\ast(\boldsymbol{\state}_\timestep)$ can be selected measurably (e.g., the first minimizer in a fixed ordering of $S_\nagents$); write $\refVar^{\ast, i}_\timestep \coloneqq \refVar^{\sigma_\timestep^\ast(i)}$ for the target assigned to agent $i$.
Given that agents are now assigned to targets, we can now consider the single-agent optimal control problem from $\state_\timestep^i$ to the corresponding target $\refVar^{\ast, i}$.
Specifically, for $\varepsilon>0$, there exists a universally measurable $\frac{\varepsilon}{\horizon}$-optimal action selector $\action_\timestep^{\varepsilon}\colon \statespace\times\refSpace \to \actionspace$, satisfying, for all $(\state,\refVar) \in \statespace\times\refSpace$:
\begin{equation*}
    \expectedvalue{
    \noise
    }{
    \stagecostsmall{\timestep}(\state, \action_\timestep^{\varepsilon}(\state,\refVar), \noise)
    +
    \costtogosmall{\nexttimestep}(\dynamics{\state}{\action_\timestep^{\varepsilon}(\state,\refVar), \noise}{\timestep}, \refVar)
    }
    \leq
    \costtogosmall{\timestep}(\state, \refVar) 
    +
    \frac{\varepsilon}{\horizon}.
\end{equation*}

At this point, we define the separation policy $\pi^\varepsilon_{\mathrm{sep},\timestep}$ at time $\timestep$. An agent at $\state_\timestep^{i}$ executes the action $\action_\timestep^{\varepsilon}(\state_\timestep^i,\refVar_\timestep^{\ast,i})$.
Informally, the optimal transport problem indicates the target, whereas the single-agent optimal control problem determines the optimal action to reach that target. 
This policy is deterministic, Markovian (it depends only on $\boldsymbol{\state}_\timestep$), and universally measurable, being the composition of $\action_\timestep^\varepsilon$ with the assignment map $\boldsymbol{\state}_\timestep \mapsto (\state_\timestep^i, \refVar_\timestep^{\ast, i})$; hence $\pi^\varepsilon_{\mathrm{sep}} \in \Pi$.

At this point, it suffices to verify that this policy yields the bound on the cost. In particular, we show that the expected cost-to-go function under this policy, denoted by $\costtogo{\timestep}^{\nagents,\varepsilon}(\boldsymbol{\state}_\timestep, \refPVar^\nagents)$, is upper bounded as follows: 
\begin{equation*}
    \costtogo{\timestep}^{\nagents,\varepsilon}(\boldsymbol{\state}_\timestep, \refPVar^\nagents)
    \leq
    \kantorovich{\costtogosmall{\timestep}}{\fleet_\timestep^\nagents}{\refPVar^\nagents}{}
    +
    \frac{\horizon-\timestep}{\horizon}\varepsilon,
\end{equation*}
which gives the desired upper bound. We proceed by induction. The equality at time $t=\horizon$ is trivial, as the terminal cost does not depend on the policy but only on the terminal state. Suppose then the statement holds at time $t+1$. By the policy-cost recursion (using the independence of the noise across time) and the induction hypothesis inside the expectation,
\begin{align*}
    \costtogo{\timestep}^{\nagents,\varepsilon}(\boldsymbol{\state}_\timestep, \refPVar^\nagents)
    &\leq
    \expectedvalue{\boldsymbol{\noise}_\timestep}{
        \frac{1}{\nagents}\sum_{i=1}^\nagents
        \stagecostsmall{\timestep}(\state_\timestep^i,\action_\timestep^i,\noise_\timestep^i)
        +
        \kantorovich{\costtogosmall{\nexttimestep}}{\fleet_{\nexttimestep}^\nagents}{\refPVar^\nagents}{}
    }
    +
    \frac{\horizon-\timestep-1}{\horizon}\varepsilon
    \\
    &=
    \frac{1}{\nagents}
    \expectedvalue{\boldsymbol{\noise}_\timestep}{
        \sum_{i=1}^\nagents
        \stagecostsmall{\timestep}(\state_\timestep^i,\action_\timestep^i,\noise_\timestep^i)
        +
        \min_{\sigma_{\nexttimestep}\in S_\nagents}\sum_{i=1}^\nagents\costtogosmall{\nexttimestep}(\dynamics{}{}{\timestep}(\state_\timestep^i,\action_\timestep^i,\noise_\timestep^i),\refVar^{\sigma_{\nexttimestep}(i)})
    }
    \\
    &\hspace{2cm}+
    \frac{\horizon-\timestep-1}{\horizon}\varepsilon
    \\
    &\leq
    \frac{1}{\nagents}
    \sum_{i=1}^\nagents
    \expectedvalue{\noise_\timestep^i}{
        \stagecostsmall{\timestep}(\state_\timestep^i,\action_\timestep^i,\noise_\timestep^i)
        +
        \costtogosmall{\nexttimestep}(\dynamics{}{}{\timestep}(\state_\timestep^i,\action_\timestep^i,\noise_\timestep^i),\refVar^{\sigma_\timestep^\ast(i)})
    }
    +
    \frac{\horizon-\timestep-1}{\horizon}\varepsilon
    \\
    &\leq
    \frac{1}{\nagents}
    \sum_{i=1}^\nagents
    \costtogosmall{\timestep}(\state_\timestep^i,\refVar^{\sigma_\timestep^\ast(i)})+\frac{\varepsilon}{\horizon}+
    \frac{\horizon-\timestep-1}{\horizon}\varepsilon
    \\
    &=
    \frac{1}{\nagents}
    \min_{\sigma\in S_\nagents}
    \sum_{i=1}^\nagents
    \costtogosmall{\timestep}(\state_\timestep^i,\refVar^{\sigma(i)})+
    \frac{\horizon-\timestep}{\horizon}\varepsilon,
\end{align*}
where the first inequality combines the policy-cost recursion with the inductive assumption, the second inequality follows since $\sigma_\timestep^\ast$ is a potentially suboptimal permutation for the assignment at time $\nexttimestep$, the last inequality follows from the $\frac{\varepsilon}{\horizon}$-optimality of the action selector $\action_\timestep^\varepsilon$ used by the policy $\pi^\varepsilon_{\mathrm{sep},\timestep}$, and the last equality follows from the definition of $\sigma_\timestep^\ast$.
Note that the expectation step above only uses the marginal law $\noise_\timestep^i \sim \noisedist{\timestep}(\cdot \st \state_\timestep^i, \action_\timestep^i)$ of each agent's noise together with the linearity of expectation; it is therefore valid under all noise models considered in this work (individual, local, and global), regardless of the correlation of the noise across agents.
This gives
\begin{equation*}
    \costtogo{\timestep}^{\nagents,\varepsilon}(\boldsymbol{\state}_\timestep, \refPVar^\nagents)
    \leq
     \kantorovich{\costtogosmall{\timestep}}{\fleet_\timestep^\nagents}{\refPVar^\nagents}{}
     +
     \frac{\horizon-\timestep}{\horizon}\varepsilon,
\end{equation*}
which completes the induction. 

Since $\varepsilon>0$ is arbitrary, and $\costtogo{\timestep}^\nagents$ is the infimum of the expected cost over all admissible policies, letting $\varepsilon\downarrow 0$ gives
\begin{equation*}
    \costtogo{\timestep}^\nagents(\boldsymbol{\state}_\timestep, \refPVar^\nagents)
    \leq
     \kantorovich{\costtogosmall{\timestep}}{\fleet_\timestep^\nagents}{\refPVar^\nagents}{}.
\end{equation*}
\end{proof}

\section{Quantifying the tightness of the optimal transport upper bound}
\label{section:bound_quantification_v4}

The inequality in the proof of \sepbound{} emerges from the fact that we remove the
dependence of the target assignment on the realized noise. More precisely, we first apply
\cref{lemma:marginalinf} to remove the transport plan dependence on the noise in
the term $(\ast)$ of~\eqref{eq:inductionstep}, and then apply \cref{lemma:infexp}
to allow the transport plan to be chosen before the noise is realized.
In this appendix, we quantify the loss introduced by these two steps.

Throughout the appendix, we retain
\cref{assumption:measurable_field,assumption:barycenter_condition} and write
$\empnoisedist{\tau}
=\big(\empnoisedist{\tau}^{\state,\action}\big)_{(\state,\action)}$ for the
realized noise field. Recall that \cref{assumption:barycenter_condition} reads
\begin{equation*}
    \mathbb E_{\distoverdist{\tau}}\left[
        \int_{\noisespace}\varphi(\noise)
        \dd\empnoisedist{\tau}^{\state,\action}(\noise)
    \right]
    =
    \int_{\noisespace}\varphi(\noise)
    \dd\noisedist{\tau}(\noise\st\state,\action)
\end{equation*}
for every bounded Borel function $\varphi$ and almost every
$(\state,\action)$. By truncation and monotone convergence, the same identity
holds for every nonnegative Borel $\varphi$, which is the form we use below.

\begin{assumption}[Lipschitz single-agent cost-to-go]
\label{assumption:bq4:cost_to_go_lipschitz}
Equip $\statespace$ with a metric $d_{\statespace}$. We assume that for every
$\tau\in\{1,\ldots,\horizon\}$ and for some $L_\tau<\infty$,
\begin{equation}
\label{eq:bq4:cost_to_go_lipschitz}
    \left|
        \costtogosmall{\tau}(\state,\refVar)
        -\costtogosmall{\tau}(\state',\refVar)
    \right|
    \leq
    L_\tau d_{\statespace}(\state,\state')
\end{equation}
for all $\state,\state'\in\statespace$ and $\refVar\in\refSpace$.
\end{assumption}

We need the constant to be uniform in $\refVar$ because the target assigned
to an agent may itself be reassigned once the noise has been observed. The
assumption holds, for instance, whenever all the underlying spaces are finite and
the single-agent cost-to-go functions are finite-valued.

For each $\tau<\horizon$, we measure the spread of the transition by
\begin{equation}
\label{eq:bq4:pairwise_dispersion}
    \dispersion{\tau}
    \coloneqq
    \sup_{(\state,\action)\in\statespace\times\actionspace}
    \int_{\noisespace\times\noisespace}
    d_{\statespace}\!\left(
        \dynamics{\state}{\action,\noise}{\tau},
        \dynamics{\state}{\action,\tilde\noise}{\tau}
    \right)
    \dd\noisedist{\tau}(\noise\st\state,\action)
    \dd\noisedist{\tau}(\tilde\noise\st\state,\action),
\end{equation}
that is, by the expected distance between two conditionally independent next
states generated from the same state-action pair. We assume throughout that
$\dispersion{\tau}$, together with the expectations and transport costs appearing
in the comparison below, is finite. This is the case, for example, when
$\noise\mapsto\dynamics{\state}{\action,\noise}{\tau}$ is Lipschitz uniformly in
$(\state,\action)$, the nominal kernels have uniformly bounded first moments, and
$\kantorovich{\costtogosmall{\tau}}{\fleet}{\refPVar}{}<\infty$, or more simply
when all the underlying spaces are compact and data are continuous.

For the following analysis, it is convenient to write the
recursion~\eqref{eq:dp:population} in operator form.
For a next-stage function $F$ and
$\rho_\tau^\lambda\coloneqq
\lambda\otimes\empnoisedist{\tau}^{\state,\action}$, we hence set
\begin{equation*}
\begin{aligned}
 (\mathcal T_\tau F)(\fleet,\refPVar)
 \coloneqq \inf_{\lambda\in M(\fleet)}
 \mathbb E_{\distoverdist{\tau}}\!\left[
   \int_{\statespace\times\actionspace\times\noisespace}
   \stagecostsmall{\tau}(\state,\action,\noise)
   \dd\rho_\tau^\lambda(\state,\action,\noise)
   +F\!\left(
     \pushforward{\dynamics{}{}{\tau}}{\rho_\tau^\lambda},\refPVar
   \right)
 \right],
\end{aligned}
\end{equation*}
so that $\costtogo{\tau}=\mathcal T_\tau\costtogo{\tau+1}$
by~\eqref{eq:dp:population}.
The population Bellman operator $\mathcal T_\tau$ satisfies the monotonicity and constant-shift
properties that are standard for dynamic programming
operators~\cite[Ch.~8]{bertsekas1996stochastic}. More precisely, the following lemma holds.

\begin{lemma}[Monotonicity and constant shift of the population Bellman operator]
\label{lemma:bq4:operator_properties}
Let $F,\widetilde F\colon\probabilityspace{\statespace}\times
\probabilityspace{\refSpace}\to\realsBar$ be universally measurable and bounded
below, and let $c\in\reals$. Then
\begin{enumerate}[label=(\roman*),leftmargin=*]
\item $F\leq\widetilde F$ pointwise implies
$\mathcal T_\tau F\leq\mathcal T_\tau\widetilde F$ pointwise;
\item $\mathcal T_\tau[F+c]=\mathcal T_\tau[F]+c$.
\end{enumerate}
\end{lemma}

\begin{proof}
Both properties are immediate from the definition of $\mathcal T_\tau$: the
next-stage function enters the right-hand side additively and monotonically,
while a constant depends neither on the realized next population nor on
$\lambda$, so it factors out of the expectation and out of the infimum. The
argument uses only the structure of the definition.
\end{proof}
Finally, we also recall that
\begin{equation*}
    \kantorovich{\costtogosmall{\tau}}{\fleet}{\refPVar}{}
    =\inf_{\plan{\gamma}\in\plans{\fleet}{\refPVar}}
    \int_{\statespace\times\refSpace}
    \costtogosmall{\tau}(\state,\refVar)
    \dd\plan{\gamma}(\state,\refVar).
\end{equation*}
Here, $\plan{\gamma}$ assigns the mass in the current state distribution
$\fleet$ to the target distribution $\refPVar$, and
$\costtogosmall{\tau}(\state,\refVar)$ is the single-agent cost associated with
each state--target pair. This assignment is made before the noise at time
$\tau$ is observed. In contrast,
$\mathcal T_\tau[\kantorovich{\costtogosmall{\tau+1}}{}{}{}]$ assigns targets
after the realized noise has determined the next population. The difference
in \eqref{eq:bq4:bellman_defect} is the cost of not being able to use that
realized noise when assigning targets.

\begin{lemma}[One-step Bellman defect]
\label{lemma:bq4:bellman_defect}
Under
\cref{assumption:measurable_field,assumption:barycenter_condition,%
assumption:bq4:cost_to_go_lipschitz}, for every
$\tau<\horizon$, $\fleet\in\probabilityspace{\statespace}$, and
$\refPVar\in\probabilityspace{\refSpace}$, it holds that
\begin{equation}
\label{eq:bq4:bellman_defect}
    0
    \leq
    \kantorovich{\costtogosmall{\tau}}{\fleet}{\refPVar}{}
    -\mathcal T_\tau\!\left[
      \kantorovich{\costtogosmall{\tau+1}}{}{}{}
    \right](\fleet,\refPVar)
    \leq
    L_{\tau+1}\dispersion{\tau}.
\end{equation}
\end{lemma}

\begin{proof}
Throughout this proof we abbreviate the state-action pair as $x=(\state,\action)$. We start by showing that the two
quantities in \eqref{eq:bq4:bellman_defect} have the same expected stage cost.
As a consequence, we only need to compare how the next-stage single-agent cost-to-go
$\costtogosmall{\tau+1}(\dynamics{\state}{\action,\noise}{\tau},\refVar)$
is minimized in the two transport problems. In $\kantorovich{\costtogosmall{\tau}}{\fleet}{\refPVar}{}$,
the target assignment depends only on $x$, whereas in $\mathcal T_\tau\!\left[ \kantorovich{\costtogosmall{\tau+1}}{}{}{} \right](\fleet,\refPVar)$
it may also depend on the realized noise.
The second minimization is therefore over a larger set
of assignments and cannot have a larger value. In the following proof, we show that the
difference between the two values is at most
$L_{\tau+1}\dispersion{\tau}$.

\medskip
\noindent\textbf{Step 1: Define the two next-stage transport problems.}
Fix $\lambda\in M(\fleet)$. If the target is
assigned using $x$ alone, the expected next-stage cost-to-go is
\begin{equation*}
    A_\tau(\lambda)
    \coloneqq
    \inf_{\plan{\gamma}\in\plans{\lambda}{\refPVar}}
    \int_{(\statespace\times\actionspace)\times\refSpace}
    \int_{\noisespace}
    \costtogosmall{\tau+1}
    \bigl(\dynamics{\state}{\action,\tilde\noise}{\tau},\refVar\bigr)
    \dd\noisedist{\tau}(\tilde\noise\st x)
    \dd\plan{\gamma}(x,\refVar).
\end{equation*}
Here, the plan $\plan{\gamma}$ assigns a target $\refVar$ to each
$x=(\state,\action)$, before the noise $\tilde\noise$ is drawn from the nominal
kernel. Hence, this target assignment does not depend on $\tilde\noise$.

On the other hand, if the noise field is observed before the target is assigned, the transport
plan may couple the target with the full pair $(x,\noise)$. For a fixed
realization of the noise field, the smallest next-stage cost-to-go is then
\begin{equation*}
    A_\tau(\lambda;\empnoisedist{\tau})
    \coloneqq
    \inf_{\plan{\gamma}'\in
        \plans{\lambda\otimes\empnoisedist{\tau}^{x}}{\refPVar}}
    \int_{(\statespace\times\actionspace)\times\noisespace\times\refSpace}
    \costtogosmall{\tau+1}
    \bigl(\dynamics{\state}{\action,\noise}{\tau},\refVar\bigr)
    \dd\plan{\gamma}'(x,\noise,\refVar).
\end{equation*}
Indeed, since the first marginal of $\plan{\gamma}'$ is
$\lambda\otimes\empnoisedist{\tau}^{x}$, the transport plan $\plan{\gamma}'$ assigns a target
after both $x$ and its realized noise $\noise$ are known.

\medskip
\noindent\textbf{Step 2: Decompose the transport cost.}
We now verify that these two quantities are exactly the next-stage terms
that appear in \eqref{eq:bq4:bellman_defect}. We begin with
$\kantorovich{\costtogosmall{\tau}}{\fleet}{\refPVar}{}$. Define
\begin{equation*}
    c_\tau(\state,\action,\refVar)
    \coloneqq
    \int_{\noisespace}
      \stagecostsmall{\tau}(\state,\action,\noise)
      +\costtogosmall{\tau+1}
       \bigl(\dynamics{\state}{\action,\noise}{\tau},\refVar\bigr)
    \dd\noisedist{\tau}(\noise\st\state,\action).
\end{equation*}
By the single-agent Bellman recursion we have
\begin{equation*}
    \costtogosmall{\tau}(\state,\refVar)
    =\inf_{\action\in\actionspace}
      c_\tau(\state,\action,\refVar),
\end{equation*}
and plugging it into the definition of the transport cost gives
\begin{equation*}
    \kantorovich{\costtogosmall{\tau}}{\fleet}{\refPVar}{}
    =\inf_{\plan{\gamma}_1\in\plans{\fleet}{\refPVar}}
      \int_{\statespace\times\refSpace}
      \inf_{\action\in\actionspace}
      c_\tau(\state,\action,\refVar)
      \dd\plan{\gamma}_1(\state,\refVar).
\end{equation*}

The function $c_\tau$ is lower semianalytic under the conventions of
Appendix~\ref{appendix:preliminary_lemmata}, therefore we can apply
\cref{lemma:global_to_pointwise}. In particular, for every fixed
$\plan{\gamma}_1\in\plans{\fleet}{\refPVar}$,
\begin{equation*}
    \int_{\statespace\times\refSpace}
      \inf_{\action\in\actionspace}
      c_\tau(\state,\action,\refVar)
      \dd\plan{\gamma}_1(\state,\refVar)
    =
  \inf_{\{\alpha^{\state,\refVar}\}\in\Lambda(\actionspace)}
  \int_{\statespace\times\refSpace}\int_{\actionspace}
    c_\tau(\state,\action,\refVar)
    \dd\alpha^{\state,\refVar}(\action)
    \dd\plan{\gamma}_1(\state,\refVar).
\end{equation*}

Here, $\alpha^{\state,\refVar}$ is a Borel kernel that chooses an action
for each state--target pair. Now, the pair
$(\plan{\gamma}_1,\alpha^{\state,\refVar})$ defines a measure
$\plan{\gamma}$ on
$(\statespace\times\actionspace)\times\refSpace$ by
\begin{equation*}
    \dd\plan{\gamma}(\state,\action,\refVar)
    =\dd\alpha^{\state,\refVar}(\action)
     \dd\plan{\gamma}_1(\state,\refVar),
\end{equation*}
and we can rewrite its marginals in the form used by the population Bellman problem.
To do so, let $\lambda$ be the $(\state,\action)$-marginal of
$\plan{\gamma}$. Since the state marginal of $\plan{\gamma}_1$ is $\fleet$,
the state marginal of $\lambda$ is also $\fleet$, and therefore
$\lambda\in M(\fleet)$. Combining this with the fact that the target marginal of $\plan{\gamma}$ is
$\refPVar$, we obtain $\plan{\gamma}\in\plans{\lambda}{\refPVar}$.
Conversely, from every pair consisting of $\lambda\in M(\fleet)$
and $\plan{\gamma}\in\plans{\lambda}{\refPVar}$, we can obtain a
coupling $\plan{\gamma}_1\in\plans{\fleet}{\refPVar}$ by taking the
$(\state,\refVar)$-marginal of $\plan{\gamma}$, and a Borel action kernel
$\alpha^{\state,\refVar}$ by disintegrating $\plan{\gamma}$ with respect
to $(\state,\refVar)$.
Therefore, optimizing over $(\plan{\gamma}_1,\alpha^{\state,\refVar})$
is equivalent to optimizing over $\lambda\in M(\fleet)$ and
$\plan{\gamma}\in\plans{\lambda}{\refPVar}$.

The transport problem can therefore be written as
\begin{equation*}
\begin{aligned}
\kantorovich{\costtogosmall{\tau}}{\fleet}{\refPVar}{}
=\inf_{\lambda\in M(\fleet)}
  \inf_{\plan{\gamma}\in\plans{\lambda}{\refPVar}}
  \int_{(\statespace\times\actionspace)\times\refSpace}
  c_\tau(\state,\action,\refVar)
  \dd\plan{\gamma}(\state,\action,\refVar).
\end{aligned}
\end{equation*}
The outer infimum chooses an admissible state-action distribution
$\lambda$, while the inner infimum assigns its mass to the target distribution
$\refPVar$.

We can now expand the integral of $c_\tau$. Since the stage cost does not
depend on the target $\refVar$, its integral depends only on the
$(\state,\action)$-marginal $\lambda$. We denote it as
\begin{equation*}
    G_\tau(\lambda)
    \coloneqq
    \int_{\statespace\times\actionspace}\int_{\noisespace}
    \stagecostsmall{\tau}(\state,\action,\noise)
    \dd\noisedist{\tau}(\noise\st x)\dd\lambda(x).
\end{equation*}
Therefore
\begin{equation*}    \int_{(\statespace\times\actionspace)\times\refSpace}
    c_\tau(\state,\action,\refVar)\dd\plan{\gamma}(x,\refVar)
    =
    G_\tau(\lambda)
    +\int_{(\statespace\times\actionspace)\times\refSpace}
    \int_{\noisespace}
    \costtogosmall{\tau+1}
    \bigl(\dynamics{\state}{\action,\tilde\noise}{\tau},\refVar\bigr)
    \dd\noisedist{\tau}(\tilde\noise\st x)
    \dd\plan{\gamma}(x,\refVar).
\end{equation*}

For a given $\lambda$, the first term is fixed. Minimizing the second
term over $\plan{\gamma}\in\plans{\lambda}{\refPVar}$ gives
$A_\tau(\lambda)$, and minimizing again over $\lambda\in M(\fleet)$ gives
\begin{equation}
\label{eq:bq4:surrogate_decomposition}
    \kantorovich{\costtogosmall{\tau}}{\fleet}{\refPVar}{}
    =\inf_{\lambda\in M(\fleet)}\left\{
    G_\tau(\lambda)
    +A_\tau(\lambda)\right\}.
\end{equation}

\medskip
\noindent\textbf{Step 3: Decompose the population Bellman term.}
We now turn to the second quantity in
\eqref{eq:bq4:bellman_defect}. From the definition of $\mathcal T_\tau$,
\begin{align*}
\mathcal T_\tau\!\left[
  \kantorovich{\costtogosmall{\tau+1}}{}{}{}
\right]&(\fleet,\refPVar)\\
&=
\inf_{\lambda\in M(\fleet)}
\mathbb E_{\distoverdist{\tau}}\!\left[
\int_{\statespace\times\actionspace\times\noisespace}
\stagecostsmall{\tau}(\state,\action,\noise)
\dd(\lambda\otimes\empnoisedist{\tau}^{x})(x,\noise)
+\kantorovich{\costtogosmall{\tau+1}}
 {\pushforward{\dynamics{}{}{\tau}}
  {\lambda\otimes\empnoisedist{\tau}^{x}}}
 {\refPVar}{}
\right].
\end{align*}

We look at the two terms in the expectation separately.
The expectation of the stage cost can be rewritten using \cref{assumption:barycenter_condition} as
\begin{equation*}
\mathbb E_{\distoverdist{\tau}}\!\left[
\int_{\statespace\times\actionspace\times\noisespace}
\stagecostsmall{\tau}(\state,\action,\noise)
\dd(\lambda\otimes\empnoisedist{\tau}^{x})(x,\noise)
\right]
=
G_\tau(\lambda).
\end{equation*}

For the transport term, we fix $\lambda\in M(\fleet)$ and a realization of the noise field.
Since the term couples the next-state distribution
$\pushforward{\dynamics{}{}{\tau}}
{\lambda\otimes\empnoisedist{\tau}^{x}}$ with the target distribution
$\refPVar$, we can apply \cref{lemma:pushandOT}.
The lemma allows us to describe the same transport
problem before applying the dynamics by coupling $(x,\noise)$ directly with a
target $\refVar$ and evaluating the cost at the resulting next state
$\dynamics{\state}{\action,\noise}{\tau}$. By doing so we directly obtain 
$A_\tau(\lambda;\empnoisedist{\tau})$. In particular, we have
\begin{align*}
\kantorovich{\costtogosmall{\tau+1}}
 {\pushforward{\dynamics{}{}{\tau}}
  {\lambda\otimes\empnoisedist{\tau}^{x}}}
 {\refPVar}{} &=
\inf_{\plan{\gamma}'\in
  \plans{\lambda\otimes\empnoisedist{\tau}^{x}}{\refPVar}}
\int_{(\statespace\times\actionspace)\times\noisespace\times\refSpace}
\costtogosmall{\tau+1}
\bigl(\dynamics{\state}{\action,\noise}{\tau},\refVar\bigr)
\dd\plan{\gamma}'(x,\noise,\refVar)\\
&=A_\tau(\lambda;\empnoisedist{\tau}).
\end{align*}

Substituting these two equalities into the definition of
$\mathcal T_\tau$ gives
\begin{equation}
\label{eq:bq4:population_decomposition}
    \mathcal T_\tau\!\left[
      \kantorovich{\costtogosmall{\tau+1}}{}{}{}
    \right](\fleet,\refPVar)
    =\inf_{\lambda\in M(\fleet)}\left\{
    G_\tau(\lambda)
    +\mathbb E_{\distoverdist{\tau}}
     \bigl[A_\tau(\lambda;\empnoisedist{\tau})\bigr]\right\}.
\end{equation}

\medskip
\noindent\textbf{Step 4: Compare the next-stage costs for a fixed
$\lambda$.}
Now, since \eqref{eq:bq4:surrogate_decomposition} 
and \eqref{eq:bq4:population_decomposition} contain the same stage-cost integrals $G_\tau(\lambda)$, it is enough to 
show that, for every $\lambda\in M(\fleet)$,
\begin{equation}
\label{eq:bq4:next_stage_comparison}
    0
    \leq
    A_\tau(\lambda)
    -
    \mathbb E_{\distoverdist{\tau}}
    \left[
        A_\tau(\lambda;\empnoisedist{\tau})
    \right]
    \leq
    L_{\tau+1}\dispersion{\tau}.
\end{equation}
Indeed, after adding the common stage-cost term, the inequalities remain valid
for every $\lambda$, and taking the infimum over $\lambda\in M(\fleet)$ gives
\eqref{eq:bq4:bellman_defect}.

\smallskip
\noindent\textbf{Upper bound.}
We first prove the upper bound. Fix a realization of the noise field and take
any $\plan{\gamma}'$ admissible for
$A_\tau(\lambda;\empnoisedist{\tau})$. Since $x=(\state,\action)$ and the $(x,\refVar)$-marginal 
of $\plan{\gamma}'$ has marginals $\lambda$ and $\refPVar$, respectively, 
it is admissible for $A_\tau(\lambda)$. 
We can therefore evaluate $A_\tau(\lambda)$ at this $(x,\refVar)$-marginal
and draw $\tilde\noise$ independently from $\noisedist{\tau}(\cdot\st x)$,
which gives
\begin{align*}
    A_\tau(\lambda)
    &\leq
    \int_{(\statespace\times\actionspace)\times\noisespace\times\refSpace}
    \int_{\noisespace}
    \costtogosmall{\tau+1}
    \bigl(\dynamics{\state}{\action,\tilde\noise}{\tau},\refVar\bigr)
    \dd\noisedist{\tau}(\tilde\noise\st x)
    \dd\plan{\gamma}'(x,\noise,\refVar)
    \\
    &\leq
    \int_{(\statespace\times\actionspace)\times\noisespace\times\refSpace}
    \costtogosmall{\tau+1}
    \bigl(\dynamics{\state}{\action,\noise}{\tau},\refVar\bigr)
    \dd\plan{\gamma}'(x,\noise,\refVar)
    \\
    &\quad+
    L_{\tau+1}
    \int_{(\statespace\times\actionspace)\times\noisespace}
    \int_{\noisespace}
    d_{\statespace}\!\left(
        \dynamics{\state}{\action,\noise}{\tau},
        \dynamics{\state}{\action,\tilde\noise}{\tau}
    \right)
    \dd\noisedist{\tau}(\tilde\noise\st x)
    \dd(\lambda\otimes\empnoisedist{\tau}^{x})(x,\noise),
\end{align*}
where the second inequality follows from \cref{assumption:bq4:cost_to_go_lipschitz}.

Looking at the obtained upper bound, we can observe that 
the first integral is the value of a fixed $\plan{\gamma}'$ in the definition of
$A_\tau(\lambda;\empnoisedist{\tau})$. The last integral, instead, does not contain
$\refVar$, and so does not depend on how $\plan{\gamma}'$ assigns targets. 
We can therefore minimize over $\plan{\gamma}'$ and obtain
\begin{equation*}
\begin{aligned}
A_\tau(\lambda)
&\leq A_\tau(\lambda;\empnoisedist{\tau})\\
&\quad+L_{\tau+1}
\int_{(\statespace\times\actionspace)\times\noisespace}
\int_{\noisespace}
d_{\statespace}\!\left(
\dynamics{\state}{\action,\noise}{\tau},
\dynamics{\state}{\action,\tilde\noise}{\tau}
\right)
\dd\noisedist{\tau}(\tilde\noise\st x)
\dd(\lambda\otimes\empnoisedist{\tau}^{x})(x,\noise).
\end{aligned}
\end{equation*}

Averaging over the noise field and applying \cref{assumption:barycenter_condition}
to replace $\empnoisedist{\tau}^{x}$ by $\noisedist{\tau}(\cdot\st x)$ in the last
integral gives
\begin{equation*}
\begin{aligned}
&\mathbb E_{\distoverdist{\tau}}\!\left[
\int_{(\statespace\times\actionspace)\times\noisespace}
\int_{\noisespace}
d_{\statespace}\!\left(
\dynamics{\state}{\action,\noise}{\tau},
\dynamics{\state}{\action,\tilde\noise}{\tau}
\right)
\dd\noisedist{\tau}(\tilde\noise\st x)
\dd(\lambda\otimes\empnoisedist{\tau}^{x})(x,\noise)
\right]\\
&\quad=
\int_{\statespace\times\actionspace}
\int_{\noisespace\times\noisespace}
d_{\statespace}\!\left(
\dynamics{\state}{\action,\noise}{\tau},
\dynamics{\state}{\action,\tilde\noise}{\tau}
\right)
\dd\noisedist{\tau}(\noise\st x)
\dd\noisedist{\tau}(\tilde\noise\st x)
\dd\lambda(x)\\
&\quad\leq\dispersion{\tau},
\end{aligned}
\end{equation*}
where the last inequality follows from the supremum in
\eqref{eq:bq4:pairwise_dispersion}. We have therefore proved
the upper bound
\begin{equation*}
    A_\tau(\lambda)
    \leq
    \mathbb E_{\distoverdist{\tau}}
    \left[A_\tau(\lambda;\empnoisedist{\tau})\right]
    +L_{\tau+1}\dispersion{\tau}.
\end{equation*}

\smallskip
\noindent\textbf{Lower bound.}
For the lower bound in \eqref{eq:bq4:next_stage_comparison}, fix any
$\plan{\gamma}\in\plans{\lambda}{\refPVar}$. After the noise field is realized,
we can couple this plan with the realized noise kernel by setting
\begin{equation*}
    \dd\plan{\gamma}_{\empnoisedist{\tau}}'
    (x,\noise,\refVar)
    \coloneqq
    \dd\plan{\gamma}(x,\refVar)
    \dd\empnoisedist{\tau}^{x}(\noise).
\end{equation*}
The resulting plan is admissible for
$A_\tau(\lambda;\empnoisedist{\tau})$ since its $(x,\noise)$-marginal is
$\lambda\otimes\empnoisedist{\tau}^{x}$, because the $x$-marginal of
$\plan{\gamma}$ is $\lambda$, and its target marginal is $\refPVar$. 
Evaluating the infimum at this plan,
averaging over the noise field, and then
applying \cref{lemma:noiseexp} with $(x,\refVar)$ as the base variable
and
$\costtogosmall{\tau+1}
(\dynamics{\state}{\action,\noise}{\tau},\refVar)$ as the cost,
gives
\begin{align*}
    \mathbb E_{\distoverdist{\tau}}
    \left[A_\tau(\lambda;\empnoisedist{\tau})\right]
    &\leq
    \mathbb E_{\distoverdist{\tau}}
    \left[
        \int_{(\statespace\times\actionspace)\times\noisespace\times\refSpace}
        \costtogosmall{\tau+1}
        \bigl(\dynamics{\state}{\action,\noise}{\tau},\refVar\bigr)
        \dd\plan{\gamma}_{\empnoisedist{\tau}}'
        (x,\noise,\refVar)
    \right]
    \\
    &\quad=
    \int_{(\statespace\times\actionspace)\times\refSpace}
    \int_{\noisespace}
    \costtogosmall{\tau+1}
    \bigl(\dynamics{\state}{\action,\tilde\noise}{\tau},\refVar\bigr)
    \dd\noisedist{\tau}(\tilde\noise\st x)
    \dd\plan{\gamma}(x,\refVar).
\end{align*}

The right-hand side corresponds exactly to the value of the fixed 
$\plan{\gamma}$ in the definition of $A_\tau(\lambda)$, and, since 
the inequality holds for every $\plan{\gamma}\in\plans{\lambda}{\refPVar}$, 
minimizing over $\plan{\gamma}$ yields
\begin{equation*}
    \mathbb E_{\distoverdist{\tau}}
    \left[A_\tau(\lambda;\empnoisedist{\tau})\right]
    \leq A_\tau(\lambda).
\end{equation*}

\medskip
\noindent\textbf{Step 5: Return to the two Bellman quantities.}
This proves \eqref{eq:bq4:next_stage_comparison}. To conclude, for every $\lambda\in M(\fleet)$,
adding the common stage-cost expectation to \eqref{eq:bq4:next_stage_comparison}
gives
\begin{equation*}
\begin{aligned}
G_\tau(\lambda)
+\mathbb E_{\distoverdist{\tau}}
  \left[A_\tau(\lambda;\empnoisedist{\tau})\right]
&\leq G_\tau(\lambda)+A_\tau(\lambda)\\
&\leq G_\tau(\lambda)
+\mathbb E_{\distoverdist{\tau}}
  \left[A_\tau(\lambda;\empnoisedist{\tau})\right]
+L_{\tau+1}\dispersion{\tau}.
\end{aligned}
\end{equation*}
Taking the infimum over $\lambda$ in all three terms and using
\eqref{eq:bq4:surrogate_decomposition} and
\eqref{eq:bq4:population_decomposition} gives exactly
\eqref{eq:bq4:bellman_defect}.
\end{proof}

Lemma~\ref{lemma:bq4:bellman_defect} bounds, for one time step, the
difference between assigning targets before and after the noise is observed.
Applying this bound successively from time $\timestep$ to time $\horizon-1$
gives the total difference stated in the following theorem.
\begin{mdframed}[hidealllines=true,backgroundcolor=blue!5,nobreak=true]
\begin{theorem}[Quantification of the separation gap]
\label{theorem:bq4:direct_bound}
Under \cref{assumption:measurable_field,assumption:barycenter_condition,%
assumption:bq4:cost_to_go_lipschitz}, for every
$\timestep\in\{0,\ldots,\horizon\}$, every
$\fleet_\timestep\in\probabilityspace{\statespace}$, and every
$\refPVar\in\probabilityspace{\refSpace}$,
\begin{equation}
\label{eq:bq4:main_bound}
    0
    \leq
    \kantorovich{\costtogosmall{\timestep}}
    {\fleet_\timestep}{\refPVar}{}
    -\costtogo{\timestep}(\fleet_\timestep,\refPVar)
    \leq
    \sum_{\tau=\timestep}^{\horizon-1}
    L_{\tau+1}\dispersion{\tau}.
\end{equation}
\end{theorem}
\end{mdframed}

\begin{proof}
We proceed by backward induction. At the horizon the two sides agree, since
$\costtogosmall{\horizon}=\terminalcostsmall$ and therefore
$\kantorovich{\costtogosmall{\horizon}}{\fleet}{\refPVar}{}
=\costtogo{\horizon}(\fleet,\refPVar)$. 
Assume next that~\eqref{eq:bq4:main_bound} holds at time $\tau+1$, pointwise in the
population and target distributions, so that
\begin{equation*}
    0
    \leq
    \kantorovich{\costtogosmall{\tau+1}}{}{}{}
    -\costtogo{\tau+1}
    \leq
    \sum_{r=\tau+1}^{\horizon-1}L_{r+1}\dispersion{r}.
\end{equation*}
Since $\mathcal T_\tau$ is monotone and commutes with the addition of
constants by \cref{lemma:bq4:operator_properties}, applying it to both sides gives
\begin{equation}
\label{eq:bq4:propagated_error}
    0
    \leq
    \mathcal T_\tau\!\left[
      \kantorovich{\costtogosmall{\tau+1}}{}{}{}
    \right]
    -\mathcal T_\tau\costtogo{\tau+1}
    \leq
    \sum_{r=\tau+1}^{\horizon-1}L_{r+1}\dispersion{r}.
\end{equation}
We can now compare $\kantorovich{\costtogosmall{\tau}}{}{}{} -\costtogo{\tau}$.
In particular, using
$\costtogo{\tau}=\mathcal T_\tau\costtogo{\tau+1}$ and \eqref{eq:bq4:bellman_defect} and
\eqref{eq:bq4:propagated_error}, we obtain
\begin{equation*}
\begin{aligned}
    0
    &\leq
    \kantorovich{\costtogosmall{\tau}}{}{}{}
    -\costtogo{\tau}
    \\
    &\leq
    \left( \kantorovich{\costtogosmall{\tau}}{}{}{}
    - \mathcal T_\tau\!\left[
      \kantorovich{\costtogosmall{\tau+1}}{}{}{}
    \right] \right)
    + \left( \mathcal T_\tau\!\left[
      \kantorovich{\costtogosmall{\tau+1}}{}{}{}
    \right]
    - \mathcal T_\tau\costtogo{\tau+1}
    \right)\\
    &\leq
    L_{\tau+1}\dispersion{\tau}
    +\sum_{r=\tau+1}^{\horizon-1}L_{r+1}\dispersion{r}
    =\sum_{r=\tau}^{\horizon-1}L_{r+1}\dispersion{r},
\end{aligned}
\end{equation*}
which is~\eqref{eq:bq4:main_bound} at time $\tau$ and completes the
induction.
\end{proof}

\begin{corollary}[Finite-agent consequence]
\label{corollary:bq4:finite_bound}
Let $\boldsymbol{\state}_\timestep
=(\state_\timestep^1,\ldots,\state_\timestep^\nagents)\in\statespace^\nagents$ be
the agent states at time $\timestep$, with empirical distribution
$\fleet_\timestep^\nagents
=\frac1\nagents\sum_{i=1}^\nagents\delta_{\state_\timestep^i}$, and let
$\refPVar^\nagents$ be an $\nagents$-atom target distribution as in
\cref{corollary:agent_problem_upper_bound}. Under the assumptions of
\cref{theorem:bq4:direct_bound}, it holds that
\begin{equation}
\label{eq:bq4:finite_bound}
    0
    \leq
    \kantorovich{\costtogosmall{\timestep}}
    {\fleet_\timestep^\nagents}{\refPVar^\nagents}{}
    -\costtogo{\timestep}^\nagents
    (\boldsymbol{\state}_\timestep,\refPVar^\nagents)
    \leq
    \sum_{\tau=\timestep}^{\horizon-1}
    L_{\tau+1}\dispersion{\tau}.
\end{equation}
\end{corollary}

\begin{proof}
For a joint noise law and costs that are invariant under permutations of agent labels, the finite-agent value depends only on the empirical state. We may therefore choose an $\varepsilon$-optimal policy whose empirical state–action measure is a feedback function of that state. This measure is admissible for the population problem and has the same cost, so the population problem is a relaxation of the agent-based one.
Together with \cref{corollary:agent_problem_upper_bound}, this gives
\begin{equation*}
    \costtogo{\timestep}
    (\fleet_\timestep^\nagents,\refPVar^\nagents)
    \leq
    \costtogo{\timestep}^\nagents
    (\boldsymbol{\state}_\timestep,\refPVar^\nagents)
    \leq
    \kantorovich{\costtogosmall{\timestep}}
    {\fleet_\timestep^\nagents}{\refPVar^\nagents}{}.
\end{equation*}
Subtracting the middle term from the transport cost and
applying~\eqref{eq:bq4:main_bound} at
$(\fleet_\timestep^\nagents,\refPVar^\nagents)$
yields~\eqref{eq:bq4:finite_bound}.
\end{proof}

\paragraph{Interpretation.}
The two sides of~\eqref{eq:bq4:bellman_defect} differ because the
population Bellman recursion may reassign targets after seeing the
realized next population, whereas the transport bound commits to an
assignment before the nominal noise is integrated. The bound in
\cref{theorem:bq4:direct_bound} gives a clear interpretation of this loss.
In particular, $L_{\tau+1}$ measures how sensitive the remaining cost is to
a displacement of an agent, while $\dispersion{\tau}$ measures how far
the noise can push apart two agents that start from the same state-action pair.
The gap therefore closes for deterministic dynamics, and more generally whenever
the optimal assignment does not depend on the realized noise.

\paragraph{Bounding the dispersion term.}
The definition of $\dispersion{\tau}$ involves the distance between next
states. This quantity can be bounded directly in terms of the noise when
$\noisespace$ carries a metric $d_{\noisespace}$ and the dynamics are
$K_\tau$-Lipschitz in their noise argument, uniformly in
$(\state,\action)$:
\begin{equation*}
    d_{\statespace}\!\left(
        \dynamics{\state}{\action,\noise}{\tau},
        \dynamics{\state}{\action,\tilde\noise}{\tau}
    \right)
    \leq
    K_\tau d_{\noisespace}(\noise,\tilde\noise).
\end{equation*}
Indeed, applying this inequality to the two independent noise draws in the
definition of $\dispersion{\tau}$ gives
\begin{equation*}
    \dispersion{\tau}
    \leq
    K_\tau
    \sup_{(\state,\action)}
    \expectedvalue{\noisedist{\tau}(\cdot\st\state,\action)^{\otimes2}}{
        d_{\noisespace}(\noise_\tau,\tilde\noise_\tau)
    }.
\end{equation*}
If $\noisespace$ is also a normed space and the conditional mean exists,
we can apply the triangle inequality through that mean and, since the two noise draws
have the same conditional distribution, obtain
\begin{equation*}
    \dispersion{\tau}
    \leq
    2K_\tau
    \sup_{(\state,\action)}
    \expectedvalue{\noisedist{\tau}(\cdot\st\state,\action)}{
        \norm{
            \noise_\tau
            -\expectedvalue{\noisedist{\tau}(\cdot\st\state,\action)}{\noise_\tau}
        }
    }.
\end{equation*}
Finally, for additive dynamics
$\dynamics{\state}{\action,\noise}{\tau}
=\widetilde f_\tau(\state,\action)+\noise$, with $\statespace$ and $\noisespace$
carrying the same norm metric, the distance between the two next states is
exactly $\norm{\noise-\tilde\noise}$, and the first estimate holds with
equality and $K_\tau=1$.

\paragraph{Tightness.}
We here show through an example that the linear dependence on $\dispersion{\tau}$ in
\cref{lemma:bq4:bellman_defect,theorem:bq4:direct_bound} cannot be improved
without further assumptions.

\begin{example}[The linear coefficient is tight]
\label{example:bq4:coinflip}
Let $\tau=\horizon-1$, $\actionspace = \{0\}$, and $\statespace=\refSpace=\{0,1\}$, equipped with the discrete metric. Set zero stage cost $\stagecostsmall{\timestep}(\state, \action, \refVar)=0$, and
the following terminal cost and dynamics:
\begin{equation*}
    \terminalcostsmall(\state,\refVar)=\lvert\state-\refVar\rvert,
    \qquad
    \dynamics{\state}{\action,\noise}{\tau}=\noise,
\end{equation*}
where $\noise\sim\mathrm{Bernoulli}(\tfrac12)$. Consider an infinite population and use the
individual noise model in Example~\ref{example:indiv_noise_inf_agents}.
Suppose that initially the population is concentrated at $\state=0$, i.e. $\fleet_0=\delta_0$, and
the target distribution is selected as $\refPVar=\tfrac12\delta_0+\tfrac12\delta_1$. 
Before the noise is observed, the only possible coupling assigns half the mass to each target. Moreover, since the next state equals $\noise$ regardless of the current state, the single-agent Bellman recursion gives
\begin{equation*}
    \costtogosmall{\tau}(0,\refVar)
    =\expectedvalue{\noise}{\lvert\noise-\refVar\rvert}
    =\tfrac12\lvert 0-\refVar\rvert+\tfrac12\lvert 1-\refVar\rvert
    =\tfrac12,
\end{equation*}
for every $\refVar\in\refSpace$.
Hence $\costtogosmall{\tau}(0,0)=\costtogosmall{\tau}(0,1)=\tfrac12$, and
\begin{equation*}
    \kantorovich{\costtogosmall{\tau}}{\fleet}{\refPVar}{}=\frac12.
\end{equation*}
After the transition, on the other hand, the realized population is exactly
$\refPVar$, and the population Bellman step matches equal states and targets at
zero cost:
\begin{equation*}
    \mathcal T_\tau\!\left[
      \kantorovich{\costtogosmall{\horizon}}{}{}{}
    \right](\fleet,\refPVar)=0.
\end{equation*}
Since $L_\horizon=1$ and two independent $\mathrm{Bernoulli}(\tfrac12)$ draws differ with probability
$1/2$, we have $\dispersion{\tau}=1/2$, so equality holds
in~\eqref{eq:bq4:bellman_defect}.
\end{example}

\section{Robustness to weakly coupled dynamics}
\label{section:weak_coupling}

In this appendix, we study what happens when the separation-based policy 
is applied to a model in which the agents are weakly coupled, that is, a system whose dynamics and stage cost may depend on the current population distribution.
Since the
separation principle of \cref{section:main_results} requires decoupled agents,
we construct the policy from a decoupled surrogate model, deploy it in the
coupled system, and bound the resulting loss of optimality.
Because the
admissible controls remain $M(\fleet)$ in both models, the following result covers soft population effects, such as 
congestion or queueing, but not hard capacity or collision constraints 
that change the set of feasible controls.

\paragraph{Models and assumptions.}
Throughout this appendix, we denote the decoupled surrogate
with a superscript $D$, so its dynamics, stage cost, single-agent 
value, and population cost-to-go are written as
\begin{equation*}
    f_\tau^{D}(\state,\action,\noise),
    \qquad
    g_\tau^{D}(\state,\action,\noise),
    \qquad
    \costtogosmall{\tau}^{D}(\state,\refVar),
    \qquad
    \costtogo{\tau}^{D}(\fleet,\refPVar).
\end{equation*}
The corresponding quantities for the coupled model are denoted by a
superscript $C$:
\begin{equation*}
    f_\tau^{C}(\state,\action,\fleet,\noise),
    \qquad
    g_\tau^{C}(\state,\action,\fleet,\noise),
    \qquad
    \costtogo{\tau}^{C}(\fleet,\refPVar).
\end{equation*}
Here, the dynamics and stage costs also depend on the population, so there is no direct counterpart of $\costtogosmall{\tau}^{D}$. Indeed, because $f_\tau^{C}$ depends on $\fleet$, the cost-to-go of a single agent cannot be defined without fixing the trajectory of the population. 
At the terminal time $\horizon$, we denote by $\costtogo{\horizon}^{C}(\fleet,\refPVar)$ the terminal objective, evaluated at the terminal fleet state $\fleet$ and target distribution $\refPVar$.
Apart from the population dependence displayed above, the two models
share the same basic setting.
The models have the same state, action, noise, and target spaces, the same
nominal noise kernel, and the same admissible controls $M(\fleet)$. We also drive 
them with the same realized noise field, which satisfies
\cref{assumption:measurable_field,assumption:barycenter_condition}, so
their next populations are 
\begin{equation}
\label{eq:wc:next_populations}
    \fleet_{\tau+1}^{C,\lambda}
    \coloneqq
    \pushforward{f_\tau^{C}(\cdot,\cdot,\fleet,\cdot)}
    {\rho_\tau^\lambda},
    \qquad
    \fleet_{\tau+1}^{D,\lambda}
    \coloneqq
    \pushforward{f_\tau^{D}}{\rho_\tau^\lambda},
\end{equation}
where $\rho_\tau^\lambda\coloneqq\lambda\otimes\empnoisedist{\tau}^{\state,\action}$
denotes the joint state-action-noise distribution for a fixed population $\fleet$ 
and a control $\lambda\in M(\fleet)$.
Finally, the terminal objective of the coupled and decoupled models is
common and denoted by
\begin{equation}
\label{eq:wc:common_terminal}
    \costtogo{\horizon}^{C}(\fleet,\refPVar)
    =
    \costtogo{\horizon}^{D}(\fleet,\refPVar)
    =
    \kantorovich{\terminalcostsmall}{\fleet}{\refPVar}{}.
\end{equation}

We next specify how far the coupled model may be from the decoupled
one at each time step.
\begin{assumption}[Bounded one-step model mismatch]
\label{assumption:wc:model_mismatch}
Fix a reference point $\state_0\in\statespace$ and assume that every population
distribution $\fleet \in \probabilityspace{\statespace}$ at which the coupled and decoupled models are compared below has finite first moment,
\begin{equation*}
    \int_{\statespace}d_{\statespace}(\state,\state_0)\dd\fleet(\state)<\infty.
\end{equation*}
For every $\tau<\horizon$, let
$\statemismatch{\tau},\delta_\tau\geq0$ be constants such that the following
hold:
\begin{enumerate}[label=(\roman*),leftmargin=*,itemsep=0.4em]
\item \emph{Dynamics mismatch.}
When the coupled and decoupled
dynamics are evaluated at the same state, action, and noise, their outputs
satisfy
\begin{equation}
\label{eq:wc:dynamics_mismatch}
    d_{\statespace}\!\left(
        f_\tau^{C}(\state,\action,\fleet,\noise),
        f_\tau^{D}(\state,\action,\noise)
    \right)
    \leq
    \statemismatch{\tau}
\end{equation}
uniformly in $\fleet\in\probabilityspace{\statespace}$ and $(\state,\action,\noise)\in\statespace\times\actionspace\times\noisespace$. 
\item \emph{Stage-cost mismatch.} 
The
expected stage costs, evaluated at the same population and the same admissible
state-action distribution, satisfy
\begin{equation}
\label{eq:wc:stage_mismatch}
    \left|
    \mathbb E_{\distoverdist{\tau}}
    \left[
        \int_{\statespace\times\actionspace\times\noisespace}
        \left(
            g_\tau^{C}(\state,\action,\fleet,\noise)
            -g_\tau^{D}(\state,\action,\noise)
        \right)
        \dd\rho_\tau^\lambda(\state,\action,\noise)
    \right]
    \right|
    \leq
    \delta_\tau
\end{equation}
uniformly in $\fleet\in\probabilityspace{\statespace}$ and $\lambda\in M(\fleet)$.
\end{enumerate}
\end{assumption}

These two constants control the two parts of one Bellman step. Indeed,
$\delta_\tau$ bounds the difference in the current stage cost, while
$\statemismatch{\tau}$ bounds the difference between the next populations. 

Note also that condition \eqref{eq:wc:stage_mismatch} is satisfied if there exists a pointwise 
bound $|g_\tau^{C}-g_\tau^{D}|\leq\delta_\tau$. Additionally, for a fixed initial population, then it is sufficient that 
\eqref{eq:wc:dynamics_mismatch} and \eqref{eq:wc:stage_mismatch} hold on the 
population states reachable under any admissible control.

These one-step bounds enter the comparison through the Bellman
recursions.
Recall the Bellman operator $\mathcal T_\tau$ introduced in
\cref{section:bound_quantification_v4}.
We now rewrite the two population recursions, for the coupled and decoupled models, 
in a common form. For $m\in\{C,D\}$, define
\begin{equation*}
\begin{aligned}
 (\mathcal T_\tau^{m}F)(\fleet,\refPVar)
 \coloneqq \inf_{\lambda\in M(\fleet)}
 \mathbb E_{\distoverdist{\tau}}\!\left[
   \int_{\statespace\times\actionspace\times\noisespace}
   g_\tau^{m}(\state,\action,\fleet,\noise)
   \dd\rho_\tau^\lambda(\state,\action,\noise)
   +F(\fleet_{\tau+1}^{m,\lambda},\refPVar)
 \right].
\end{aligned}
\end{equation*}

Here and below, we use the convention
$g_\tau^D(\state,\action,\fleet,\noise)
\coloneqq g_\tau^D(\state,\action,\noise)$, thus neither $f_\tau^D$ nor
$g_\tau^D$ depends on the population argument. With
this notation, we have the Bellman recursions
$\costtogo{\tau}^{m}=\mathcal T_\tau^{m}\costtogo{\tau+1}^{m}$. In
particular, $\mathcal T_\tau^{D}=\mathcal T_\tau$, since the decoupled stage
cost and dynamics are exactly the data $\stagecostsmall{\tau}$ and
$\dynamics{}{}{\tau}$ used to define $\mathcal T_\tau$ in
\cref{section:bound_quantification_v4}, so every result proved there for
$\mathcal T_\tau$ applies verbatim to $\mathcal T_\tau^{D}$. Additionally, 
if $\pi$ is a feedback policy with $\pi_\tau(\fleet)\in M(\fleet)$, we drop the infimum and denote the same expression evaluated at $\lambda=\pi_\tau(\fleet)$ as
\begin{equation*}
    (\mathcal T_\tau^{m,\pi}F)(\fleet,\refPVar)
 \coloneqq
 \mathbb E_{\distoverdist{\tau}}\!\left[
   \int_{\statespace\times\actionspace\times\noisespace}
   g_\tau^{m}(\state,\action,\fleet,\noise)
   \dd\rho_\tau^{\pi_\tau(\fleet)}(\state,\action,\noise)
   +F(\fleet_{\tau+1}^{m,\pi_\tau(\fleet)},\refPVar)
 \right].
\end{equation*}
The cost of this policy satisfies
\begin{equation*}
    \costtogo{\horizon}^{m,\pi}
      =\kantorovich{\terminalcostsmall}{}{}{},
    \qquad
    \costtogo{\tau}^{m,\pi}
      =\mathcal T_\tau^{m,\pi}\costtogo{\tau+1}^{m,\pi}.
\end{equation*}

The next-stage terms in these recursions may be evaluated at different
population distributions, so to compare those distributions, we use the
$1$-Wasserstein distance induced by $d_{\statespace}$.
Explicitly, for any  $\mu,\nu\in\probabilityspace{\statespace}$ 
with finite first moments, the $1$-Wasserstein distance induced 
by $d_{\statespace}$ is denoted by $W_1(\mu,\nu)$ and defined as
\begin{equation*}
W_1(\mu,\nu)
\coloneqq
\inf_{q\in\plans{\mu}{\nu}}
\int_{\statespace\times\statespace}
d_{\statespace}(\state,\widetilde\state)
\dd q(\state,\widetilde\state).
\end{equation*}

To compare the next populations in \eqref{eq:wc:next_populations}, we need an estimate for two maps applied to the same measure. We formalize this in the following lemma.
Note that this differs from \cite[Proposition~3]{aolaritei2025distributional}, which considers one map applied to two measures.

\begin{lemma}[Wasserstein bound for pushforwards of a common measure]
Let $U$ be a measurable space, let
$\mu\in\probabilityspace{U}$, and let
$F,G\colon U\to\statespace$ be measurable. Then
\begin{equation}
\label{eq:wc:pushforward_wasserstein}
    W_1(\pushforward{F}{\mu},\pushforward{G}{\mu})
    \leq
    \int_{U} d_{\statespace}\bigl(F(u),G(u)\bigr)\dd\mu(u).
\end{equation}
\end{lemma}

\begin{proof}
Consider the coupling
$q_{F,G}\coloneqq(F,G)_\#\mu$, where $(F,G)$ maps $u$ to
$(F(u),G(u))$. Its first marginal is $F_\#\mu$, since for
every Borel set $A\subseteq\statespace$,
\begin{equation*}
q_{F,G}(A\times\statespace)
=\mu\bigl(F^{-1}(A)\bigr)
=(F_\#\mu)(A).
\end{equation*}
The same argument shows that its second marginal is $G_\#\mu$, so $q_{F,G}\in\plans{F_\#\mu}{G_\#\mu}$. 
Evaluating the definition of $W_1$ at this plan and applying the change-of-variables formula
for the pushforward $q_{F,G}=(F,G)_\#\mu$ gives
\begin{equation*}
\begin{aligned}
W_1(\pushforward{F}{\mu},\pushforward{G}{\mu})
&\leq
\int_{\statespace\times\statespace}
d_{\statespace}(\state,\widetilde\state)
\dd q_{F,G}(\state,\widetilde\state)
\\
&=
\int_{U}
d_{\statespace}\bigl(F(u),G(u)\bigr)\dd\mu(u).
\end{aligned}
\end{equation*} 
This concludes the proof. 
\end{proof}
In other words, when two distributions are obtained from the
same measure through two different maps, their Wasserstein distance can be
bounded by comparing the outputs of the maps pointwise.

It remains to relate this distance between populations to the
next-stage cost. To do so, we assume that the decoupled single-agent
cost-to-go $\costtogosmall{\tau}^{D}$ satisfies
\cref{assumption:bq4:cost_to_go_lipschitz}, with constants $L_\tau$, and use
$\dispersion{\tau}$~\eqref{eq:bq4:pairwise_dispersion} for the dispersion of
the decoupled dynamics $f_\tau^{D}$.
As in Appendix~\ref{section:bound_quantification_v4}, we assume that
all transport costs appearing below are finite.

We can now proceed with the study of the robustness of the separation-based policy.
The first step is to show that the transport cost is Lipschitz in the population
distribution. This will allow us, in \cref{lemma:wc:operator_perturbation} below, 
to turn the state mismatch in \eqref{eq:wc:dynamics_mismatch} into a bound 
on the next-stage cost.

\begin{lemma}[Population sensitivity of the transport cost]
\label{lemma:wc:H_lipschitz}
Suppose \cref{assumption:bq4:cost_to_go_lipschitz} holds for
$\costtogosmall{\tau}^{D}$ with constant $L_\tau$. Then, for every
$\tau\in\{1,\ldots,\horizon\}$, every
$\fleet,\widetilde\fleet\in\probabilityspace{\statespace}$ with finite first
moments, and every $\refPVar\in\probabilityspace{\refSpace}$,
\begin{equation}
\label{eq:wc:H_lipschitz}
    \left|
        \kantorovich{\costtogosmall{\tau}^{D}}{\fleet}{\refPVar}{}
        -\kantorovich{\costtogosmall{\tau}^{D}}{\widetilde\fleet}{\refPVar}{}
    \right|
    \leq
    L_\tau W_1(\fleet,\widetilde\fleet).
\end{equation}
\end{lemma}

\begin{proof}
The proof consists of transferring a nearly optimal target assignment for $\fleet$ to
$\widetilde\fleet$. Fix $\varepsilon>0$. By the definition of the transport cost
and the Wasserstein distance, we can choose
$\plan{\gamma}_\varepsilon\in\plans{\fleet}{\refPVar}$ and
$q_\varepsilon\in\plans{\fleet}{\widetilde\fleet}$ such that
\begin{align*}
    \int_{\statespace\times\refSpace}
    \costtogosmall{\tau}^{D}(\state,\refVar)
    \dd\plan{\gamma}_\varepsilon(\state,\refVar)
    &\leq
    \kantorovich{\costtogosmall{\tau}^{D}}{\fleet}{\refPVar}{}+\varepsilon,
    \\
    \int_{\statespace\times\statespace}
    d_{\statespace}(\state,\widetilde\state)
    \dd q_\varepsilon(\state,\widetilde\state)
    &\leq
    W_1(\fleet,\widetilde\fleet)+\varepsilon.
\end{align*}
Since the two plans have the same $\state$-marginal $\fleet$, the gluing 
lemma gives a measure $r_\varepsilon$ on
$\statespace\times\statespace\times\refSpace$ with
$(\state,\refVar)$-marginal $\plan{\gamma}_\varepsilon$ and
$(\state,\widetilde\state)$-marginal $q_\varepsilon$. Note, in particular, 
that its $(\widetilde\state,\refVar)$-marginal couples $\widetilde\fleet$ with
$\refPVar$, so we can use it in the transport problem for
$\kantorovich{\costtogosmall{\tau}^{D}}{\widetilde\fleet}{\refPVar}{}$.
The first inequality below evaluates the transport cost at the
$(\widetilde\state,\refVar)$-marginal of $r_\varepsilon$, the second applies
\eqref{eq:bq4:cost_to_go_lipschitz} pointwise, and then we use the 
$(\state,\refVar)$ and $(\state,\widetilde\state)$-marginals
of $r_\varepsilon$ to separate the two integrals:
\begin{equation}
\label{eq:wc:H_lipschitz_chain}
\begin{aligned}
    \kantorovich{\costtogosmall{\tau}^{D}}
    {\widetilde\fleet}{\refPVar}{}
    &\leq
    \int_{\statespace\times\statespace\times\refSpace}
    \costtogosmall{\tau}^{D}(\widetilde\state,\refVar)
    \dd r_\varepsilon(\state,\widetilde\state,\refVar)
    \\
    &\leq
    \int_{\statespace\times\statespace\times\refSpace}
    \left[
      \costtogosmall{\tau}^{D}(\state,\refVar)
      +L_\tau d_{\statespace}(\state,\widetilde\state)
    \right]
    \dd r_\varepsilon(\state,\widetilde\state,\refVar)
    \\
    &=
    \int_{\statespace\times\refSpace}
    \costtogosmall{\tau}^{D}(\state,\refVar)
    \dd\plan{\gamma}_\varepsilon(\state,\refVar)
    +L_\tau
    \int_{\statespace\times\statespace}
    d_{\statespace}(\state,\widetilde\state)
    \dd q_\varepsilon(\state,\widetilde\state)
    \\
    &\leq
    \kantorovich{\costtogosmall{\tau}^{D}}{\fleet}{\refPVar}{}
    +\varepsilon
    +L_\tau\bigl(W_1(\fleet,\widetilde\fleet)+\varepsilon\bigr)
    \\
    &=
    \kantorovich{\costtogosmall{\tau}^{D}}{\fleet}{\refPVar}{}
    +L_\tau W_1(\fleet,\widetilde\fleet)
    +(1+L_\tau)\varepsilon,
\end{aligned}
\end{equation}
where the last inequality uses the two $\varepsilon$-optimality bounds in the
choice of $\plan{\gamma}_\varepsilon$ and $q_\varepsilon$.
Since the construction holds for all $\varepsilon>0$, letting
$\varepsilon\downarrow0$ gives
\begin{equation*}
    \kantorovich{\costtogosmall{\tau}^{D}}
      {\widetilde\fleet}{\refPVar}{}
    -\kantorovich{\costtogosmall{\tau}^{D}}{\fleet}{\refPVar}{}
    \leq L_\tau W_1(\fleet,\widetilde\fleet).
\end{equation*}
Exchanging the roles of $\fleet$ and $\widetilde\fleet$ throughout the
construction above, that is, choosing $\plan{\gamma}_\varepsilon\in
\plans{\widetilde\fleet}{\refPVar}$ that is $\varepsilon$-optimal for
$\kantorovich{\costtogosmall{\tau}^{D}}{\widetilde\fleet}{\refPVar}{}$ and
$q_\varepsilon\in\plans{\widetilde\fleet}{\fleet}$ that is
$\varepsilon$-optimal for $W_1(\widetilde\fleet,\fleet)$, and then repeating the
gluing step and~\eqref{eq:wc:H_lipschitz_chain} with $\state$ and
$\widetilde\state$ swapped, gives
\begin{equation*}
    \kantorovich{\costtogosmall{\tau}^{D}}{\fleet}{\refPVar}{}
    -\kantorovich{\costtogosmall{\tau}^{D}}
      {\widetilde\fleet}{\refPVar}{}
    \leq L_\tau W_1(\widetilde\fleet,\fleet).
\end{equation*}
Since
$W_1(\widetilde\fleet,\fleet)=W_1(\fleet,\widetilde\fleet)$,
the two bounds give~\eqref{eq:wc:H_lipschitz}.
\end{proof}

Lemma~\ref{lemma:wc:H_lipschitz} allows us to compare one Bellman step in the two models. 
To do so, we first keep the control $\lambda$ fixed, so that the only 
differences come from the stage cost and the next population, 
and then minimize over $\lambda$.

\begin{lemma}[One-step model perturbation]
\label{lemma:wc:operator_perturbation}
Under \cref{assumption:measurable_field,assumption:barycenter_condition,assumption:bq4:cost_to_go_lipschitz,assumption:wc:model_mismatch}, for
every $\tau<\horizon$, every $\fleet\in\probabilityspace{\statespace}$
with finite first moment, every
$\refPVar\in\probabilityspace{\refSpace}$, and every admissible feedback policy
$\pi$,
\begin{align}
\label{eq:wc:optimal_operator_perturbation}
    \left|
        \mathcal T_\tau^{C}\!\left[
          \kantorovich{\costtogosmall{\tau+1}^{D}}{}{}{}
        \right](\fleet,\refPVar)
        -\mathcal T_\tau^{D}\!\left[
          \kantorovich{\costtogosmall{\tau+1}^{D}}{}{}{}
        \right](\fleet,\refPVar)
    \right|
    &\leq
    \delta_\tau+L_{\tau+1}\statemismatch{\tau},
    \\
\label{eq:wc:policy_operator_perturbation}
    \left|
        \mathcal T_\tau^{C,\pi}\!\left[
          \kantorovich{\costtogosmall{\tau+1}^{D}}{}{}{}
        \right](\fleet,\refPVar)
        -\mathcal T_\tau^{D,\pi}\!\left[
          \kantorovich{\costtogosmall{\tau+1}^{D}}{}{}{}
        \right](\fleet,\refPVar)
    \right|
    &\leq
    \delta_\tau+L_{\tau+1}\statemismatch{\tau}.
\end{align}
\end{lemma}

\begin{proof}
\noindent\textbf{Step 1: Compare the next populations for a fixed
control.}
Fix $\lambda\in M(\fleet)$ and a realization of the noise field. By
\eqref{eq:wc:next_populations}, we can obtain the two next populations by
pushing forward the same state-action-noise distribution
$\rho_\tau^\lambda$, once through $f_\tau^C$ and once through $f_\tau^D$. 
We can then apply \eqref{eq:wc:pushforward_wasserstein} with
$\mu=\rho_\tau^\lambda$,
$F=f_\tau^C(\cdot,\cdot,\fleet,\cdot)$, and $G=f_\tau^D$ and obtain
\begin{equation}
\label{eq:wc:pathwise_population_mismatch}
\begin{aligned}
    W_1(\fleet_{\tau+1}^{C,\lambda},
        \fleet_{\tau+1}^{D,\lambda})
    &\leq
    \int_{\statespace\times\actionspace\times\noisespace}
    d_{\statespace}\!\left(
        f_\tau^{C}(\state,\action,\fleet,\noise),
        f_\tau^{D}(\state,\action,\noise)
    \right)
    \dd\rho_\tau^\lambda(\state,\action,\noise)
    \leq \statemismatch{\tau}.
\end{aligned}
\end{equation}
The last inequality follows directly from the uniform dynamics-mismatch bound
\eqref{eq:wc:dynamics_mismatch} in
\cref{assumption:wc:model_mismatch}.

\medskip
\noindent\textbf{Step 2: Compare the next-stage transport costs.}
For the fixed realization of the noise field selected in the previous step,
\cref{lemma:wc:H_lipschitz} applied at time $\tau+1$, together with
\eqref{eq:wc:pathwise_population_mismatch}, gives
\begin{equation*}
\left|
\kantorovich{\costtogosmall{\tau+1}^{D}}
 {\fleet_{\tau+1}^{C,\lambda}}{\refPVar}{}
-\kantorovich{\costtogosmall{\tau+1}^{D}}
 {\fleet_{\tau+1}^{D,\lambda}}{\refPVar}{}
\right|
\leq
L_{\tau+1}W_1\!\left(
\fleet_{\tau+1}^{C,\lambda},
\fleet_{\tau+1}^{D,\lambda}
\right)
\leq L_{\tau+1}\statemismatch{\tau}.
\end{equation*}

Since this bound holds for every realization of the field, we can
average it out and obtain
\begin{equation*}
\begin{aligned}
\mathbb E_{\distoverdist{\tau}}\!\left[
\left|
\kantorovich{\costtogosmall{\tau+1}^{D}}
 {\fleet_{\tau+1}^{C,\lambda}}{\refPVar}{}
-\kantorovich{\costtogosmall{\tau+1}^{D}}
 {\fleet_{\tau+1}^{D,\lambda}}{\refPVar}{}
\right|
\right]
\leq L_{\tau+1}\statemismatch{\tau}.
\end{aligned}
\end{equation*}

\medskip
\noindent\textbf{Step 3: Add the stage-cost difference.}
By \eqref{eq:wc:stage_mismatch} in \cref{assumption:wc:model_mismatch}, 
the absolute difference between the two expected stage costs is at most $\delta_\tau$. 
We now add this difference to the difference between the next-stage transport costs. 
The triangle inequality and the estimate from Step 2 give
\begin{equation}
\label{eq:wc:fixed_control_bound}
\begin{aligned}
&\Bigg|
\mathbb E_{\distoverdist{\tau}}\!\bigg[
    \int_{\statespace\times\actionspace\times\noisespace}
    \bigl(g_\tau^{C}(\state,\action,\fleet,\noise)
          -g_\tau^{D}(\state,\action,\noise)\bigr)
    \dd\rho_\tau^\lambda(\state,\action,\noise)\\
    &\qquad+\kantorovich{\costtogosmall{\tau+1}^{D}}
       {\fleet_{\tau+1}^{C,\lambda}}{\refPVar}{}
    -\kantorovich{\costtogosmall{\tau+1}^{D}}
       {\fleet_{\tau+1}^{D,\lambda}}{\refPVar}{}
\bigg]\Bigg|
\\
&\quad\leq
\left|
\mathbb E_{\distoverdist{\tau}}\!\left[
\int_{\statespace\times\actionspace\times\noisespace}
\bigl(g_\tau^{C}(\state,\action,\fleet,\noise)
      -g_\tau^{D}(\state,\action,\noise)\bigr)
\dd\rho_\tau^\lambda(\state,\action,\noise)
\right]
\right|
\\
&\qquad+
\mathbb E_{\distoverdist{\tau}}\!\left[
\left|
\kantorovich{\costtogosmall{\tau+1}^{D}}
 {\fleet_{\tau+1}^{C,\lambda}}{\refPVar}{}
-\kantorovich{\costtogosmall{\tau+1}^{D}}
 {\fleet_{\tau+1}^{D,\lambda}}{\refPVar}{}
\right|
\right]
\\
&\quad\leq \delta_\tau+L_{\tau+1}\statemismatch{\tau}.
\end{aligned}
\end{equation}

\medskip
\noindent\textbf{Step 4: Pass to the Bellman operators.}
By linearity of expectation, one of the two one-sided inequalities
implied by \eqref{eq:wc:fixed_control_bound} is
\begin{equation}
\label{eq:wc:fixed_control_upper_C}
\begin{aligned}
&\mathbb E_{\distoverdist{\tau}}\!\Bigg[
    \int_{\statespace\times\actionspace\times\noisespace}
    g_\tau^{C}(\state,\action,\fleet,\noise)
    \dd\rho_\tau^\lambda(\state,\action,\noise)
    +\kantorovich{\costtogosmall{\tau+1}^{D}}
       {\fleet_{\tau+1}^{C,\lambda}}{\refPVar}{}
\Bigg]
\\
&\quad\leq
\mathbb E_{\distoverdist{\tau}}\!\Bigg[
    \int_{\statespace\times\actionspace\times\noisespace}
    g_\tau^{D}(\state,\action,\noise)
    \dd\rho_\tau^\lambda(\state,\action,\noise)
    +\kantorovich{\costtogosmall{\tau+1}^{D}}
       {\fleet_{\tau+1}^{D,\lambda}}{\refPVar}{}
\Bigg]
+\delta_\tau+L_{\tau+1}\statemismatch{\tau}.
\end{aligned}
\end{equation}
The left- and right-hand sides are precisely the fixed-control objectives
minimized in the definitions of $\mathcal T_\tau^C$ and
$\mathcal T_\tau^D$.
Since \eqref{eq:wc:fixed_control_upper_C} holds for every admissible
$\lambda$, we can take the infimum over the controls in $M(\fleet)$ on both
sides, and obtain
\begin{equation*}
\begin{aligned}
&\mathcal T_\tau^{C}\!\left[
  \kantorovich{\costtogosmall{\tau+1}^{D}}{}{}{}
\right](\fleet,\refPVar)
\\
&\quad=
\inf_{\lambda\in M(\fleet)}
\mathbb E_{\distoverdist{\tau}}\!\Bigg[
    \int_{\statespace\times\actionspace\times\noisespace}
    g_\tau^{C}(\state,\action,\fleet,\noise)
    \dd\rho_\tau^\lambda(\state,\action,\noise)
    +\kantorovich{\costtogosmall{\tau+1}^{D}}
       {\fleet_{\tau+1}^{C,\lambda}}{\refPVar}{}
\Bigg]
\\
&\quad\leq
\inf_{\lambda\in M(\fleet)}
\mathbb E_{\distoverdist{\tau}}\!\Bigg[
    \int_{\statespace\times\actionspace\times\noisespace}
    g_\tau^{D}(\state,\action,\noise)
    \dd\rho_\tau^\lambda(\state,\action,\noise)
    +\kantorovich{\costtogosmall{\tau+1}^{D}}
       {\fleet_{\tau+1}^{D,\lambda}}{\refPVar}{}
\Bigg]
+\delta_\tau+L_{\tau+1}\statemismatch{\tau}
\\
&\quad=
\mathcal T_\tau^{D}\!\left[
  \kantorovich{\costtogosmall{\tau+1}^{D}}{}{}{}
\right](\fleet,\refPVar)
+\delta_\tau+L_{\tau+1}\statemismatch{\tau}.
\end{aligned}
\end{equation*}
Similarly, taking the infimum over $M(\fleet)$ in the reverse fixed-control 
inequality of \eqref{eq:wc:fixed_control_bound} gives the other one-sided bound
\begin{equation*}
\mathcal T_\tau^{D}\!\left[
  \kantorovich{\costtogosmall{\tau+1}^{D}}{}{}{}
\right](\fleet,\refPVar)
\leq
\mathcal T_\tau^{C}\!\left[
  \kantorovich{\costtogosmall{\tau+1}^{D}}{}{}{}
\right](\fleet,\refPVar)
+\delta_\tau+L_{\tau+1}\statemismatch{\tau}.
\end{equation*}

Combining these two inequalities proves \eqref{eq:wc:optimal_operator_perturbation}.
Finally, for the fixed-policy operators, it is enough to evaluate
the same objectives at the same control $\lambda=\pi_\tau(\fleet)$. 
Indeed, substituting this value into \eqref{eq:wc:fixed_control_bound} gives
directly \eqref{eq:wc:policy_operator_perturbation}.
\end{proof}

\paragraph{The separation-based policy.}
We now construct the separation-based policy from the decoupled model,
following the same steps as in \cref{section:algorithm} and \sepconstruct{}.
We will then evaluate this policy in the coupled model, so to bound the performance 
loss due to the model mismatch.
Briefly, the policy construction proceeds as follows: at an observed population $\fleet$, 
the policy first chooses a $\septol{\tau}/2$-optimal target assignment
$\plan{\gamma}_\tau^\theta\in\plans{\fleet}{\refPVar}$ for the decoupled
transport problem, so that
\begin{equation}
\label{eq:wc:near_optimal_transport}
    \int_{\statespace\times\refSpace}
    \costtogosmall{\tau}^{D}(\state,\refVar)
    \dd\plan{\gamma}_\tau^\theta(\state,\refVar)
    \leq
    \kantorovich{\costtogosmall{\tau}^{D}}{\fleet}{\refPVar}{}
    +\frac{\septol{\tau}}{2}.
\end{equation}
It then chooses a measurable action
$\action_\tau^\theta(\state,\refVar)$ whose single-agent cost is within
$\septol{\tau}/2$ of the decoupled optimum. Writing
$\action=\action_\tau^\theta(\state,\refVar)$, this means that
\begin{equation}
\label{eq:wc:near_optimal_action}
    \int_{\noisespace}
    \left[
    g_\tau^{D}(\state,\action,\noise)
    +\costtogosmall{\tau+1}^{D}
    \bigl(f_\tau^{D}(\state,\action,\noise),\refVar\bigr)
    \right]
    \dd\noisedist{\tau}(\noise\st\state,\action)
    \leq
    \costtogosmall{\tau}^{D}(\state,\refVar)
    +\frac{\septol{\tau}}{2}.
\end{equation}

Since this inequality holds for every $(\state,\refVar)\in\statespace\times\refSpace$, the target plan and
the selected actions together induce the following population control
\begin{equation}
\label{eq:wc:separation_control}
    \pi_{\mathrm{sep},\tau}^{\theta}(\fleet)
    \coloneqq
    \pushforward{
        \bigl(
            \projectionFromTo{\statespace\times\refSpace}{\statespace},
            \action_\tau^\theta
        \bigr)
    }{\plan{\gamma}_\tau^\theta}
    \in M(\fleet).
\end{equation}
Note that the control $\pi_{\mathrm{sep},\tau}^{\theta}(\fleet)$ is a
measurable map from the population space to the set of admissible controls, 
as required by the definition of a feedback policy. 
The construction ensures that the policy is well-defined and can be implemented in practice, 
as it relies on measurable selections of near-optimal target assignments and actions.
The two errors of size $\septol{\tau}/2$ add up to the per-stage tolerance $\septol{\tau}$.
If, additionally, the target plan and the actions can
both be chosen as measurable exact minimizers, then the construction holds when we 
take $\septol{\tau}=0$.

Finally, recall that $\rho_\tau^\lambda$ was
defined immediately after \eqref{eq:wc:next_populations}, for every
$\lambda\in M(\fleet)$, as the corresponding state-action-noise
distribution. Since
$\pi_{\mathrm{sep},\tau}^\theta(\fleet)\in M(\fleet)$, we can apply that
definition with
$\lambda=\pi_{\mathrm{sep},\tau}^\theta(\fleet)$, obtaining $\rho_\tau^{\pi_{\mathrm{sep},\tau}^\theta(\fleet)}$.

Before evaluating this separation-based policy in the coupled model, we show that
when working in the decoupled model, one Bellman step under this policy costs at 
most the decoupled transport surrogate plus $\septol{\tau}$. 
This is the estimate that will enter the backward argument of the main theorem.

\begin{lemma}[One-step bound for the separation-based policy]
\label{lemma:wc:policy_residual}
Under
\cref{assumption:measurable_field,assumption:barycenter_condition}, the separation-based 
policy defined by
\eqref{eq:wc:near_optimal_transport}--\eqref{eq:wc:separation_control} satisfies,
for every $\tau<\horizon$, every $\fleet\in\probabilityspace{\statespace}$, and
every $\refPVar\in\probabilityspace{\refSpace}$,
\begin{equation}
\label{eq:wc:policy_residual}
    \mathcal T_\tau^{D,\pi_{\mathrm{sep}}^{\theta}}\!\left[
      \kantorovich{\costtogosmall{\tau+1}^{D}}{}{}{}
    \right](\fleet,\refPVar)
    \leq
    \kantorovich{\costtogosmall{\tau}^{D}}{\fleet}{\refPVar}{}
    +\septol{\tau}.
\end{equation}
\end{lemma}

\begin{proof}
\noindent\textbf{Step 1: Lift the policy to a measure on $(\state,\refVar,\noise)$.}
We start by fixing $\fleet$ and a realization of the noise field and, to obtain an
admissible target assignment at the next time, we keep the target attached to
each state while the state moves according to its selected action and noise.
More precisely, consider the measure
\begin{equation*}
    \dd\plan{\gamma}_\tau^\theta(\state,\refVar)
    \dd\empnoisedist{\tau}^{\state,
      \action_\tau^\theta(\state,\refVar)}(\noise)
\end{equation*}
on $(\state,\refVar,\noise)$. It
generates the same state, action, and noise distribution
$\rho_\tau^{\pi_{\mathrm{sep},\tau}^\theta(\fleet)}$ as the
separation-based policy. More precisely, for every nonnegative measurable
function $h$ on
$\statespace\times\actionspace\times\noisespace$,
\begin{equation}
\label{eq:wc:policy_lift_identity}
\begin{aligned}
&\int_{\statespace\times\actionspace\times\noisespace}
h(\state,\action,\noise)
\dd\rho_\tau^{\pi_{\mathrm{sep},\tau}^\theta(\fleet)}
  (\state,\action,\noise)
\\
&\quad=
\int_{\statespace\times\refSpace}\int_{\noisespace}
h\bigl(\state,\action_\tau^\theta(\state,\refVar),\noise\bigr)
\dd\empnoisedist{\tau}^{\state,
  \action_\tau^\theta(\state,\refVar)}(\noise)
\dd\plan{\gamma}_\tau^\theta(\state,\refVar).
\end{aligned}
\end{equation}
Indeed, \eqref{eq:wc:separation_control} says that applying
$(\state,\refVar)\mapsto
(\state,\action_\tau^\theta(\state,\refVar))$ to
$\plan{\gamma}_\tau^\theta$ gives the state-action distribution chosen by
the policy. Adding the realized noise kernel then gives
$\rho_\tau^{\pi_{\mathrm{sep},\tau}^\theta(\fleet)}$, which proves the
identity.

\noindent\textbf{Step 2: Build a feasible next-stage transport plan.}
We now push the measure on $(\state,\refVar,\noise)$ forward under
the following map:
\begin{equation*}
    (\state,\refVar,\noise)
    \longmapsto
    \bigl(
      f_\tau^{D}(\state,
        \action_\tau^\theta(\state,\refVar),\noise),
      \refVar
    \bigr).
\end{equation*}
Its first marginal is
$\fleet_{\tau+1}^{D,\pi_{\mathrm{sep},\tau}^\theta(\fleet)}$ by
\eqref{eq:wc:next_populations} and the identity
\eqref{eq:wc:policy_lift_identity}, while its second
marginal is $\refPVar$, because the map leaves the target coordinate unchanged.
The pushforward is therefore an admissible coupling in the transport
problem
\begin{equation*}
    \kantorovich{\costtogosmall{\tau+1}^{D}}
    {\fleet_{\tau+1}^{D,\pi_{\mathrm{sep},\tau}^\theta(\fleet)}}{\refPVar}{}.
\end{equation*}

Hence, this coupling handles the next-stage term of the Bellman objective, it remains
to show that the coupling also accounts for the current stage cost. To do so, take
$h=g_\tau^D$ in \eqref{eq:wc:policy_lift_identity} and write the stage cost
using the same measure on $(\state,\refVar,\noise)$:
\begin{equation*}
\begin{aligned}
&\int_{\statespace\times\actionspace\times\noisespace}
g_\tau^D(\state,\action,\noise)
\dd\rho_\tau^{\pi_{\mathrm{sep},\tau}^\theta(\fleet)}
  (\state,\action,\noise)
\\
&\quad=
\int_{\statespace\times\refSpace}\int_{\noisespace}
g_\tau^D\bigl(
\state,\action_\tau^\theta(\state,\refVar),\noise
\bigr)
\dd\empnoisedist{\tau}^{\state,
\action_\tau^\theta(\state,\refVar)}(\noise)
\dd\plan{\gamma}_\tau^\theta(\state,\refVar).
\end{aligned}
\end{equation*}
Thus, both parts of the Bellman objective can be bounded using one
integral with respect to the measure constructed above.

\medskip
\noindent\textbf{Step 3: Bound the cost of this coupling.}
We now combine the two parts of the Bellman objective and use the near-optimality of the
target assignment and the actions to bound the total cost.
Start by observing that the next-stage transport cost
$\kantorovich{\costtogosmall{\tau+1}^{D}}{\fleet_{\tau+1}^{D,\pi_{\mathrm{sep},\tau}^\theta(\fleet)}}{\refPVar}{}$
is the infimum over all admissible
couplings, so evaluating it at the coupling just constructed gives an upper
bound. Consequently, for the fixed realization of the noise field in the 
previous step, and writing $\action=\action_\tau^\theta(\state,\refVar)$, 
the Bellman objective of the separation-based policy is bounded by
\begin{equation*}
\begin{aligned}
&\int_{\statespace\times\actionspace\times\noisespace}
g_\tau^D(\state,\action,\noise)
   \dd\rho_\tau^{\pi_{\mathrm{sep},\tau}^\theta(\fleet)}
   (\state,\action,\noise)
+\kantorovich{\costtogosmall{\tau+1}^{D}}
 {\fleet_{\tau+1}^{D,\pi_{\mathrm{sep},\tau}^\theta(\fleet)}}{\refPVar}{}
\\
&\quad\leq
\int_{\statespace\times\refSpace}\int_{\noisespace}
\left[
 g_\tau^{D}(\state,\action,\noise)
 +\costtogosmall{\tau+1}^{D}\bigl(
   f_\tau^{D}(\state,\action,\noise),\refVar
 \bigr)
\right]
\dd\empnoisedist{\tau}^{\state,\action}(\noise)
\dd\plan{\gamma}_\tau^\theta(\state,\refVar).
\end{aligned}
\end{equation*}
To average this bound over the noise field, fix $\fleet$ and
$\plan{\gamma}_\tau^\theta$. The universally measurable selector
$\action_\tau^\theta$ has a Borel version
$b:\statespace\times\refSpace\to\actionspace$ that agrees with it
$\plan{\gamma}_\tau^\theta$-almost everywhere. Such a version follows
by applying \cite[Lemma~7.28(c)]{bertsekas1996stochastic} to the
selector's Dirac kernel and inverting the Dirac embedding on the
resulting full-measure set. Replacing the selector by $b$
changes neither the control nor the integrals. With
$\action=b(\state,\refVar)$, the cost in square brackets is nonnegative
and lower semianalytic. The nominal kernel and the law of the realized
kernel are Borel after this substitution on a
$\plan{\gamma}_\tau^\theta$-full set, where
\cref{assumption:measurable_field,assumption:barycenter_condition} hold. Extending the kernels off that set, we may apply
\cref{lemma:noiseexp} with base measure $\plan{\gamma}_\tau^\theta$ on
$\statespace\times\refSpace$ and a singleton action space. Since the
coupling and selector were fixed before the noise realization, taking
expectations and recognizing the Bellman operator gives
\begin{equation*}
\begin{aligned}
&\mathcal T_\tau^{D,\pi_{\mathrm{sep}}^{\theta}}\!\left[
  \kantorovich{\costtogosmall{\tau+1}^{D}}{}{}{}
\right](\fleet,\refPVar)
\\
&\quad=
\mathbb E_{\distoverdist{\tau}}\!\Bigg[
\begin{aligned}[t]
\int_{\statespace\times\actionspace\times\noisespace}
g_\tau^D(\state,\action,\noise)
\dd\rho_\tau^{\pi_{\mathrm{sep},\tau}^\theta(\fleet)}
  (\state,\action,\noise)
+
\kantorovich{\costtogosmall{\tau+1}^{D}}
 {\fleet_{\tau+1}^{D,\pi_{\mathrm{sep},\tau}^\theta(\fleet)}}{\refPVar}{}
\end{aligned}
\Bigg]
\\
&\quad\leq
\mathbb E_{\distoverdist{\tau}}\!\Bigg[
\int_{\statespace\times\refSpace}\int_{\noisespace}
\left[
 g_\tau^{D}(\state,\action,\noise)
 +\costtogosmall{\tau+1}^{D}\bigl(
   f_\tau^{D}(\state,\action,\noise),\refVar
 \bigr)
\right]
\dd\empnoisedist{\tau}^{\state,\action}(\noise)
\dd\plan{\gamma}_\tau^\theta(\state,\refVar)
\Bigg]
\\
&\quad=
\int_{\statespace\times\refSpace}\int_{\noisespace}
\left[
 g_\tau^{D}(\state,\action,\noise)
 +\costtogosmall{\tau+1}^{D}\bigl(
   f_\tau^{D}(\state,\action,\noise),\refVar
 \bigr)
\right]
\dd\noisedist{\tau}(\noise\st\state,\action)
\dd\plan{\gamma}_\tau^\theta(\state,\refVar),
\end{aligned}
\end{equation*}
where the last equality is exactly the application of \cref{lemma:noiseexp}
with the expression in square brackets as the cost.

By \eqref{eq:wc:near_optimal_action}, the inner integral is at most
$\costtogosmall{\tau}^{D}(\state,\refVar)+\septol{\tau}/2$. Therefore,
\begin{equation*}
\begin{aligned}
    \mathcal T_\tau^{D,\pi_{\mathrm{sep}}^{\theta}}\!\left[
      \kantorovich{\costtogosmall{\tau+1}^{D}}{}{}{}
    \right](\fleet,\refPVar)
    \leq
    \int_{\statespace\times\refSpace}
    \left[
      \costtogosmall{\tau}^{D}(\state,\refVar)
      +\frac{\septol{\tau}}{2}
    \right]
    \dd\plan{\gamma}_\tau^\theta(\state,\refVar)
    \leq
    \kantorovich{\costtogosmall{\tau}^{D}}{\fleet}{\refPVar}{}
    +\septol{\tau},
\end{aligned}
\end{equation*}
where the last inequality uses the near-optimality of the target assignment in
\eqref{eq:wc:near_optimal_transport} to contribute another
$\septol{\tau}/2$. This is exactly \eqref{eq:wc:policy_residual}.
\end{proof}

The previous lemma controls the cost of the chosen policy relative to
the decoupled transport surrogate. To compare that surrogate with the
decoupled optimal Bellman step, we also use the result proved in
Appendix~\ref{section:bound_quantification_v4}. In particular, 
applying \cref{lemma:bq4:bellman_defect} to the decoupled model gives
\begin{equation}
\label{eq:wc:separation_defect}
    0
    \leq
    \kantorovich{\costtogosmall{\tau}^{D}}{\fleet}{\refPVar}{}
    -\mathcal T_\tau^{D}\!\left[
      \kantorovich{\costtogosmall{\tau+1}^{D}}{}{}{}
    \right](\fleet,\refPVar)
    \leq
    L_{\tau+1}\dispersion{\tau}.
\end{equation}

We now have the three one-step estimates needed to derive the full performance
bound. (i) Lemma~\ref{lemma:wc:policy_residual} controls the separation-based
policy in the decoupled model, (ii)
Lemma~\ref{lemma:wc:operator_perturbation} transfers a Bellman step between the
coupled and decoupled models, and (iii) \eqref{eq:wc:separation_defect} compares the
decoupled transport surrogate with the decoupled optimal Bellman step. The
theorem below accumulates these errors over the remaining time steps.

\begin{mdframed}[hidealllines=true,backgroundcolor=blue!5,nobreak=true]
\begin{theorem}[Performance of separation under weak coupling]
\label{theorem:wc:performance}
Consider the coupled and decoupled models defined above. Assume that the field satisfies
\cref{assumption:measurable_field,assumption:barycenter_condition}, that the
decoupled single-agent cost-to-go satisfies
\cref{assumption:bq4:cost_to_go_lipschitz} with constants $L_\tau$, and that
\cref{assumption:wc:model_mismatch} holds with constants
$\statemismatch{\tau}$ and $\delta_\tau$. Let
$\pi_{\mathrm{sep}}^\theta$ be the policy defined in
\eqref{eq:wc:near_optimal_transport}--\eqref{eq:wc:separation_control}. Then,
for every $\tau\in\{0,\ldots,\horizon\}$,
$\fleet\in\probabilityspace{\statespace}$ with finite first moment, and
$\refPVar\in\probabilityspace{\refSpace}$,
\begin{equation}
\label{eq:wc:main_performance_bound}
    0
    \leq
    \costtogo{\tau}^{C,\pi_{\mathrm{sep}}^{\theta}}(\fleet,\refPVar)
    -\costtogo{\tau}^{C}(\fleet,\refPVar)
    \leq
    \sum_{r=\tau}^{\horizon-1}
    \left[
        \septol{r}
        +L_{r+1}\dispersion{r}
        +2\delta_r
        +2L_{r+1}\statemismatch{r}
    \right].
\end{equation}
When the transport and action minima are attained measurably, we can take
$\septol{r}=0$ for every $r$. In this case, at the initial population
$\bar\fleet_0$, the bound becomes
\begin{equation}
\label{eq:wc:exact_performance_bound}
    \costtogo{0}^{C,\pi_{\mathrm{sep}}^{0}}
    (\bar\fleet_0,\refPVar)
    -\costtogo{0}^{C}(\bar\fleet_0,\refPVar)
    \leq
    \sum_{r=0}^{\horizon-1}L_{r+1}\dispersion{r}
    +2\sum_{r=0}^{\horizon-1}
    \left(\delta_r+L_{r+1}\statemismatch{r}\right).
\end{equation}
\end{theorem}
\end{mdframed}

\begin{proof}
We use the decoupled transport cost as a common reference. First, we
bound from above the cost incurred by the separation-based policy in the coupled
model. Second, we bound the coupled optimal value from below using the same
reference. Finally, we subtract the second estimate from the first to obtain the
claimed performance bound. Both estimates are proved backward in time. At
$\tau=\horizon$, the policy value, the coupled optimum, and
$\kantorovich{\costtogosmall{\horizon}^{D}}{}{}{}$ all equal the common
terminal cost in~\eqref{eq:wc:common_terminal}, so both inductions have the
same base case. Explicitly, for every $(\fleet,\refPVar)$,
\begin{equation*}
\costtogo{\horizon}^{C,\pi_{\mathrm{sep}}^{\theta}}(\fleet,\refPVar)
=\costtogo{\horizon}^{C}(\fleet,\refPVar)
=\kantorovich{\costtogosmall{\horizon}^{D}}{\fleet}{\refPVar}{}
=\kantorovich{\terminalcostsmall}{\fleet}{\refPVar}{}.
\end{equation*}
Since the sums in \eqref{eq:wc:policy_upper} and
\eqref{eq:wc:optimal_lower} are empty at $\tau=\horizon$, these equalities
directly prove both base cases.

In both inductions that follow, we use the monotonicity and constant-shift properties of the Bellman operators (\cref{lemma:bq4:operator_properties}). 
Indeed, the lemma is stated for $\mathcal T_\tau^{D}=\mathcal T_\tau$, and since its proof uses only
the structure of the definition, it also applies verbatim to
$\mathcal T_\tau^{C}$ and to the fixed-policy operators
$\mathcal T_\tau^{m,\pi}$.

\medskip
\noindent\textbf{Step 1: Upper bound for the separation-based policy.}
We first show that the policy cost in the coupled model is at most the
decoupled transport cost plus the near-optimality and model-mismatch errors.
More precisely, for every $(\fleet,\refPVar)$,
\begin{equation}
\label{eq:wc:policy_upper}
\begin{aligned}
    \costtogo{\tau}^{C,\pi_{\mathrm{sep}}^{\theta}}
    (\fleet,\refPVar)
    \leq
    \kantorovich{\costtogosmall{\tau}^{D}}{\fleet}{\refPVar}{}
    +\sum_{r=\tau}^{\horizon-1}
    \left(\septol{r}+\delta_r+L_{r+1}\statemismatch{r}\right).
\end{aligned}
\end{equation}
Assume, by backward induction, that~\eqref{eq:wc:policy_upper} holds at
time $\tau+1$ for every population. The induction hypothesis says that, from
any such population, the cost of following the separation-based policy is at
most the corresponding decoupled transport cost plus the errors accumulated
from times $\tau+1$ to $\horizon-1$.

Now fix $(\fleet,\refPVar)$ and apply the induction hypothesis at the realized next population
$\fleet_{\tau+1}^{C,\pi_{\mathrm{sep},\tau}^{\theta}(\fleet)}$. We obtain
\begin{align*}
    \costtogo{\tau}^{C,\pi_{\mathrm{sep}}^{\theta}}(\fleet,\refPVar)
    &=
    \mathbb E_{\distoverdist{\tau}}\!\left[
    \int_{\statespace\times\actionspace\times\noisespace}
    g_\tau^C(\state,\action,\fleet,\noise)
    \dd\rho_\tau^{\pi_{\mathrm{sep},\tau}^{\theta}(\fleet)}
      (\state,\action,\noise)
    +
    \costtogo{\tau+1}^{C,\pi_{\mathrm{sep}}^{\theta}}\!\left(
      \fleet_{\tau+1}^{C,\pi_{\mathrm{sep},\tau}^{\theta}(\fleet)},
      \refPVar
    \right)
    \right]
    \\
    &\leq
    \mathbb E_{\distoverdist{\tau}}\!\left[
    \int_{\statespace\times\actionspace\times\noisespace}
    g_\tau^C(\state,\action,\fleet,\noise)
    \dd\rho_\tau^{\pi_{\mathrm{sep},\tau}^{\theta}(\fleet)}
      (\state,\action,\noise)
    +
    \kantorovich{\costtogosmall{\tau+1}^{D}}
      {\fleet_{\tau+1}^{C,\pi_{\mathrm{sep},\tau}^{\theta}(\fleet)}}
      {\refPVar}{}
    \right]\\
    &\qquad
    +\sum_{r=\tau+1}^{\horizon-1}
      \left(\septol{r}+\delta_r+L_{r+1}\statemismatch{r}\right)
    \\
    &=
    \mathcal T_\tau^{C,\pi_{\mathrm{sep}}^{\theta}}\!\left[
      \kantorovich{\costtogosmall{\tau+1}^{D}}{}{}{}
    \right](\fleet,\refPVar)
    +\sum_{r=\tau+1}^{\horizon-1}
      \left(\septol{r}+\delta_r+L_{r+1}\statemismatch{r}\right)
    \\
    &\leq
    \mathcal T_\tau^{D,\pi_{\mathrm{sep}}^{\theta}}\!\left[
      \kantorovich{\costtogosmall{\tau+1}^{D}}{}{}{}
    \right](\fleet,\refPVar)
    +\delta_\tau+L_{\tau+1}\statemismatch{\tau}
    +\sum_{r=\tau+1}^{\horizon-1}
      \left(\septol{r}+\delta_r+L_{r+1}\statemismatch{r}\right)
    \\
    &\leq
    \kantorovich{\costtogosmall{\tau}^{D}}{\fleet}{\refPVar}{}
    +\sum_{r=\tau}^{\horizon-1}
      \left(\septol{r}+\delta_r+L_{r+1}\statemismatch{r}\right).
\end{align*}
The second inequality is \eqref{eq:wc:policy_operator_perturbation}
in \cref{lemma:wc:operator_perturbation}, which costs
$\delta_\tau+L_{\tau+1}\statemismatch{\tau}$, while the last inequality is
\eqref{eq:wc:policy_residual} in \cref{lemma:wc:policy_residual}, 
which contributes $\septol{\tau}$. 
After adding these current errors to the errors from later times, we obtain
\eqref{eq:wc:policy_upper} at time $\tau$ and prove the upper bound.

\medskip
\noindent\textbf{Step 2: Lower bound for the coupled optimum.}
We now compare the coupled optimal value with the same decoupled
transport cost. Here the transport cost provides a lower bound after we
subtract the separation and model-mismatch errors:
\begin{equation}
\label{eq:wc:optimal_lower}
    \costtogo{\tau}^{C}(\fleet,\refPVar)
    \geq
    \kantorovich{\costtogosmall{\tau}^{D}}{\fleet}{\refPVar}{}
    -\sum_{r=\tau}^{\horizon-1}
    \left[
        L_{r+1}\dispersion{r}
        +\delta_r
        +L_{r+1}\statemismatch{r}
    \right].
\end{equation}
Now fix $(\fleet,\refPVar)$. Since the induction hypothesis is a pointwise lower
bound on $\costtogo{\tau+1}^{C}$, applying it inside $\mathcal T_\tau^{C}$ and using~\cref{lemma:bq4:operator_properties} gives
\begin{align*}
    \costtogo{\tau}^{C}(\fleet,\refPVar)
    &=
    \mathcal T_\tau^{C}\!\left[
      \costtogo{\tau+1}^{C}
    \right](\fleet,\refPVar)
    \\
    &\geq
    \mathcal T_\tau^{C}\!\left[
      \kantorovich{\costtogosmall{\tau+1}^{D}}{}{}{}
    \right](\fleet,\refPVar)
    -\sum_{r=\tau+1}^{\horizon-1}
      \left[L_{r+1}\dispersion{r}
            +\delta_r+L_{r+1}\statemismatch{r}\right]
    \\
    &\geq
    \mathcal T_\tau^{D}\!\left[
      \kantorovich{\costtogosmall{\tau+1}^{D}}{}{}{}
    \right](\fleet,\refPVar)
    -\delta_\tau-L_{\tau+1}\statemismatch{\tau}
    -\sum_{r=\tau+1}^{\horizon-1}
      \left[L_{r+1}\dispersion{r}
            +\delta_r+L_{r+1}\statemismatch{r}\right]
    \\
    &\geq
    \kantorovich{\costtogosmall{\tau}^{D}}{\fleet}{\refPVar}{}
    -\sum_{r=\tau}^{\horizon-1}
      \left[L_{r+1}\dispersion{r}
            +\delta_r+L_{r+1}\statemismatch{r}\right].
\end{align*}
Here, the second inequality is
\eqref{eq:wc:optimal_operator_perturbation} in \cref{lemma:wc:operator_perturbation}, 
which contributes $\delta_\tau+L_{\tau+1}\statemismatch{\tau}$.
Rearranging
\eqref{eq:wc:separation_defect} gives
\begin{equation*}
\mathcal T_\tau^{D}\!\left[
  \kantorovich{\costtogosmall{\tau+1}^{D}}{}{}{}
\right](\fleet,\refPVar)
\geq
\kantorovich{\costtogosmall{\tau}^{D}}{\fleet}{\refPVar}{}
-L_{\tau+1}\dispersion{\tau},
\end{equation*}
which is the last inequality in the display. This proves
\eqref{eq:wc:optimal_lower} at time $\tau$.

\medskip
\noindent\textbf{Step 3: Combine the two bounds.}
Subtracting the obtained lower bound~\eqref{eq:wc:optimal_lower} from
the upper bound~\eqref{eq:wc:policy_upper} cancels the common decoupled 
transport cost and gives
\begin{equation*}
\begin{aligned}
\costtogo{\tau}^{C,\pi_{\mathrm{sep}}^{\theta}}(\fleet,\refPVar)
-\costtogo{\tau}^{C}(\fleet,\refPVar)
&\leq
\sum_{r=\tau}^{\horizon-1}
\left(\septol{r}+\delta_r+L_{r+1}\statemismatch{r}\right)
+\sum_{r=\tau}^{\horizon-1}
\left(L_{r+1}\dispersion{r}
      +\delta_r+L_{r+1}\statemismatch{r}\right)
\\
&\quad=
\sum_{r=\tau}^{\horizon-1}
\left[
\septol{r}+L_{r+1}\dispersion{r}
+2\delta_r+2L_{r+1}\statemismatch{r}
\right].
\end{aligned}
\end{equation*}
That corresponds exactly to the upper bound in \eqref{eq:wc:main_performance_bound}. 
To conclude the proof, it is enough to note that the zero lower bound is immediate because
$\pi_{\mathrm{sep}}^{\theta}$ is admissible in the coupled model and
$\costtogo{\tau}^{C}$ is the infimum of the costs over all admissible policies there.
Finally, if exact minimizers can be selected measurably,
setting $\tau=0$, $\fleet=\bar\fleet_0$, and $\septol{r}=0$ gives
\eqref{eq:wc:exact_performance_bound}.
\end{proof}

\paragraph{Interpretation.}
The three terms in the bound correspond to the three comparisons used in
the proof. The tolerance $\septol{r}$ comes from the near-optimal target plan
and action used to define the separation-based policy in 
\eqref{eq:wc:near_optimal_transport}--\eqref{eq:wc:separation_control}. 
The term
$L_{r+1}\dispersion{r}$ is already present in \cref{lemma:bq4:bellman_defect} 
for the decoupled problem and bounds
the difference between assigning targets before the noise is observed and
optimizing the population Bellman step after the resulting population is
known. Finally, $\delta_r+L_{r+1}\statemismatch{r}$ is the cost of replacing
one coupled Bellman step by its decoupled counterpart, as bounded in \cref{lemma:wc:operator_perturbation}.
This last term appears twice because the proof makes that replacement 
once for the deployed policy and once for the coupled optimal value.

\paragraph{Bounding the model mismatch.}
Note that Assumption~\ref{assumption:wc:model_mismatch} only quantifies the 
model mismatch in terms of the abstract mismatch $\statemismatch{\tau}$. 
To use the theorem in a concrete model, it is often more convenient to relate the 
quantity to the strength of the population interaction. 
This is the case, for example, when the coupled dynamics are Lipschitz 
in the population distribution:
\begin{equation}
\label{eq:wc:dynamics_lipschitz_population}
    d_{\statespace}\!\left(
        f_\tau^{C}(\state,\action,\fleet,\noise),
        f_\tau^{C}(\state,\action,\widetilde\fleet,\noise)
    \right)
    \leq
    \popLip{\tau} W_1(\fleet,\widetilde\fleet),
\end{equation}
uniformly in $(\state,\action,\noise)$. In this case, indeed, the constant 
$\popLip{\tau}$ measures how strongly the population affects the dynamics. 
If, in addition, the decoupled surrogate is obtained by freezing that
population argument at a reference profile $\refProfile{\tau}$:
\begin{equation*}
    f_\tau^{D}(\state,\action,\noise)
    =f_\tau^{C}
      (\state,\action,\refProfile{\tau},\noise),
\end{equation*}
and it holds that $W_1(\fleet,\refProfile{\tau})\leq R_\tau$ at every 
population where the two models are compared, then we can apply
\eqref{eq:wc:dynamics_lipschitz_population} with
$\widetilde\fleet=\refProfile{\tau}$. By the definition of $f_\tau^D$, this
allows us to bound the general model mismatch in 
\cref{assumption:wc:model_mismatch} by the more informative
$\statemismatch{\tau}\leq\popLip{\tau}R_\tau$. 
Under this class of models, \eqref{eq:wc:exact_performance_bound} becomes
\begin{equation*}
    \costtogo{0}^{C,\pi_{\mathrm{sep}}^{0}}(\bar\fleet_0,\refPVar)
    -\costtogo{0}^{C}(\bar\fleet_0,\refPVar)
    \leq
    \sum_{\tau=0}^{\horizon-1}L_{\tau+1}\dispersion{\tau}
    +2\sum_{\tau=0}^{\horizon-1}
    \left(\delta_\tau+L_{\tau+1}\popLip{\tau}R_\tau\right).
\end{equation*}

If, in addition, the stage costs coincide, so that $\delta_\tau=0$, then
\begin{equation}
\label{eq:wc:lipschitz_performance}
    \costtogo{0}^{C,\pi_{\mathrm{sep}}^{0}}(\bar\fleet_0,\refPVar)
    -\costtogo{0}^{C}(\bar\fleet_0,\refPVar)
    \leq
    \sum_{\tau=0}^{\horizon-1}L_{\tau+1}\dispersion{\tau}
    +2\sum_{\tau=0}^{\horizon-1}
    L_{\tau+1}\popLip{\tau}R_\tau.
\end{equation}
The second sum in \eqref{eq:wc:lipschitz_performance} corresponds 
to the additional error due to weak coupling: it grows linearly in $\popLip{\tau}$ and vanishes when the dynamics stop depending on the population, in which case \eqref{eq:wc:lipschitz_performance} reduces to the bound of \cref{theorem:bq4:direct_bound}.

\section{Additional numerical results}
\label{app:additional_experiments}

This appendix provides the implementation details and complementary results for the experiments in Section~\ref{section:results}. We first describe the grid-world experiments and report the additional performance metrics used in our evaluation (Appendix~\ref{appendix:results:performance}). We then give the details of the scalability study (Appendix~\ref{appendix:results:scaling}) and examine how the origin--destination pairs covered during training affect transfer across population sizes (Appendix~\ref{appendix:results:pair-frequency}). Finally, we describe the Manhattan road network (Appendix~\ref{appendix:results:manhattan}) and report the computational resources used in the experiments.

\subsection{Additional details for the grid-world experiments in Section~\ref{section:results:benchmarking}}
\label{appendix:results:performance}

\paragraph{Environment.}
We use an $11\times 30$ grid with horizon $\horizon=100$ and $\nagents=7$ agents. At the beginning of each episode, the agents are placed uniformly at random in the first column, while the $\nagents$ targets are sampled uniformly from the last ten columns.
The action space is $\actionspace=\{\texttt{up},\texttt{down},\texttt{left},\texttt{right},\texttt{stay}\}$. At each time step, the selected action is executed with probability $1-p$ and is replaced uniformly by one of the other four actions with probability $p$. We consider $p \in \{0,0.1,0.25\}$ under the individual, local, and global noise models defined in Appendix~\ref{section:noise}. Specifically, under individual noise, each agent receives an independent noise draw, under local noise, agents with the same state-action pair share the same draw, while under global noise, the entire population shares one draw. We finally clip the resulting positions of the agents to make them remain in the grid after each transition.

An agent becomes inactive after reaching a relevant target and enters an absorbing state. In the case of a MARL agent, a relevant target for an agent is any previously unoccupied target, in the case of $\famalg$, instead, each agent is assigned a target $z$ before the rollout, and the agent becomes inactive when it reaches that specific target.\footnote{This target-dependent stopping rule is an experimental convention and is not part of the target-independent motion map $\dynamics{}{}{\timestep}(\state,\action,\noise)$ in Section~\ref{section:problem_setting}. The simulator also holds an inactive agent at its arrival position and masks subsequent actions, perturbations, and stage costs.} 
For $\famalg$ with a fixed assignment, both the routing policy and the stopping rule depend only on each agent’s own state and target. Conditional on the initial states and assignments, the trajectories are independent under individual noise, but may be correlated under local or global noise.

\paragraph{Reward structure.}
An active agent incurs a stage cost of $1$ on every transition until it does not reach a target. At the horizon, the MARL training objective also includes the empirical optimal transport cost $\kantorovich{c}{\fleet_\horizon}{\refPVar}{}$, where $c$ is the Manhattan distance. In the target-conditioned single-agent problems used by $\famalg$, the corresponding terminal penalty is the Manhattan distance from the agent to its assigned target. At evaluation, we report travel time, reach rate, and empirical terminal optimal transport cost separately.

\paragraph{Baselines.} We compare $\famalg$ with MAPPO, IPPO, HAPPO, VDN, QMIX, and MFQ. All baselines use the same episode budget and training seeds.
All neural policies have two hidden layers with 64 units and $\tanh$ activations. The separation-based neural methods are trained as goal-conditioned single-agent controllers. For $\algname{PPO}$, $\algname{A2C}$, and $\algname{DDQN}$, the policy receives the input
\begin{equation*}
  \left(\frac{r}{H_g},\ \frac{c}{W_g},\ \frac{r_z}{H_g},\ \frac{c_z}{W_g},\ \frac{\timestep}{\horizon-1},\frac{\lVert\state-z\rVert_1}{H_g+W_g}\right),
\end{equation*}
where $\state=(r,c)$ is the current agent position, $z=(r_z,c_z)$ is its assigned target, $\timestep$ is the time step, and $H_g=10$ and $W_g=29$ are the maximum row and column indices. The separation policy therefore selects an action without observing the full population. Coordination between agents occurs only through solving the optimal transport problem, whose solution selects $z$.

MAPPO, IPPO, HAPPO, QMIX, VDN, and MFQ instead receive target-relative local observations. For agent $i$, this observation has dimension $5+3\nagents$:
\begin{equation*}
  \left(\frac{r_i}{H_g},\ \frac{c_i}{W_g},\ \frac{i}{\nagents-1},\ \frac{\timestep}{\horizon-1},\ a_i,
  \frac{\Delta r_{i1}}{H_g},\frac{\Delta c_{i1}}{W_g},b_1,\ldots,
  \frac{\Delta r_{i\nagents}}{H_g},\frac{\Delta c_{i\nagents}}{W_g},b_{\nagents}\right).
\end{equation*}
Here $a_i$ indicates whether agent $i$ is active, $\Delta r_{ij}=r_{z_j}-r_i$ and $\Delta c_{ij}=c_{z_j}-c_i$ are the coordinates of target $j$ relative to agent $i$, and $b_j$ indicates whether target $j$ remains unoccupied. 
Thus, each agent observes the target set as an ordered list of relative offsets and availability flags. 
MAPPO also uses a centralized critic state of dimension $6\nagents+1$. 
This state contains all normalized agent positions, all active-agent flags, the time step, all normalized target positions, and all target-availability flags. 
HAPPO uses the same centralized critic state, but has one non-shared actor per agent and updates the actors sequentially. 
QMIX passes the global state to its monotone mixing network, whose embedding dimension is 32. 
IPPO uses a local critic, VDN uses additive value decomposition, and MFQ augments each agent's action value with the global mean action distribution of the currently active agents. 
All neural methods are optimized with Adam~\cite{kingma2015adam}.

\paragraph{Training and evaluation.}
For each algorithm, noise model, and noise probability, we train five policies using seeds $0$--$4$ and evaluate each policy on $50$ rollout instances. 
Within a given condition, evaluation seed $s\in\{0,\ldots,49\}$ fixes the starts, targets, and initial state of the transition-noise generator for every algorithm. 

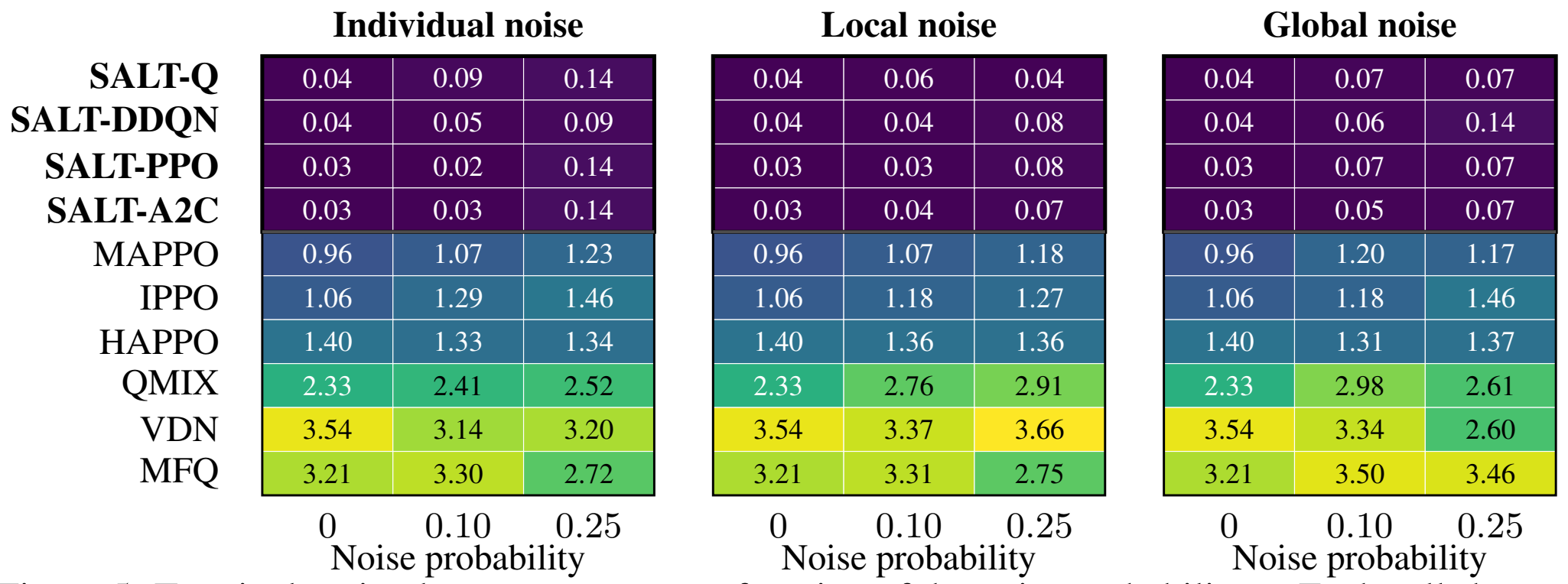
\begin{figure}[ht]
    \begin{tikzpicture}[x=1.13cm,y=0.38cm]
  \node[anchor=east,font=\small\bfseries] at (-0.22,9.5) {SALT-Q};
  \node[anchor=east,font=\small\bfseries] at (-0.22,8.5) {SALT-DDQN};
  \node[anchor=east,font=\small\bfseries] at (-0.22,7.5) {SALT-PPO};
  \node[anchor=east,font=\small\bfseries] at (-0.22,6.5) {SALT-A2C};
  \node[anchor=east,font=\small] at (-0.22,5.5) {MAPPO};
  \node[anchor=east,font=\small] at (-0.22,4.5) {IPPO};
  \node[anchor=east,font=\small] at (-0.22,3.5) {HAPPO};
  \node[anchor=east,font=\small] at (-0.22,2.5) {QMIX};
  \node[anchor=east,font=\small] at (-0.22,1.5) {VDN};
  \node[anchor=east,font=\small] at (-0.22,0.5) {MFQ};

  \node[anchor=south,font=\bfseries\small] at (1.500,10.1) {Individual noise};
  \Cell{0.000}{9}{68}{2}{86}{white}{0.04}
  \Cell{1.000}{9}{70}{8}{92}{white}{0.09}
  \Cell{2.000}{9}{71}{13}{96}{white}{0.14}
  \Cell{0.000}{8}{68}{2}{86}{white}{0.04}
  \Cell{1.000}{8}{69}{4}{87}{white}{0.05}
  \Cell{2.000}{8}{70}{8}{92}{white}{0.09}
  \Cell{0.000}{7}{68}{1}{84}{white}{0.03}
  \Cell{1.000}{7}{68}{1}{84}{white}{0.02}
  \Cell{2.000}{7}{71}{13}{96}{white}{0.14}
  \Cell{0.000}{6}{68}{2}{86}{white}{0.03}
  \Cell{1.000}{6}{68}{2}{86}{white}{0.03}
  \Cell{2.000}{6}{71}{13}{96}{white}{0.14}
  \Cell{0.000}{5}{58}{84}{140}{white}{0.96}
  \Cell{1.000}{5}{54}{92}{141}{white}{1.07}
  \Cell{2.000}{5}{49}{104}{142}{white}{1.23}
  \Cell{0.000}{4}{54}{92}{141}{white}{1.06}
  \Cell{1.000}{4}{47}{108}{142}{white}{1.29}
  \Cell{2.000}{4}{42}{119}{142}{white}{1.46}
  \Cell{0.000}{3}{44}{115}{142}{white}{1.40}
  \Cell{1.000}{3}{46}{111}{142}{white}{1.33}
  \Cell{2.000}{3}{46}{111}{142}{white}{1.34}
  \Cell{0.000}{2}{42}{176}{127}{white}{2.33}
  \Cell{1.000}{2}{50}{182}{122}{black}{2.41}
  \Cell{2.000}{2}{61}{188}{116}{black}{2.52}
  \Cell{0.000}{1}{234}{229}{26}{black}{3.54}
  \Cell{1.000}{1}{160}{218}{57}{black}{3.14}
  \Cell{2.000}{1}{170}{220}{50}{black}{3.20}
  \Cell{0.000}{0}{173}{220}{48}{black}{3.21}
  \Cell{1.000}{0}{189}{223}{38}{black}{3.30}
  \Cell{2.000}{0}{88}{199}{101}{black}{2.72}
  \draw[black,line width=0.55pt] (0.000,0) rectangle (3.000,10);
  \draw[black,line width=0.95pt] (0.000,6) rectangle (3.000,10);
  \draw[black!70,line width=0.75pt] (0.000,6) -- (3.000,6);
  \node[anchor=north,font=\small] at (0.500,-0.1) {$0$};
  \node[anchor=north,font=\small] at (1.500,-0.1) {$0.10$};
  \node[anchor=north,font=\small] at (2.500,-0.1) {$0.25$};
  \node[overlay,anchor=north,font=\small] at (1.500,-0.8) {Noise probability};

  \node[anchor=south,font=\bfseries\small] at (4.950,10.1) {Local noise};
  \Cell{3.450}{9}{68}{2}{86}{white}{0.04}
  \Cell{4.450}{9}{69}{4}{87}{white}{0.06}
  \Cell{5.450}{9}{68}{2}{86}{white}{0.04}
  \Cell{3.450}{8}{68}{2}{86}{white}{0.04}
  \Cell{4.450}{8}{68}{2}{86}{white}{0.04}
  \Cell{5.450}{8}{70}{7}{90}{white}{0.08}
  \Cell{3.450}{7}{68}{1}{84}{white}{0.03}
  \Cell{4.450}{7}{68}{2}{86}{white}{0.03}
  \Cell{5.450}{7}{70}{7}{90}{white}{0.08}
  \Cell{3.450}{6}{68}{2}{86}{white}{0.03}
  \Cell{4.450}{6}{68}{2}{86}{white}{0.04}
  \Cell{5.450}{6}{69}{5}{89}{white}{0.07}
  \Cell{3.450}{5}{58}{84}{140}{white}{0.96}
  \Cell{4.450}{5}{54}{93}{141}{white}{1.07}
  \Cell{5.450}{5}{50}{100}{142}{white}{1.18}
  \Cell{3.450}{4}{54}{92}{141}{white}{1.06}
  \Cell{4.450}{4}{50}{100}{142}{white}{1.18}
  \Cell{5.450}{4}{47}{107}{142}{white}{1.27}
  \Cell{3.450}{3}{44}{115}{142}{white}{1.40}
  \Cell{4.450}{3}{45}{113}{142}{white}{1.36}
  \Cell{5.450}{3}{45}{113}{142}{white}{1.36}
  \Cell{3.450}{2}{42}{176}{127}{white}{2.33}
  \Cell{4.450}{2}{94}{201}{98}{black}{2.76}
  \Cell{5.450}{2}{119}{209}{83}{black}{2.91}
  \Cell{3.450}{1}{234}{229}{26}{black}{3.54}
  \Cell{4.450}{1}{202}{225}{31}{black}{3.37}
  \Cell{5.450}{1}{253}{231}{37}{black}{3.66}
  \Cell{3.450}{0}{173}{220}{48}{black}{3.21}
  \Cell{4.450}{0}{189}{223}{38}{black}{3.31}
  \Cell{5.450}{0}{92}{200}{99}{black}{2.75}
  \draw[black,line width=0.55pt] (3.450,0) rectangle (6.450,10);
  \draw[black,line width=0.95pt] (3.450,6) rectangle (6.450,10);
  \draw[black!70,line width=0.75pt] (3.450,6) -- (6.450,6);
  \node[anchor=north,font=\small] at (3.950,-0.1) {$0$};
  \node[anchor=north,font=\small] at (4.950,-0.1) {$0.10$};
  \node[anchor=north,font=\small] at (5.950,-0.1) {$0.25$};
  \node[overlay,anchor=north,font=\small] at (4.950,-0.8) {Noise probability};

  \node[anchor=south,font=\bfseries\small] at (8.400,10.1) {Global noise};
  \Cell{6.900}{9}{68}{2}{86}{white}{0.04}
  \Cell{7.900}{9}{69}{5}{89}{white}{0.07}
  \Cell{8.900}{9}{69}{5}{89}{white}{0.07}
  \Cell{6.900}{8}{68}{2}{86}{white}{0.04}
  \Cell{7.900}{8}{69}{5}{89}{white}{0.06}
  \Cell{8.900}{8}{71}{13}{96}{white}{0.14}
  \Cell{6.900}{7}{68}{1}{84}{white}{0.03}
  \Cell{7.900}{7}{69}{5}{89}{white}{0.07}
  \Cell{8.900}{7}{69}{5}{89}{white}{0.07}
  \Cell{6.900}{6}{68}{2}{86}{white}{0.03}
  \Cell{7.900}{6}{69}{4}{87}{white}{0.05}
  \Cell{8.900}{6}{69}{5}{89}{white}{0.07}
  \Cell{6.900}{5}{58}{84}{140}{white}{0.96}
  \Cell{7.900}{5}{49}{102}{142}{white}{1.20}
  \Cell{8.900}{5}{51}{99}{141}{white}{1.17}
  \Cell{6.900}{4}{54}{92}{141}{white}{1.06}
  \Cell{7.900}{4}{50}{100}{142}{white}{1.18}
  \Cell{8.900}{4}{42}{119}{142}{white}{1.46}
  \Cell{6.900}{3}{44}{115}{142}{white}{1.40}
  \Cell{7.900}{3}{46}{109}{142}{white}{1.31}
  \Cell{8.900}{3}{45}{113}{142}{white}{1.37}
  \Cell{6.900}{2}{42}{176}{127}{white}{2.33}
  \Cell{7.900}{2}{129}{211}{77}{black}{2.98}
  \Cell{8.900}{2}{74}{193}{109}{black}{2.61}
  \Cell{6.900}{1}{234}{229}{26}{black}{3.54}
  \Cell{7.900}{1}{197}{224}{33}{black}{3.34}
  \Cell{8.900}{1}{72}{193}{110}{black}{2.60}
  \Cell{6.900}{0}{173}{220}{48}{black}{3.21}
  \Cell{7.900}{0}{226}{228}{24}{black}{3.50}
  \Cell{8.900}{0}{218}{227}{25}{black}{3.46}
  \draw[black,line width=0.55pt] (6.900,0) rectangle (9.900,10);
  \draw[black,line width=0.95pt] (6.900,6) rectangle (9.900,10);
  \draw[black!70,line width=0.75pt] (6.900,6) -- (9.900,6);
  \node[anchor=north,font=\small] at (7.400,-0.1) {$0$};
  \node[anchor=north,font=\small] at (8.400,-0.1) {$0.10$};
  \node[anchor=north,font=\small] at (9.400,-0.1) {$0.25$};
  \node[overlay,anchor=north,font=\small] at (8.400,-0.8) {Noise probability};
\end{tikzpicture}
    \caption{Terminal optimal transport cost as a function of the noise probability $p$. Each cell shows the pooled arithmetic mean under the training and evaluation protocol described above. Figure~\ref{fig:gridworld_combined_metrics} shows the corresponding distributions. Brighter cells indicate higher cost.}
    \label{fig:OTcost}
\end{figure}

\begin{figure}[ht]
    \includegraphics[width=\linewidth]{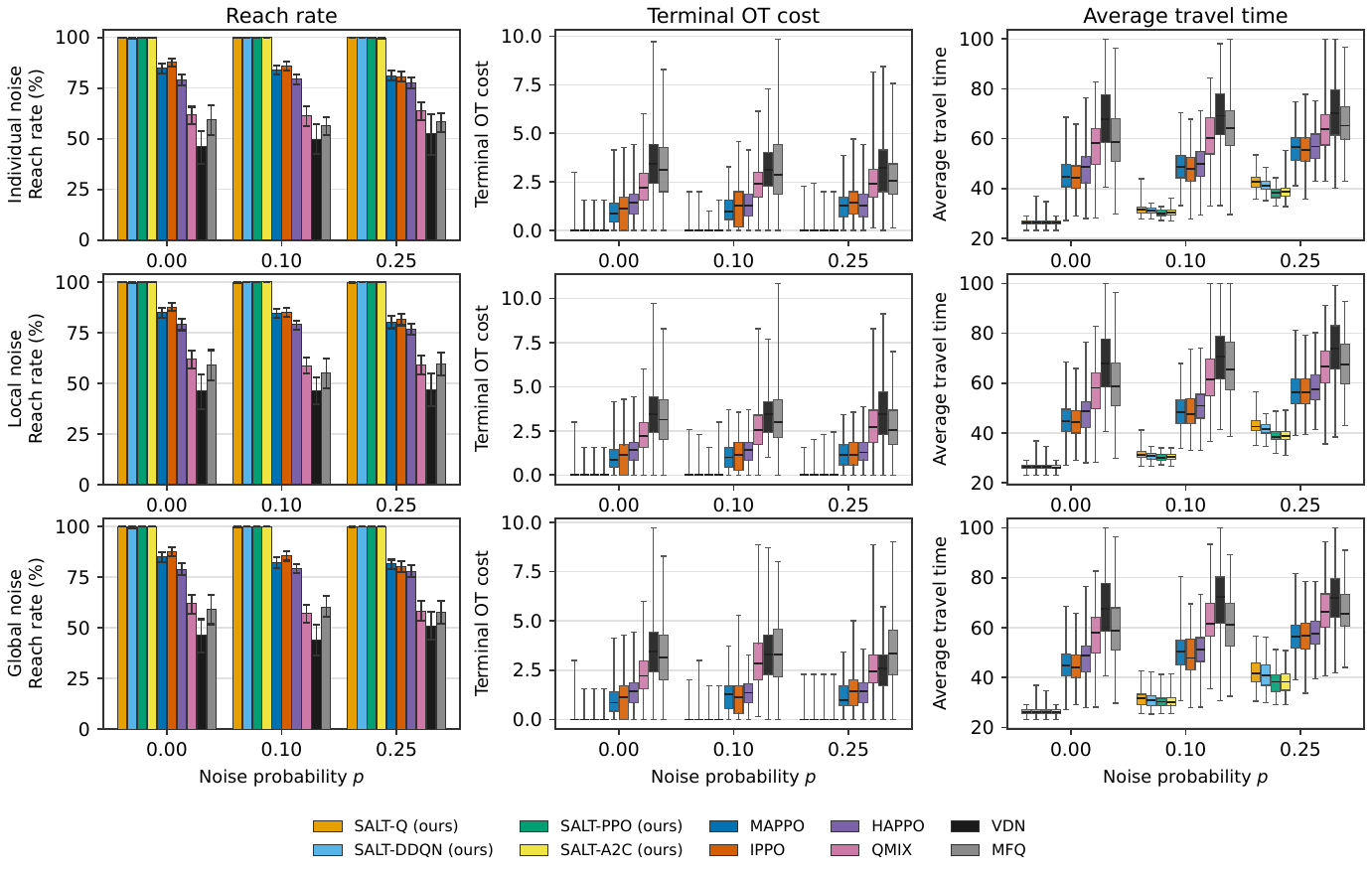}
    \caption{Reach rate, terminal optimal transport cost, and average travel time under the evaluation protocol described above. Reach-rate bars show pooled means and two-sided $95\%$ crossed two-way bootstrap confidence intervals. The remaining panels show the pooled outcome distributions. Boxes span the interquartile range, center lines are medians, and whiskers show the observed minimum and maximum.}
    \label{fig:gridworld_combined_metrics}
\end{figure}

For each metric, we pool the $50$ outcomes from each training seed, and report the arithmetic mean of the resulting $250$ outcomes in
the heatmaps in Figures~\ref{fig:TravelTimesBenchmark}, \ref{fig:OTcost}, and \ref{fig:ReachRate}. 
We then summarize all the results and include the corresponding confidence intervals in Figure~\ref{fig:gridworld_combined_metrics}.
In particular, the reach-rate bars in Figure~\ref{fig:gridworld_combined_metrics} show the pooled means together with two-sided $95\%$ confidence intervals,
while the terminal-cost and travel-time boxplots show the median and interquartile range (from the boxes), and observed minimum and maximum of the pooled outcomes (from the whiskers).

\paragraph{Hyperparameters.}
Tables~\ref{tab:gridworld-training-settings},~\ref{tab:separation-hyperparameters}, and~\ref{tab:marl-baseline-hyperparameters} report all hyperparameters used in the grid-world experiments.
\begin{table}[h]
\centering
\caption{Shared grid-world training budget and evaluation settings for Section~\ref{section:results:benchmarking}.}
\label{tab:gridworld-training-settings}
\begin{tabular}{ll}
\toprule
Quantity & Value \\
\midrule
Grid / horizon & $11 \times 30$, $\horizon=100$ \\
Agents / targets & $\nagents=7$ \\
Training seeds & $0$--$4$ (5 independent runs) \\
Evaluation instances & 50 per training seed, 250 pooled \\
Noise probabilities & $0, 0.1, 0.25$ \\
Noise modes & individual, local, global \\
Training episodes & 320{,}000 \\
Environment interactions & $320{,}000 \times 7 \times 100 = 2.24\times10^8$ per run \\
Batch episodes & 64 \\
Update batches for neural methods & 5{,}000 \\
Step cost & 1 \\
Goal bonus & 0 \\
Evaluation target region & columns 20--29 \\
Evaluation rematching for $\famalg$ & every time step \\
\bottomrule
\end{tabular}
\end{table}

\begin{table}[h]
\centering
\caption{Hyperparameters used for $\famalg$ in Section~\ref{section:results:benchmarking}.}
\label{tab:separation-hyperparameters}
\resizebox{\textwidth}{!}{%
\begin{tabular}{lcccc}
\toprule
Hyperparameter & $\algname{Q}$ & $\algname{DDQN}$ & $\algname{PPO}$ & $\algname{A2C}$ \\
\midrule
Discount $\gamma$ & 0.98 & 0.98 & 0.98 & 0.98 \\
Learning rate
& 0.25 $\to$ 0.02
& $3\times10^{-4}$
& \makecell{actor $3\times10^{-4}$\\critic $10^{-3}$}
& \makecell{actor $3\times10^{-4}$\\critic $10^{-3}$} \\
Learning-rate decay & 20{,}000 episodes & -- & -- & -- \\
Hidden size & -- & 64 & 64 & 64 \\
GAE $\lambda$ & -- & -- & 0.95 & 0.95 \\
PPO clip / epochs & -- & -- & 0.2 / 4 & -- \\
Entropy coefficient & -- & -- & 0.01 & 0.01 \\
Maximum gradient norm & -- & 0.5 & 0.5 & 0.5 \\
$\epsilon$-exploration & 0.5 $\to$ 0.05 & 0.5 $\to$ 0.05 & 0.5 $\to$ 0.05 & 0.5 $\to$ 0.05 \\
$\epsilon$-decay & 160{,}000 episodes & $1.12\times10^8$ interactions & $1.12\times10^8$ interactions & $1.12\times10^8$ interactions \\
Replay capacity & -- & 500{,}000 & -- & -- \\
Replay warmup / minibatch & -- & 5{,}000 / 256 & -- & -- \\
Learner steps per batch & -- & 64 & -- & -- \\
Target update period & -- & 500 learner steps & -- & -- \\
\bottomrule
\end{tabular}
}
\end{table}

For $\algname{PPO}$ and $\algname{A2C}$, the $\epsilon$ schedule in Table~\ref{tab:separation-hyperparameters} controls random exploration of the target assignment, while action exploration is induced by the stochastic actor. $\algname{Q}$ and $\algname{DDQN}$ use the same schedule for random assignments and $\epsilon$-greedy action exploration.

\begin{table}[h]
\centering
\caption{Hyperparameters used for MARL baselines in Section~\ref{section:results:benchmarking}.}
\label{tab:marl-baseline-hyperparameters}
\resizebox{\textwidth}{!}{%
\begin{tabular}{lcccccc}
\toprule
Hyperparameter & MAPPO & IPPO & HAPPO & QMIX & VDN & MFQ \\
\midrule
Discount $\gamma$ & 0.98 & 0.98 & 0.98 & 0.98 & 0.98 & 0.98 \\
Learning rate
& \makecell{actor $3\times10^{-4}$\\critic $10^{-3}$}
& \makecell{actor $3\times10^{-4}$\\critic $10^{-3}$}
& \makecell{actor $3\times10^{-4}$\\critic $10^{-3}$}
& $3\times10^{-4}$
& $3\times10^{-4}$
& $3\times10^{-4}$ \\
Hidden size & 64 & 64 & 64 & 64 & 64 & 64 \\
GAE $\lambda$ & 0.95 & 0.95 & 0.95 & -- & -- & -- \\
PPO clip / epochs & 0.2 / 4 & 0.2 / 4 & 0.2 / 4 & -- & -- & -- \\
Entropy coefficient & 0.01 $\to$ 0.001 & 0.01 $\to$ 0.001 & 0.01 $\to$ 0.001 & -- & -- & -- \\
Maximum gradient norm & 0.5 & 0.5 & 0.5 & 0.5 & 0.5 & 0.5 \\
Actor minibatch & 4{,}096 & 4{,}096 & 4{,}096 & -- & -- & -- \\
Critic minibatch & 1{,}024 & 4{,}096 & 1{,}024 & -- & -- & -- \\
Replay capacity & -- & -- & -- & 100{,}000 & 100{,}000 & 100{,}000 \\
Minimum replay size & -- & -- & -- & 2{,}048 & 2{,}048 & 2{,}048 \\
Replay batch size & -- & -- & -- & 512 & 512 & 512 \\
Updates per batch & -- & -- & -- & 8 & 8 & 8 \\
Target update period & -- & -- & -- & 50 batches & 50 batches & 50 batches \\
Exploration & -- & -- & -- & $\epsilon$: 0.5 $\to$ 0.05 & $\epsilon$: 0.5 $\to$ 0.05 & \makecell{$\tau$: 5 $\to$ 0.1\\$\epsilon$: 0.3 $\to$ 0.02} \\
Exploration decay & -- & -- & -- & 160{,}000 episodes & 160{,}000 episodes & 320{,}000 episodes \\
\bottomrule
\end{tabular}
}
\end{table}

\paragraph{Additional performance metrics.}

\begin{figure}[ht]
    \begin{tikzpicture}[x=1.13cm,y=0.38cm]
  \node[anchor=east,font=\small\bfseries] at (-0.22,9.5) {SALT-Q};
  \node[anchor=east,font=\small\bfseries] at (-0.22,8.5) {SALT-DDQN};
  \node[anchor=east,font=\small\bfseries] at (-0.22,7.5) {SALT-PPO};
  \node[anchor=east,font=\small\bfseries] at (-0.22,6.5) {SALT-A2C};
  \node[anchor=east,font=\small] at (-0.22,5.5) {MAPPO};
  \node[anchor=east,font=\small] at (-0.22,4.5) {IPPO};
  \node[anchor=east,font=\small] at (-0.22,3.5) {HAPPO};
  \node[anchor=east,font=\small] at (-0.22,2.5) {QMIX};
  \node[anchor=east,font=\small] at (-0.22,1.5) {VDN};
  \node[anchor=east,font=\small] at (-0.22,0.5) {MFQ};

  \node[anchor=south,font=\bfseries\small] at (1.500,10.1) {Individual noise};
  \Cell{0.000}{9}{68}{1}{84}{white}{100}
  \Cell{1.000}{9}{68}{1}{84}{white}{99.9}
  \Cell{2.000}{9}{68}{1}{84}{white}{99.9}
  \Cell{0.000}{8}{68}{1}{84}{white}{99.8}
  \Cell{1.000}{8}{68}{1}{84}{white}{100}
  \Cell{2.000}{8}{68}{1}{84}{white}{100}
  \Cell{0.000}{7}{68}{1}{84}{white}{100}
  \Cell{1.000}{7}{68}{1}{84}{white}{100}
  \Cell{2.000}{7}{68}{1}{84}{white}{100}
  \Cell{0.000}{6}{68}{1}{84}{white}{100}
  \Cell{1.000}{6}{68}{1}{84}{white}{100}
  \Cell{2.000}{6}{68}{1}{84}{white}{99.8}
  \Cell{0.000}{5}{56}{88}{140}{white}{84.9}
  \Cell{1.000}{5}{54}{92}{141}{white}{83.9}
  \Cell{2.000}{5}{48}{105}{142}{white}{81.1}
  \Cell{0.000}{4}{62}{73}{137}{white}{87.7}
  \Cell{1.000}{4}{59}{82}{139}{white}{85.9}
  \Cell{2.000}{4}{47}{107}{142}{white}{80.7}
  \Cell{0.000}{3}{44}{114}{142}{white}{79}
  \Cell{1.000}{3}{45}{113}{142}{white}{79.4}
  \Cell{2.000}{3}{42}{119}{142}{white}{77.7}
  \Cell{0.000}{2}{59}{187}{117}{black}{61.9}
  \Cell{1.000}{2}{64}{189}{114}{black}{61.3}
  \Cell{2.000}{2}{47}{180}{124}{black}{63.7}
  \Cell{0.000}{1}{229}{228}{25}{black}{46.2}
  \Cell{1.000}{1}{186}{222}{40}{black}{49.9}
  \Cell{2.000}{1}{152}{216}{62}{black}{52.7}
  \Cell{0.000}{0}{82}{197}{105}{black}{59.3}
  \Cell{1.000}{0}{110}{206}{88}{black}{56.5}
  \Cell{2.000}{0}{90}{200}{100}{black}{58.4}
  \draw[black,line width=0.55pt] (0.000,0) rectangle (3.000,10);
  \draw[black,line width=0.95pt] (0.000,6) rectangle (3.000,10);
  \draw[black!70,line width=0.75pt] (0.000,6) -- (3.000,6);
  \node[anchor=north,font=\small] at (0.500,-0.1) {$0$};
  \node[anchor=north,font=\small] at (1.500,-0.1) {$0.10$};
  \node[anchor=north,font=\small] at (2.500,-0.1) {$0.25$};
  \node[overlay,anchor=north,font=\small] at (1.500,-0.8) {Noise probability};

  \node[anchor=south,font=\bfseries\small] at (4.950,10.1) {Local noise};
  \Cell{3.450}{9}{68}{1}{84}{white}{100}
  \Cell{4.450}{9}{68}{1}{84}{white}{99.9}
  \Cell{5.450}{9}{68}{1}{84}{white}{99.9}
  \Cell{3.450}{8}{68}{1}{84}{white}{99.8}
  \Cell{4.450}{8}{68}{1}{84}{white}{100}
  \Cell{5.450}{8}{68}{1}{84}{white}{100}
  \Cell{3.450}{7}{68}{1}{84}{white}{100}
  \Cell{4.450}{7}{68}{1}{84}{white}{100}
  \Cell{5.450}{7}{68}{1}{84}{white}{100}
  \Cell{3.450}{6}{68}{1}{84}{white}{100}
  \Cell{4.450}{6}{68}{1}{84}{white}{100}
  \Cell{5.450}{6}{68}{1}{84}{white}{100}
  \Cell{3.450}{5}{56}{88}{140}{white}{84.9}
  \Cell{4.450}{5}{55}{90}{140}{white}{84.4}
  \Cell{5.450}{5}{46}{109}{142}{white}{80.2}
  \Cell{3.450}{4}{62}{73}{137}{white}{87.7}
  \Cell{4.450}{4}{57}{86}{140}{white}{85}
  \Cell{5.450}{4}{49}{103}{142}{white}{81.4}
  \Cell{3.450}{3}{44}{114}{142}{white}{79}
  \Cell{4.450}{3}{44}{114}{142}{white}{78.9}
  \Cell{5.450}{3}{40}{125}{142}{white}{76.6}
  \Cell{3.450}{2}{59}{187}{117}{black}{61.9}
  \Cell{4.450}{2}{86}{198}{103}{black}{58.7}
  \Cell{5.450}{2}{82}{197}{105}{black}{59.1}
  \Cell{3.450}{1}{229}{228}{25}{black}{46.2}
  \Cell{4.450}{1}{226}{228}{24}{black}{46.5}
  \Cell{5.450}{1}{223}{227}{24}{black}{46.7}
  \Cell{3.450}{0}{82}{197}{105}{black}{59.3}
  \Cell{4.450}{0}{127}{211}{78}{black}{54.9}
  \Cell{5.450}{0}{80}{196}{106}{black}{59.3}
  \draw[black,line width=0.55pt] (3.450,0) rectangle (6.450,10);
  \draw[black,line width=0.95pt] (3.450,6) rectangle (6.450,10);
  \draw[black!70,line width=0.75pt] (3.450,6) -- (6.450,6);
  \node[anchor=north,font=\small] at (3.950,-0.1) {$0$};
  \node[anchor=north,font=\small] at (4.950,-0.1) {$0.10$};
  \node[anchor=north,font=\small] at (5.950,-0.1) {$0.25$};
  \node[overlay,anchor=north,font=\small] at (4.950,-0.8) {Noise probability};

  \node[anchor=south,font=\bfseries\small] at (8.400,10.1) {Global noise};
  \Cell{6.900}{9}{68}{1}{84}{white}{100}
  \Cell{7.900}{9}{68}{1}{84}{white}{99.9}
  \Cell{8.900}{9}{68}{1}{84}{white}{99.9}
  \Cell{6.900}{8}{68}{1}{84}{white}{99.8}
  \Cell{7.900}{8}{68}{1}{84}{white}{100}
  \Cell{8.900}{8}{68}{1}{84}{white}{100}
  \Cell{6.900}{7}{68}{1}{84}{white}{100}
  \Cell{7.900}{7}{68}{1}{84}{white}{100}
  \Cell{8.900}{7}{68}{1}{84}{white}{100}
  \Cell{6.900}{6}{68}{1}{84}{white}{100}
  \Cell{7.900}{6}{68}{1}{84}{white}{100}
  \Cell{8.900}{6}{68}{1}{84}{white}{100}
  \Cell{6.900}{5}{56}{88}{140}{white}{84.9}
  \Cell{7.900}{5}{50}{100}{142}{white}{82.2}
  \Cell{8.900}{5}{49}{103}{142}{white}{81.4}
  \Cell{6.900}{4}{62}{73}{137}{white}{87.7}
  \Cell{7.900}{4}{58}{84}{140}{white}{85.5}
  \Cell{8.900}{4}{46}{109}{142}{white}{80.2}
  \Cell{6.900}{3}{44}{114}{142}{white}{79}
  \Cell{7.900}{3}{45}{113}{142}{white}{79.4}
  \Cell{8.900}{3}{42}{118}{142}{white}{77.9}
  \Cell{6.900}{2}{59}{187}{117}{black}{61.9}
  \Cell{7.900}{2}{103}{204}{92}{black}{57}
  \Cell{8.900}{2}{90}{200}{100}{black}{58.3}
  \Cell{6.900}{1}{229}{228}{25}{black}{46.2}
  \Cell{7.900}{1}{253}{231}{37}{black}{44}
  \Cell{8.900}{1}{176}{221}{47}{black}{50.6}
  \Cell{6.900}{0}{82}{197}{105}{black}{59.3}
  \Cell{7.900}{0}{74}{193}{109}{black}{60.1}
  \Cell{8.900}{0}{96}{202}{96}{black}{57.7}
  \draw[black,line width=0.55pt] (6.900,0) rectangle (9.900,10);
  \draw[black,line width=0.95pt] (6.900,6) rectangle (9.900,10);
  \draw[black!70,line width=0.75pt] (6.900,6) -- (9.900,6);
  \node[anchor=north,font=\small] at (7.400,-0.1) {$0$};
  \node[anchor=north,font=\small] at (8.400,-0.1) {$0.10$};
  \node[anchor=north,font=\small] at (9.400,-0.1) {$0.25$};
  \node[overlay,anchor=north,font=\small] at (8.400,-0.8) {Noise probability};
\end{tikzpicture}
    \caption{Reach rate as a function of the noise probability $p$. Each cell shows the pooled arithmetic mean under the training and evaluation protocol described above. Figure~\ref{fig:gridworld_combined_metrics} reports the corresponding $95\%$ confidence intervals. Darker cells indicate higher reach rate.}
    \label{fig:ReachRate}
\end{figure}

In addition to average travel time, we report two further metrics.
Figure~\ref{fig:OTcost} reports the empirical optimal transport cost between the final agent positions and the targets, that is, the minimum average Manhattan distance obtained by matching agents to targets.
Figure~\ref{fig:ReachRate} reports the fraction of agents that reach a target before the horizon.
More precisely, let
\begin{equation*}
\fleet_\horizon^\nagents = \frac1\nagents \sum_{i=1}^\nagents \delta_{\state_\horizon^i},
\qquad
\refPVar^\nagents = \frac1\nagents \sum_{j=1}^\nagents \delta_{\refVar^j}.
\end{equation*}
With this notation, the empirical terminal optimal transport cost is
\begin{equation*}
\kantorovich{d}{\fleet_\horizon^\nagents}{\refPVar^\nagents}{}
=
\min_{\sigma \in S_\nagents}
\frac1\nagents \sum_{i=1}^\nagents
d\!\left(\state_\horizon^i,\refVar^{\sigma(i)}\right),
\end{equation*}
where $S_\nagents$ is the set of permutations of $\{1,\ldots,\nagents\}$, $\state^i_\horizon$ is the final position of agent $i$, $\refVar^j$ is target $j$, $d(\cdot,\cdot)$ is the Manhattan distance, and $\sigma$ assigns agents to targets. 
This cost is zero exactly when every target is occupied and, unlike the reach rate, it also distinguishes near misses from terminal configurations that remain far from the target set.

The three metrics complement each other in describing performance and should therefore be read together. 
Average travel time tells us how long the agents travel, but an unsuccessful trajectory may have the same travel time as a slow but successful one. 
The reach rate tells us how many agents actually reach a target, but treats all unsuccessful agents in the same way. 
The terminal optimal transport cost adds the remaining information by distinguishing final configurations that are close to full target coverage from those that remain far from it. 
Figure~\ref{fig:gridworld_combined_metrics} reports the three metrics together. To interpret it recall that 
higher values indicate better performance for the reach rate, while lower values indicate better performance for terminal optimal transport cost and average travel time. 
We can observe that across all three metrics, the $\famalg$ variants achieve higher reach rates, lower terminal costs, and shorter travel times than the MARL baselines.

\subsection{Additional details for the scalability experiment}
\label{appendix:results:scaling}

We next study how the number of environment interactions required for training changes with the population size. 
For this experiment, we use square grids and individual noise with probability $p=0.1$.
Each agent starts from an independently sampled row of the first column, while the targets are sampled independently and uniformly over the full grid.

To avoid confounding population size with target density, we scale the grid area approximately in proportion to $\nagents$, 
the resulting grid sizes are reported in Table~\ref{tab:scalability-grid-sizes}. 
We select PPO as the single-agent learner in $\famalg$ and compare the resulting method with MAPPO.
The hyperparameters are the same as in Tables~\ref{tab:separation-hyperparameters} and~\ref{tab:marl-baseline-hyperparameters}.
\begin{table}[h]
\centering
\caption{Population sizes and grid sizes used in the scalability experiment.}
\label{tab:scalability-grid-sizes}
\resizebox{\textwidth}{!}{%
\begin{tabular}{lcccccccc}
\toprule
Agents $\nagents$ & 2 & 4 & 6 & 8 & 16 & 32 & 64 & 128 \\
\midrule
Grid dimensions & $10\times10$ & $14\times14$ & $17\times17$ & $20\times20$ & $28\times28$ & $40\times40$ & $57\times57$ & $80\times80$ \\
Horizon & 20 & 28 & 34 & 40 & 56 & 80 & 114 & 160 \\
Maximum planned batches & 781{,}250 & 279{,}018 & 153{,}186 & 97{,}656 & 34{,}877 & 12{,}207 & 4{,}283 & 1{,}526     \\
\bottomrule
\end{tabular}
}
\end{table}
For each population size in Table \ref{tab:scalability-grid-sizes}, we choose the maximum number of batches so that the 
per-agent-transition budget is approximately $2\times10^9$.
We record a reach-rate milestone at the first of $10$ consecutive batches whose $64$-episode reach rate meets the stated threshold in Figure~\ref{fig:ReachRateMilestones}.

\subsection{Origin--destination coverage under population-size transfer}
\label{appendix:results:pair-frequency}

The transportation cost used by $\famalg$ is learned from target-conditioned trajectories. 
Consequently, an origin--destination (OD) pair selected only a few times during training may therefore have a less accurate cost-to-go estimate than a pair selected more often,
and this can affect performance at evaluation especially when the population size differs from the one used during training. 
Indeed, as training progresses, a larger population gives the assignment layer more opportunities to select low-cost matchings, 
whereas some OD pairs may then be selected less often and become underrepresented in the training data. 
With fewer agents, however, we have a limited number of available matchings, and some unfavorable assignments cannot be avoided. 
As a result, the OD pairs encountered at evaluation may differ from those for which the transportation cost was learned accurately.

Here, we examine this effect. To do so, we use a grid world in which the start and target distributions 
are known and remain the same during training and evaluation, and keep track of the visitation frequency of each OD pair during training.

\paragraph{Setup.}
We use a $30\times30$ grid with horizon $\horizon=100$, unit stage cost, individual action noise with probability $p=0.1$, and uniform start and target distributions over the full grid.
We compare training with $N=20$ and $N=5$ and evaluate the performance across fleets of $N\in\{1,5,10,20,50\}$ agents.
Table~\ref{tab:pair-frequency-training-settings} reports the training settings for the two population sizes.

\begin{table}[h]
\centering
\caption{Training settings for the origin--destination coverage experiment.}
\label{tab:pair-frequency-training-settings}
\begin{tabular}{lcc}
\toprule
Quantity & $N=20$ & $N=5$ \\
\midrule
Training seeds & $0$--$4$ & $0$--$4$ \\
Nominal interaction budget & $2\times10^9$ agent steps & $2\times10^9$ agent steps \\
Assignment observations & $20{,}000{,}000$ & $20{,}000{,}000$ \\
Population episodes & $1{,}000{,}000$ & $4{,}000{,}000$ \\
Learning rate & $0.25\to0.05$ & $0.25\to0.05$ \\
Learning-rate decay & $800{,}000$ episodes & $3{,}200{,}000$ episodes \\
Action exploration & $0.5\to0.05$ & $0.5\to0.05$ \\
Action-exploration decay & $500{,}000$ episodes & $2{,}000{,}000$ episodes \\
Assignment exploration $\varepsilon_{\mathrm{match}}$ & $0.5\to0$ & $0.5\to0$ \\
Assignment-exploration decay & $800{,}000$ episodes & $3{,}200{,}000$ episodes \\
\bottomrule
\end{tabular}
\end{table}

Since an episode produces four times as many assignment observations when $N=20$ as when $N=5$, the decay periods for $N=5$ are four times longer. The schedules are therefore the same when measured in assignment observations.

At the beginning of each training episode, follow the assignment exploration rule of Algorithm~\ref{alg:separation}, with $\varepsilon_{\mathrm{match}}$ corresponding to $\alpha_k$. 
With probability $\varepsilon_{\mathrm{match}}$, we choose a uniformly random one-to-one agent--target matching, we solve the empirical optimal transport problem using the current Q-based cost-to-go estimate.
To measure the coverage of the training data, we record how often each OD pair is selected:
\begin{equation*}
    n(\state,\refVar)
    =
    \#\{\text{training episodes in which an agent starting at $s$ is assigned to $z$}\}.
\end{equation*}
Although the learned matching concentrates on lower-cost pairs, assignment exploration mitigates this and gives nonzero support to at least $809{,}801$ of the $810{,}000$ possible OD pairs. 
The grid-world Q-table is initialized at zero.

For each trained Q-table, we evaluate fleets of $N\in\{1,5,10,20,50\}$ agents for $5{,}000$ episodes. 
During evaluation, we compute the initial optimal transport assignment from the learned Q-table and keep it fixed for the entire rollout, 
so to better relate training coverage to evaluation performance.

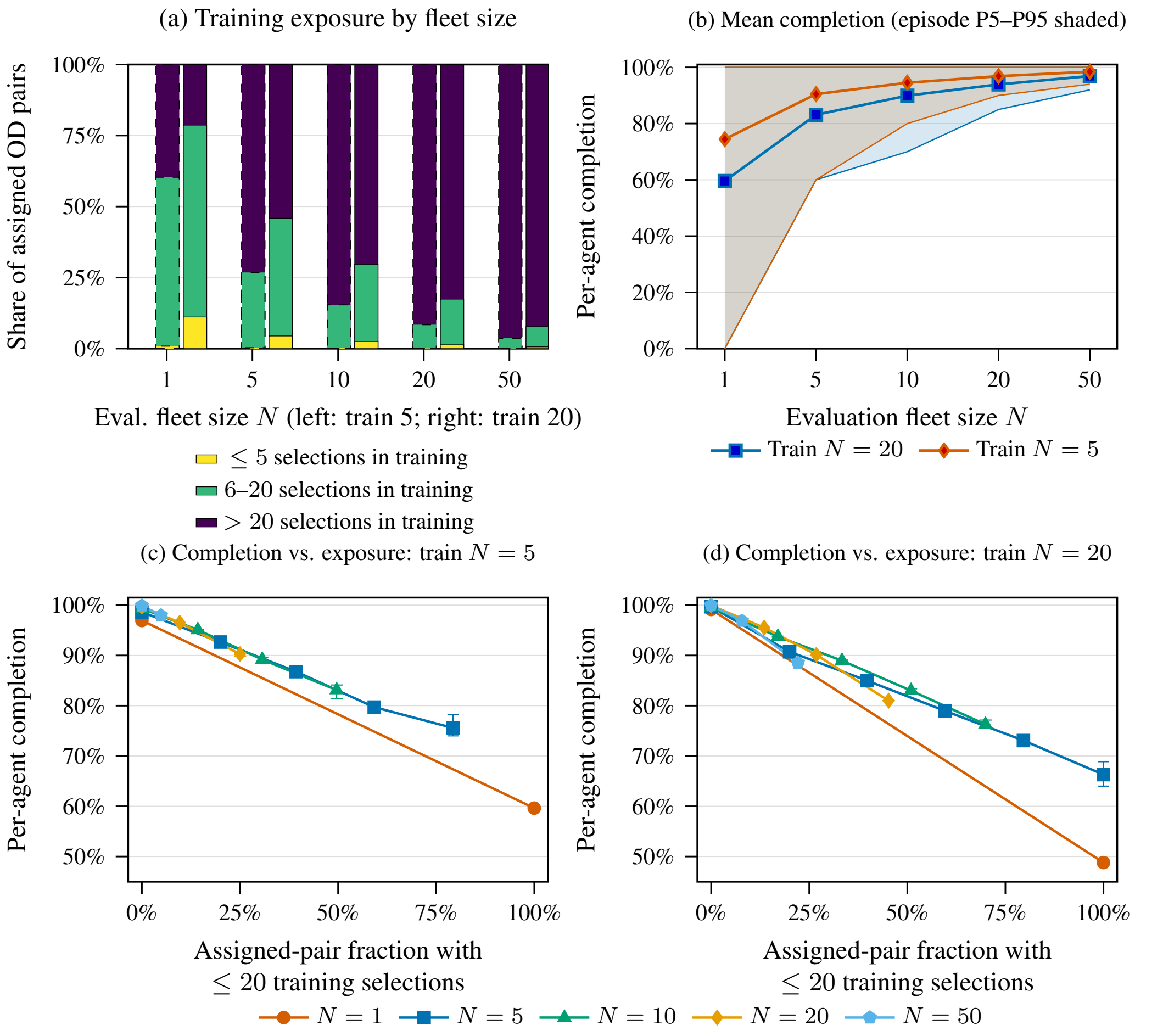
\begin{figure*}[t]
    \centering

\definecolor{pfrare}{RGB}{253,231,37}
\definecolor{pfmoderate}{RGB}{53,183,121}
\definecolor{pffrequent}{RGB}{68,1,84}
\definecolor{pfblue}{RGB}{0,114,178}
\definecolor{pforange}{RGB}{230,159,0}
\definecolor{pflblue}{RGB}{86,180,233}
\definecolor{pfgreen}{RGB}{0,158,115}
\definecolor{pfvermillion}{RGB}{213,94,0}

\pgfplotsset{
  pair frequency axis/.style={
    axis lines=box,
    axis line style={black, line width=0.65pt},
    tick align=outside,
    tick pos=left,
    tick style={black, line width=0.55pt},
    title style={font=\small, yshift=0.35ex},
    label style={font=\small},
    tick label style={font=\footnotesize},
    legend style={font=\footnotesize, draw=none, fill=none,
                  /tikz/every even column/.append style={column sep=3pt}},
    grid style={draw=gray!25, line width=0.35pt},
    ymajorgrids=true,
    unbounded coords=jump,
    scale only axis,
  },
}

\begin{tikzpicture}
\begin{groupplot}[
  group style={group size=2 by 2, horizontal sep=18mm, vertical sep=30mm},
  pair frequency axis,
  width=0.362\linewidth,
  height=0.245\linewidth,
]

\nextgroupplot[
  title={(a) Training exposure by fleet size},
  ybar,
  bar width=8pt,
  xmin=0.55, xmax=5.45,
  ymin=0, ymax=1,
  xtick={1,2,3,4,5},
  xticklabels={1,5,10,20,50},
  ytick={0,0.25,0.5,0.75,1},
  yticklabels={0\%,25\%,50\%,75\%,100\%},
  xlabel={Eval. fleet size $N$ (left: train 5; right: train 20)},
  ylabel={Share of assigned OD pairs},
  legend columns=1,
  legend style={at={(0.5,-0.31)},anchor=north,row sep=0pt},
]
\addplot+[draw=black, line width=0.25pt, fill=pffrequent, forget plot]
  table[col sep=comma,x=t20_x,y=t20_total_share] {figures/pair_frequency_transfer.csv};
\addplot+[draw=black, dashed, line width=0.35pt, fill=pffrequent, forget plot]
  table[col sep=comma,x=t5_x,y=t5_total_share] {figures/pair_frequency_transfer.csv};
\addplot+[draw=black, line width=0.25pt, fill=pfmoderate, forget plot]
  table[col sep=comma,x=t20_x,y=t20_low_mid_share] {figures/pair_frequency_transfer.csv};
\addplot+[draw=black, dashed, line width=0.35pt, fill=pfmoderate, forget plot]
  table[col sep=comma,x=t5_x,y=t5_low_mid_share] {figures/pair_frequency_transfer.csv};
\addplot+[draw=black, line width=0.25pt, fill=pfrare, forget plot]
  table[col sep=comma,x=t20_x,y=t20_low_share] {figures/pair_frequency_transfer.csv};
\addplot+[draw=black, dashed, line width=0.35pt, fill=pfrare, forget plot]
  table[col sep=comma,x=t5_x,y=t5_low_share] {figures/pair_frequency_transfer.csv};
\addlegendimage{legend image code/.code={
  \path[fill=pfrare,draw=black,line width=0.25pt]
    (0cm,-0.05cm) rectangle (0.25cm,0.05cm);}}
\addlegendentry{$\leq5$ selections in training}
\addlegendimage{legend image code/.code={
  \path[fill=pfmoderate,draw=black,line width=0.25pt]
    (0cm,-0.05cm) rectangle (0.25cm,0.05cm);}}
\addlegendentry{$6$--$20$ selections in training}
\addlegendimage{legend image code/.code={
  \path[fill=pffrequent,draw=black,line width=0.25pt]
    (0cm,-0.05cm) rectangle (0.25cm,0.05cm);}}
\addlegendentry{$>20$ selections in training}

\nextgroupplot[
  title={(b) Mean completion (episode P5--P95 shaded)},
  title style={font=\footnotesize},
  xmin=0.7, xmax=5.3,
  ymin=0, ymax=1.01,
  xtick={1,2,3,4,5},
  xticklabels={1,5,10,20,50},
  ytick={0,0.2,0.4,0.6,0.8,1},
  yticklabels={0\%,20\%,40\%,60\%,80\%,100\%},
  xlabel={Evaluation fleet size $N$},
  ylabel={Per-agent completion},
  legend columns=2,
  legend style={at={(0.5,-0.28)},anchor=north},
]
\addplot[name path=t20p05, draw=pfblue, line width=0.45pt, forget plot]
  table[col sep=comma, x=fleet_index,y=t20_completion_p05] {figures/pair_frequency_transfer.csv};
\addplot[name path=t20p95, draw=pfblue, line width=0.45pt, forget plot]
  table[col sep=comma,x=fleet_index,y=t20_completion_p95] {figures/pair_frequency_transfer.csv};
\addplot[fill=pfblue, fill opacity=0.16, draw=none, forget plot]
  fill between[of=t20p05 and t20p95];
\addplot+[pfblue, line width=0.9pt, mark=square*, mark size=2pt]
  table[col sep=comma, x=fleet_index,y=t20_completion_mean] {figures/pair_frequency_transfer.csv};
\addlegendentry{Train $N=20$}
\addplot[name path=t5p05, draw=pfvermillion, line width=0.45pt, forget plot]
  table[col sep=comma,x=fleet_index,y=t5_completion_p05] {figures/pair_frequency_transfer.csv};
\addplot[name path=t5p95, draw=pfvermillion, line width=0.45pt, forget plot]
  table[col sep=comma,x=fleet_index,y=t5_completion_p95] {figures/pair_frequency_transfer.csv};
\addplot[fill=pfvermillion, fill opacity=0.16, draw=none, forget plot]
  fill between[of=t5p05 and t5p95];
\addplot+[pfvermillion, line width=0.9pt, mark=diamond*, mark size=2.2pt]
  table[col sep=comma,x=fleet_index,y=t5_completion_mean] {figures/pair_frequency_transfer.csv};
\addlegendentry{Train $N=5$}

\nextgroupplot[
  title={(c) Completion vs. exposure: train $N=5$},
  title style={font=\footnotesize},
  xmin=-0.035, xmax=1.035,
  ymin=0.45, ymax=1.015,
  xtick={0,0.25,0.5,0.75,1},
  xticklabels={0\%,25\%,50\%,75\%,100\%},
  ytick={0.5,0.6,0.7,0.8,0.9,1},
  yticklabels={50\%,60\%,70\%,80\%,90\%,100\%},
  xlabel={\shortstack{Assigned-pair fraction with\\$\leq20$ training selections}},
  ylabel={Per-agent completion},
  legend columns=5,
  legend to name=pairfrequencyfleetlegend,
]
\addplot+[pfvermillion, line width=0.85pt, mark=*, mark size=1.9pt,
  mark options={fill=pfvermillion,draw=pfvermillion},
  error bars/.cd, y dir=both, y explicit]
  table[col sep=comma,x=t5_n1_x,y=t5_n1_mean,y error minus=t5_n1_errminus,y error plus=t5_n1_errplus] {figures/pair_frequency_transfer.csv};
\addlegendentry{$N=1$}
\addplot+[pfblue, line width=0.85pt, mark=square*, mark size=1.9pt,
  mark options={fill=pfblue,draw=pfblue}, error bars/.cd, y dir=both, y explicit]
  table[col sep=comma,x=t5_n5_x,y=t5_n5_mean,y error minus=t5_n5_errminus,y error plus=t5_n5_errplus] {figures/pair_frequency_transfer.csv};
\addlegendentry{$N=5$}
\addplot+[pfgreen, line width=0.85pt, mark=triangle*, mark size=2.1pt,
  mark options={fill=pfgreen,draw=pfgreen}, error bars/.cd, y dir=both, y explicit]
  table[col sep=comma,x=t5_n10_x,y=t5_n10_mean,y error minus=t5_n10_errminus,y error plus=t5_n10_errplus] {figures/pair_frequency_transfer.csv};
\addlegendentry{$N=10$}
\addplot+[pforange, line width=0.85pt, mark=diamond*, mark size=2.1pt,
  mark options={fill=pforange,draw=pforange}, error bars/.cd, y dir=both, y explicit]
  table[col sep=comma,x=t5_n20_x,y=t5_n20_mean,y error minus=t5_n20_errminus,y error plus=t5_n20_errplus] {figures/pair_frequency_transfer.csv};
\addlegendentry{$N=20$}
\addplot+[pflblue, line width=0.85pt, mark=pentagon*, mark size=2pt,
  mark options={fill=pflblue,draw=pflblue}, error bars/.cd, y dir=both, y explicit]
  table[col sep=comma,x=t5_n50_x,y=t5_n50_mean,y error minus=t5_n50_errminus,y error plus=t5_n50_errplus] {figures/pair_frequency_transfer.csv};
\addlegendentry{$N=50$}

\nextgroupplot[
  title={(d) Completion vs. exposure: train $N=20$},
  title style={font=\footnotesize},
  xmin=-0.035, xmax=1.035,
  ymin=0.45, ymax=1.015,
  xtick={0,0.25,0.5,0.75,1},
  xticklabels={0\%,25\%,50\%,75\%,100\%},
  ytick={0.5,0.6,0.7,0.8,0.9,1},
  yticklabels={50\%,60\%,70\%,80\%,90\%,100\%},
  xlabel={\shortstack{Assigned-pair fraction with\\$\leq20$ training selections}},
  ylabel={Per-agent completion},
]
\addplot+[pfvermillion, line width=0.85pt, mark=*, mark size=1.9pt, forget plot,
  mark options={fill=pfvermillion,draw=pfvermillion},
  error bars/.cd, y dir=both, y explicit]
  table[col sep=comma,x=t20_n1_x,y=t20_n1_mean,y error minus=t20_n1_errminus,y error plus=t20_n1_errplus] {figures/pair_frequency_transfer.csv};
\addplot+[pfblue, line width=0.85pt, mark=square*, mark size=1.9pt, forget plot,
  mark options={fill=pfblue,draw=pfblue},
  error bars/.cd, y dir=both, y explicit]
  table[col sep=comma,x=t20_n5_x,y=t20_n5_mean,y error minus=t20_n5_errminus,y error plus=t20_n5_errplus] {figures/pair_frequency_transfer.csv};
\addplot+[pfgreen, line width=0.85pt, mark=triangle*, mark size=2.1pt, forget plot,
  mark options={fill=pfgreen,draw=pfgreen},
  error bars/.cd, y dir=both, y explicit]
  table[col sep=comma,x=t20_n10_x,y=t20_n10_mean,y error minus=t20_n10_errminus,y error plus=t20_n10_errplus] {figures/pair_frequency_transfer.csv};
\addplot+[pforange, line width=0.85pt, mark=diamond*, mark size=2.1pt, forget plot,
  mark options={fill=pforange,draw=pforange},
  error bars/.cd, y dir=both, y explicit]
  table[col sep=comma,x=t20_n20_x,y=t20_n20_mean,y error minus=t20_n20_errminus,y error plus=t20_n20_errplus] {figures/pair_frequency_transfer.csv};
\addplot+[pflblue, line width=0.85pt, mark=pentagon*, mark size=2pt, forget plot,
  mark options={fill=pflblue,draw=pflblue},
  error bars/.cd, y dir=both, y explicit]
  table[col sep=comma,x=t20_n50_x,y=t20_n50_mean,y error minus=t20_n50_errminus,y error plus=t20_n50_errplus] {figures/pair_frequency_transfer.csv};

\end{groupplot}
\node[anchor=north] at ($(group c1r2.south east)!0.5!(group c2r2.south west)+(0,-1.25cm)$)
  {\pgfplotslegendfromname{pairfrequencyfleetlegend}};
\end{tikzpicture}
\caption{OD-pair coverage and population-size transfer after training $\algname{Q}$ with $N=20$ or $N=5$ agents at the same nominal interaction budget. (a)~Share of nontrivial evaluation assignments grouped by training count; within each evaluation fleet, the left and right bars correspond to training with $N=5$ and $N=20$. (b)~Overall per-agent completion; lines show pooled episode means and shaded regions the episode-level fifth--ninety-fifth percentiles. (c--d)~Per-agent completion conditioned on the fraction of assignments satisfying $n(s,z)\leq20$, shown separately for the two training fleets. Points are means and error bars span the minimum and maximum over five training seeds. The $x$ coordinate is the mean realized fraction within each bin.}
    \label{fig:pair-frequency-transfer}
\end{figure*}

When classifying coverage, we exclude the trivial pairs for which $s=z$ and group every remaining pair 
$(\state,\refVar)$ according to the number of times it was selected during training: 
$n(\state,\refVar)\leq5$, $6\leq n(\state,\refVar)\leq20$, or $n(\state,\refVar)>20$. 
In Figures~\ref{fig:pair-frequency-transfer}c--d, the horizontal coordinate is the fraction of nontrivial assignments in an episode that satisfy $n(\state,\refVar)\leq20$. 
We divide this fraction into the bins $0\%$, $(0,20\%]$, $(20,40\%]$, $(40,60\%]$, $(60,80\%]$, and $(80,100\%]$ 
and a bin is shown only if each of the five training seeds contributes at least $20$ evaluation episodes. 
In this case completion corresponds to the fraction of agents that reach their assigned target by the horizon.

\paragraph{Results.}
Figure~\ref{fig:pair-frequency-transfer}a reports the percentage of evaluation OD pairs in each of the three training-count ranges described above.
For both training fleets, the shares satisfying $n(s,z)\leq5$ and $6\leq n(s,z)\leq20$ decrease as the evaluation fleet grows. 
We can see that, with the same interaction budget, training with $N=5$ agents gives broader coverage with respect to training with $N=20$ agents. 
For example, for an evaluation fleet of size $N=1$, the share satisfying $n(s,z)>20$ increases from $21.3\%$ after training with $N=20$ to $39.7\%$ after training with $N=5$. 
At the same time, the share satisfying $n(s,z)\leq5$ decreases from $11.1\%$ to $0.9\%$.

Figure~\ref{fig:pair-frequency-transfer}b shows how this difference in coverage impacts the completion rate. 
Training with $N=20$ leads to completion rates that increase from $59.6\%$ for an evaluation fleet of size $N=1$ to $96.9\%$ for $N=50$. 
Training with $N=5$ raises these values to $74.5\%$ and $98.4\%$, respectively, and also improves completion at every intermediate fleet size. 
The shaded areas show the episode-level fifth--ninety-fifth percentiles after pooling the evaluation episodes from all five training seeds.

Figures~\ref{fig:pair-frequency-transfer}c--d examine the relation more directly by plotting completion against the fraction of evaluation assignments satisfying $n(s,z)\leq20$. 
For each fleet size and under both training regimes, completion decreases as this fraction increases. 
For an evaluation fleet of size $N=1$, the assignment layer has no alternative target. In this case, in episodes involving an under-covered OD pair, the agent reaches its target in $48.8\%$ of cases after training with $N=20$, 
compared with $59.6\%$ after training with $N=5$. A similar difference is present for $N=5$: when $20\%$ of the assignments are under-covered, completion increases from $90.7\%$ to $92.7\%$, while when their mean fraction is approximately $60\%$, it increases from $79.0\%$ to $79.7\%$. 
The minimum--maximum bars show that these trends are consistent across the five training seeds.

These results show that the OD-pair coverage obtained during training has an important impact on the transfer of $\famalg$ across fleet sizes. 
When the evaluation fleet is smaller than the training fleet, the assignment layer may select pairs that were underrepresented in the training data, leading to worse performance. 
At a fixed interaction budget, training with fewer agents improves coverage and reduces this mismatch. However, it requires proportionally more population episodes and therefore a longer training time. 
We believe that adapting the fleet size during training may help balance these two effects, and point it out as a promising direction for future work.

\subsection{Towards a realistic, time-varying urban logistics task}
\label{appendix:results:manhattan}

\paragraph{Manhattan environment and dynamics.}
We model southern Manhattan as a directed road graph extracted from OpenStreetMap \cite{OpenStreetMap}, whose data are available under the Open Data Commons Open Database License (ODbL). 
We restrict the graph to $38$ taxi zones in southern Manhattan, retain its largest strongly connected component, and remove self-loops. 
The resulting graph contains $\lvert V\rvert=2087$ nodes and $4255$ directed edges. 
Each node in the graph represents an intersection, while each directed edge represents an admissible road segment. 
The nominal edge travel times are computed from the road length and a road-type-dependent speed limit.

To model variable traffic conditions, we include traffic signals and stochastic congestion noise.
We model edge-level traffic signals using a $90$-second cycle whose green-phase duration depends on the road type.
Primary roads are green for $50\%$ of the cycle, whereas secondary and residential roads are green for $30\%$. 
The remaining road types (motorway, trunk, tertiary, and unclassified roads) are treated as always green.
Finally, the signal offset of each intersection is randomly initialized. 
The impact of traffic signals is that an agent may have to wait at an intersection if it arrives during a red phase.
More precisely, when an agent traverses edge $e$ at time $t$, its realized transition time is the sum of the travel time and the waiting time determined by the signal phase.

To represent time-varying traffic conditions, we also add stochastic congestion noise. 
When edge $e$ is traversed, its travel time is multiplied by
\begin{equation*}
    1 + u,
    \qquad
    u \sim \operatorname{Unif}(0,\eta_e),
\end{equation*}
where $\eta_e = 0.1$ for primary roads, $\eta_e = 0.15$ for secondary roads, and $\eta_e=0$ otherwise. 
A new value of $u$ is drawn for every agent-edge traversal.
We exclude operationally unrealistic immediate U-turns from the urban routing task, in both training and evaluation. 
After a vehicle traverses $u\to v$, the reverse action $v\to u$ is masked whenever $v$ has another admissible successor.

\paragraph{Time discretization and Q-table.}
The Manhattan simulator evolves in event time, thus, one decision step corresponds to traversing one outgoing road segment, rather than advancing time by a fixed one-second increment. 
The tabular Q-learner instead uses a discrete time coordinate with resolution $\Delta t = 1 \text{ second}$.
Since all traffic lights follow a $90$-second cycle, we record in the time coordinate of the Q-table the traffic phase instead of unbounded physical time. 
Thus, the Q-table contains $T_{\mathrm{disc}} = 90/\Delta t = 90$ time slices. 
At physical time $t$, the corresponding table index is
\begin{equation*}
    k(t)
    =
    \operatorname{clip}
    (\operatorname{floor}(t/\Delta t), 0, 89).
\end{equation*}

More precisely, we consider a target-conditioned Q-table, whose entries are indexed by the current node $v$, the request node $z$, the traffic-phase index $k$, and the outgoing action $\action$:
$Q(v,z,k,\action)$.
For the Manhattan graph, this gives a table of shape
\begin{equation*}
    \lvert V\rvert \times \lvert V\rvert \times T_{\mathrm{disc}}
    \times \max_v \lvert\mathrm{out}(v)\rvert
    =
    2087 \times 2087 \times 90 \times 4,
\end{equation*}
where $\lvert\mathrm{out}(v)\rvert$ is the out-degree of node $v$. 
Invalid padded actions are masked during both action selection and Bellman backups.

The physical edge travel times remain continuous, while the one-second grid is used
only to index the Q-table. 
If $T_{v,a}$ is the nominal travel time of edge
$(v,a)$ and $u$ is its congestion perturbation, the realized edge travel time is
$\bar T_{v,a}=T_{v,a}(1+u)$.
Once the edge has been traversed, the simulator computes the traffic-signal waiting time from the arrival phase. 
For an edge with period $P_{v,a}$, offset $o_{v,a}$, and green duration $g_{v,a}$, this phase is
\begin{equation*}
    \phi(t) = (t + \bar T_{v,a} + o_{v,a}) \bmod P_{v,a},
\end{equation*}
and the corresponding waiting time is
\begin{equation*}
    W_{v,a}(t)=
    \begin{cases}
    0, & \phi < g_{v,a},\\
    P_{v,a}-\phi, & \phi \ge g_{v,a}.
    \end{cases}
\end{equation*}
The next physical time is therefore
\begin{equation*}
    t' = t + \bar T_{v,a} + W_{v,a}(t),
\end{equation*}
and to obtain the next Q-table index, we round $t'$ to the discrete grid and reduce it modulo the common traffic-signal period:
\begin{equation*}
    k'
    =
    \operatorname{round}(t'/\Delta t)
    \bmod T_{\mathrm{disc}}.
\end{equation*}
The next lookup therefore uses the resulting phase index $k'$. The reward is based on the same realized delay $\bar T_{v,a}+W_{v,a}(t)$, expressed in minutes.

\paragraph{Task setup.}
We sample the starting positions from the three fixed depot nodes $\{587,1226,1972\}$, which have the highest betweenness centrality, while each request is represented by a destination node $z$ sampled uniformly from the graph.
During training, we assign the requests at the beginning of each population episode, by solving an epsilon-greedy optimal transport problem,
and keep the assignment fixed for the entire training episode, as required to learn the target-conditioned subproblem.
At evaluation, instead, the online-assignment $\famalg$ policy recomputes the optimal transport problem from the agents' current nodes and traffic phases at every routing step, 
and then takes one greedy Q-learning action. 
Figure~\ref{fig:manhattan_travel_times} compares this policy with a static nominal shortest-path baseline. 
The baseline computes a single distance-based optimal transport assignment at $t=0$ and follows nominal shortest paths without accounting for the signal phase or congestion.

\paragraph{Learning algorithm.}
For the Manhattan experiment, we use tabular Q-learning with learning rate $0.1$ and no discount factor. 
Action exploration is $\epsilon$-greedy, with $\epsilon$ decreasing linearly from $1.0$ to $0.01$ over $90\%$ of the total training steps.
The training budget is $5{,}000{,}000$ population episodes with $20$ agents. 
The maximum shortest-path length in the processed graph is $66$ edges, thus during both training and the reported final evaluation, 
we cap each trajectory at twice this value, or $132$ road traversals. 
To provide a warm start to the learning algorithm, we initialize the Q-table using nominal shortest-path travel-time estimates, 
hence ignoring congestion noise or traffic-signal phases.

\paragraph{Reward.}
At each step, the stage reward of an active agent is the negative realized delay in minutes:
\begin{equation*}
    \reward_\timestep
    =
    -\frac{\bar T_{v,\action} + W_{v,\action}(t)}{60}.
\end{equation*}
Inactive agents enter an absorbing state and receive zero reward thereafter. If the $132$-traversal cap is reached before the target, the terminal Bellman target also accounts for the remaining distance: it includes $-d_{\mathrm{SP}}(v,z)/60$, the negative nominal shortest-path travel time from the current node $v$ to the target $z$, rather than an arbitrary fixed timeout penalty.

\begin{figure}[t]
    \centering
\begin{tikzpicture}
\begin{groupplot}[
  group style={group size=3 by 1, horizontal sep=0.30cm},
  scale only axis,
  width=0.280\linewidth,
  height=0.23\linewidth,
  xmin=0.5,
  xmax=5.5,
  ymin=-52,
  ymax=68,
  axis lines=box,
  axis line style={black, line width=0.75pt},
  tick align=outside,
  tick pos=left,
  tick style={black, line width=0.75pt},
  xtick={1,2,3,4,5},
  xticklabels={1,5,10,20,100},
  minor x tick num=1,
  minor tick style={draw=none},
  ytick={-40,-20,0,20,40,60},
  yticklabels={-40\%,-20\%,0\%,20\%,40\%,60\%},
  xlabel={Number of agents},
  label style={font=\small},
  tick label style={font=\small},
  title style={font=\small\bfseries},
  xmajorgrids=true,
  xminorgrids=true,
  grid style={draw=gray!35, solid, line width=0.35pt},
  ymajorgrids=false,
]
\nextgroupplot[
  title={Static SP},
  ylabel={Relative improvement},
]
\foreach \yy in {-40,-20,20,40,60}{%
  \addplot[forget plot, draw=gray!35, dashed, line width=0.35pt]
    coordinates {(0.5,\yy) (5.5,\yy)};
}
\addplot[forget plot, draw=black, dashed, line width=0.55pt]
  coordinates {(0.5,0) (5.5,0)};
\addplot[boxplotstyle,boxplot prepared={draw position=1,lower whisker=-45.056668,lower quartile=0.000000,median=14.845996,upper quartile=30.696553,upper whisker=63.335937}] coordinates {};
\addplot[boxplotstyle,boxplot prepared={draw position=2,lower whisker=-7.833956,lower quartile=13.416065,median=21.213559,upper quartile=28.012151,upper whisker=48.222901}] coordinates {};
\addplot[boxplotstyle,boxplot prepared={draw position=3,lower whisker=0.013008,lower quartile=15.631962,median=21.060194,upper quartile=26.106960,upper whisker=40.971681}] coordinates {};
\addplot[boxplotstyle,boxplot prepared={draw position=4,lower whisker=6.442881,lower quartile=17.435774,median=21.224713,upper quartile=25.000285,upper whisker=36.163655}] coordinates {};
\addplot[boxplotstyle,boxplot prepared={draw position=5,lower whisker=14.088794,lower quartile=19.528262,median=21.325891,upper quartile=23.248151,upper whisker=28.268776}] coordinates {};

\nextgroupplot[
  title={SP reassignment: $K=5$},
  yticklabels=\empty,
]
\foreach \yy in {-40,-20,20,40,60}{%
  \addplot[forget plot, draw=gray!35, dashed, line width=0.35pt]
    coordinates {(0.5,\yy) (5.5,\yy)};
}
\addplot[forget plot, draw=black, dashed, line width=0.55pt]
  coordinates {(0.5,0) (5.5,0)};
\addplot[boxplotstyle,boxplot prepared={draw position=1,lower whisker=-46.272494,lower quartile=0.000000,median=14.876033,upper quartile=30.857100,upper whisker=63.335937}] coordinates {};
\addplot[boxplotstyle,boxplot prepared={draw position=2,lower whisker=-8.108462,lower quartile=13.631282,median=21.096616,upper quartile=28.207813,upper whisker=48.227216}] coordinates {};
\addplot[boxplotstyle,boxplot prepared={draw position=3,lower whisker=0.045840,lower quartile=15.602658,median=21.007125,upper quartile=26.037174,upper whisker=40.739420}] coordinates {};
\addplot[boxplotstyle,boxplot prepared={draw position=4,lower whisker=6.011871,lower quartile=17.358175,median=21.276624,upper quartile=24.984450,upper whisker=35.790956}] coordinates {};
\addplot[boxplotstyle,boxplot prepared={draw position=5,lower whisker=14.380838,lower quartile=19.469888,median=21.334565,upper quartile=23.184261,upper whisker=28.631967}] coordinates {};

\nextgroupplot[
  title={SP reassignment: $K=10$},
  yticklabels=\empty,
]
\foreach \yy in {-40,-20,20,40,60}{%
  \addplot[forget plot, draw=gray!35, dashed, line width=0.35pt]
    coordinates {(0.5,\yy) (5.5,\yy)};
}
\addplot[forget plot, draw=black, dashed, line width=0.55pt]
  coordinates {(0.5,0) (5.5,0)};
\addplot[boxplotstyle,boxplot prepared={draw position=1,lower whisker=-45.098039,lower quartile=0.000000,median=15.287837,upper quartile=30.712980,upper whisker=63.335937}] coordinates {};
\addplot[boxplotstyle,boxplot prepared={draw position=2,lower whisker=-8.085830,lower quartile=13.393350,median=21.216805,upper quartile=28.062262,upper whisker=49.178215}] coordinates {};
\addplot[boxplotstyle,boxplot prepared={draw position=3,lower whisker=0.395711,lower quartile=15.723301,median=21.141540,upper quartile=25.998122,upper whisker=40.651231}] coordinates {};
\addplot[boxplotstyle,boxplot prepared={draw position=4,lower whisker=5.969855,lower quartile=17.404979,median=21.258387,upper quartile=25.034800,upper whisker=36.338365}] coordinates {};
\addplot[boxplotstyle,boxplot prepared={draw position=5,lower whisker=14.233334,lower quartile=19.467540,median=21.374464,upper quartile=23.290520,upper whisker=28.623129}] coordinates {};
\end{groupplot}
\end{tikzpicture}
    \caption{Relative travel-time improvement of $\algname{Q}$ against nominal shortest-path routing with a static assignment (left), reassignment every five routing steps (center), or reassignment every ten routing steps (right). 
    The evaluation samples and boxplot conventions match Figure~\ref{fig:manhattan_travel_times}.}
    \label{fig:manhattan_reassignment_baselines}
\end{figure}
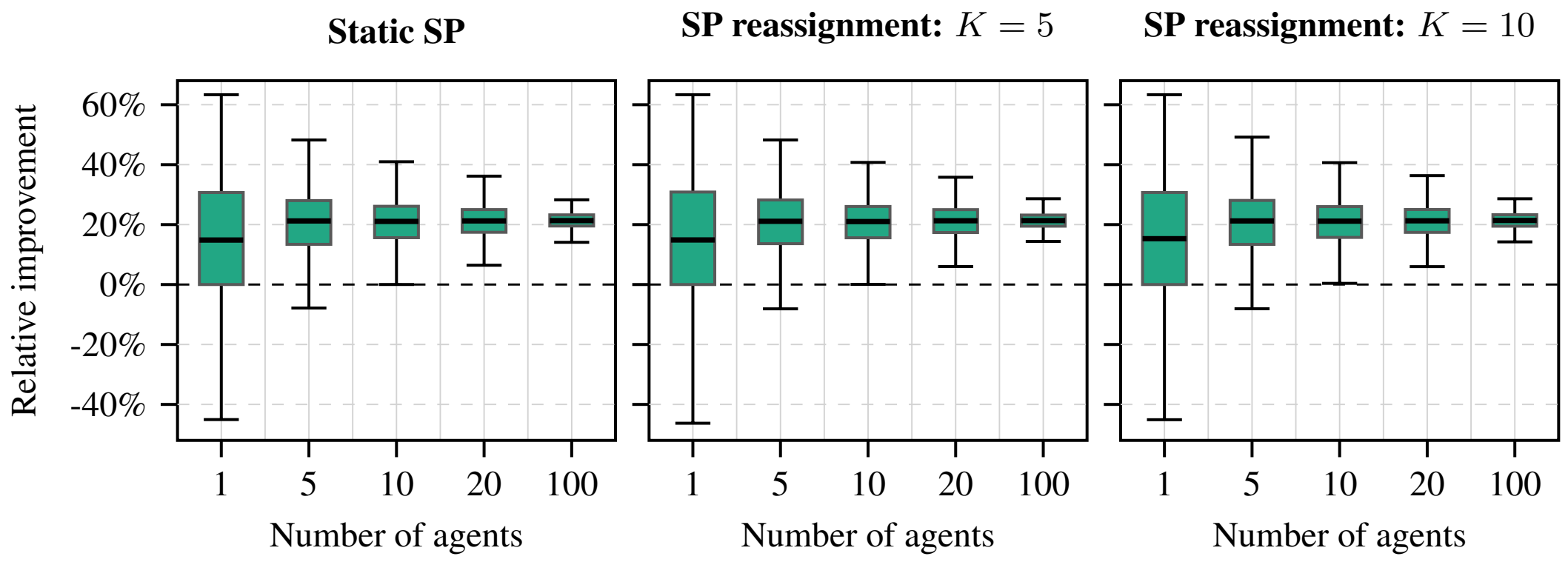

\paragraph{Evaluation and logging.}
We report results for $8$ training seeds ($0$--$7$). For each seed and each fleet size $N\in\{1,5,10,20, 100\}$, we draw $500$ evaluation instances, giving $4000$ outcomes per fleet size.
Within each evaluation run, $\algname{Q}$ and all shortest-path baselines share starts, requests, and the fixed traffic-signal configuration, while the separate rollouts use independent congestion-noise streams. 
We then report the continuous travel-time improvements over the baseline as $100 (T_{\mathrm{baseline}}-T_{\algname{Q}})/T_{\mathrm{baseline}}$ per instance.
Figures~\ref{fig:manhattan_travel_times} and~\ref{fig:manhattan_reassignment_baselines} show medians and interquartile ranges after pooling outcomes across training seeds. Here, whiskers follow the Tukey $1.5$-interquartile-range rule and outliers remain in the underlying samples but are not represented in the plots.

We evaluate the performance of $\famalg$ over three shortest-path baselines differing in the reassignment frequency they use.
Static shortest path computes the optimal transport assignment once at the beginning of the episode and keep the assignment fixed until the end of the episode itself.
$K$-reassignment shortest path instead recomputes the optimal transport assignment every $K$ routing steps from the vehicles' current nodes to the still-unserved requests, served requests instead remain locked to the vehicles that already reached them.
Between reassignments, routing follows the same nominal shortest paths as the static baseline. 
In \cref{fig:manhattan_reassignment_baselines} we show the performance of $5$-reassignment and $10$-reassignment shortest path.
As visible, the reassignment improvement is minimal. 
To measure it directly, define $G_K(N)=100(\bar T_{\mathrm{static}}-\bar T_{\mathrm{reassign},K})/\bar T_{\mathrm{static}}$, where bars denote means over the available continuous-time outcomes. 
For $N=(5,10,20,100)$, we obtain $G_5=(-0.0014,\ 0.0008,\ 0.0031,\ 0.0086)\%$ and $G_{10}=(-0.0480,\ -0.0243,\ -0.0406,\ -0.0045)\%$. 
The marginal differences are extremely small and can be traced back to the stochasticity of the environment. 
Therefore, we can conclude that for $\nagents =1$ reassignment cannot change the matching, 
and the small aggregate differences can only be due to different stochastic congestion realizations.
Therefore, we can conclude that periodic target reassignment has essentially no practical improvement to nominal shortest-path routing in this experiment, 
and that the substantial advantage in Figure~\ref{fig:manhattan_travel_times} is not an artifact of comparing online SALT assignment against a static assignment heuristic.

\subsection{Compute resources}

We ran all reported experiments on the Clariden supercomputer at the Swiss National Supercomputing Centre (CSCS). 
Each Clariden compute node contains four NVIDIA GH200 modules, each with a 72-core Grace CPU and $96$~GB of HBM3 memory. 
An independent rank requested one GH200 and $72$ CPU cores. 

For the grid-world experiments, we used PyTorch~\cite{paszke2019pytorch} with CUDA~\cite{nickolls2008scalable} and assigned one independent rank to each module.
Within a grid-world condition, the recorded training times summed across methods range from $3.9$ to $4.9$ hours, with a mean of $4.2$ hours, 
and the $50$ final rollouts per method were generated in the same jobs and required little additional time relative to training. 
The eight-condition scalability sweep used the same allocation and at most $96$ additional active GPU-hours under the 12-hour job limit.

For the Manhattan experiments, we used JAX~\cite{jax2018github} with CUDA and the same one-GH200, 72-core allocation for each training seed. 
We ran four training seeds on each node. 
Training one policy for $5{,}000{,}000$ episodes required approximately $77.5$ active GPU-hours.

\end{document}